\documentclass[aps, rmp, showpacs, nofootinbib,10pt, longbibliography,twocolumn, amsfonts, amsmath, amssymb, superscriptaddress, floatfix]{revtex4-2}
\usepackage[colorlinks,bookmarks=false,citecolor=blue,linkcolor=blue,urlcolor=black]{hyperref}

\usepackage[T1]{fontenc}
\usepackage{epigraph}
\usepackage{color}
\usepackage{xcolor}

\usepackage{tabularx}
 \usepackage{array,multirow,graphicx}
 \usepackage{float}
 
\usepackage{graphicx}
\usepackage{epsfig}
\usepackage{ulem}
\usepackage{braket}
\usepackage{dsfont}

\def\be{\begin{equation}}
\def\ee{\end{equation}}
\def\bea{\begin{eqnarray}}
\def\eea{\end{eqnarray}}

\def\bi{\begin{itemize}}
\def\ei{\end{itemize}}

\DeclareMathAlphabet{\mathcalnew}{OMS}{cmsy}{m}{n}

\begin{document}

\title{Many-Body Localization in the Age of Classical Computing}

\author{ Piotr Sierant}
\email{piotr.sierant@icfo.eu}
\address{ \mbox{ICFO-Institut de Ciencies Fotoniques, The Barcelona Institute of
Science and Technology,} \mbox{ Castelldefels (Barcelona) 08860, Spain}  }

\author{Maciej Lewenstein}
\address{ \mbox{ICFO-Institut de Ciencies Fotoniques, The Barcelona Institute of
Science and Technology,} \mbox{ Castelldefels (Barcelona) 08860, Spain}  }
\address{ \mbox{ICREA, Pg. Llu\'is Companys 23,} \mbox{ 08010 Barcelona, Spain} }

\author{Antonello Scardicchio}
 \address{\mbox{Condensed Matter and Statistical Physics Group, Abdus Salam International Centre of Theoretical Physics,} \mbox{ Strada Costiera 11,
 34151 Trieste, Italy } }

\author{Lev Vidmar}
\address{\mbox{Department of Theoretical Physics, Jo\v{z}ef Stefan Institute} \mbox{1000 Ljubljana, Slovenia} }
\address{ \mbox{Department of Physics, Faculty of Mathematics and Physics, University of Ljubljana, SI-1000 Ljubljana, Slovenia}}

 \author{Jakub Zakrzewski}
 \address{ 
\mbox{Instytut Fizyki Teoretycznej, Wydzia\l{} Fizyki, Astronomii i Informatyki Stosowanej, Uniwersytet Jagiello\'nski,} \mbox{ \L{}ojasiewicza 11, PL-30-348 Krak\'ow, Poland} }
\address{\mbox{Mark Kac Complex Systems Research Center, Jagiellonian University in Krak\'ow,} \mbox{Krak\'ow, Poland} }

\date{\today}
\begin{abstract}
Statistical mechanics provides a framework for describing the physics of large, complex many-body systems using only a few macroscopic parameters to determine the state of the system. 
For isolated quantum many-body systems, such a description is achieved via the eigenstate thermalization hypothesis (ETH), which links thermalization, ergodicity and quantum chaotic behavior. However, tendency towards thermalization is not observed at finite system sizes and evolution times in a robust many-body localization (MBL) regime found numerically and experimentally in the dynamics of interacting many-body systems at strong disorder.
Although the phenomenology of the MBL regime is well-established, the central question remains unanswered: under what conditions does the MBL \textit{regime} give rise to an MBL \textit{phase}, in which the thermalization does not occur even in the \textit{asymptotic} limit of infinite system size and evolution time? 

This review focuses on recent numerical investigations aiming to clarify the status of the MBL phase, and it establishes the critical open questions about the dynamics of disordered many-body systems. The last decades of research have brought an unprecedented new variety of tools and indicators to study the breakdown of ergodicity, ranging from spectral and wave function measures, observable matrix elements, and unitary quantum dynamics, to transport and quantum information measures. We give a comprehensive overview of these approaches and attempt to provide a unified understanding of their main features.
We emphasize general trends towards ergodicity with increasing length and time scales, which exclude naive single-parameter scaling hypothesis, necessitate the use of more refined scaling procedures, and prevent unambiguous extrapolations of numerical results to the asymptotic limit.
Providing a concise description of numerical methods for studying ETH and MBL, we explore various approaches to tackle the question of the MBL phase. 
Persistent finite size drifts towards ergodicity consistently emerge in quantities derived from eigenvalues and eigenvectors of disordered many-body systems.
The drifts are related to continuous inching towards ergodicity and non-vanishing transport observed in the dynamics of many-body systems, even at strong disorder. These phenomena impede the understanding of microscopic processes at the ETH-MBL crossover. Nevertheless, the abrupt slowdown of dynamics with increasing disorder strength provides premises suggesting the proximity of the MBL phase.
This review concludes that the questions about thermalization and its failure in disordered many-body systems remain a captivating area open for further explorations.

\end{abstract}

\maketitle

\setlength\afterepigraphskip{-10pt}
\epigraph{\it  \flushright In memory of Dominique Delande.}

\tableofcontents

\section{Introduction}


Since the XIX century and thanks to the works of Ludwig Boltzmann, James Clerk Maxwell, and Josiah Willard Gibbs, physicists have understood some of the reasons behind the universal and successful applicability of thermodynamics. It was recognized that most of the physical systems undergo a {\it thermalization} process in their dynamics, and that their long time behaviour is very accurately described by probability distributions on the phase space, termed statistical ensembles \cite{Pathria16,Huang08}. Time averages of physical observables can then be calculated as ensemble averages with respect to the corresponding statistical ensembles: the statement is commonly referred to as the ergodic hypothesis. Technically, the ergodic hypothesis states that the long-time averages of few-body observables are the same for (almost all) initial states. Hence, these averages can only depend on the conserved quantities of the system, which, in a generic non-integrable system, are just macroscopic observables such as energy, angular momentum, total magnetization, total charge, {\it etc.}

It should be stressed that, while it is widely believed that the ergodic hypothesis holds for the overwhelming majority of many-body systems, the examples of systems in which this statement can be proven are rather rare. 
Systems that undergo second order phase transitions approach, at the critical point,  the equilibrium state very slowly, with equilibration times that grow algebraically in the number of microscopic constituents of the system. Typically, however, we term such situations as critical slowing down rather than ergodicity breaking. The situation is much different for disordered systems, such as spin glasses, introduced by \cite{Edwards75, Sherrington75}. Below the freezing temperature, systems in spin glass phase are trapped in a non-ergodic set of states: the system may fluctuate between several states, but cannot transition to all states of equivalent energy. This was understood within the theory of replica symmetry breaking~\cite{Parisi79,Parisi80}, instrumental for the 2021 Nobel Prize in Physics to G. Parisi \cite{Parisi23}. While this solution is exact for the Sherrington-Kirkpatrick model~\cite{Talagrand02}, the ergodicity of the Edwards-Anderson model on a finite-dimensional lattice is a more complex problem \cite{Stein13}.

Diffusion, which arises in classical systems due to numerous collisions between the constituent particles~\cite{Einstein05, Smoluchowski06}, is one of the standard mechanisms of thermalization in such systems. However, a phenomenon that is also abundant in classical systems is {\it anomalous diffusion} \cite{Volpe16}. The recent progress in single particle tracking in biological settings \cite{Manzo15} provides one example of anomalous diffusion that can be characterized down to the level of trajectories of individual classical particles \cite{Munoz21}. The situation is fundamentally different in the realm of quantum mechanics: diffusion of particles may be inhibited due to presence of disorder, leading to the phenomenon of Anderson localization~\cite{Anderson58}, recognized by the Nobel Prize in Physics for P. Anderson in 1977, and recently observed also in systems of ultracold atoms \cite{Billy08, Roati08}.

Thermalization of quantum systems is typically discussed in terms of quench experiments in which systems are prepared in certain out-of-equilibrium states and then evolve in time. Such a scenario was considered in \cite{Rigol08}, which demonstrated that a non-integrable isolated quantum many-body system relaxes to an equilibrium state described by a microcanonical ensemble determined, in turn, by a few variables related to global conserved quantities mentioned before. This process of thermalization must be visible at the level of individual many-body eigenstates, according to the eigenstate thermalization hypothesis (\textbf{ETH}) proposed in the seminal papers \cite{Deutsch91, Srednicki94}. 
In this sense, an Anderson insulator, as a single particle system, is manifestly non-ergodic since the particles remain close to their original positions rather than propagating throughout the system. While the conditions for the occurrence of Anderson localization in systems of non-interacting quantum particles are well understood, it is of fundamental importance to understand the effects of disorder in a more generic situation -- in the presence of interactions between particles.

It was proposed \cite{Basko06, Gornyi05, Oganesyan07} that a sufficiently strong disorder, even in presence of interactions, may inhibit thermalization of quantum many-body systems leading to the phenomenon of many-body localization (\textbf{MBL}). Rather than relaxing to the equilibrium state determined by global conserved quantities, MBL systems retain memory of certain details of the initial state that is preserved in local observables, the so-called local integrals of motion LIOMs, see \cite{Imbrie17} and references therein, for an indefinite period of time. Such a behavior would result in a transition between an ergodic and MBL phases observable even at infinite temperature, as conjectured in \cite{Pal10}.

The existence of MBL transition is of fundamental importance for our understanding of non-equilibrium phenomena in many-body systems, the robustness of diffusive transport, and the applicability of statistical physics. The aim of the present review is to discuss the current status of the phenomenon of MBL and of the putative ETH-MBL phase transition. A particular emphasis will be put on the numerical results which are of paramount importance for our understanding of MBL.
The interpretation of many numerical results, as we shall argue, remains to this day inconclusive, despite the intensive efforts of the community interested in non-equilibrium phenomena in quantum many-body systems spanning over the last 15 years. 
An abrupt slow down of many-body dynamics with increasing disorder strength makes it possible to distinctly observe an MBL \textit{regime}, in which finite size systems fail to thermalize even over extended periods of time. However, the question of whether the observed MBL regime actually corresponds to an MBL \textit{phase} -- a dynamical phase of matter in which the thermalization does not occur at \textit{arbitrarily long} times and across \textit{all} system sizes -- remains unresolved.
The difficulties in the interpretation of the numerical results and understanding the status of the MBL phase are tied to the exponential growth of many-body Hilbert space with system size $L$, which severely limits system sizes and time scales accessible to present day classical computers and supercomputers. 
Similar limitations emerge in state-of-the-art experiments, in which perfect isolation from environment can only be maintained for a finite time interval.

The present review is organized as follows. We begin with a concise historical overview of the development of the MBL field in Sec.~\ref{sec:persp}. Next, in Sec.~\ref{sec:thermANDcounter}, we provide a definition of MBL and discuss its key phenomenological features. In Sec.~\ref{sec:status} we describe the current status of the field, highlighting open questions and qualitatively characterizing the challenges associated with interpreting numerical results in strongly disordered interacting systems. We hope that this Section may serve not only as a roadmap of challenges in the field of MBL, but also can be used as a reference point for understanding the difficulties in the interpretation of numerical results for generic complex systems. Sec.~\ref{sec:num} offers a brief exposition of numerical methods typically employed in studies of MBL. In Sec.~\ref{sec:numerical}, we provide a quantitative discussion of the ETH-MBL crossover found in the exact diagonalization studies of disordered spin chains, and emphasize the system size drifts, disallowing an unambiguous interpretation of the results. Sec.~\ref{sec:dynamics} discusses the present understanding of time evolution of disordered many-body systems, highlighting the slow but persistent trends towards thermalization found even at strong disorder. We briefly discuss the relation of thermalization and transport in many-body systems in Sec.~\ref{sec:trans}, and, in Sec.~\ref{sec:aval}, we outline the microscopic mechanisms responsible for the emergence of ergodicity with decrease of the disorder strength, in particular, the quantum avalanche mechanism and many-body resonances. Finally, in Sec.~\ref{sec:ext} we consider various extensions of the MBL phenomenology, while in Sec.~\ref{sec:exp}, we discuss experiments on MBL, highlighting the remaining challenges.

\section{MBL: historical perspective in short}
\label{sec:persp}

In 1958 P. W. Anderson studied the motion of a quantum mechanical particle in a disordered medium from the perspective of its transport properties, and found that interference effects between multiple-scattering  paths lead to inhibition of diffusion \cite{Anderson58}. Such a system becomes an insulator, its eigenstates are \textit{localized} in position space, hence the name of Anderson localization. The ensuing transitions between metallic and localized phases of disordered electronic systems became subject of intense research in decades following the Anderson's work, as reviewed, e.g., in \cite{Kramer93, Evers08}. Several works \cite{Altshuler79, Fleishman80, Giamarchi88, Altshuler97} attempted to understand transport properties of interacting many-body systems in presence of disorder, a question posed already in the seminal work of Anderson. The culminating point of those studies came when \cite{Basko06, Gornyi05} put forward perturbative arguments that interacting electrons in static random potentials undergo a metal-insulator transition at a nonzero critical temperature.

A seemingly unrelated field of research was conceived in the context of statistical physics and thermalization.
In 1955, E. Wigner proposed that statistics of spacings between the lines of heavy atomic nuclei may be modelled by random matrices \cite{Wigner55} of appropriate symmetry class, as introduced later by \cite{Dyson62a}. The observations for the heavy atomic nuclei, which are examples of strongly interacting many-body systems, were followed, among others, 
by the Berry conjecture, which suggested that quantum states associated with irregular stochastic classical motion appear to be Gaussian random function~\cite{berry_77}, and by the Bohigas–Giannoni–Schmit conjecture, which asserts that the spectral statistics of quantum systems whose classical counterparts exhibit chaotic behaviour are described by random matrix theory \cite{Casati80, Bohigas84}.
These early contributions started the field of quantum chaos \cite{Haakebook}. Repulsion of energy levels, characteristic for random matrices, was predicted for electrons in random impurity potentials \cite{Altshuler86}, providing a link between quantum chaos and the physics of disordered quantum systems. 

Those results were followed by studies of many-body systems with tools that rely on the random matrix theory \cite{Montambaux93, Poilblanc93, Berkovits99, Georgeot00, Santos05} and even focused on entanglement in disordered many-body systems \cite{Santos04, DeChiara06}. However, the link between quantum chaotic behavior and thermalization of isolated many-body systems, understood in terms of eigenstate thermalization hypothesis (\textbf{ETH}) developed by \cite{Deutsch91, Srednicki94}, was established in the context of systems that do not have semi-classical limit, e.g., spin-1/2 models, only later, in \cite{Rigol08}\footnote{See \cite{Jensen85} for an early study of applicability of statistical mechanics to dynamics of spin-1/2 chains.}.
The latter will be further discussed in Sec.~\ref{sec:eth}.

The developments in the initially unrelated fields of Anderson localization, quantum chaos and thermalization of many-body systems, set the stage for studies of many-body localization (\textbf{MBL}), which focus on properties of disordered interacting many-body quantum systems at high temperature. The studies of MBL aim at understanding the impact of interactions and disorder on the dynamics of many body systems, and they typically employ tools of random matrix theory and measures of entanglement from quantum information theory \cite{Amico08, Frerot23} to characterize eigenvalues and eigenvectors of many-body systems. Research on MBL, commenced with the numerical study \cite{Oganesyan07}, gained significant momentum when it was postulated that the interplay of disorder and interactions leads to a transition between ergodic and MBL phases that manifests itself in singular changes of many-body eigenstates at arbitrary energies \cite{Pal10}, even at infinite temperature\footnote{Throughout this review, the notions of \textit{infinite} and \textit{high} temperature are used interchangeably and, unless otherwise noted, mean the energy density corresponding to the infinite temperature.}. Interest in MBL is additionally driven by the remarkable recent progress in experiments on synthetic quantum matter which allow to directly observe whether and how the many-body systems approach thermal equilibrium \cite{Schreiber15, Smith16}. But, despite the experimental progress, numerical simulations on classical computers, that are the focal point of this review, constitute the main source of our understanding of the phenomenon of MBL.

\section{Thermalization and its counterexamples}
\label{sec:thermANDcounter}

In this Section, we set the stage for further discussions of MBL. We start with a brief discussion of the notion of ergodicity of isolated quantum many-body systems and introduce the eigenstate thermalization hypothesis (\textbf{ETH}). Then, we proceed to describe the rough features of MBL systems and survey the most widely studied models of MBL. For a detailed reviews of ETH, we refer the reader to \cite{Dalessio16}, whereas for previous discussions of the MBL phenomenology, we refer the reader to \cite{Nandkishore15, Alet18, Abanin19}.

\subsection{Quantum thermalization}
\label{sec:eth}

Statistical mechanics aims to describe complex many-body systems in terms of a few macroscopic quantities. By way of the ergodic hypothesis and mixing property, one obtains a drastic simplification, going from deterministic equations describing the position and momenta of classical particles to the statistical properties of such observables. An analogous role for interacting quantum many-body systems is played by the ETH, where the deterministic Schr\"{o}dinger equation evolution is traded for the statistical properties of matrix elements of operators. 

We consider an isolated quantum many-body system governed by a Hamiltonian $\hat H$ with a 
complete set of eigenstates $|m\rangle$ corresponding to non-degenerate eigenvalues $E_m$, an initial state $|\psi\rangle$ that is not an eigenstate of $\hat H$, and an 
observable $\hat{A}$ which is few-body, i.e., acting on number of particles (or spins) much smaller than the total number of constituents of the system. We set $\hbar=1$.
The Schr\"{o}dinger equation implies that the average value of $\hat{A}$ after time $\tau$ reads
\begin{equation}
 A(\tau)  = \sum_{m} |c_m|^2 A_{mm} + \sum_{m,n\neq m}
 \mathrm{e}^{-\mathrm{i} (E_m-E_n) \tau} c^*_m c_n A_{mn},
 \label{eq:A1}
\end{equation}
where $A_{mn} = \langle m |\hat A | n \rangle$ and $c_n = \braket{n| \psi}$. The time average of $A(\tau)$ over time $t$ is given by
\begin{equation}
 \bar{A}(t) = \frac{1}{t} \int_0^{t}d\tau\, A(\tau) \stackrel{t \to \infty }{\longrightarrow} \sum_m |c_m|^2 A_{mm}.
  \label{eq:A2}
\end{equation}
Hence, the question about the the thermalization of the system, i.e., about the long time behavior of the observables, is related to properties of the diagonal matrix elements $A_{mm}$.
Moreover, the second term in \eqref{eq:A1} sets the time scale of the approach to the equilibrium value, $\lim_{t\to \infty}\bar{A}(t)$, which is determined by the properties of the off-diagonal matrix elements $A_{mn}$. This time scale can be, in some cases, exponentially long in the system size as it is determined by the differences of energies in the middle of many-body spectrum\footnote{ In many cases, the approach to the equilibrium value is much faster, see, e.g.,~\cite{Reimann16fast}. }.

Early works \cite{Deutsch91, Srednicki94} paved the way to formulation of 
ETH  as an ansatz for both diagonal and off-diagonal matrix elements of
observables in the eigenbasis of $\hat H$ \cite{Srednicki99}\footnote{See Sec.~III of~\cite{Polkovnikov11} and Sec.~4.2 of~\cite{Dalessio16} for survey of works reporting results consistent with Eq.~\eqref{eq:ETH1} obtained in the semi-classical limit of quantum systems whose classical counterpart is chaotic, including~\cite{Peres84, Feingold84, Feingold86}. See also \cite{Tasaki98} for Boltzmann distribution derivation using typicality arguments.},
\begin{equation}
 A_{mn} = \mathcal{A}(\bar E) \delta_{mn} + \mathrm{e}^{-S(\bar E)/2}f_{\mathcal A} (\bar E, \omega_{mn}) R_{mn},
 \label{eq:ETH1}
\end{equation}
where  $\bar E = (E_m+E_n)/2$ is the mean energy, $\omega_{mn} = E_m-E_n$ is the energy difference, 
and $R_{mn}$ is a random variable
with zero mean and unit variance.
The function $S(\bar E)$ is the  thermodynamic entropy at energy $\bar E$, i.e., the logarithm of the density of states \cite{Burke23}, and $\mathcal{A}(\bar E)$ and $f_{\mathcal A} (\bar E, \omega)$ are smooth functions of their arguments. 
Within the random matrix theory ({\bf RMT}), $\mathcal{A}(\bar E) = {\rm const}$ and $f_{\mathcal A} (\bar E, \omega) = 1$.
A nontrivial dependence on $\bar E$ in $\mathcal{A}(\bar E)$ can be attributed to the presence of conserved operators, including the Hamilton operator and its higher powers \cite{Hamazaki18, Mierzejewski_2020}, while a nontrivial dependence on $\omega$ in $f_{\mathcal A} (\bar E, \omega)$ contains information on the nature of transport and the fine structure of response functions \cite{Dalessio16}. Indeed, in quantum many-body systems, the function $f_{\mathcal A} (\omega)$ (we skip the $\bar E$ dependence here) typically interpolates between the 
characteristic \textit{plateau} $f_{\mathcal A} (\omega) = \mathrm{const}$ at small frequency $\omega$ and a large frequency decay $f_{\mathcal A} (\omega) \stackrel{\omega \to \infty}{\longrightarrow} 0$. The change of the behavior of $f_{\mathcal A} (\omega) $ between the two limits 
defines the Thouless energy $\omega_{\mathrm{Th}}$, and the relaxation processes of physical observables are expected to be completed at the Thouless time $t_{\mathrm{Th}} \propto \omega_{\mathrm{Th}}^{-1}$. The scaling of $t_{\mathrm{Th}}$ with system size $L$ may determine the type of dynamics and transport in the system, for instance, for diffusive dynamics one expects $t_{\mathrm{Th}} \propto L^{2} $.
The notions of Thouless time and Thouless energy in disordered many-body systems are discussed more quantitatively in Secs.~\ref{sec:numerical} and~\ref{sec:dynamics}, respectively.

Expanding \eqref{eq:ETH1} for the diagonal matrix elements around the mean energy $\braket{E}=\braket{\psi|\hat{H}|\psi}$, and assuming that the energy variance $\delta E^2$ in the initial state $\ket{\psi}$ is sufficiently small, one arrives at 
\cite{Rigol08} 
\begin{equation}
 \lim_{t\to \infty}\bar{A}(t) = \frac{1}{\mathcal N_{\langle E \rangle, \delta E}}
 \sum_{|E_m - \langle E \rangle| < \delta E} A_{mm} \equiv \mathrm{Tr} \left[ \hat{\rho}_{MC} \hat{A} \right]\;,
 \label{eq:ETH2}
\end{equation}
with corrections that typically decay algebraically with the number of lattice sites.
Equation~\eqref{eq:ETH2} shows that the equilibrium value of a few-body observable $\hat{A}$ is an average over a microcanonical ensemble of states $ \hat{\rho}_{MC}$, and $\mathcal N_{\langle E \rangle, \delta E}$ determines the number of states in this ensemble. This is the notion of quantum ergodicity: the long time average of the observable $\hat{A}$ is independent of the specific arrangement of the coefficients $|c_m|^2$, and is the same as the average of $A_{mm}$ over the entire energy shell $|E_m - \langle E \rangle| < \delta E$. This closely parallels the classical statistical ensembles that include all microstates compatible with macroscopic constraints with equal probability. 

It is nowadays well established that the ETH represents a sufficient criterion for the onset of thermalization~\cite{Dalessio16}.
The systems that satisfy ETH typically exhibit diffusive transport; to which extent the ETH can be extended to systems with non-diffusive types of transport, in particular when the system is open to the boundary, is discussed in Sec.~\ref{sec:open}.
Other recent activities in the studies of ETH include, among others, the role of the matrix elements correlations for higher point correlation functions~\cite{Foini19, pappalardi2022eigenstate}, and the modification of the ETH in the vicinity of the ergodicity breaking transition point~\cite{kliczkowski_swietek_24}.

\subsection{Ergodicity breaking: single particle systems and beyond}

While ETH has been observed numerically in a wide class of many-body systems, there is no proof of it for a generic, local Hamiltonian, and provable counterexamples of fine-tuned, integrable, Hamiltonians can be constructed. In this sense, the situation is similar to the classical mechanics' ergodic hypothesis, which is widely believed to be true for generic Hamiltonian dynamics (for a sufficiently large number of degrees of freedom), but could only be proved for some realistic systems~\cite{anosov1967some, sinai1970dynamical}. 

Anderson localization \cite{Anderson58} is often referred to as a counter-example to ETH, which has {\it no counterpart} in classical mechanics. 
However, this interpretation requires some care, since the Anderson model, invoked to model Anderson localization on a lattice, is a quadratic model,\footnote{Quadratic Hamiltonians are sums of bilinear forms of creation and annihilation operators, see Eq.~\eqref{eq:Hand} for one example.}
and it can be proven that many-body eigenstates of quadratic models can not exhibit ETH in the many-particle Hilbert space~\cite{lydzba_mierzejewski_23}. However, recent work has formulated the notion of {\it single-particle eigenstate thermalization}~\cite{lydzba_zhang_21}\footnote{
We note that observables in integrable systems relax to the values predicted by the generalized Gibbs ensemble~\cite{Rigol07GGE} in a process called generalized thermalization~\cite{Cassidy11, Vidmar16}.}, which can be seen as the analogue of~\eqref{eq:ETH1}, defined for the matrix elements in single-particle eigenstates of quadratic models, and is a natural consequence of the Berry conjecture~\cite{berry_77}.
Within this framework, Anderson insulator should be seen as an example of the breakdown of single-particle eigenstate thermalization.

Consider a lattice system consisting of $N$ sites where {\it non-interacting} particles move (this is also known as one-electron approximation). The Hamiltonian of the system is given as 
\begin{equation}
    \hat H = \sum_{\braket{i,j}} \left( \hat{c}^\dag_i \hat{c}_j + \mathrm{h.c.} \right) +
    \sum_i \epsilon_i \hat{c}^\dag_i \hat{c}_i,
    \label{eq:Hand}
\end{equation}
where $\hat{c}^\dag_i$ ($\hat{c}_i$) are fermionic creation (anihilation) operators creating particle at site $i$, the sum over neighboring lattice sites is denoted by $\braket{i,j}$, and $\epsilon_i$ are on-site energies.
In case the on-site energies $\epsilon_i$ are random uncorrelated numbers, the model in Eq.~\eqref{eq:Hand} is referred to as the Anderson model. If the variance of $\epsilon_i$ is large enough (with respect to the hopping rate), it turns out that each particle stays confined close to its original position, and the system avoids the process of reaching thermal equilibrium. It goes without saying that there is no conduction in the system and the system is an {\it insulator}. In this case, the single-particle version of the ETH ansatz \eqref{eq:ETH1} does not apply~\cite{lydzba_zhang_21}, and the long-time value of an observable such as site occupation is not recovered by the microcanonical average \eqref{eq:ETH2}. This gives rise to the absence of transport and several mathematical proofs of localization at large disorder exist
\cite{Goldshtein77, Frohlich83, Aizenman93, Bourgain05}\footnote{It is curious, however, that no proof of {\it absence} of localization at small disorder exist for the Anderson model except for the Bethe lattice \cite{aizenman2006stability}.}. 

In spite of the above arguments, it could not be ruled out that interactions would destroy the one-electron picture of Anderson localization and a small conductivity will be born, possibly even exponentially small in the inverse of the interaction strength (measured in units of the local density of single-particle states). In fact, a mechanism in which a small interaction between electrons and phonons can induce a non-zero conductivity via the phonon-assisted hopping was already pointed out by Mott \cite{Mott90}. However, crucially, the phonons considered by Mott are supposed to be a good, ergodic, bath not perturbed by the presence of localized electrons. But can a bath of {\it localized} electrons provide the necessary frequencies for a conduction probe electron to move for macroscopic distances? In other words, could electrons act like Mott's phonon bath? Fleishman and Anderson \cite{Fleishman80} already in the 1980's proposed, on the basis of perturbation theory (and some general intuition), that, at small enough temperatures and weak interactions, localization would be preserved.

\subsection{MBL phase: a definition}
Twenty six years after Fleishman and Anderson, 
\cite{Basko06, Gornyi05} presented a 
thorough analysis of the perturbation theory, where they found that at fixed interaction strength, there is a critical temperature below which conductivity vanishes (the increase in temperature opens more channels and therefore even a small interaction is sufficient to delocalize the system). 
Their analysis of perturbation theory (in the Keldysh formalism) of course is not exact. Some diagrams have to be thrown away for the remaining ones to be tractable, and the ones that get re-summed are essentially those without loops (they are called rainbow diagrams), an approximation which is called by the authors Self-Consistent Born Approximation (SCBA, which, when further supplemented by the negligence of the real part of the self-energies, gets renamed Imaginary SCBA, or ImSCBA). This is 
analogous to the ``upper bound" condition of \cite{Anderson58} and has been studied, under the name of Forward Scattering Approximation, for a variety of one-electron problems on lattices and also for some many-body problems \cite{pietracaprina2016forward}. For the one-electron Anderson problem there is no doubt that this approximation provides an {\it overestimate} of the localization critical disorder, namely, the system could be localized even for smaller disorder but cannot be delocalized for larger disorder. This can be explained in a number of ways (see, for example, \cite{parisi2019anderson} for one based on the properties of the Abou-Chacra, Anderson and Thouless integral equation \cite{abou1973selfconsistent}) and in the one-electron problem, as said before, it is beyond doubt. If this upper bound property would transfer {\it tout-court} to the many-body problem, there would be no doubt that an MBL phase would exists on the basis of Basko et al.\ analytic work. Unfortunately, this has not been proved yet, and the stability of MBL phase, defined in the following, remains an open question.

Subsequent work in lattice systems mostly focused on the infinite temperature states, 
and, importantly, 
emphasized the formulation of the MBL as a phase of matter that avoids thermalization \cite{Pal10,Oganesyan07,Nandkishore15}.
Following these works, 
the dynamics generated by Hamiltonian $\hat{H}$ of an isolated many-body system is non-ergodic if
there exists a local\footnote{A local observable acts non-trivially at a finite number of lattice sites independent of the system size $L$.} observable $\hat{A}$ for which 
\begin{equation}
 A_{\infty}  \equiv  \lim_{L\to\infty} \lim_{t\to\infty}  \bar{A}(t)\neq \mathrm{Tr} \left[ \hat{\rho}_{MC} \hat{A} \right].
 \label{eq:MBL1}
\end{equation}
An isolated interacting many-body system is said to be in an MBL \textit{phase} when the non-ergodic behavior is caused by the disorder and occurs for a robust class of initial states and observables.
This is the definition of MBL that we employ in the present review. Conversely, if the thermalization occurs for
all local observables $\hat{A}$, i.e., when \eqref{eq:ETH2} is satisfied, we say that the system in an \textit{ergodic} phase.
In systems in which the MBL phase exists, it is separated from the ergodic phase by an \textit{MBL phase transition}.  
We note that our definition pertains to observables $\hat{A}$ that are local in real space, i.e., the space in which the sites of the considered many-body system reside\footnote{This aspect of the definition can be readily extended to include different types of few-body observables, for instance, observables local in quasi-momentum space.}.
Moreover, 
the MBL \textit{phase} is defined as a robust ergodicity breaking phenomenon for disordered systems and  
arbitrary physically relevant\footnote{Studies of thermalization predominantly focus on product or low-entangled initial states.} initial states.
Hence, our definition does not apply to fine-tuned settings such as integrable models \cite{Vidmar16} or many-body scars \cite{Serbyn21}.

In ergodic many-body systems, the ETH ansatz \eqref{eq:ETH1} is satisfied, in the large system size limit, by all eigenstates~\cite{Kim14testing}. In contrast, a \textit{weak} ETH allows for (i) polynomial decay with system size $L$ of the fluctuations of matrix elements, as in integrable models \cite{Ikeda13Finite, Alba15ethINT, lydzba_swietek_24}, and (ii) the existence of special eigenstates, dubbed outliers, which do not approach the corresponding microcanonical averages, such as quantum scars, c.f. Sec.~\ref{subsec:clean}. 
The weak ETH is not expected to apply in the MBL phase. Instead, in the MBL phase, one anticipates that the fluctuations of matrix elements of local observables do not decay with $L$, making the ETH invalid for the vast majority of eigenstates. In that case, Eq.~\eqref{eq:MBL1} is satisfied, in the limit $t \to \infty$ and $L \to \infty$, for the physically relevant initial states.

The definition of the MBL phase \eqref{eq:MBL1} contains the double limit of infinite time, $t\to \infty$, and infinite system size, $L \to \infty$. The presence of both limits is essential to define the MBL phase with properties sharply different from the features of the ergodic phase. This leads to a question about the order of limits. Exact diagonalization studies provide access to eigenvalues and eigenvectors of many-body systems, thus naturally allowing to take the limit of infinite time, $t\to \infty$, first, as demonstrated in Eq.~\eqref{eq:A2}. This leads to a system size dependent average value of the considered observable, $A_{\infty}(L) = \lim_{t\to \infty} \bar A(t)$ (the $L$ dependence is now explicitly indicated), which can be subsequently extrapolated to the large system size limit as $A_{\infty} = \lim_{L\to \infty} A_{\infty}(L)$, and compared with the microcanonical ensemble prediction, $\mathrm{Tr} \left[ \hat{\rho}_{MC} \hat{A} \right]$. In the context of disordered 1D many-body systems with local Hamiltonians, which are the focus of this review, we implicitly assume that the limit $L\to \infty$ in Eq.~\eqref{eq:MBL1} is taken in the usual manner, i.e., by keeping the parameters of the Hamiltonian, such as the interaction amplitude, disorder strength and tunneling amplitude, fixed. Alternative choices, allowing for scaling of the microscopic parameters with the system size $L$, may also lead to interesting dynamical phenomena, such as, e.g., in the presence of long-range couplings, as we highlight in Sec.~\ref{subsec:unconv}.

The approach of sending $t \to \infty$ first, 
natural when eigenvalues and eigenvectors of $\hat{H}$ are computed, is impractical in experimental and numerical studies of time evolution of many-body systems\footnote{In a solid-state physics experimental context, one usually assumes the limit $L\to\infty$ first, assuming the system to be effectively in the thermodynamic limit. One then is interested in the time evolution of local perturbation acting on a spatial region $\Omega$ of size $O(1)$ of an initially homogeneous state, as for example the creation of a particle on top of an homogeneous state or a particle-hole excitation created by a photon or a neutron scattered. If the local perturbation eventually ({\it i.e.} in the limit $t\to\infty$) escapes the region $\Omega$, then the system is not localized (neither Anderson nor MBL).}. Let us consider, as an example, an ergodic system with diffusive dynamics, in which local observables attain their long time saturation values after the so-called Thouless time $t_{Th} = L^2/D$, where $D$ is the diffusion constant. To correctly assess that such a system is ergodic, one has to probe properties of local observables at times $t = a L^2$, where $a$ is a sufficiently large constant assuring that the diffusive relaxation processes are finished. In contrast, probing the local observables at smaller times $t \propto L$ may yield values of local observables different than their long time saturation values leading to an incorrect conclusion about lack of thermalization in the system. The time scale required to correctly assess whether the system is ergodic or MBL increases with the slowing down of the dynamics: for instance, for an ergodic system with subdiffusive dynamics characterized by a dynamical exponent $z > 2$, one has to probe the dynamics at times scaling at least as $t \propto L^{z}$. 

The above examples show that in order to assess whether a system reaches the thermal equilibrium or not, one has to probe time evolution at times scales that diverge with system size $L$, and that the precise relation between $t$ and $L$ is fixed by the type of dynamics of the system. This property, coupled with a dramatic slow down of the dynamics of many-body systems with increasing the disorder strength, constitutes one of the main difficulties in assessing whether systems are in the ergodic or the MBL phase.\footnote{Sending $L\to \infty$ followed by the $t\to\infty$ limit may not be the most natural approach in studies of ETH and MBL that aim to explore whether system thermalizes for generic (physically realizable) conditions. In particular, initial states of domain wall type involve lengthscales that are divergent in the $L\to \infty$ limit. This leads to divergent timescales of observables located away from the domain walls, preventing one from verifying \eqref{eq:ETH2} even for a diffusive system.  }

MBL was originally defined as an insulating phase of an interacting electronic system at sufficiently low temperature \cite{Basko06}, where conductivity and diffusion coefficients vanish. In principle, conductivity can vanish for a number of reasons, and diffusion may be substituted by sub-diffusion (in particular in one-dimensional systems), leaving the final equilibrium state to be described by 
the microcanonical ensemble, see also Appendix~\ref{app:dif-sub-Gibbs}. However, it was noted already in 2012 (Oganesyan, private communications) that the vanishing of conductivity in the MBL phase may be due to the existence of local integrals of motion for MBL systems (\textbf{LIOMs}). These were properly conjectured in \cite{Serbyn13a} where also details of the implications for the MBL phenomenology was given (e.g., the slow growth of entanglement already observed in \cite{Znidaric08}), and shortly after, the perturbation series analysis of \cite{Basko06}\footnote{Ref.~\cite{Gornyi17} identified a class of terms that parametrically enhance delocalization threshold with respect to the predictions of~\cite{Basko06, Gornyi05}.} was adapted by \cite{Ros15} to show that, indeed, LIOMs should exist in the MBL phase when the same approximations of the ImSCBA are used. In parallel, numerical algorithms to find and characterize LIOMs were proposed \cite{Chandran15,Rademaker16,Mierzejewski18}. In the meanwhile, a detailed mathematical analysis suggesting stability of the MBL phase of a one-dimensional disordered spin chain was published by \cite{Imbrie16, Imbrie16a}, see also~\cite{Imbrie17} for a parallel discussion of this work and of other perturbative approaches to MBL. Starting from the analysis of \cite{Imbrie16a}, refined, and purged from a hypothesis about absence of level attraction, the authors of the recent paper~\cite{deRoeck24absence} claim a proof of the existence of LIOMs in a fraction of samples which is exponentially small in system size. Although this is {\it not yet} a proof of the existence of an MBL phase, using this result, the authors can prove the absence of diffusion in a strongly disordered interacting quantum spin-1/2 chain, namely the disordered quantum Ising model introduced in Eq.~(\ref{eq:TFIM}), leading to the vanishing of DC conductivity and of any diffusion coefficient.

Our definition of the MBL phase \eqref{eq:MBL1} is not formulated in terms of the transport properties and it does not explicitly rely on the vanishing of the DC conductivity of the system. Those properties, however, are characteristic features of a localized system as described in the next Section, and can be derived from the existence of local conserved operators, such as those that will be introduced in Eq.~\eqref{eq:LIOM}. While the perturbative treatment of \cite{Basko06} assumes that the given interacting system can be smoothly connected to an Anderson localized non-interacting model, our definition allows for MBL in a larger class of many-body systems, including for example systems which do not have a natural non-interacting limit. And, while the ergodicity breaking condition \eqref{eq:MBL1} explicitly involves the energy of the initial state, present day studies of MBL typically focus on the limit of infinite temperature analyzing eigenstates in the middle of many-body spectrum. Let us also note that the term ``many-body localization'' to refer to Anderson localization beyond the one-electron approximation was already in use in the 1980's, the earliest occurrence we could found is in a paper by W.\ McMillan dated 1981 \cite{McMillan1981}. That paper is a generalization of the ``gang of four'' renormalization group (\textbf{RG}) analysis \cite{Abrahams79}, including both conductivity and interaction in a {\it two-parameter scaling} set of RG equations (the author seems unaware of the paper \cite{Fleishman80} that was published shortly before that). As a side remark, the failure of a simple one-parameter scaling theory for describing the MBL transition was recognized in recent works and is discussed in more detail in Sec.~\ref{sec:fss}.

Nevertheless, we stress that a precise definition of MBL is still a matter of debate, as also highlighted in Sec.~\ref{sec:open}.
For example, while the above discussion implicitly assumes the necessity of LIOMs in the MBL phase, this requirement may be too restrictive.
Another, stronger form of ergodicity breaking is localization in Fock space graph, also referred to as Fock space localization \cite{DeTomasi21}\footnote{Note that Fock space localization considered here is not entirely identical to the notion of Fock space localization invoked in~\cite{Basko06}, see also Sec.~\ref{subsec:wave-fun}.\label{footnoteFSL}}.
In our definition here, MBL includes, but is not limited to, Fock space localization, since \eqref{eq:MBL1} may be valid also when many-body eigenstates are not localized in the Fock space \cite{Luca13, Mace19Multifractal}.

\subsection{Modeling and phenomenology of MBL}
\label{subsec:modelingANDphenom}

One-dimensional disordered XXZ spin-1/2 chain is a many-body system that has received a lot of attention in the investigations of ergodicity breaking, becoming a paradigmatic model in studies of MBL. Its Hamiltonian reads
\begin{equation}
 \hat H = \sum_{i=1}^{L} J_i\left(
 \hat{S}^x_i\hat{S}^x_{i+1} + \hat{S}^y_i\hat{S}^y_{i+1}
+ \Delta \hat{S}^z_i \hat{S}^z_{i+1}  \right)
 +  \sum_{i=1}^{L}h_i \hat{S}^z_i,
 \label{Hxxz}
\end{equation}
where $\hat S^{\alpha}_i=\hat{\sigma}^{\alpha}_i/2$ are spin 1/2 operators ($\alpha=x,y,z$), periodic boundary conditions are assumed ($\hat{S}^{\alpha}_{L+1} \equiv \hat{S}^{\alpha}_{1}$), and $J_i=1$ sets the energy scale. Disorder is introduced to the system via random magnetic fields, $h_i$, which are taken as independent random variables, each distributed uniformly in the interval $[-W,W]$, where $W$ is the disorder strength.
The Jordan-Wigner transformation \cite{Jordan28} allows to map the Hamiltonian \eqref{Hxxz} to a system of spinless fermions with annihilation operators given by $\hat{c}_j = (\prod_{i=1}^{j-1} \hat{S}^z_i) \hat{S}^-_j$ where $ \hat{S}^-_j = \hat{S}^x_j-i\hat{S}^y_j$. Under this mapping, the term  $\hat{S}^z_i \hat{S}^z_{i+1}$ becomes a density-density interaction term $\hat{n}_i \hat{n}_{i+1}$ in the fermionic language up to an energy shift, where $\hat{n}_i=\hat{c}^\dag_j\hat{c}_j$ is the fermionic number operator. For concreteness, one usually fixes $\Delta =1$, however, the non-equilibrium properties of the system in presence of disorder, $W>0$, are qualitatively the same as long as $\Delta$ is of the order of unity.

Early studies \cite{Santos04, DeChiara06} observed that as $W$ increases, the system starts to exhibit signatures of non-ergodicity. Those numerical results were later interpreted in terms of a transition to the MBL phase \cite{Oganesyan07, Pal10}. Since then, various features of the strongly disordered XXZ spin-1/2 chain, in particular the slow dynamics and the low-entanglement structure of many-body eigenstates, have been interpreted as features of a stable MBL phase as a dynamical phase of matter.
In the latter, the system retains memory about the details of the initial state, which may be probed by density correlation functions \cite{Schreiber15, Luschen17}. The slow dynamics of MBL phase is associated with a logarithmic increase of entanglement entropy for initial product states \cite{Znidaric08, Bardarson12, Serbyn13a, Iemini16signatures}, as well as with dephasing of local observables \cite{Serbyn14} and with local perturbations spreading along a logarithmic light cone \cite{Fan17, He17, Yichen17, Swingle17slow}. 
Spectrum of the MBL system has Poisson statistics \cite{Oganesyan07}, similarly to Anderson localized systems \cite{Shklovskii93}. Highly excited many-body eigenstates are typically multifractal (rather than localized) in the computational basis~\cite{Mace19Multifractal}, and are characterized by an area-law entanglement \cite{Serbyn13b, Bauer13}, similar to ground states of local Hamiltonians \cite{Eisert10}.

The LIOMs were proposed \cite{Serbyn13b, Huse14, Abanin19} to describe features of systems in the MBL phase. In the absence of interactions, i.e., for $\Delta=0$, the single particle eigenstates $\phi_\alpha$ of \eqref{Hxxz} define a set of occupation number operators,
\begin{equation}\label{eq:expsingle}
 \hat{n}_\alpha= \sum_{i,j} \phi_{\alpha}^{*}(i) \phi_{\alpha}(j) \hat{c}^{\dag}_i \hat{c}_j.
\end{equation}
The operators $\hat{n}_\alpha$ form a set of mutually commuting conserved quantities, which are quasilocal due to the exponential decay of the single particle eigenstate amplitude $\phi_\alpha(i)$ from the center of mass.
The picture of LIOMs  assumes that at strong disorder $W$ and in presence of interactions, $\Delta>0$, it is possible to construct quasilocal operators that are conserved by \eqref{Hxxz} as dressed versions of \eqref{eq:expsingle}, which can be expressed within the following ansatz \cite{Ros15},
\begin{equation}\label{eq:LIOM}
 \hat{I}_\alpha= \hat{n}_\alpha + \sum_{N \geq 1} \sum_{\substack{\mathcalnew{I} \neq \mathcalnew{J}\\
 |\mathcalnew{I}|=N=|\mathcalnew{J}|}} \mathcalnew{A}^{(\alpha)}_{\mathcalnew{I,J}} (\hat{O}_{\mathcalnew{I,J}}+ \hat{O}^\dag_{\mathcalnew{I,J}}),
\end{equation}
where $\mathcalnew{I}=(\beta_1, \cdots, \beta_N)$ and $\mathcalnew{J}=(\gamma_1, \cdots, \gamma_N)$ are sets of indices labeling the single particle states, and 
\begin{equation}\label{eq:normOrd}
 \hat{O}_{\mathcalnew{I,J}}= \prod_{\beta \in \mathcalnew{I}} \hat{c}^\dag_\beta \prod_{\gamma \in \mathcalnew{J}} \hat{c}_\gamma
\end{equation}
is a normal ordered operator (an ordering between the single particle indices is also assumed).
The LIOM $\hat{I}_\alpha$ is a quasilocal operator due to an exponential decay of the coefficients $\mathcalnew{A}^{(\alpha)}_{\mathcalnew{I,J}}$ away from the centre of state $\phi_\alpha(i)$.

When the above construction of a set of LIOMs succeeds, the system may be characterized by a complete set of conserved operators, $[\hat H, \hat{I}_\alpha]=0$.
Conservation of LIOMs immediately implies that the long-time average of a local observable $A$ is determined not only by the average energy, $\langle E \rangle$, and its spread, $\delta E$,  as in \eqref{eq:ETH2}, but also by the average values of LIOMs in the initial state, $\braket{\psi | \hat{I}_\alpha | \psi}$, consistently with the definition of MBL phase \eqref{eq:MBL1}, implying the memory of the initial state and the absence of transport. Spectrum of $\hat{H}$ is fully determined by LIOMs, which explains the Poisson statistics of eigenvalues as well as, via the quasi-locality of $\hat{I}_\alpha$, the area-law entanglement of eigenstates of MBL systems. The exponential decay of couplings between the LIOMs corresponding to different single particle eigenstates 
explains the slow spreading of entanglement and other information measures in the MBL phase.

The construction of the quasi-local unitary transformation $\hat{U}$, which maps the original particle number operators $\hat{n}_i$ to the LIOMs $\hat{I}_{\alpha}$, is a non-trivial task. It has been attempted for specific interacting many-body systems using perturbation theory \cite{Ros15} (with assumptions equivalent to those made in \cite{Basko06}), or numerical approaches. All of those methods are either approximate, or allow to construct LIOMs in relatively small systems. Hence, while they provide useful information about the MBL regime in a specific considered system, they do not answer the question about the fate of the system in the double limit  $t\to \infty$, $L \to \infty$. Of a different nature is the multiscale analysis of \cite{Imbrie16,deRoeck24absence}, in which LIOMs are constructed by a {\it superconvergent} method {\it a la} Kolmogorov, in which a local unitary is built iteratively to order $\lambda, \lambda^2, \lambda^4,...$ in the dimensionless interaction strength $\lambda$. The unitary in question rotates the number operators $\hat{n}_\alpha$ into $\hat{I}_\alpha$ for which $[\hat{H},\hat{I}_\alpha]=0$, and one needs to prove convergence of the procedure and locality of the result. These papers have not been vetted yet by the mathematical physics community, and we therefore refrain from making strong statements based on their results.

Another important step in modeling of MBL, going beyond perturbation theory in the interaction, was achieved by the so-called avalanche mechanism \cite{DeRoeck17}, which is discussed in more details in Sec.~\ref{sec:aval}.
The main idea of the avalanche mechanism is to describe the interplay between a small ergodic seed, which exists with probability one in regions of small disorder, and the localized qubits.
Denoting the Hilbert space of the ergodic seed and of the localized qubits, respectively, by $\mathcal H_{\mathrm{seed}}$ and $\mathcal H_{\mathrm{loc}}$, the Hamiltonian of the corresponding toy model acting on $\mathcal H_{\mathrm{seed}} \otimes \mathcal{H}_{\mathrm{loc}}$ reads~\cite{DeRoeck17}
\begin{eqnarray} \label{eq:Havalanche}
    \hat H = 
\hat{R}\otimes\mathds{1}+g_0\sum_{j=1}^{L}\alpha^j \, \hat{S}^{x}_{n(j)} \otimes \hat{S}^{x}_{j}+\sum_{j=1}^L h_j \,\mathds{1} \otimes\hat{S}^z_j \;,
\end{eqnarray}
where the first term on the r.h.s.~acts non-trivially only within the small ergodic seed, the second term describes the interaction between the spin-1/2 
at site $j$ with a randomly selected 
spin $n(j)$ within the ergodic seed ($\alpha$ is the coupling parameter and $n(j)$ for different $j$ are uncorrelated), and the third term describes the impact of random magnetic field on the localized spins. 
The actual relevance of the model in \eqref{eq:Havalanche} for the description of the disordered spin chains, and systems in higher-dimensional lattices, is still unclear. However, a recent study~\cite{Suntajs22} argued that these types of models, dubbed quantum sun models, are interesting {\it per se}, since they allow for an accurate numerical determination of the ergodicity breaking transition point that agrees to high precision with the analytical prediction based on the hybridization condition~\cite{DeRoeck17}.
Throughout this review, we will refer in several occasions to the quantum sun models, such as the one in Eq.~\eqref{eq:Havalanche}, as a reference point of the ergodicity breaking transition that is adequately described by the state-of-the-art numerical simulations on classical computers.

\subsection{Challenges to MBL}
\label{subsec:chall}
In spite of the rich characterization of properties of the random-field XXZ spin-1/2 chains discussed above, we stress that a conclusion that they  indicate presence of an MBL \textit{phase} at strong disorder is, however, only an interpretation of numerical computations performed for finite evolution times and system sizes. The premises suggesting possibilities of alternative interpretations were apparent in the numerical results, but were often dismissed as merely indicating presence of strong finite size effects at the transition \cite{Oganesyan07, Luitz15, Serbyn15, Devakul15}. In particular, \cite{Bera17}, followed by \cite{Weiner19}, observed a slow "creep" towards thermalization close to the purported MBL transition in the dynamics of disordered spin-1/2 chains with increasing system sizes and time scales.

The interpretation in terms of a stable MBL phase was widely accepted until the year 2019. In that year, a paper \cite{Suntajs19} (see also \cite{Suntajs20e}) suggested that the numerical results of spectral statistics for strongly disordered spin-1/2 chains can be interpreted in terms of an abrupt slowing down of dynamics with increasing disorder strength $W$, which, however, may not lead to the breakdown of ergodicity and the onset of MBL phase. Instead, \cite{Suntajs19} conjectured that in the $t\to \infty$, $L \to \infty$ limit the system remains ergodic at any disorder strength $W$, in consistence with the ETH \eqref{eq:ETH2}. This interpretation first triggered criticism focusing on implications of the arguments of \cite{Suntajs19} to the numerical results for models in which presence of localized phase is well established, such as Anderson model in higher dimensions \cite{Sierant20thouless} or on random regular graphs~\cite{Abanin21}. Nevertheless, as a consequence of those works, it is now accepted that the status of MBL as of a stable dynamical phase of matter is not clear \cite{Panda20}.

Arguments about a possible instability of the MBL phase to non-perturbative effects were later also formulated using complementary measures, such as those indicating an unbounded number entropy growth~\cite{Kiefer20, Kiefer21}, dynamical constraints to localization~\cite{Sels20obstruction}, and the fragility of the LIOMs 
placed in contact with large ergodic systems~\cite{Sels21dilute}.
All the contributions discussed above lead to an intense debate about the status of MBL phase, with various approaches leading to disparate and sometimes conflicting predictions that we aim to review in more details in the next Sections.
As a working definition in studies of finite systems, which feature a certain degree of non-ergodicity but one or both of the limits $t\to \infty$, $L \to \infty$ can not be properly taken into account, it was proposed to refer to them as being in the MBL \textit{regime}~\cite{Morningstar22, Sierant22challenges}.

\section{Status of MBL}
\label{sec:status}

Although significant progress has been made in understanding of the phenomenology of MBL, the status of MBL as a dynamical phase of matter remains uncertain. This is largely due to the inherent complexity of the MBL problem which lacks a well-established theoretical framework and is based mainly on numerical simulations on classical computers.  It is challenging to draw definite conclusions from the results of such simulations which, due to the limited system sizes (typically going from 6 to 20 or 22 qubits for spin-1/2 chains) or evolution times (larger systems can be simulated for a hundred or so cycles), can be extrapolated to the asymptotic limit of $t\to \infty $ and $L\to \infty$ in various ways.
While certain theories for the mechanism of the ETH-MBL transition have been proposed, as we outline in Sec.~\ref{sec:aval}, their relevance for the physics of numerically and experimentally studied prototype models of MBL has not yet been demonstrated.

\subsection{Main open questions}
\label{sec:open}

Below, we list the questions about MBL which, in our view, remain open in spite of over the decade of research on interacting and disordered many-body systems. The rest of this review will be devoted to analyzing the attempts to answer those questions and illustrating the emerging problems, with a particular emphasis on numerical calculations on classical computers. 

\begin{figure}
    \centering
    \includegraphics[width=1\columnwidth]{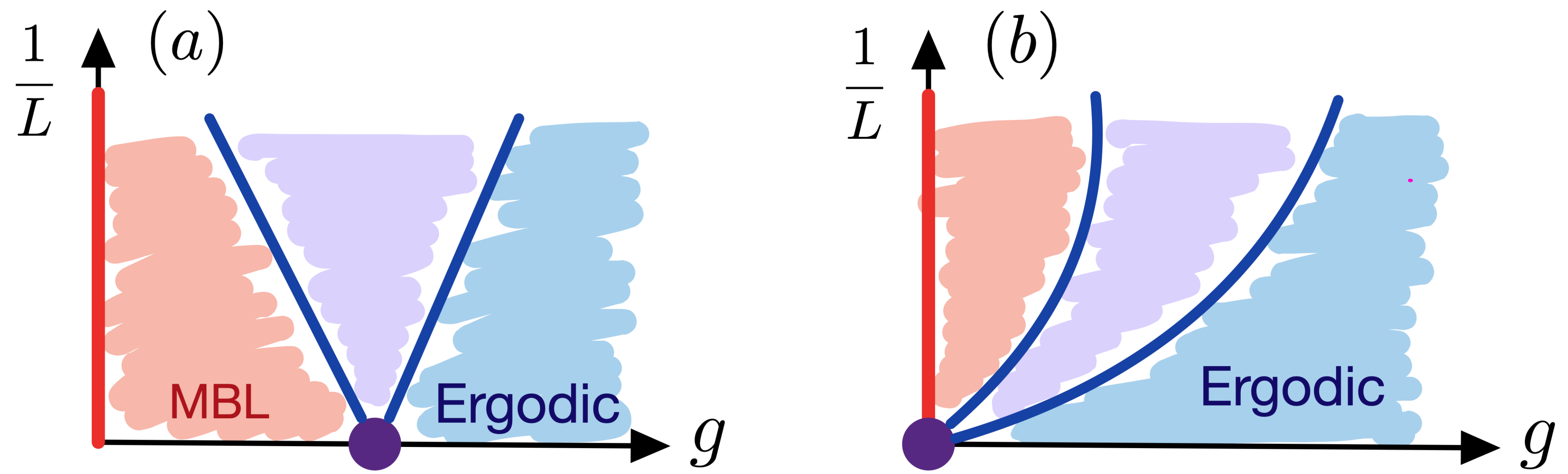}
    \caption{Finite-size scaling of the crossover from MBL to ergodic regime.
    The number of lattice sites is denoted by $L$ and $g$ is a parameter of the theory (in case of the disordered XXZ spin-1/2 chain from Eq.~\eqref{Hxxz}, $g=1/W$).
    (a) A scenario that gives rise to a stable MBL phase in the thermodynamic limit $L\to\infty$, (b) a scenario in which the MBL regime at finite $L$ evolves into an ergodic phase at $L\to\infty$, in that case the MBL in the system is only a \textit{prethermal} phenomenon and the thermalization occurs at long times and large system sizes.
\label{fig:sketch}
}
\end{figure}

\noindent\textit{1. Definition of MBL.} 
Broadly speaking, MBL can be defined 
as lack of thermalization of isolated many-body system in the presence of disorder and interactions. To make the definition of MBL phase precise, one has to take the double limit of infinitely long time evolution, $t\to \infty$, and of the infinite system size $L\to \infty$, as in Eq.~\eqref{eq:MBL1}. 
This leads to a question about establishing a relation between $t$ and $L$ that allows one to assess in a practical way, both numerically and experimentally, whether a given system thermalizes. Another question about the definition of MBL is whether the existence of LIOMs is equivalent to the onset of MBL, as defined by \eqref{eq:MBL1}. If LIOMs can be constructed for a given system, their presence implies that \eqref{eq:MBL1} is satisfied. Conversely, does it mean that all systems satisfying \eqref{eq:MBL1} possess LIOMs?
Moreover, what is the relation of the Fock-space localization to MBL? Is the former a stronger and distinct form of ergodicity breaking?

\noindent \textit{2. Understanding of the MBL phase.}  If we define the MBL phase through the breakdown of ETH \eqref{eq:MBL1}, the central question is whether such a phase of matter actually exists.
Two possible scenarios for approaching the limit $L\to\infty$ are sketched in Fig.~\ref{fig:sketch}.
The mathematical scheme indicating robustness of the MBL phase was proposed, under certain assumptions, several years ago \cite{Imbrie16}. A recent, subsequent analysis~\cite{deRoeck24absence}, made under weaker assumptions, rules out diffusion at strong disorder in certain quantum spin-1/2 chains, but does not exclude subdiffusive transport.
However, as non-experts in mathematical physics, we cannot assess the validity and mathematical rigor of  schemes. 
Another question that remains partially open is the demonstration of the signatures of the MBL phase and its extent in numerical computations for specific many-body systems. Such a demonstration necessarily involves extrapolating the results to the asymptotic limit $t\to \infty$ and $L\to \infty$. Devising an extrapolation scheme that would be convincing and widely accepted is still to be achieved. Additional challenges involve identifying quantities and settings well-suited for studies of the MBL phase, and also determining what features of the dynamics of disordered many-body systems are smoking-gun evidence for MBL phase rather than just an illustration of slow dynamics in the presence of strong disorder.

\noindent
\textit{3. Relationship between thermalization and transport.} 
The ETH is often associated with diffusive transport. There is, however, some evidence for 
other types of transport in certain systems fulfilling ETH, see Sec.~\ref{sec:trans}. Is it possible to find a system in which ETH and subdiffusion actually coexist? In such a system the smooth function $f_A(E, \omega)$ from Eq.~\eqref{eq:ETH1} may have a plateau at $\omega < \omega_{\rm Th} \sim L^{-z}$ that shrinks in a subdiffusive manner, with $z>2$ and $\omega_{\rm Th}$ denotes the Thouless energy. Conversely: could there be a non-vanishing transport in a system in the MBL phase\footnote{In presence of LIOMs there is no transport.
Hence, a non-vanishing transport in the MBL phase would exclude the existence of LIOMs.
}?

\noindent\textit{4. Mechanism of the transition.} If the MBL phase exists, understanding of the physical mechanism of the ETH-MBL transition is a significant challenge  to be addressed. Various effective toy models of the ETH-MBL transition based on strong disorder renormalization group approaches have been proposed. Demonstration of their relevance for physics of particular disordered many-body systems is an open question.
Another open question is the relevance of the avalanche mechanism for physics of disordered many-body systems, and the role of many-body resonances in the emergence of ergodic dynamics. Finding predictions of the corresponding effective models that are testable and falsifiable in numerical and experimental approaches is another open problem. This concerns, in particular, development of finite-size scaling approaches that can be employed to numerical results for disordered many-body systems. 

\textit{5. Understanding of slow dynamics at finite times. } The questions about the MBL phase necessarily concern the thermodynamic limit $t, L\to \infty$. However, the practical significance of this asymptotic limit is restricted, since many realistic physical scenarios pertain to systems of finite size inspected over a finite interval of time. This demonstrates the importance of finding descriptions of the MBL \textit{regime} characterized by a very slow dynamics of strongly disordered interacting many-body systems, irrespective of whether some parts of this regime become eventually ergodic in the asymptotic limit or not. Can we understand whether the slow dynamics leads to thermalization or gives rise to the MBL phase without the need to wait for extremely long times?
In terms of spectral statistics, this questions may be phrased whether there exist certain universal properties beyond the standard random-matrix-theory predictions, which characterize the transition point in finite systems.

\textit{6. Models of MBL and beyond.} Disordered XXZ spin-1/2 chain \eqref{Hxxz} was very often the subject of numerical studies of MBL. Consideration of alternative disordered systems in which studies of MBL may be easier, and establishing ways of comparing ergodicity breaking among different systems, are possible directions for further research. A related question is whether the ETH-MBL transition is system specific, and whether there exist different types of ergodicity breaking transitions.
This includes the question about a general classification of non-ergodic phenomena in interacting many-body systems, regardless of whether the strong disorder is the source of ergodicity breaking or not.

\subsection{Difficulties in numerical approaches to MBL}
\label{subsec:difficulties}

In the following, we discuss features commonly found in numerical studies of ergodicity breaking in disordered many-body systems. We emphasize the ambiguity in interpreting numerical results and the difficulties in extrapolating results to the asymptotic limit of infinite system size $L$ and evolution time $t$, required for the understanding of the MBL phase and of the ETH-MBL transition.

One way to numerically study MBL is to compute eigenvalues and eigenvectors of Hamiltonian $\hat H$ of a disordered many-body system and use them to calculate a quantity $\overline X$ that indicates the breakdown of ergodicity in the system. For concreteness, we assume that the quantity $\overline X$ is normalized in such a way that it is equal to $1$ if the system follows ETH and is equal to $0$ in a large MBL system. Calculating $\overline X$ as a function of disorder strength $W$ for system size $L$ and taking average over disorder realizations, one observes a \textit{crossover}\footnote{We differentiate between the notions of the crossover and phase transition. Crossover is a gradual change of the considered property of the system observed at finite system size $L$. Phase transition is a singular change of the considered property of the system in the thermodynamic  limit $L\to\infty$. Phase transition can be inferred from investigation of the system at finite $L$ if the results change systematically with $L$ according to a certain scaling form.} between the ergodic ($\overline X \to 1$) and MBL ($\overline X \to 0$) regimes. The goal is to understand what is the fate of this crossover in the $L\to \infty$ limit and whether it indicates a \textit{phase transition} between ergodic and MBL phases in the thermodynamic limit.

\begin{figure}
    \centering
    \includegraphics[width=1\columnwidth]{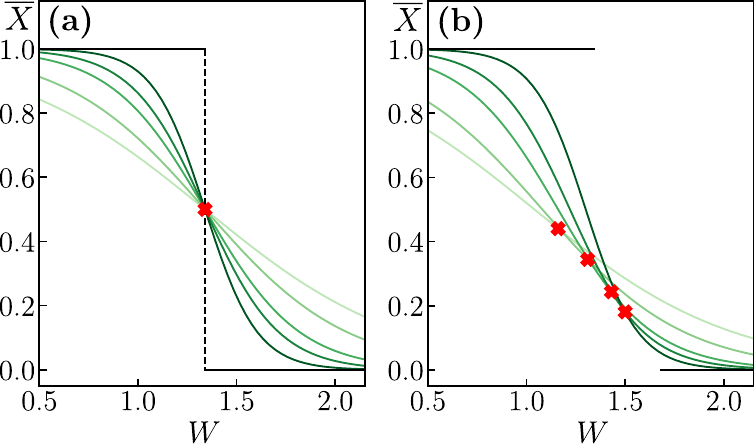}
    \caption{A quantity $\overline X$, derived from eigenvectors (or eigenvalues) of Hamiltonian of disordered many-body system is plotted as a function of disorder strength $W$. It is equal to $\overline X =1$ deep in the ergodic regime ($W\lesssim0.5$), and tends to $0$ in the MBL regime; the darker lines correspond to larger system size $L$. In an ideal scenario \textbf{(a)}, the curves for different $L$ cross at disorder strength $W^*$ (denoted by the red cross) which is independent of $L$ and estimates the critical disorder strength $W_C$. In scenario \textbf{(b)}, encountered in investigations of 1D disordered many-body systems, the crossing point $W^*$ drifts to larger disorder strengths with increasing $L$, the value of $W_C$ remains uncertain. 
\label{fig:flow}
}
\end{figure}

The fate of the crossover observed in numerical investigations at finite system sizes is clear in a scenario presented in Fig.~\ref{fig:flow}(a). There, we observe an ergodic regime at $W<W^*$, in which $\overline X$ approaches the ergodic value, $\overline X \to 1$, with increasing system size $L$, and an MBL regime at $W> W^*$, in which $\overline X $ monotonously decays to $0$ with  the increase of $L$. The crossing point, $W^*$, gives a good estimate of the critical disorder strength $W_C$ of the ETH-MBL transition at which $\overline X$ changes from $1$ to $0$ in the thermodynamic limit $L\to \infty$. This scenario is, however, to a good approximation only realized in certain toy models of ergodicity breaking such as the quantum sun models~\cite{Luitz17bath, Suntajs22}.

In contrast, investigations of 1D disordered many-body systems with local Hamiltonians often yield results schematically illustrated in Fig.~\ref{fig:flow}(b). The crossing point, $W^*(L)$, drifts with increasing system size $L$ to larger disorder strength $W$. Thus, there exists a clear ergodic regime for $W<W^*(L)$, where $\overline X$ monotonically approaches the ergodic value, $\overline X \to 1$, with increasing $L$.
However, the drift of the crossing point makes it challenging to identify the extent (or even the existence) of the MBL phase.
For $W>W^*(L)$, the value of $\overline X$ is decreasing with system size $l$ as long as $l<L$, but this trend gets reversed for $l>L$. 
Thus, the crossing point $W^*(L)$ provides a lower bound for the critical disorder strength $W_C$ at a given $L$.
The behavior of $W^*(L)$ can be highly non-universal and may depend on the specific properties of the system under investigation.
While it is expected that $\lim_{L\to\infty} W^*(L) = W_C$ if there is an MBL transition, it is also possible for $W^*(L)$ to increase indefinitely with system size $L$. In the latter case, although the ETH-MBL crossover is observed at any finite $L$, the system hosts only an ergodic phase in the thermodynamic limit. Therefore, the determination of the critical disorder strength requires a careful extrapolation of the results for $W^*(L)$ to the thermodynamic limit. 
While here we only discussed a particular extrapolation scheme, different extrapolations based on the results in the same range of system sizes may yield conflicting results, which is the source of difficulties in understanding of the status of the MBL phase. 

\begin{figure}
    \centering
    \includegraphics[width=1\columnwidth]{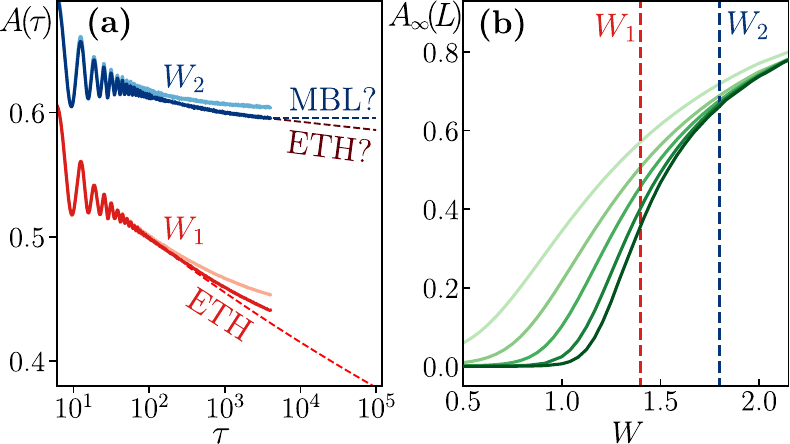}
    \caption{Observing the putative ergodic-MBL transition in time domain. \textbf{(a)} Average value $A(\tau)$ of a local observable $\hat{A}$, darker lines correspond to larger system size $L$. At disorder strength $W_1$, one observes a slow decay of $A(\tau)$ which persists to longer times $\tau$ with increasing $L$; this indicates the decay $A(\tau) \to 0$ in the limit $\tau \to \infty$ and $L \to \infty$. Hence, the ETH is valid in the thermodynamic limit at $W_1$. When $W$ is increased to $W_2>W_1$, the decay of $A(\tau)$ slows down considerably, but the increase of $L$ is accompanied with more persistent decay of $A(\tau)$. The system is in the MBL phase if $0<A_{\infty}\equiv \lim_{L\to\infty} \lim_{t\to \infty} \bar{A}(t)$. The value of $A_\infty(L)$ is accessible only for small $L$ and is decreasing with $L$ at fixed $W$, see \textbf{(b)}, where the darker lines correspond to larger $L$. 
    \label{fig:flow2}
}
\end{figure}

Examination of time evolution of an average value $A(\tau)$ of a local observable $\hat{A}$
is a different way of investigating MBL
While relevant for numerical simulations, it is especially important due to its experimental relevance.  
In the presence of interactions, $A(\tau)$ averaged over disorder realizations, after an initial transient, follows a non-oscillatory time dependence, as shown in Fig.~\ref{fig:flow2}(a). Hence, it suffices to consider $A(\tau)$, rather than the long time average $\overline A(t)$, c.f.~Eq.~\eqref{eq:ETH2}, to assess the ergodicity of the system. 
Let us assume that the observable and the initial state of a disordered many-body system are chosen in such a way that $A(\tau=0) = 1$ and that  $\lim_{t\to \infty}\bar{A}(t) =0$ if the system is ergodic \eqref{eq:ETH2}. Then, according to \eqref{eq:MBL1}, a non-zero value of $A(\tau)$ in the double limit of $L\to\infty$ and $\tau\to\infty$ signifies that the system is in the MBL phase. 

A typical situation encountered in numerical (and in experimental) studies of disordered many-body systems is illustrated in Fig.~\ref{fig:flow2}(a). At small disorder strengths, $W \ll W_1$, the quantity $A(\tau)$ quickly approaches $0$ consistently with the thermalization of the system. While the approach to the thermal value slows down with increasing the disorder strength, the decay of $A(\tau)$ persists to longer times with increasing the system size. Hence, it is clear that $A(\tau)$ vanishes in the double limit $L\to\infty $ and $\tau \to \infty$ at $W=W_1$, and that the system is in the ergodic phase. 
The fate of the system becomes unclear upon further increase of the disorder strength $W$. For $W_2>W_1$, at times available in numerical simulations, the average value $A(\tau)$ is significantly larger than $0$. This could be interpreted in terms of an onset of MBL in the system. However, even at the larger disorder strength $W=W_2$, the increase of system size $L$ leads to a stronger decay of $A(\tau)$ that persists to longer times.
Thus, in absence of additional information, the value of $A(\tau)$ in the $L\to\infty$, $\tau\to\infty$ limit remains obscure, preventing one from deciding whether the system is in an MBL phase at $W=W_2$, or whether the ergodicity is restored at sufficiently large length and time scales. 

The behavior of the system is further illustrated in Fig.~\ref{fig:flow2}(b), in which we sketch the the infinite time average $A_\infty(L) = \lim_{t\to\infty} \frac{1}{t}\int_{0}^{t} d\tau A(\tau)$ as a function of disorder strength $W$. At smaller $W$, we see a clear approach  $A_\infty(L) \to 0$, and hence, the system is in the ergodic phase. The decrease of $A_\infty(L)$, however, slows down when $W$ is increased. Already at $W=W_1$ the limitation in accessible system sizes prevents one from clearly seeing whether the system thermalizes or not. The monotonous decay of $A_\infty(L)$ with $L$ persists to larger disorder strengths. While the decrease of $A_\infty(L)$ at $W=W_2$ within the interval of available system sizes is very small compared to the value of $A_\infty(L)$, it is clearly observable. This prevents one from unambiguously concluding whether at $W=W_2$ the system is in the MBL phase, $A_\infty(L) >0$, or whether it thermalizes, $A_\infty(L) \to 0$, in the limit $L\to \infty$.

In Sec.~\ref{sec:numerical}, we will provide concrete examples of quantities that follow the above described patterns leading to difficulties in pin-pointing the MBL transition. Investigations of MBL so far show that also quantities more complex than the quantity $\overline X$ share similar non-trivial trends with the increase of time and length scales. This reflects the generic feature of numerical simulations of disordered many-body systems that reveal a persistent tendency towards thermalization in the system which strengthens with the increase of time and system size. In addition, in the presence of strong disorder, the dynamics become extremely slow. Thus, the question of the existence and position of the MBL phase is inevitably tied to a formidable task of distinguishing between very slow and completely frozen dynamics. While the overall features of ergodicity breaking indicators outlined in the above description are shared by the known observables and many-body systems, the question about identification of more favourable (easier to analyze) models and observables remains open.

\subsection{Remarks on finite-size scaling} \label{sec:fss}

Here, we discuss the main challenges associated with finite-size scaling analysis of numerical results at the putative transition between ergodic and MBL phases in 1D systems in presence of interactions and disorder. Given the difficulties in interpreting numerical results at the MBL crossover as indicative of a true phase transition, a proposed finite-size scaling analysis of the results must undergo careful scrutiny. Nevertheless, a successful finite-size scaling analysis of numerical results would help solving both the problem of existence of MBL phase as well as would allow us to understand the critical properties of the MBL transition. In this regard, we draw upon the Anderson transition on lattices with various geometries as a reference point for finite-size scaling analysis at the presumed MBL transition.

\subsubsection{Single parameter scaling}
Let us reconsider the quantity $\overline X$ that indicates the breakdown of ergodicity in the system, and examine its dependence on the system size $L$ and disorder strength $W$, denoted as $\overline X = \overline X(L,W)$.

A large body of numerical works starting from \cite{Pal10}, through the more accurate \cite{Luitz15} and continuing to these days, treated the ergodicity breaking transition of MBL as a `usual' phase transition, i.e., as if it can be described by a simple fixed point of a renormalization group analysis \cite{Abrahams79,Cardy96} with a single relevant parameter. We will review this concept, assumptions, and results in the following but we anticipate that the results are unsatisfactory. In particular, the critical exponents which best fit the data are in contradiction with general bounds (to be discussed below), and the irrelevant corrections are usually large, with a considerable flow of the apparent fixed point with system size.

In a single-parameter scaling hypothesis the results for $\overline X(L,W)$ sufficiently close to the transition (i.e., when all the {\it irrelevant} parameter flows can be neglected) can be interpreted as depending solely only on the product of a scaling variable $\chi(w)$ and the system size $L$,
\begin{equation}
\overline X(L,W) = f_0(\chi(w) L^{1/\nu}),
    \label{eq:scal1}
\end{equation}
where $f_0$ is a certain scaling function, $w \equiv (W-W_C)/W_C$ is the dimensionless disorder, and the scaling variable, generally written as $\chi(w) = \sum_n a_n w^n$, can be assumed to be proportional to $w$, $\chi(w) \sim w$. The critical exponent $\nu$ is given by the linearization of the beta function, see Eq.~\eqref{eq:beta_function}, of a relevant variable close to the critical point $w=0$. We remind the reader that if one considers $\overline  X$ as an renormalization group ``charge" (like a coupling or a simple function of couplings in the Hamiltonian), one can use the RG flow to organize different finite size systems on the same orbit \cite{Cardy96}. This is done by means of a beta function
\begin{equation} \label{eq:beta_function}
    \beta\equiv\frac{d\ln \overline  X}{d\ln L}.
\end{equation}
The one-parameter scaling gives rise to the hypothesis that $\beta$ is solely a function of $\overline X$, i.e., $\beta=\beta(\overline X)$. Assuming a simple critical fixed point with the behavior $\beta(\overline  X)\simeq s(\overline  X-\overline 
 X_0)$ close to the fixed point, then integrating the equation we have $\ln(|\overline  X(L)-\overline  X_0|/|\overline  X(L_0)-\overline  X_0|)=\overline  X_0 s\ln(L/L_0)$. Hence, $|\overline  X(L)-\overline  X_0|\simeq |X(L_0)-X_0|(L/L_0)^{1/\nu}$, with $\nu=1/(s \overline  X_0)$ the critical exponent. Since at finite $L_0$ everything is analytic in $w$, we must have $|\overline  X(L_0)-\overline  X_0|\propto \chi(w),$ where $\chi(w)$ is an analytic function of $w$. In the generic case we expect $\chi(w)\simeq w+O(w^2)$, and therefore we can write $\overline  X(L)= \overline X_0+c(L/\xi)^{1/\nu}$, where $\xi$ is a correlation length that exhibits a power-law divergence, 
\begin{equation}
\xi_{P} = \xi_{\pm} |w|^{-\nu}.
    \label{eq:NU}
\end{equation}
$\xi_{\pm}$ are constants that may be different for $w>0$ and $w<0$. So, under the assumption $w = \chi(w)$, Eq.~\eqref{eq:scal1} implies that $\overline X(L,W)$ depends only on the ratio of the system size and the correlation length, $\overline X(L,W) = f( L/\xi_P )$.

The Anderson localization transition in three (and higher) dimensional lattices serves as an example of a critical phenomenon captured by the scaling hypothesis \eqref{eq:scal1}-\eqref{eq:NU}. Numerical studies have demonstrated that this scaling hypothesis describes multiple physical characteristics of the 3D Anderson model in the vicinity of the localization transition \cite{MacKinnon81,Rodriguez10, Lopez12, Slevin14, Prelovsek21, Suntajs21}. The theoretical basis for this success is the {\it one parameter scaling theory} of the Anderson transition put forward in \cite{Abrahams79}.

However, it turns out that applying the single parameter scaling hypothesis \eqref{eq:scal1}-\eqref{eq:NU} to the results at the putative ETH-MBL transition presents significant difficulties. Under a natural assumption that the function $f_0(x)$, defined in \eqref{eq:scal1}, is continuous and monotonous in the vicinity of the transition, the single parameter scaling hypothesis {\it by construction} fails to reproduce the characteristic, significant drift of the crossing point observed in results at the ETH-MBL crossover. Indeed, in this scenario, the equality $\overline X(L_1, W^*) = \overline X(L_2, W^*)$ holding for two system sizes $L_1 \neq L_2$ at the crossing point $W^*$, implies, through \eqref{eq:scal1}-\eqref{eq:NU}, that $W^* = W_C$. Consequently, while $W^* = W_C$ for the behavior schematically represented in Fig.~\ref{fig:flow}(a), the single parameter scaling is not well suited for the ETH-MBL crossover, c.f. Fig.~\ref{fig:flow}(b). Nevertheless, if one disregards the drift of the crossing point $W^*$, for instance by restricting the results to a few largest system sizes available, the finite size scaling ansatz \eqref{eq:scal1}-\eqref{eq:NU} can yield a visually appealing data collapses at the ETH-MBL crossover. Unfortunately, the resulting critical exponent $\nu$ obtained in this approach is close to $1$ \cite{Pal10, Kjall14, Luitz15}, in apparent contradiction with the Harris-Chayes criterion $\nu \geq 2$  \cite{Harris74, Chayes86} applicable to 1D systems \cite{Chandran15a, Khemani17}.

Several phenomenological approaches \cite{Goremykina19, Dumitrescu19} proposed that the ETH-MBL transition may be of the Berezinskii–Kosterlitz–Thouless (BKT) origin~\cite{Kosterlitz16}, suggesting to replace \eqref{eq:NU} with the following behavior of the correlation length, 
\begin{equation}
    \xi_{BKT} = \exp \left( b_{\pm}/\sqrt{w} \right),
    \label{eq:BKT}
\end{equation}
where $b_{\pm}$ are constants that may be different on the two sides of the transition. The possibility of BKT scaling of results at the ETH-MBL crossover in the disordered XXZ spin-1/2 chain~\eqref{Hxxz} was explored in~\cite{Suntajs20, Laflorencie20}, see also~\cite{Hopjan21}. While the BKT scaling form is not directly contradicting the Harris-Chayes criterion (since it can be thought of as a particular $\nu \to \infty$ limit of \eqref{eq:NU}), its application to the numerical results is associated with the following difficulties. 

The BKT scaling \eqref{eq:BKT} is characterized by a much slower scaling of the results with system size when compared to the power-law scenario \eqref{eq:NU}. To demonstrate this, let us consider fixing a specific value $x_0$ of our ergodicity breaking indicator, 
and define a system size dependent disorder strength $W_{x_0}(L)$ such that $\overline X(L, W_{x_0}(L))=x_0$. Assuming that the system undergoes a transition at which $\overline X$ exhibits a step-like behavior in the $L\to\infty$ limit, we observe that for $x_0$ smaller (larger) than the value of $\overline X$ corresponding to the critical point, $W_{x_0}(L)$ converges with increasing $L$ to the critical disorder strength $W_C$ from 
above (below), compare Fig.~\ref{fig:flow}. For the power-law scenario, from \eqref{eq:scal1} and \eqref{eq:NU}, we find that
\begin{equation}
   W_{x_0}(L) = W_C \pm c L^{-1/\nu},
    \label{eq:Wx01}
\end{equation}
where $c = (W_C^{\nu} \, \xi_{\pm} f^{-1}(x_0))^{1/\nu}$ and the sign depends on the value of $x_0$. In contrast, the BKT scenario implies through \eqref{eq:BKT} that 
\begin{equation}
   W_{x_0}(L) = W_C \pm  \frac{a}{(\ln (L/L_0))^2},
    \label{eq:Wx02}
\end{equation}
where $a =W_C b_{\pm}^2 $ and $L_0 = 1/(f^{-1}(x_0))$. 

A comparison of \eqref{eq:Wx01} and \eqref{eq:Wx02} reveals that $W_{x_0}$ changes with the system size much more slowly in the BKT scenario than in the power-law case. Consequently, it is challenging to distinguish the two scenarios in numerically generated results. This point is exemplified by a renormalization-group inspired toy model of MBL transition proposed in \cite{Dumitrescu17}. The results of that approach were initially interpreted in terms of a power-law scaling with exponent $\nu \approx 3$ \cite{Dumitrescu17}, and only a refined analysis of numerical results for the same model \cite{Dumitrescu19} demonstrated a BKT scaling at the transition. Those difficulties arose in spite of the fact that numerical results were available for this toy model in a wide interval of system sizes ranging up to $L \approx 10^5$. In fact, similar problems are encountered in analyses of the BKT scaling at quantum critical points, see~\cite{Lacki21} for one example. These examples show that distinguishing between the BKT and power-law scaling for the disordered many-body systems, in which numerical results are generated for system sizes that span less than two orders of magnitude, is less likely successful. This, in turn, demonstrates the importance of understanding of the mechanism of the transition whose predictions can be then compared  with numerics, see Sec.~\ref{sec:aval}.

Since the understanding of this subject is still evolving, it is important to mention that for the Anderson transition on random graphs (which is not in the universality class of any finite-dimensional Anderson transition), it has been pointed out in \cite{vanoni2023renormalization} (see also \cite{Sierant23RRG,Garcia-Mata22}) that the one-parameter scaling theory of \cite{Abrahams79} needs to be amended. The RG equations of \cite{Abrahams79} on infinite dimensional lattices are inherently different from the ones on finite-dimensional lattices. Among other things, this implies that the critical point is localized itself (the level statistics is Poisson, and the critical eigenstates are localized \cite{Tikhonov19}). Since the localized fixed point and the critical point cannot be distinguished, the linearization of the beta function close to the critical point that should give rise to the critical exponent $\nu$ and to the usual one-parameter scaling behavior does not work {\it tout-court}. When one considers the whole beta function~\eqref{eq:beta_function}, together with the now non-negligible {\it irrelevant parameters} flow at the critical point, the whole picture can be mapped to the BKT transition described before. It could therefore be the case that the MBL transition is in the same universality class as the Anderson transition in infinite dimensions (something already implied by the \cite{Altshuler97} and reiterated in \cite{Luca13,de2014anderson,vanoni2023renormalization}). Work to establish this connection, following \cite{vanoni2023renormalization, altshuler2024renormalization}, and to re-analyze the existing results from this new point of view, is in progress.

\subsubsection{Sub-leading corrections to the scaling}

As pointed out in the previous subsection, the currently available numerical results for the disordered many-body systems do not fall within the regime of applicability of the single parameter scaling hypothesis due to the finite size drifts at the ETH-MBL crossover.  Hence, the scaling ansatz \eqref{eq:scal1} has to be modified in order to account for those effects. Moreover, even if there exists a regime in which the finite size drifts become negligible, it is probable that it will continue to stay beyond the reach of exact numerical methods on classical (super)computers in the foreseeable future, see also Sec.~\ref{sec:num}. 

In this situation, one needs to include some {\it irrelevant parameters} flow, as one needs to do it in Anderson transitions on random graphs to get precise values for the exponents. In fact, numerical results for the Anderson transitions on lattices of dimension 4 and higher \cite{Tarquini17}\footnote{With increasing dimension $d$, the contribution from the irrelevant parameters increases, and finally, in the $d\to \infty$ limit, the leading irrelevant exponent becomes marginal in the RG sense~\cite{altshuler2024renormalization}.} or on random regular graphs~\cite{de2014anderson,Tikhonov16,Sierant23RRG}, exhibit a noticeable drift of the crossing point, which may appear similar to the numerical results at the ETH-MBL crossover shown in Fig.~\ref{fig:flow}(b). A finite size scaling analysis of such results involves introduction of sub-leading corrections to the scaling ansatz \cite{Aharony83, Efetov83, Slevin99}. The starting point, suggested by renormalization group equations~\cite{Cardy96}, and which goes beyond the analysis of \cite{Abrahams79}, reads
\begin{equation}
    \overline X(L,W) = F(w L^{1/ \nu}, \psi L^y) ,
    \label{eq:scal2}
\end{equation}
where $w=(W-W_C)/W_C$ is the relevant scaling variable, $\psi$ is an irrelevant scaling variable and $y$ is the irrelevant critical exponent of the smallest absolute value. Performing the Taylor expansion of the two-parameter function $F$ to first order in the irrelevant variable, we get
\begin{equation}
    \overline X(L,W) = f_0(w L^{1/ \nu}) + \psi L^y f_1(w L^{1/ \nu}),
    \label{eq:scal3}
\end{equation}
where $f_0(w L^{1/ \nu})= F(w L^{1/ \nu}, 0)$ and $f_1(w L^{1/ \nu})= \partial_{x} F(w L^{1/ \nu}, x)|_{x=0}$. Comparing \eqref{eq:scal3} with \eqref{eq:scal1}, we see that the irrelevant variable yields a correction to the scaling through the term $\psi L^y f_1(w L^{1/ \nu})$. In contrast to the single parameter scaling form \eqref{eq:scal1}, the scaling ansatz \eqref{eq:scal3} describes a drift of the crossing point $W^*$ with increasing system size $L$. The relation between $W^*$ and $L$ is determined by the exponents $\nu$ and $y$, as well as the functional forms of the functions $f_0$ and $f_1$; see \cite{Sierant23RRG} for an example of Anderson transition on random graphs. Thus, the scaling ansatz \eqref{eq:scal3} is more suited for numerical results at the ETH-MBL crossover as exemplified in \cite{Sierant22floquet}, which predicted $\nu \approx 2$  for 1D interacting disordered Floquet system, in consistence with the Harris-Chayes criterion. However, the scaling form \eqref{eq:scal3} contains multiple free parameters while the numerical results for many-body systems are available only in a limited range of system sizes, which severely restricts the predictive power of a finite scaling analysis with \eqref{eq:scal3}. This is illustrated by the scaling analysis of Anderson localization transition on random regular graphs. While the exact knowledge of the critical disorder strength makes the analysis more constrained, the presently available numerical results alone are insufficient to clearly determine the value of the critical exponent $\nu$, see \cite{Pino20, Sierant23RRG, Tikhonov19, Garcia-Mata20, Garcia-Mata22, Biroli22}.

A different way to account for the drifts of the results at the MBL crossover was considered in \cite{Suntajs20}. In that approach one employs the ansatz $\overline X(L,W) = f( L/\chi_P )$, with an additional assumption that the transition point itself has a functional dependence on the system size $L$,
\begin{equation}
W_C \rightarrow \widetilde{W}_{C}(L),
\label{eq:WcL}
\end{equation}
which leads to a correlation length that depends not only on the disorder strength $W$, for instance through \eqref{eq:NU} or \eqref{eq:BKT}, but also on the system size $L$, $\chi=\chi(L, W)$. 
Different scenarios for the transition can be described by different scaling of $\widetilde{W}_{C}(L)$
with the system size $L$.
Conventional scenario assumes that the transition occurs at a finite critical disorder strength in the thermodynamic limit $\widetilde{W}_{C}(L) \stackrel{L\to\infty}{\longrightarrow} W_C$. However, Eq.~\eqref{eq:WcL} also introduces possibilities of unconventional scaling, in which the critical point diverges in a certain fashion with system size, $\widetilde{W}_{C}(L) \stackrel{L\to\infty}{\longrightarrow} \infty$, or that the regime at small $W$ shrinks to a single point in the large system size limit, $\widetilde{W}_{C}(L) \stackrel{L\to\infty}{\longrightarrow} 0$.

\subsubsection{Unconventional scaling limits}
\label{subsec:unconv}

A particular example of an unconventional scaling limit was proposed in the study \cite{Suntajs20}, which, by comparing collapses of results obtained with a few scaling ansatzes, found that results at the ETH-MBL crossover in the XXZ spin chain were best described by the BKT scenario \eqref{eq:BKT}, but with a transition point that increases linearly with system size, $\widetilde{W}_{C}(L) = a_0 + a_1 L$, where $a_0, a_1$ are constants. Taken literally, that result would imply that there is no transition to an MBL phase in the commonly understood sense, as $\widetilde{W}_{C}(L) \to \infty$ in the thermodynamic limit. However, the implications of this result remain to be understood. One possibility, highlighted in \cite{Suntajs20}, is that the linear drift of  $\widetilde{W}_{C}(L)$ with $L$ is a feature of numerical results at available system sizes $L\approx 20$, and that it is replaced by a milder system size dependence at larger $L$, so that $\widetilde{W}_{C}(L) \stackrel{L \to \infty}{\longrightarrow} W_C$, where $W_C$ is a finite critical disorder strength. Interestingly, a recent analysis of the ETH-MBL crossover \cite{Sierant21constraint} in constrained spin chains~\cite{Chen18}, which exhibit a specific interplay of constraints and interactions, reported a linear drift of $\widetilde{W}_{C}(L)$ with $L$ that persists to very large values of $L$.

Since the introduction of the scaling theory of localization~\cite{Abrahams79}, it was understood that the 2D Anderson model remains localized at any disorder strength $W$. Nevertheless, early numerical studies of the model, e.g.,~\cite{Edwards72, Licciardello75}, found a regime of disorder strengths at which eigenstates of the model appeared to be delocalized. 
In contrast, a diagrammatic approach of \cite{Vollhardt80Diagrammatic}, relevant for a weak disorder, predicts the breakdown of transport due to interference corrections at a lengthscale which, for asymptotically small $W$, must behave like $\xi \propto e^{a/W^2}$, where $a$ is a constant. The lengthscale $\xi$ is identified, within the self-consistent theory of localization, with the localization length of the 2D Anderson model~\cite{lee1985disordered}.
This is due to the fact that the $\beta$ function has a {\it double zero} rather than a simple zero at the critical point\footnote{At simple zero $\beta(\overline  X) = 0$ and $d \beta(\overline  X)/d \overline  X \neq 0$. At double zero $\beta(\overline X) = 0$ and $d \beta(\overline X)/d\overline X =0$.}. This is the way the RG captures the effects of divergent series of weak-localization corrections.
Numerical analysis of 2D Anderson model, relying on a direct extraction of the localization length~\cite{MacKinnon81}, or, recently, on a study of the average gap ratio $\overline r$~\cite{Suntajs23}, suggests a single-parameter scaling with the length scale $\xi' \propto e^{a'/W^{\mu'}}$, with $\mu'\approx 1$ for the accessible system sizes. While a gradual adherence of the numerical results to the perturbative prediction (which becomes exact as $W\to0$) can be observed~\cite{Lee13honeycomb}, the Anderson model in 2D is an important reference point for studies of the localization transitions. It shows that data collapses over even seemingly wide intervals of $L$ may yield the unconventional value of the exponent $\mu'=1$. The delocalized-localized crossover can be interpreted, via \eqref{eq:WcL}, as an unconventional transition which is getting sharper with increasing $L$, but eventually the delocalized phase shrinks to a point, $\widetilde{W}_{C}(L) \stackrel{ L \to \infty }{\longrightarrow} 0$. In some sense, explained by the work of \cite{vanoni2023renormalization}, the Anderson model in 2D shares a similarity with the Anderson model on random regular graphs, since in both cases the critical point coincides with one of the ``gaussian fixed points" (localized or ergodic). 
In 2D it is the ergodic fixed point\footnote{For 2D Anderson model, at the ergodic fixed point the fractal dimension $D_1$ (see Sec.~\ref{subsec:wave-fun}) is equal to unity, and the $\beta$ function can be expanded around the $D_1=1$ fixed point as $\beta(D_1) = -(1-D_1)^2$
\cite{altshuler2024renormalization}.}, in the random regular graphs it is the localized fixed point. In both cases larger than power-law critical lengths emerge since a `naive' analysis based on a simple zero of the beta function must fail. Both cases need to be treated using a more complex scaling analysis, and this is a cautionary tale for the studies of MBL as well.

\begin{table*}[ht!]
    \centering
    \begin{tabular}{ccccccccccc}
        \hline \hline
         Quantity & & Measures & & \multicolumn{2}{c}{Ergodic regime} &  &  \multicolumn{2}{c}{MBL regime}  & & Section \\
        \cline{1-1}  \cline{3-3} \cline{5-6} \cline{8-8} \cline{10-11}
         & &  &  &  &   & & \\
Average gap ratio $\overline r$ & & {\raisebox{1.3ex}[0pt]{  \small{ short-range } }} & &$\overline r \stackrel{L \gg1 }{\longrightarrow } \overline r_{\mathrm{GOE}} = 0.5307(1)$ & & &$\overline r \stackrel{L \gg1 }{\longrightarrow } \overline r_{\mathrm{PS}} = 0.386(1)$ & & & \ref{subsubsec:gapratio} \\
& & {\raisebox{1.7ex}[0pt]{ \small{ level statistics} }} & &  & & &  & & & \\
Spectral form factor $K(t)$ & & {\raisebox{1.3ex}[0pt]{ \small{ long-range} }} & &$K(t) = K_{\mathrm{GOE}}(t)$ for $t>t_{\mathrm{Th}}$ & & &$K(t) = K_{\mathrm{PS}}(t)=1$,  $t_{\mathrm{Th}} \geq t_{\mathrm{H} }$ & & & \ref{subsec:thouless} \\
& & {\raisebox{1.7ex}[0pt]{ \small{level statistics} }} & &  & & &  & & & \\
Participation entropy $S_q$ & & {\raisebox{1.3ex}[0pt]{ \small{wave function} }} & &$S_q = \log_2(\mathcal N) + c_q$, $c_q < 0$& & &$S_q = D_q \log_2(\mathcal N) + c_q$, $D_q < 1$ & & & \ref{subsec:wave-fun} \\
& & {\raisebox{1.7ex}[0pt]{ \small{delocalization} }} & &  & & &  & & & \\
Entanglement entropy $S_A$ & & {\raisebox{1.3ex}[0pt]{ \small{eigenstate} }} & & \small{ $S_A = S_{RMT} \propto L_A$} \small{(volume-law)}& & & $S_A = \mathrm{const} $  (area-law)& & & \ref{subsec:quant_inf} \\
& & {\raisebox{1.7ex}[0pt]{ \small{entanglement} }} & &  & & &  & & & \\
Spectral function $f^2(\omega)$ & & {\raisebox{1.3ex}[0pt]{ \small{off-diagonal} }} & & $f^2(\omega) = \mathrm{const}$ for $\omega < \omega_{\mathrm{Th}}$& & & $f^2(\omega) = C_0 \delta(\omega)$$+$$f^2_{\mathrm{reg}}(\omega) $, $C_0$$>$$0$& & & \ref{subsec:spectralFUN} \\
& & {\raisebox{1.7ex}[0pt]{ \small{matrix elements} }} & &  & & &  & & & \\
Fidelity susceptibility $\chi_{\mathrm{typ}}$ & & {\raisebox{1.3ex}[0pt]{ \small{eigenstate} }} & & $\chi_{\mathrm{typ}} \sim \mathcal N$ & & & $\chi_{\mathrm{typ}}/\mathcal{N} \stackrel{ L \gg 1}{\longrightarrow } 0$ & & & \ref{subsec:sensitivity} \\
& & {\raisebox{1.7ex}[0pt]{ \small{sensitivity} }} & &  & & &  & & & \\ \hline
& & & &  & & &  & & & \\
Density correlation $C(t)$ & & {\raisebox{1.3ex}[0pt]{ \small{relaxation} }} & & $C(t)\stackrel{t \to \infty}{\longrightarrow } 0$ for $L\gg 1$ & & & $C(t)\stackrel{t \to \infty}{\longrightarrow } C_0 >0$ for $L\gg 1$ & & & \ref{sec:density} \\
& & {\raisebox{1.7ex}[0pt]{ \small{dynamics} }} & &  & & &  & & &\\
 {\raisebox{1.3ex}[0pt]{ \small{Configurational entangle-} }} & & {\raisebox{1.3ex}[0pt]{ \small{entanglement} }} & & $S_c(t) \propto t$& & & $S_c(t) \propto \log_2(t)$ & & & \ref{sec:entanglementSC} \\  {\raisebox{1.7ex}[0pt]{ \small{ment entropy $S_c(t)$} }} 
& & {\raisebox{1.7ex}[0pt]{ \small{generation} }} & &  & & &  & & &\\
 Number entropy $S_n(t)$ & & {\raisebox{1.3ex}[0pt]{ \small{particle number} }} & & $S_n(t) \propto \log_2(t)$& & & $S_n(t) \propto \mathrm{const}$& & & \ref{sec:entanglementSN} \\  
& & {\raisebox{1.7ex}[0pt]{ \small{fluctuations} }} & &  & & &  & & &\\
 {\raisebox{1.3ex}[0pt]{ \small{Nested commutator} }} & & {\raisebox{1.3ex}[0pt]{ \small{operator} }} & & $||\mathcal{L}^n \hat{A}||\propto (n/\ln(n))^n$& & & $||\mathcal{L}^n \hat{A}|| \leq (\mathrm{const})^ n$ & & & \ref{subsec:operatorGR} \\  {\raisebox{1.7ex}[0pt]{ \small{norm $||\mathcal{L}^n \hat{A}||$} }} 
& & {\raisebox{1.7ex}[0pt]{ \small{spreading} }} & &  & & &  & & & \\ \hline
& & & &  & & &  & & & \\
Spin conductivity $\sigma(\omega)$ & & {\raisebox{1.3ex}[0pt]{ \small{spin} }} & & $\sigma(\omega) \stackrel{\omega \ll 1}{=} \sigma_0 + \zeta \omega^{1-2/z}$, $z\geq 2$ & & & $\sigma(\omega) \stackrel{\omega \ll 1}{=} 0$ & & & \ref{subsec:spinCON} \\
& & {\raisebox{1.7ex}[0pt]{ \small{transport} }} & &  & & &  & & &\\
 {\raisebox{1.3ex}[0pt]{ \small{Two-point density} }} & & {\raisebox{1.3ex}[0pt]{ \small{density } }} & & $\Delta x(t) \propto t^{1/z}$ with $z \geq2 $& & & $\Delta x(t) =\mathrm{const} $ & & & \ref{subsec:unitTEV} \\  {\raisebox{1.7ex}[0pt]{ \small{correlator $\Phi(x,t)$} }} 
& & {\raisebox{1.7ex}[0pt]{ \small{spreading} }} & &  & & &  & & &\\
Spin current $j$ & & {\raisebox{1.3ex}[0pt]{ \small{spin} }} & & $ j\propto L^{1-z}$ with $z\geq 2$ & & & $ j\propto \exp( -\gamma L)$ with $\gamma >0$  & & & \ref{subsec:NESS} \\
& & {\raisebox{1.7ex}[0pt]{ \small{transport} }} & &  & & &  & & &\\ \hline \hline
    \end{tabular}
    \caption{ Summary of the quantities employed as ergodicity breaking indicators and their behavior in ergodic and MBL regimes, together with the number of the Section in which they are defined and considered. Here and throughout, we consider systems defined on one-dimensional lattices consisting of $L$ sites, $L_A$ is size of subsystem $A$ which defines a bipartition of the system, $t$ is time (measured in tunneling times of the system), $\mathcal N$ is the dimension of the relevant sector of the many-body Hilbert space, $t_{\mathrm{Th}} \,(\omega_{\mathrm{Th}})$ is the Thouless time (energy), $t_{\mathrm{H}}$ ($\omega_{\mathrm{H}}$) is the Heisenberg time (energy), $z$ is the dynamical critical exponent. The persistence of the behavior associated with the ergodic (MBL) regime in the asymptotic limits $t \to \infty$ and $L \to \infty$ over an extended range of disorder strengths $W$ indicates the existence of an ergodic (MBL) phase. 
    \label{table1}
    } 
\end{table*}

Disordered spin-1/2 chains with power-law interactions provide a further example in which an unconventional scaling limit allows for interesting ergodicity breaking phenomena~\cite{Tikhonov18, Gopalakrishnan19}. The presence of power-law interactions characterized by an exponent $a$ leads to an absence of MBL phase in that model, $\widetilde{W}_{C}(L) \stackrel{L\to\infty}{\longrightarrow} \infty$, when the thermodynamic limit is taken in the usual manner, i.e., by keeping the tunneling matrix elements, disorder and interaction strengths fixed. 
The thermodynamic limit can be taken in unconventional way. Ref.~\cite{Tikhonov18} observes that the variable $w=(W-\widetilde{W}_{C}(L))/\widetilde{W}_{C}(L)$ allows for tuning the disordered spin-1/2 chains with power-law interactions across 
a transition in the $L\to\infty$ limit, see also Sec.~\ref{subsec:longrange}. Similarly, changing $w$ leads to sharp transition between ergodic and MBL phases in the thermodynamic limit in the avalanche model considered by~\cite{Gopalakrishnan19}.

\subsection{Conclusion}

Numerical investigations of non-equilibrium phenomena in many-body systems over the last two decades allowed us to understand and characterize the MBL regime arising in systems of finite size $L$ at sufficiently large disorder strength $W$. However, in spite of the abundant efforts of the community, the fate of the system in the asymptotic limits $t \to \infty$ and $L \to \infty$ remains unclear. The persistent drifts of the ergodicity breaking indicator towards the ETH behavior with increasing time and length scales, pose difficulties in discerning the slow dynamics from a complete arrest of the dynamics in the limit $L,t \to \infty$ that defines the MBL phase. Consequently, the fundamental questions about MBL presented in Sec.~\ref{sec:open} remain open. 

We have outlined, without resorting to concrete quantities or models, the features of numerical results for disordered many-body systems, highlighting the finite size drifts towards the ETH regime that prevent the unambiguous interpretation of the results. We have argued that single parameter scaling is insufficient to capture the system size dependence of the results at the ETH-MBL crossover, and that sub-leading corrections to the scaling have to be incorporated, either via irrelevant scaling variables or by the use of unconventional scaling functions. 

The Anderson models of various dimensionality provide examples of various types of critical scalings in disordered (and non-interacting) systems: i) single parameter scaling describes accurately the Anderson localization transition in 3D, ii) the Anderson model in 2D hosts a broad delocalized regime at finite $L$, which shrinks to a point $W=0$ in the thermodynamic limit, iii) the Anderson model on random regular graphs, which constitutes a particular realization of the $D \to \infty$ limit, features strong finite size drifts at the delocalized-localized crossover, which necessitate introduction of sub-leading corrections to the scaling. The finite size drifts for the latter model resemble, to a certain degree, the behavior at the ETH-MBL crossover in interacting many-body systems. Even though the known critical disorder strength $W_c$ for Anderson localization transition on random regular graphs simplifies the finite size analysis, the present-day numerical investigations of the Anderson model on random regular graphs arrive at inconsistent results, differing, for instance, by the value of the critical exponent $\nu$~\cite{Garcia-Mata22, Sierant23RRG}. This illustrates the formidable challenge of comprehending the critical features that may emerge at the ETH-MBL crossover in the interacting disordered many-body systems.

In the following Section \ref{sec:num}, we briefly outline numerical approaches typically employed in studies of ergodicity and MBL. A reader interested in physics of disordered quantum many-body may skip Sec.~\ref{sec:num}, and proceed directly to Sec.~\ref{sec:numerical}, in which we review numerical investigations of the ETH-MBL crossover in disordered many-body systems, providing concrete examples of the quantities that follow the general patterns described in the present Section. 
Subsequently, we focus on unitary dynamics of disordered many-body systems in Sec.~\ref{sec:dynamics} and its relation to transport in Sec.~\ref{sec:trans}.   
In Table~\ref{table1}, we summarize the most important quantities employed in these considerations.

\section{Numerical methods for MBL}
\label{sec:num}
This Section aims to equip the reader with intuitions about system sizes and time scales that can be probed with the discussed numerical methods rather than to provide an in-depth description of various numerical algorithms. The phenomena of quantum thermalization and ergodicity breaking rely on properties of highly excited eigenstates of investigated many-body systems, as visible already from the ETH ansatz~\eqref{eq:ETH1}. Consequently, exact diagonalization methods that enable the computation of eigenstates at arbitrarily chosen energies constitute one of the essential tools in studies of MBL. Alternatively,  questions of thermalization may be answered by investigations of the time evolution of many-body systems. Therefore, we will discuss numerical tools that rely on the Hamiltonian matrix's sparsity to numerically calculate the system's exact time evolution. Finally, we will mention tensor network approaches that approximate eigenstates and compute the time evolution of many-body systems, provided that the quantum states in question are not too strongly entangled.

\subsection{Exact diagonalization}
\label{subsec:ED}

Exact diagonalization (\textbf{ED}) is a numerical method in which a certain number of eigenvalues $E_m$ and eigenvectors $\ket{m}$ of Hamiltonian $\hat{H}$ of a many-body system is calculated. To that end, one needs to specify the matrix $\hat{H}$ corresponding to the Hamiltonian $\hat{H}$ in a selected many-body basis, as described, e.g.,  in an introductory fashion in \cite{Zhang10, Jung20}. 
Once the Hamiltonian matrix $\hat{H}$ is calculated, one may employ \textit{a full exact diagonalization} routine, using one of the available linear algebra packages, for instance, the \path{LAPACK} package~\cite{Anderson99}, to compute \textit{all} eigenvalues $E_m$ and eigenvectors $\ket{m}$ of $\hat{H}$. Subsequently, the questions about thermalization \eqref{eq:A1} in the many-body system described by the Hamiltonian $\hat H$ can be answered. However, understanding whether the investigated system is in the MBL phase \eqref{eq:MBL1}, or whether it follows the ETH, requires taking the large system size limit $L$. The computational resources required by the full ED quickly become prohibitive with increasing $L$ due to the exponential growth of the Hilbert space dimension $\mathcal N$ with the system size $L$, impeding the extrapolation of the results to $L \to \infty$. The typical matrix size accessible for full ED calculations in present-day computational studies is of the order of $\mathcal N \sim 10^5$. Hence, full ED calculations for the disordered XXZ spin chain \eqref{Hxxz} in the $\sum_i \hat{S}^z_i =0 $ sector are restricted to $L \leq 18$. Note that investigations of MBL usually involve disorder average, which makes studies of larger system sizes infeasible.

An observation that enables the exploration of system sizes beyond the reach of full ED is that the Hamiltonians of quantum many-body systems can often be written as sparse matrices for an appropriately chosen basis. The number of non-zero off-diagonal entries per row in such sparse matrices is much smaller than the matrix size $\mathcal N$. For example, the Hamiltonian of the disordered XXZ spin chain \eqref{Hxxz}, written in the eigenbasis of $\hat{S}^z_i$ operators, contains no more than $L$ non-zero off-diagonal elements per row. By retaining the non-vanishing elements of the sparse Hamiltonian matrix $\hat{H}$ solely, it becomes possible to significantly reduce the computation time associated with matrix-vector multiplication in comparison to a dense matrix (which, by definition,  has $O({\mathcal N})$ non-zero entries per row). This opens up the possibility of using the Lanczos algorithm~\cite{Lanczos50}, which employs repeated matrix-vector multiplications to compute eigenvectors corresponding to eigenvalues at the edges of the spectrum of $\hat{H}$. The Lanczos algorithm's convergence slows considerably as the number of requested eigenvectors at the spectrum edges increases. The Lanczos algorithm needs to be paired with a so-called \textit{spectral transformation} to make it practical to calculate the eigenvectors in the middle of the spectrum of $\hat{H}$ that are of interest in studies of MBL. 

Shift-and-invert method relies on a transformation of the Hamiltonian matrix $\hat{H}$ via 
\begin{equation}
    \hat{H} \rightarrow \hat{R}_{\sigma}(\hat{H}) = (\sigma-\hat{H})^{-1},
    \label{eq:shift}
\end{equation}
where $\sigma$ is a chosen energy target. The spectral transformation \eqref{eq:shift} has two important properties: i) the eigenvectors $\ket{m}$ of $\hat{H}$ are still eigenvectors of the transformed matrix $R_{\sigma}(\hat{H})$, ii) the eigenvalues are mapped according to $E_m \rightarrow  R_{\sigma}(E_m)$, which, by the construction of the function, ensures that the eigenvalues in the vicinity of the energy target $\sigma$ become the eigenvalues at the spectrum edges of the transformed matrix $ R_{\sigma}(\hat{H})$. The two properties imply that the Lanczos algorithm for the transformed matrix $R_{\sigma}(\hat{H})$ converges to eigenvalues of $\hat{H}$ close to $\sigma$. Shift-and-invert method proved to be fruitful in studies of MBL~\cite{Luitz15}, extending to $L\leq 26$ the range of system sizes for which calculation of eigenvectors in the middle of the spectrum of \eqref{Hxxz} is possible. The main difficulty of the shift-and-invert method is the presence of the inverse operator in \eqref{eq:shift}. For interacting many-body systems, the Lanczos iteration with $R_{\sigma}(\hat{H})$ can be performed efficiently by a massively parallel $LU$ decomposition~\cite{Amestoy01,Hernandez05,petsc}. However, the sparsity pattern of the Hamiltonian matrix $\hat{H}$ is not preserved during the $LU$ decomposition, which leads to a huge memory overhead, and was identified as the main bottle-neck of the shift-and-invert method~\cite{Pietracaprina18}.

Polynomial spectral transformations~\cite{Saad11}, which replace the complicated function \eqref{eq:shift} by a polynomial of $\hat{H}$, constitute  an alternative to the shift-and-invert method. Calculation of eigenvectors in the middle of the spectrum of a large sparse matrix such as \eqref{Hxxz} can be efficiently carried out with an algorithm POLFED~\cite{Sierant20polfed} that relies on the following polynomial spectral transformation,  
\begin{eqnarray}
\hat{H} \rightarrow P^K_{\sigma}(\hat{H}) = \frac{1}{D}\sum_{k=0}^K c^{\sigma}_k T_k(\hat{H}),
\label{polyTransf}
\end{eqnarray}
where $T_k(x)$ denotes $k$-th Chebyshev polynomial, 
the coefficients read $c^{\sigma}_k = \sqrt{4-3 \delta_{0,k}} \cos(n \arccos{\sigma})$, and $D$ is a normalization factor. For sufficiently high and appropriately chosen order $K$, the polynomial $P^K_{\sigma}(E)$ has a sharp peak around the energy target $\sigma$ since the coefficients $c^{\sigma}_n$ are obtained from expanding a Dirac delta function centered at this $\sigma$ value. Hence, the Lanczos algorithm for  $P^K_{\sigma}(\hat{H})$ converges to eigenvectors of $\hat{H}$ with energies around the chosen energy target. Notably, besides basic linear algebra operations, the Lanczos algorithm employs only the multiplication of vectors by $P^K_{\sigma}(\hat{H})$. The multiplication of vectors by $P^K_{\sigma}(\hat{H})$ reduces to repeated matrix-vector multiplications by the matrix $\hat{H}$ and linear algebra operations. Hence, the sparse structure of $\hat{H}$ is preserved by the POLFED algorithm, which significantly reduces the procedure's memory cost. For instance, POLFED requires $\approx 100$ GB of RAM to calculate $2000$ eigenvectors in the middle of the spectrum of~\eqref{Hxxz} for $L=24$, easily fitting into a single computing node of a present-day supercomputer. The shift-and-invert method requires $\approx 10^4$ GB of RAM for this task. At the same time, the total CPU time used by the two methods is of similar order; see \cite{Sierant20polfed} and \cite{Pietracaprina18} for further benchmarks. In passing, we note that combining a spectral transformation based on a second-order polynomial of $\hat{H}$ with appropriate preconditioning methods provides another exact diagonalization approach to MBL \cite{Beeumen22}.

Finally, the polynomial spectral transformations can also be used for efficient calculations of eigenvectors for large Floquet operator $\hat{U}_F$~\cite{Luitz21}. 
In that case, analogously to the POLFED algorithm, one considers a high order polynomial of the Floquet operator,
 \begin{eqnarray}
  g_K(\hat{U}_F) = \sum_{m=0}^K e^{-i m \phi_{\mathrm {tg} }} \hat{U}_F^m,
  \label{eqsup1}
 \end{eqnarray}
 where $\phi_\mathrm{th}$ is the target eigenphase and the order of polynomial $K$ is fixed as $K=f \frac{\mathcal  N}{N_{\mathrm{ev}}}$, where $N_{\mathrm{ev}}$  is the number of requested eigenvectors, the Hilbert space dimension is  $\mathcal N=2^L$, and the factor $f=1.46$ was obtained from an optimization of the performance of the algorithm. 
The Arnoldi iteration~\cite{Arnoldi51} for the operator $g_K(\hat{U}_F)$ yields $ N_{\mathrm{ev}}$ eigenvectors $\ket{\psi_n}$ corresponding to the dominant eigenvalues of $g_K(\hat{U}_F)$ (which is no longer a unitary operator). The latter, by construction, correspond to eigenvectors of $\hat{U}_F$ with eigenphases closest to $\phi_\mathrm{tg}$. The convergence is reached after approximately $ \alpha N_{\mathrm{ev}}$ steps of the iteration, where $\alpha \approx 2.1$. 
 This algorithm is especially effective when the matrix-vector multiplication by the operator $\hat{U}_F$ can be efficiently implemented. This is the case for the Kicked Ising model, as described in Sec.~\ref{sec:floq}, or for various types of quantum circuits, one example was considered by~\cite{Morningstar22}. For these cases, on present day computers, the algorithm enables study of eigenphases and eigenvectors of Floquet operators for $L=20$ or more (corresponding to matrix size $\mathcal N > 10^6$), while the full exact diagonalization of Floquet operators is feasible up to $L\approx 15$ and  $\mathcal N \approx 3 \cdot 10^4$.
 In general, efficient matrix vector products can be formulated with a matrix product operator formulation of $\hat{U}_F$~\cite{Zhang17floq}. 

\subsection{Time evolution methods}
\label{subsec:numtevol}

Exact diagonalization approaches allow us to study eigenvectors of $\hat H$, which encode, via \eqref{eq:A2}, the limit of infinite times, $t \to \infty$. Moreover, the complete information about the eigenvalues and eigenvectors enables calculation of the time evolved state $\ket{ \psi(t)}$, and consequently, the time evolution of any observable $\hat{A}$ with \eqref{eq:A1}. The memory complexity of the full ED is $O(\mathcal N^2)$, while the space needed to store the time evolved state $\ket{ \psi(t)}$  scales linearly with $\mathcal N$. This observation, together with the sparsity of the Hamiltonian matrix $\hat{H}$, motivates the search for more effective time evolution algorithms.

The efficient algorithm for time propagation of the Schr\"{o}dinger equation was proposed in \cite{Tal‐Ezer84}. The algorithm is based on an expansion of the time evolution operator $\hat{U}(\Delta t) = e^{-i H \Delta t}$ over a time step $\Delta t$ into Chebyshev polynomials $T_k(x)$ of the Hamiltonian matrix
 \begin{equation}
     \hat{U}( \Delta t) \approx \mathrm{e}^{-\mathrm{i}b \Delta t} \left( J_0(a \Delta t) + 2\sum_{k=1}^N (-i)^k J_k(a \Delta t) T_k \left( \mathcal{H} \right) \right),
 \label{eqcheby}
\end{equation}
where $a=(E_{\rm max} - E_{\rm min})/2$, $b=(E_{\rm max} + E_{\rm min})/2$ and $E_{\rm min}$ ($E_{\rm max}$) is the energy of the ground state (the highest excited eigenstate) of $\hat{H}$. The Hamiltonian is rescaled as $\mathcal{H} = \frac{1}{a}(H-b)$, and $J_k(t)$ is the Bessel function of the order $k$. The order of expansion, $N \ll \mathcal N$, needed to assure convergence of \eqref{eqcheby} up to any desired precision, can be determined numerically for a given time step $\Delta t$, see, for instance, \cite{Sierant22challenges}. An application of $\hat{U}(\Delta t)$ allows to update the state of the system $\ket{\psi{(t_0+\Delta t)}} =\hat{U}(\Delta t) \ket{\psi{(t_0)}}$ and, upon repetition, results in the time-evolved state $\ket{\psi(t)}$ with $t>t_0$. Since, for a fixed order $N$, $\hat{U}( \Delta t)$ is simply a polynomial of the matrix $\hat{H}$, the calculation of $\hat{U}(\Delta t) \ket{\psi{(t_0)}}$ reduces to multiple multiplications of vectors by matrix $\hat{H}$ and linear algebra operators. Hence, the efficiency of the method relies crucially on the sparsity of the matrix $\hat{H}$. This Chebyshev time evolution algorithm allows to calculate the time-evolved state $\ket{\psi(t)}$ couple of times faster than Suzuki-Trotter-based schemes~\cite{Dobrovitski03}, see also \cite{Fehske08}.

A closely related method, called Krylov time evolution, was proposed in \cite{Nauts83}. A direct truncation of the Taylor expansion of the time evolution operator $\hat{U}( \Delta t) \ket{\psi{(t_0)}} = \sum_k \frac{(-i \Delta t)^k }{k!} H^k \ket{\psi{(t_0)}}$ to order $m$ is numerically unstable \cite{Moler03} and allows to calculate the updated state $\ket{\psi{(t_0+\Delta t)}}$ only for very small time steps $\Delta t$. However, this Taylor expansion indicates that the updated state $\ket{\psi{(t_0+\Delta t)}}$ is a linear combination of vectors from the Krylov space $\mathcal{K}_{m}$ spanned by  $\{ \ket{\psi\left(t_0\right)},\hat{H}\ket{\psi\left(t_0\right)},\hat{H}^{2}\ket{\psi\left(t_0\right)},\dots,\hat{H}^{m-1}\ket{\psi\left(t_0\right)} \}$, which can be generated, to a given order $m$, by the Arnoldi algorithm~\cite{Arnoldi51}. Denoting by ${V_{m}}\in\mathbb{C}^{\mathcal{N}\times m}$ the matrix that maps the vectors of the canonical basis $\{\vec{e}_i\}$ in $\mathbb{C}^m$ to the naturally ordered vectors of the Krylov space, defines the step of the Krylov time evolution
\begin{equation}
\ket{\psi{(t_0+\Delta t)}} = \mathrm{e}^{-\mathrm{i}\hat{H}\Delta t}\ket{\psi\left(t_0\right)}\approx{V_{m}}\mathrm{e}^{-\mathrm{i}{V_{m}^{\dagger}} H{V_{m}}\Delta t}\vec{e_{1}}.\label{eq:Krylov-timeevolution}
\end{equation}
The dimension of the matrix ${V_{m}^{\dagger}} HV_{m}\in\mathbb{C}^{m\times m}$ is \emph{small}, $m\ll\mathcal{N}$, and its exponential can be straightforwardly computed with the full ED algorithm, while the dimension of the Krylov space, $m$, is continuously increased until the updated state $\ket{\psi{(t_0+\Delta t)}}$ converges to the desired precision~\cite{Luitz17b}.

Both Chebyshev and Krylov time evolution algorithms utilize the sparse structure of the Hamiltonian matrix and allow for an efficient, numerically exact, calculation of the time-evolved state $\ket{ \psi(t)}$ under a many-body Hamiltonian $\hat H$. The range of system sizes available to those methods extends significantly beyond the capabilities of full ED. Computations performed on single nodes of present-day supercomputers allow to reach Hilbert space dimension $\approx 10^9$ \cite{Luitz17, Varma17}, corresponding to $L=32$ for the XXZ spin chain \eqref{Hxxz}, and time scales of the order of $10^3$ tunneling times~\cite{Sierant23slow}. Implementation of the methods that employs internode communication allows to tackle time evolution for  $\mathcal N \approx 10^{10}$, which corresponds to $L=36$ for the XXZ spin chain \cite{Brenes19}. For specifically arranged many-body dynamics in which a finite number of subsystems interact only sporadically in time, but otherwise evolve independently, hybrid Schrödinger-Feynman techniques may offer an advantage~\cite{Richter24simulating} over the Chebyshev and Krylov time evolution algorithms.

\subsection{Tensor network and other approaches}
\label{subsec:tenNET}

The numerical methods discussed so far allow to compute eigenstates or time evolution of many-body systems in a numerically exact manner and rely on storing the states of interest $\ket{\psi}$ as a vector in $\mathcal N$ dimensional Hilbert space. The exponential growth of $\mathcal N$ with system size $L$ restricts the scope of those methods, entirely precluding their usefulness for size of the order of $100$ sites, that are relevant to present-day experiments with synthetic quantum matter.

Tensor networks \cite{Orus19} represent the state $\ket{\psi}$ of a many-body system as a contraction of a network of tensors~\cite{Ran20}. The core idea of the tensor networks is that in certain circumstances the dimension of the tensors can be bounded in such a way that the number of parameters necessary to encode the state grows only polynomially with $L$, much slower than the dimension $\mathcal N$ of the Hilbert space. For instance, states of one-dimensional many-body systems can be written as matrix product states (MPS)~\cite{Schollwoeck11}. The required dimension $\chi$ of the involved tensors grows polynomially with $L$ for ground states of non-critical many-body systems that follow the area-law of entanglement~\cite{Eisert10}. 

The tensor network representations of many-body states enable efficient time evolution algorithms~\cite{Xie19} such as Time Evolving Block Decimation (TEBD) \cite{Vidal03,Vidal04}, which is essentially equivalent to time-dependent density matrix renormalization group algorithm~\cite{White04,Daley04}, or its consecutive improvement, the Time-Dependent Variational Principle (TDVP) \cite{Haegeman11, Koffel12, Haegeman16, Paeckel19, Goto19}. Starting from a given state represented in a matrix product form, those algorithms produce an approximation of the time evolved state $\ket{\psi_{\chi}(t)}$ for a given bond dimension $\chi$. Careful monitoring of the changes of the state $\ket{\psi_{\chi}(t)}$ with the bond dimension $\chi$ allows to verify whether the results of such simulations are converged with $\chi$. The convergence occurs when $\chi$ is sufficiently large and a further increase of $\chi$ does not lead to appreciable changes of the quantities of interest. The tensor network algorithms allow to compute time evolution of many-body systems comprising of hundreds of lattice sites in one dimension \cite{Sierant18, Doggen18, Kloss18, Zakrzewski18,Chanda20m,Chanda20t} and even to tackle time evolution of quasi-1D or small two-dimensional systems~\cite{Doggen20}.

The growth of entanglement under coherent evolution of interacting many-body systems is the main limiting factor of those methods. The bigger the entanglement entropy, the larger are the values of $\chi$ needed to faithfully represent the time evolved state. For instance, if the entanglement entropy follows a volume-law and scales proportionally to the subsystem size of a one-dimensional system, the required $\chi$ scales exponentially with $L$, implying that there is essentially no gain over the exact representation of the state as a vector in the $\mathcal N$ dimensional Hilbert space. This, in turn, means that for a given bond dimension $\chi$ there exists a time $t_\chi$ such that for $t<t_\chi$, the state $\ket{\psi_{\chi}(t)}$ is, for a given observable and with a desired accuracy, a sufficiently good approximation of $\ket{\psi(t)}$.
The entanglement spreads ballistically~\cite{Kim13} in typical chaotic many-body systems, which implies that the time $t_\chi$, for bond dimensions realistic for present-day calculations, is of the order of $10$ tunneling times. However, an increase of disorder strength implies emergence of MBL phenomenology and a significant slow-down of the dynamics, which enables to calculate time evolution of systems of few hundreds of lattice sites up to time scales of the order of few thousand tunneling times~\cite{Sierant22challenges}. This nicely parallels the claim of the slow, logarithmic in time entanglement entropy growth for the MBL-type of dynamics \cite{Znidaric08,Bardarson12}.

Light cone renormalization group \cite{Enss12LighConeRG} is a notable approach that allows to study 1D many-body systems directly in the thermodynamic limit $L\to\infty$. This method relies on the fact that the unitary time evolution of an operator can be restricted to an effective light cone due to the Lieb-Robinson bounds~\cite{Lieb1972}. The required translational invariance can be assured by enlarging the on-site Hilbert space and considering discrete disorder models, see also Sec.~\ref{sec:otherDisorder}, enabling studies of MBL directly in the thermodynamic limit~\cite{Andraschko14Purification, Enss17infinite}. Discrete disorder models allow for using of numerical linked cluster expansion~\cite{Rigol14NLCE} for studies of quench experiments in ergodic and MBL regimes in the $L \to \infty$ limit~\cite{Tang15}.

The slow-down of dynamics and the associated area-laws conjectured for eigenstates of systems in the MBL phase motivate another avenue of investigation of strongly disordered systems. The approaches of 
\cite{Yu17, Khemani16, Lim16, Villalonga18} merge the ideas of spectral transformation with the machinery of density matrix renormalization group ~\cite{Schollwoeck05}. The former allow to target the putative area-law eigenstates in the middle of the spectra of strongly disordered many-body systems, while the latter construct tensor network representation of the targeted eigenstates. Those approaches open a promising avenue towards investigation of the MBL phase. However, they are biased towards less entangled states and may miss certain subtle effects that herald the onset of thermalization in considered systems. 

Another type of tensor-network based approaches to MBL was opened by \cite{Pekker17encoding, Pollmann16}, who proposed that the full Hamiltonian of an MBL system can be represented as a low-depth quantum circuit. This approach was demonstrated to be useful for investigation of two-dimensional systems at experimentally relevant time scales \cite{Wahl17, Wahl19, Venn22, Li24} and even enables to make predictions about three-dimensional systems \cite{Chertkov21}. However, the range of time scales and system sizes at which those methods are reliable is yet to be understood.

\subsection{Outlook}

The full ED algorithm provides access to all eigenvalues and eigenvectors of the Hamiltonian matrix and constitutes a basis of our understanding of ETH and MBL physics in isolated many-body systems. However, the exponential increase of Hilbert space dimension $\mathcal N$ with system size $L$ limits the scope of the full ED. Systematic investigations of thermalization of quantum many-body systems require access to eigenstates in the middle of the many-body spectra for as large $L$ as possible. Methods that employ spectral transformations such as the shift-and-invert approach or the POLFED algorithm extend the range of $L$ for which eigenstates in the middle of the spectrum can be calculated. 

A typical quench scenario in which a many-body system is initialized in a certain out-of-equilibrium state and evolves under Hamiltonian $\hat{H}$, can be investigated with full ED. However, the sparsity of many-body Hamiltonians makes Chebyshev and Krylov time propagation schemes much more efficient tools for computation of time evolution of many-body systems. 

The ED methods and Chebyshev/Krylov time propagation schemes provide numerically exact results, which makes them especially useful in studies of the crossover between ETH and MBL regimes. Hence, in the following sections, we will focus on the results obtained with those methods. Nevertheless, the tensor network approaches enable access to much larger system sizes and, provided that their convergence is carefully monitored, yield very useful insights into thermalization of many-body systems.

Finally, we would like to briefly touch upon other numerical methods relevant for studies of ETH and MBL. Algorithms aimed at a direct calculation of LIOMs in MBL systems include approaches based on time evolution of local operators \cite{Chandran15, Mierzejewski18}, displacement operators \cite{Rademaker16, Ortuno19}, flow equations \cite{Pekker17, Thomson18, Kelly20, Thomson21flow, Thomson23, Lu2024}, or direct numerical optimization \cite{Brien16, Peng19, Johns19, Adami22}. It was argued in~\cite{Thomson2023unraveling} that the flow equation techniques combined with a method of scrambling transforms yield a method for exploration of intermediate-scale time evolution in many-body systems.
Microcanonical Lanczos method (MCLM) introduced in~\cite{Jaklic94} combines Lanczos algorithm and random sampling to investigate finite-temperature static and dynamical quantities of many-body systems and can be used to study transport properties, see, e.g.,~\cite{Mierzejewski10, Barisic10}. The transport properties can also be studied in an open system approach, involving the Gorini-Kossakowski-Sudarshan-Lindblad (GKSL) equation~\cite{Gorini76, Lindblad76},
which can be simulated
with tensor network algorithm that calculates the non-equilibrium steady state of the open system~\cite{Prosen09}.

Recently, machine learning approaches have found new applications to quantum sciences~\cite{Dawid2022modern}, see also earlier works~\cite{Mehta19, Carrasquilla20}.
While algorithms for an automated distinction between the ETH and MBL behaviors were proposed~\cite{Huembeli19, Doggen18}, a thorough analysis~\cite{Theveniaut19} indicates a large sensitivity of the quantitative results for the ETH-MBL transition with respect to the network structure and training parameters, preventing one from reaching a definite conclusions about the transition. Nevertheless, machine learning approaches can be useful for tasks such as the analysis of details of time evolution of disordered systems~\cite{Szoldra21, Kotthoff21} or classification of families of non-ergodic eigenstates of many-body Hamiltonians~\cite{Szoldra22}. In addition, unsupervised machine learning approaches have been successful in identifying phases of quantum many-body systems~\cite{RodriguezNieva19, MendesSantos21, HsinYuan22} and have been recently employed in the context of Anderson transition~\cite{Vanoni24analysis}.

\section{Phenomenology of the ETH-MBL crossover}
\label{sec:numerical}

This Section focuses on the exact diagonalization results for the disordered XXZ model and its close cousins, such as the $J_1$-$J_2$ model that we define below. We emphasize the ambiguities in interpretations of numerical results and the difficulty in extrapolating results to the limit of infinite system size $L$. We use Eq.~\eqref{eq:MBL1} as the definition of MBL, and we assume the standard picture of MBL phenomenology, outlined in Sec.~\ref{subsec:modelingANDphenom}, as the guiding principle of our analysis. We provide examples of concrete ergodicity breaking indicators that follow the generic trends described in Sec.~\ref{subsec:difficulties}, illustrating the relevance of the open questions from Sec.~\ref{sec:open}.

\subsection{Spectral statistics}
\label{subsec:spectral}
As postulated by Wigner, Dyson, Porter and other pioneers of random matrix theory, the statistical properties of the energy levels of complex systems are well approximated by random matrices \cite{Mehtabook,Haakebook}. Quantum chaotic systems exhibit level repulsion \cite{Bohigas84}, with the probability distribution of
level spacing 
depending only on the global symmetries of the Hamiltonian. 
For example, the Gaussian Orthogonal Ensemble (\textbf{GOE}) applies to time-reversal invariant systems such as \eqref{Hxxz}, see \cite{Mehtabook, Haakebook}. In contrast, systems in the MBL regime, whose Hamiltonians in finite systems commute with the set of LIOMs\footnote{The LIOMs, i.e., the quasi-local operators which commute with the Hamiltonian, can be numerically constructed in the MBL regime at finite $L$, irrespective of their status in the $L\to\infty$ limit.} \eqref{eq:LIOM}, have uncorrelated energy levels with Poisson statistics. The ease of accessing the eigenvalues $E_n$ through exact diagonalization studies of $\hat{H}$, coupled with the clear distinction between ergodic and MBL systems, makes investigating spectral statistics a natural choice in numerical studies of MBL. 

\subsubsection{Gap ratio}
\label{subsubsec:gapratio}
While earlier studies considered level spacing statistics, they suffered from the procedure of level unfolding~\cite{Haakebook}. 
That procedure is avoided considering
a gap ratio, introduced in \cite{Oganesyan07}. The gap ratio is a dimensionless quantity defined as 
\begin{equation}
    r_n = \frac{ \min\{\delta_n, \delta_{n-1} \} }{ \max\{\delta_n, \delta_{n-1} \} },
    \label{eq:er}
\end{equation}
where $\delta_n = E_{n+1}-E_{n}$ is the gap between two adjacent eigenvalues in the spectrum of the system. Studies of MBL typically concern the middle of the spectrum, where the density of states is the largest. Averaging $r_n$ over eigenvalues in this energy range and over different disorder realizations yields an average gap ratio, denoted by $\overline r$. Comparing the average gap ratio, $\overline r$, to the values $\overline r_{GOE} \approx 0.5307$ and $\overline r_{PS}=2\ln2 - 1\approx 0.3863$, which are characteristic for the GOE \cite{Atas13} 
and Poisson level statistics, respectively, is for most physical systems a way of assessing whether the system is ergodic or not\footnote{This remark concerns a generic case. However, the lack of ergodicity associated with the existence of a particular class of eigenstates, for instance with quantum scar states~\cite{Turner18}, may not be detectable with the probes of the level statistics.}.

\begin{figure}
    \centering
    \includegraphics[width=1\columnwidth]{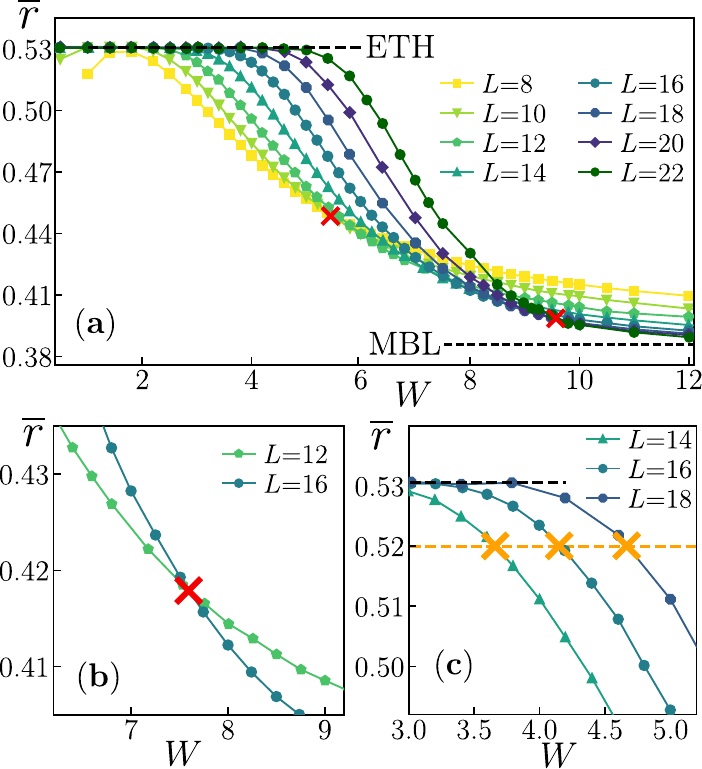}
    \caption{ETH-MBL crossover in the $J_1$-$J_2$ spin-1/2 chain. The average gap ratio $\overline r$ for system size $L$, shown in \textbf{(a)}, changes upon a variation of disorder strength $W$ between the GOE value $\overline r_{GOE}\approx 0.5307$ in the ETH regime and Poisson value $\overline r_{PS}\approx 0.3863$ in the MBL regime. The crossing point of $\overline r(W)$ curves, denoted in red, drifts with increasing disorder strength. Panel \textbf{(b)} illustrates the extraction of the $L$ dependent crossing point $W^*(L)$ for $L=14$ and $\Delta L=2$. Panel \textbf{(c)} shows how to extract the disorder strength $W^T(L)$ at which system creases to follow the ETH prediction.
\label{fig:ER}
}
\end{figure}

The average gap ratio, $\overline r$, calculated for disordered many-body systems, is the prototypical quantity featuring the trends illustrated in Fig.~\ref{fig:flow}(b). Here, we present results for the $J_1$-$J_2$ model with the Hamiltonian 
\begin{equation}
\hat{H}_{J_1-J_2} = \hat{H}_{\textit{XXZ} } +\sum_{i=1}^{L} \left(
 \hat{S}^x_i\hat{S}^x_{i+2} + \hat{S}^y_i\hat{S}^y_{i+2}
+ \Delta \hat{S}^z_i \hat{S}^z_{i+2}\right), 
\label{eq:J1J2}
\end{equation}
where $ \hat{H}_{\textit{XXZ} } $ is the Hamiltonian of the disordered XXZ spin chain, see Eq.~\eqref{Hxxz}, $\Delta=0.55$, and periodic boundary conditions are assumed.
In contrast to the disordered XXZ spin chain, however, the limit $W\to 0$ is not Bethe ansatz integrable, which simplifies the analysis in the weak disorder limit.
The Hamiltonian $\hat{H}_{J_1-J_2}$, similarly to $\hat{H}_{\textit{XXZ} }$, conserves the $z$ component of total spin, $\hat{S}^z = \sum_{i=1}^{L} \hat{S}^z_i$. We follow the standard choice in the literature on MBL and consider the sector with $\hat{S}^z=0$, which is the largest symmetry sector of the model\footnote{ A meaningful analysis of level statistics is possible only when the global symmetries of the Hamiltonian are resolved. Before the resolution of the symmetries, the spectrum of an ergodic system is a superposition of independent GOE/GUE spectra with statistical properties approaching the Poisson case with the increase of the number of symmetry sectors, see~\cite{Giraud22Symmetries} for a recent discussion of this well known problem. }. Moreover, the results depend on the energy density at which the gap ratio is calculated~\cite{Luitz15}. In the following, to extract $\overline r$, we average \eqref{eq:er} over less than $5\%$ of energies in the middle of the spectrum of the system. 
The average gap ratio $\overline r$ for the $J_1$-$J_2$ model is shown in Fig.~\ref{fig:ER}(a). While we here present the concrete example of the  $J_1$-$J_2$ model, analogous trends are found in the results for XXZ spin chain and other disordered many-body systems, see Sec.~\ref{subsec:otherModels}.
Note, however, that the quantitative values of disorder strengths in the crossover regime are larger in the $J_1$-$J_2$ model compared to the XXZ model.

For a sufficiently weak disorder in the $J_1$-$J_2$ model, $W \lessapprox 6$, there is a clear trend towards $\overline r_{GOE}$ predicted for a system that follows ETH. In contrast, for the strong disorder, $W \gtrapprox 10$, we observe a monotonous with $L$ decrease of $\overline r$ towards the Poisson value $\overline r_{PS}$, consistent with an emergent localization of the system. While the slope of the curves $\overline r(W)$ in the intermediate region increases with system size, the results do not follow the simple-to-interpret trend illustrated in Fig.~\ref{fig:flow}(a). On the contrary, there is a significant drift of the crossing point of the $\overline r(W)$ curves. While this problem was noted in the early paper \cite{Oganesyan07}, the authors observed that the drift of the crossing point slows down with the increase of the system size, providing a hint of the presence of the MBL phase. The presence of the MBL phase, based on a more detailed study of the XXZ spin chain \cite{Pal10}, became for over a decade the prevailing interpretation of the results for strongly disordered many-body systems. One of the deficiencies of this interpretation was highlighted in \cite{Suntajs19}, where it was shown  that the results for different $L$ collapse upon a rescaling $ \overline r(W) \to  \overline r(W/L)$ close to the ETH regime, i.e., for $W$ at which $\overline r(W) >\overline r_{GOE} - \Delta \overline  r$ (where $\Delta \overline r \ll 1$ is a constant). The latter rescaling, taken literally, implies that the value of $\overline r$ remains fixed if the disorder strength increases proportionally to the system size $L$, suggesting that there is no MBL phase and $ \overline r(W) \to \overline r_{GOE}$ at any $W$ for $L\to \infty$.

The two interpretations are contradictory. To quantify the two types of trends observed at the ETH-MBL crossover, \cite{Sierant20polfed} introduced two system size-dependent disorder strengths, see Figs.~\ref{fig:ER}(b) and \ref{fig:ER}(c): 
\begin{itemize}
    \item the crossing point $W^*(L)$ of $\overline r(W)$ curves for system of size $L-\Delta L$ and $L+\Delta L$ (where $\Delta L \ll L$);
    \item the disorder strength $W^T(L)$ of deviation from the ETH behavior, at which  $\overline r( W^T(L) ) = \overline r_{GOE} - \Delta \overline  r$ for a given $L$ and where $\Delta \overline  r$ is a fixed small constant.
\end{itemize}
Notably, the disorder strengths $W^T(L)$ and $W^*(L)$ provide one realization of the critical "fan", proposed as the phase diagram in variables $L$ and $W$ for disordered many-body systems~\cite{Khemani17a}. For $W<W^T(L)$, the system is in the ergodic regime, and the average gap ratio $\overline r(W)$ is close to $\overline r_{GOE}$. In contrast, for $W>W^*(L)$, the value of $\overline r(W)$ decreases upon an increase of the system size $L$ signifying the MBL regime. Finally, there is a \textit{crossover regime} $W^T(L)<W<W^*(L)$ with an increase of $\overline r$ with system size $L$ at fixed $W$. Since the analysis is performed at a given finite system size $L$, we refer to each of the intervals of $W$ as different regimes.

The interplay between the ergodic, crossover, and MBL regimes, or equivalently, the system size dependence of $W^*(L)$ and $W^T(L)$, determines the system's fate in the thermodynamic limit. For instance, for the results in the ideal case discussed in Fig.~\ref{fig:flow}(a), one would expect $W^*(L) = W_C$, independently of $L$, and $W^T(L) \to W_C$ as $L \to \infty$, at a rate determined by the critical properties of the transition, c.f.~\eqref{eq:Wx01} and \eqref{eq:Wx02}.

Consistently with the earlier studies, a large-scale numerical study \cite{Sierant20polfed} found that the results for the disordered XXZ and $J_1$-$J_2$ models are not following the trends from Fig.~\ref{fig:flow}(a). Both  $W^T_{\overline r}(L)$ and $W^*_{\overline r}(L)$ increase monotonously with system size. The boundary of the ergodic regime drifts, with a good approximation, linearly with system size: 
\begin{equation}
    W^T(L)\sim L.
    \label{eq:WT}
\end{equation}
The $L$ dependence of the crossing point may be approximated by 
\begin{equation}
W^*(L)\sim W_\infty- \mathrm{const}/L,
\label{eq:W*}
\end{equation}
compatibly with the observation that its drift slows down with increasing $L$\footnote{There are no strong arguments in favor of the $1/L$ fit in Eq.~\eqref{eq:W*}, and $W^*(L)$ could be fitted with a different system size dependence within the available range of $L$. In particular, a logarithmic in $L$ fit would yield $W^*(L)\stackrel{L\to\infty}{\longrightarrow} \infty$. }. The trends visible in Fig.~\ref{fig:ER}(a) clearly imply that $W^T(L) < W^*(L)$, hence, the scalings \eqref{eq:WT}, \eqref{eq:W*} are incompatible with each other in the $L \to \infty$ limit. This indicates the following possible Scenarios for the disordered spin chains:
\begin{enumerate}
\item The scaling \eqref{eq:W*} prevails (possibly with certain sub-leading terms) in the large $L$ limit, and both $W^T(L)$ and $W^*(L)$ converge in the thermodynamic limit to a finite critical disorder strength $W_C$.
\item The scaling \eqref{eq:WT} is obeyed in the large $L$ limit, and the crossing points $W^*(L)$ adapt to the behavior of $W^T(L)$ so that both disorder strengths diverge, $W^*(L) \to \infty$, $W^T(L) \to \infty$, in the thermodynamic limit $L\to \infty$.
\end{enumerate}
Finally, it is possible that neither of the scalings \eqref{eq:WT}, \eqref{eq:W*}, even after modifications including the putative sub-leading terms, persists in the $L\to\infty$ limit. This can never be entirely excluded solely based on analysis of numerical data at finite $L$. Nevertheless, the numerical results presented in Fig.~\ref{fig:ER}(a) do not provide any substantial premises suggesting quantitative changes of the trends at larger system sizes that are out of the reach of present-day classical supercomputers. This leaves out scenarios 1. and 2., which cannot be unambiguously differentiated by the analysis of presently available numerical results. Scenario 1.~implies the existence of an MBL phase and a transition between ergodic and MBL phases at a finite disorder strength. 
(For the disordered XXZ spin-1/2 chain, an extrapolation of the scaling \eqref{eq:W*} to $L\to \infty$ limit yields $W_C=W_\infty \approx 5.4$, larger than $W_C \approx 3.7$ proposed in \cite{Luitz15}.)
In contrast, scenario 2.~predicts that there is no MBL phase and that the disordered spin chains may become ergodic at any $W$ if the system size is sufficiently large. 

While the above perspective on the numerical results for the disordered spin chains appears grim, it may be possible to distinguish Scenarios 1.~and 2.~in the not-so-distant future. Extrapolation of the scalings \eqref{eq:W*} and \eqref{eq:WT} shows that they become incompatible at system size $L_0 \approx 50$ for the disordered XXZ spin-1/2 chain. The exact calculation of eigenstates in the middle of the spectrum for $L=50$ is far beyond the capabilities of present-day supercomputers. However, the signatures of the breakdown of  \eqref{eq:W*} or of \eqref{eq:WT} may be observable at significantly smaller system sizes, only slightly beyond $L$ studied so far. 

In passing, let us note that a crossover between delocalized and localized phases of the Anderson model on random regular graphs shows phenomenology that shares certain similarities with the ETH-MBL crossover \cite{Tikhonov16, Tikhonov21}. Analysis of the crossover with disorder strengths $W^T(L)$ and $W^*(L)$ shows a sub-linear scaling of $W^T(L)$ with $L$ and indicates that extrapolation of the $W^*(L)$ allows to reproduce accurately the critical disorder strength for Anderson localization transition on random regular graphs \cite{Sierant23RRG}. While those results provide a non-trivial test of the method of analysis of the delocalized-localized crossover with disorder strengths $W^T(L)$ and $W^*(L)$, the Anderson model on random regular graphs, as a single-particle model, is crucially different from the problem of MBL in disordered many-body systems. Finally, we would like to note that the deviations from the linear shift of $W^T(L)$ with $L$, which constitute another premise of the existence of an MBL phase in the thermodynamic limit, were found in disordered Floquet models~\cite{Sierant22floquet}, see Sec.~\ref{sec:floq}.

\subsubsection{Thouless time}
\label{subsec:thouless}

\begin{figure}
    \centering
    \includegraphics[width=1\columnwidth]{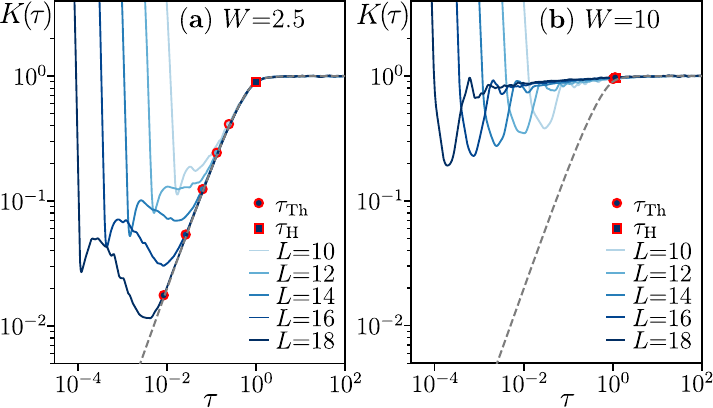}
    \caption{Spectral form factor $K(\tau)$ in the $J_1$-$J_2$ spin-1/2 chain. (a) Deep in the ergodic regime, for $W=2.5$, the SFF of the $J_1$-$J_2$ model has the ramp-plateau structure following the GOE prediction \eqref{eq:sffGOE} (denoted by gray dashed line) for $\tau > t_{Th}/t_H$, defining the Thouless time $t_{Th}$. The points $\tau_{Th} = t_{Th}/t_H$ are denoted by red circles, while the red square denotes $\tau_\mathrm{H}=1$. (b) Deep in the MBL regime of $J_1$-$J_2$ model, for $W=10$, the SFF is slowly tending, with an increase of $L$, to the result $K_{PS}(\tau)=1$ for Poisson level statistics. Convergence of SFF to $K_{PS}(\tau)$ for $L\to\infty$ at $W>W_c$ would imply the presence of the MBL phase in the system.
\label{fig:SFF}
}
\end{figure}

The gap ratio \eqref{eq:er} probes the statistical properties of energy levels $E_n$ at the scale of the average level spacing $\overline \delta$, i.e., the smallest energy scale in the many-body spectrum. Consequently, $\overline r$ reflects the properties of the system at the so called Heisenberg time, $t_H = 1/\overline \delta$, beyond which the dynamics start to be quasiperiodic. Therefore, the average gap ratio is accurately capturing whether the thermalization of the system occurs even at the largest physically relevant time scale $t_H$. The latter increases proportionally to the Hilbert space dimension, i.e., exponentially with system size $L$, e.g., as $t_H \propto 2^{L}$ for spin-1/2 chains.

The spectral form factor (\textbf{SFF}) is a Fourier transform of a two-point spectral correlation function. It quantifies long-range correlations between eigenvalues of many-body systems, and it provides insight into the dynamical properties at timescales below the Heisenberg time $t_H$. Investigations of the SFF played a significant role in our understanding of the ETH-MBL crossover.
If we denote the unfolded\footnote{Spectral unfolding is a procedure of setting the density of eigenvalues to unity \cite{Haakebook,Gomez02}, see also Sec.~\ref{sec:other_measures_stat}.}
eigenvalues of the considered quantum system by $\{\epsilon_1,\epsilon_2,\ldots,\epsilon_{\mathcal{N}} \}$, the averaged SFF is defined as 
     \begin{equation}
 K(\tau)  = \frac{1}{Z}\left \langle \left| \sum_{j=1}^{\mathcal N} g( \epsilon_j) \mathrm{e}^{-i \epsilon_j \tau} \right|^2 \right \rangle,
 \label{eq:Kt}
\end{equation}
where the average is taken over disorder realizations. The filter function $g(\epsilon)$ in Eq.~(\ref{eq:Kt}) is a Gaussian function with the mean in the middle of the unfolded spectrum and with the variance proportional to the variance of $\{\epsilon_1,\epsilon_2,\ldots,\epsilon_{\mathcal{N}} \}$, with the proportionality factor $\frac{3}{10}$, following the choice of \cite{Suntajs19}. The filter $g(\epsilon)$ reduces  
the influence of the spectrum's edges, 
while $Z$ is a normalization factor ensuring that $K(\tau) \stackrel{ \tau \to \infty }{\longrightarrow} 1$.
The SFF is known analytically for the Gaussian ensembles of random matrices. In particular, for the GOE, it reads~\cite{Mehtabook}:
\begin{equation}
   K_{GOE}(\tau) = \begin{cases}
 2 \tau - \tau \log(2\tau+1) &\text{for $ 0  < \tau \leqslant 1$} \\
2- \tau \log\left( \frac{2\tau+1}{2\tau-1}\right) &\text{for $\tau > 1$}.
\end{cases}
\label{eq:sffGOE}
\end{equation}
The above formula implies that, to a good approximation, the SFF for GOE matrices consists of a \textit{linear ramp} $K_{GOE}(\tau)\approx 2\tau$ for $ 0  < \tau \leqslant 1$ and a \textit{plateau} $K_{GOE}(\tau)\approx1$ at $\tau>1$. The linear ramp reflects the long-range correlations between \textit{all pairs} of eigenvalues of a GOE matrix. At the same time, the plateau at $\tau >1$ corresponds to the onset of quasiperiodic dynamics beyond the Heisenberg time. Note that the physical time $t$ and the parameter $\tau$ are linked via $t = \tau \,t_H $ due to the rescaling of spectral density associated with the unfolding procedure.

Remarkably, the SFF calculated for systems fulfilling the ETH also possesses the structure of a linear ramp and a plateau~\cite{Cotler17, Chen18spectral, Gharibyan18, Chan18, Suntajs20e, Berkovits21, Cipolloni2023,Buijsman2023longrange}, an example for the $J_1$-$J_2$ model is shown in Fig.~\ref{fig:SFF}(a). However, the SFF for quantum many-body systems follows the RMT predictions only for sufficiently large $\tau$. The onset of the adherence to the RMT predictions occurs at a time scale called a \textit{Thouless time} $t_{Th}$.
By definition, for time-reversal invariant systems, $K(\tau)=K_{GOE}(\tau)$ at $\tau > t_{Th}/t_H$, up to a fixed tolerance. In the following, we adopt the choice of \cite{Suntajs20e} and determine the Thouless time as $t_{Th}=\tau_{Th} t_H$, where $\tau_{Th}$ is the smallest number such that $\ln\left( K(\tau) / K_{GOE}(\tau)\right) < \epsilon$ for every $\tau > \tau_{Th}$, where $\epsilon$ is a small threshold (we put $\epsilon=0.05$). The Thouless time, defined in this manner, is inversely proportional to an energy scale $\omega_{\mathrm{Th}}$ up to which the eigenvalue correlations resemble Gaussian RMT correlations, see also Sec.~\ref{subsec:matrix_elements}. 

The Thouless time is usually understood as the longest physically relevant relaxation time in the system. Its notion was initially introduced in the context of Anderson localization in non-interacting models as a time scale at which a particle reaches the boundary of the system~\cite{Edwards72, Thouless74}. The 3D Anderson model analysis shows that the Thouless time, extracted from the SFF, follows the expected scalings. Consistently with the presence of diffusion in the delocalized phase of the 3D Anderson model,~\cite{Sierant20thouless, Suntajs21} find a diffusive scaling $t_{\rm Th} \propto L^2$ below the critical disorder strength\footnote{In contrast, a sub-diffusive scaling of $t_{Th}$ is found in the delocalized phase of Anderson model on random regular graphs~\cite{Colmenarez22SubdiffusiveRRG}.}. 
This scaling is slower than the scaling of $t_H$, which in the 3D Anderson model scales as $t_H \propto L^3$.
At the critical point, however, it was found~\cite{Sierant20thouless, Suntajs21} that $t_{\rm Th}\propto t_H \propto L^3$, and hence the scale invariance (i.e., independence of $L$) of the ratio $t_H/t_{\rm Th}$ represents an adequate criterion for the detection of the critical point.
In the localized phase, the SFF of the 3D Anderson model tends, with increasing $L$, to the SFF for Poisson level statistics, $K_{PS}(\tau) = 1$\footnote{For Poisson level statistics,~$K_{PS}(\tau) = 1$ at sufficiently large $\tau$, see~\cite{Prakash21SFF} for a detailed study of $K_{PS}(\tau)$ at arbitrary $\tau$.}, and hence the notion of $t_{\rm Th}$ extracted from the SFF becomes less meaningful.

A natural goal is then to extend this framework to interacting systems, in which $t_H$ increases exponentially with $L$.
Does this imply that $t_{\rm Th}$ should also scale exponentially with $L$ close to the transition?
\cite{Suntajs22} showed evidence that this is indeed the case in the quantum sun model, which is similar to the model in Eq.~\eqref{eq:Havalanche}.
Specifically, it was shown that in the ergodic phase when the short-range spectral statistics complies with the RMT predictions, $t_{\rm Th}$ increases exponentially with $L$ (however, with a smaller rate than $t_H$).
At the critical point, the ratio $t_H/t_{\rm Th}$ then becomes scale invariant, suggesting that this criterion to detect the critical point is useful also for interacting systems.

A nontrivial question is whether the scale invariance of $t_H/t_{\rm Th}$ also emerges in disordered spin chains such as the $J_1$-$J_2$ model.
Deep in the ergodic regime of the latter model, see Fig.~\ref{fig:SFF}(a), $\tau_{\rm Th}$ decreases significantly with the increase of system size $L$, which, given the exponential scaling of $t_H$, amounts to approximately a diffusive scaling of the Thouless time, $t_{\rm Th} \propto L^2$. 
On the other hand, $t_{\rm Th}$ increases approximately exponentially with disorder $W$~\cite{Suntajs20e}.
As a consequence, at disorder values that are not extremely large (see, e.g., the results in Fig.~\ref{fig:SFF}(b) at $W=10$), $t_{\rm Th}$ in finite systems becomes larger than (or comparable to) $t_H$.  
Investigation of the system size $L$ and disorder strength $W$ dependence of the Thouless time in the disordered XXZ and $J_1$-$J_2$ models led to the following  two-parameter scaling hypothesis for the Thouless time~\cite{Suntajs20e}, 
    \begin{equation}
     t_{Th} =t_0 \mathrm{e}^{W/\Omega }L^2,
     \label{eq:ThSc}
    \end{equation}
where $t_0$ and $\Omega$ are constants. This equation indicates a diffusive scaling of the Thouless time, with the diffusion constant (that can be interpreted as the inverse of the prefactor in \eqref{eq:ThSc}) decreasing exponentially with disorder strength $W$\footnote{This scaling is inconsistent with the recent mathematical arguments of~\cite{deRoeck24absence} for the disordered quantum Ising chain, which rule out diffusive transport at sufficiently large disorder strengths $W$. \label{redFootnote}}. At any finite system size $L$, the Thouless time \eqref{eq:ThSc} is upper-bounded by the Heisenberg time $t_{\mathrm{Th}} < t_\mathrm{H} = e^{c L}$. Hence,  there exists, at any finite $L$, a disorder strength $W_{\mathrm{Th}}^T \sim L$ (up to sub-leading logarithmic corrections) at which the scaling \eqref{eq:ThSc} breaks down. When $t_{\rm Th}$ becomes close to $t_H$, the spectral correlations do not adhere to GOE predictions even at the smallest energy scales, corresponding to time scales of the order of $t_H$. Consequently, if $t_{\rm Th}\gtrsim t_H$, the short-range level statistics, as captured, e.g., by the average gap ratio $\overline r$, deviate from the GOE behavior. Hence, $W_{\mathrm{Th}}^T$ defined here is closely related to the disorder strength $W^T(L)$ extracted as the point at which the average gap ratio deviates from the GOE value~\eqref{eq:WT}.

Importantly, the hypothesis \eqref{eq:ThSc} would imply that the disordered XXZ and $J_1$-$J_2$ spin chains follow the predictions of GOE at disorder strength $W < W_{\mathrm{Th}}^T \propto L$. The linear divergence of $W_{\mathrm{Th}}^T$  in the $L\to \infty$ limit would, therefore, mean that the disordered spin chains are ergodic at any disorder strength $W$ in the thermodynamic limit and that there is no MBL phase, as defined by \eqref{eq:MBL1}.

Nevertheless, the scaling~\eqref{eq:ThSc} of $t_{\mathrm{Th}}$ with system size $L$ and disorder strength $W$ highlights a fundamental difficulty in understanding of the MBL phase. The Thouless time $t_{\mathrm{Th}}$ increases, to a good approximation, exponentially with the disorder strength $W$, consistently with the dramatic slow down of the many-body dynamics upon the increase of $W$. The slow down of the dynamics hinders the differentiation between a very slow approach to thermal equilibrium and ultimately arrested dynamics in the $t,L\to \infty$ limit, consistent with the definition of the MBL phase \eqref{eq:MBL1}. 
In finite systems, the scaling~\eqref{eq:ThSc} is not expected to hold when the values of $t_{\rm Th}$ become comparable to those of $t_H$~\cite{Suntajs20e}.   
A subsequent analysis~\cite{Sierant20thouless, Sierant20polfed} focused on the presence of deviations from the scaling~\eqref{eq:ThSc}, arguing that the diffusive scaling $t_{\rm Th} \propto L^2$ may break down at sufficiently large $L$ and $W$. These observations suggest that the scaling \eqref{eq:ThSc}, while approximately valid for disordered spin chains, may not be valid in the asymptotic regime $L \to \infty$ at arbitrary large disorder. Hence, \eqref{eq:ThSc} per se does not rule out the existence of the MBL phase, similarly to the gap ratio result \eqref{eq:WT}.
Moreover, when the diffusive scaling in~\eqref{eq:ThSc} is replaced by a subdiffusive dependence $t_{\rm Th} \propto L^{z}$ (with $z>2$), the analysis of the results in much less constrained~\cite{Suntajs20e}, and presently available system sizes do not allow for unambiguous conclusions regarding both the nature of the dynamics at larger values of $W$ as well as the presence of the MBL phase.

In the ergodic phase of generic Floquet models without any local conservation laws, the Thouless time 
scales logarithmically with system size \cite{Kos18, Chan18}. This finding is supported numerically by transfer matrix calculations~\cite{Garratt21Pairing, Garratt21MBLasSymmetryBreaking}, but the logarithmic system size dependence is non-trivial to observe in the exact diagonalization investigations, see~\cite{Sonner21ThoulessEnergy}.
A protocol to measure SFF with the randomized measurements in quantum simulator settings was recently proposed~\cite{Joshi22probingSFF} and executed for small quantum systems in~\cite{Dong2024measSFF}.

\subsubsection{Other measures of spectral statistics} \label{sec:other_measures_stat}
A more traditional approach to study quantum chaos, manifested by the onset of level repulsion between eigenvalues of complex systems, is to consider the level spacing distribution $P(s)$, exemplified by the investigations of heavy nuclei \cite{Guhr98}, hydrogen atom in magnetic field \cite{Delande86,Friedrich89} or quantum-chaotic spin chains \cite{Avishai02, Santos10}. For systems following the predictions of random matrix theory, the level spacing distribution is approximated by the Wigner surmise,
\begin{equation}
    P_\beta(s) = A_{\beta} s^{\beta} e^{-B_{\beta} s^2},
    \label{eq:spacRMT}
\end{equation} 
where $A_\beta$, $B_\beta$ are constants and 
$\beta=1,2,4$, depending on the symmetry class of the model~\cite{Dyson62a, Dyson62b, Dyson62c}. For instance, for time reversal symmetric systems, such as \eqref{Hxxz}, $\beta=1$ and $A_1 = \pi/2$, $B_1=\pi/4$. Emergence of the level repulsion ensures that $P_\beta(s) \stackrel{ s\to 0}{\longrightarrow }0$. In contrast, the level spacing distribution for Poisson statistics reads
\begin{equation}
    P_{PS}(s) = e^{-s}.
       \label{eq:spacPS}
\end{equation} 
The evolution of the level spacing distribution from \eqref{eq:spacRMT} to \eqref{eq:spacPS} at the crossover between ETH and MBL regimes in the disordered XXZ spin-1/2 chain has been studied in a number of works
\cite{Chavda14, Serbyn16, Bertrand16, Buijsman18, Sierant19level, Kjall18}. Consistently with the trends visible in the  results for the average gap ratio, the level spacing distribution changes significantly with system size $L$ at a fixed disorder strength $W$~\cite{Sierant20model}. Moreover, to extract the level spacing distribution from eigenvalues of a many-body system, one has to perform a procedure of the so-called unfolding, which fixes the density of energy levels to unity. The unfolding procedure is not uniquely defined and may be performed in several distinct ways~\cite{Gomez02, Berkovits21}, which may lead to misleading results, i.e., the probes of quantum chaotic behavior, especially related to long-range spectral statistics, may strongly depend on the employed unfolding procedure~\cite{Gomez02}. For that reason, the analysis of the average gap ratio \eqref{eq:er} is simpler and more commonly employed.

Higher order spacing ratios are tools to assess properties of level statistics at larger energy scales, or equivalently, at time scales smaller than $t_H$. One possible definition (among several conventions~\cite{Harshini18, Kota18, Bhosale19,Rao20higher-order,Corps21long-range}) of the spacing ratio of order $k$ reads
\begin{equation}
    r^{(k)}_n = \frac{ \min\{ \delta^{(k)}_{n+k},\delta^{(k)}_{n} \} }{ \max\{\delta^{(k)}_{n+k},\delta^{(k)}_{n} \} },
    \label{eq:erK}
\end{equation}
where $\delta^{(k)}_n = E_{n+k}-E_{n}$. Equation~\eqref{eq:erK} generalizes the definition in Eq.~\eqref{eq:er}, and was used, with $k>1$, to investigate level statistics across the ETH-MBL crossover in various disordered many-body systems~\cite{Buijsman18, Sierant20model}. Moreover, the analysis of higher order gap ratios provides means of assessing the RMT properties of level statistics without resolving symmetries of the Hamiltonian~\cite{Tekur20symmetry}. Linear drifts with system size, analogous to the behavior of $W_{\mathrm{Th}}^T$, have also been observed in the analysis of higher order spacing rations at the ETH-MBL crossover in~\cite{Sierant20model}. 

We note that the number variance $\Sigma^{2}$, a quantity directly related to two-point correlation function of energy levels \cite{Mehtabook}, was employed in the studies of long-range spectral correlations at the ETH-MBL crossover~\cite{Bertrand16,Sierant19level}. The number variance $\Sigma^{2}$, together with the approach proposed in~\cite{Corps20Thouless}, provide alternative routes for the definition of Thouless time based on the properties of long-range spectral statistics.

\subsubsection{Overview} Investigation of level statistics is the simplest method of analysis of exact diagonalization results since it does not require any additional ingredients besides the Hamiltonian. Yet, the method is powerful, as the adherence of level statistics to RMT predictions signifies applicability of the ETH, while the flow towards Poisson level statistics indicates a breakdown of ergodicity. 

The average gap ratio $\overline r$ reflects the properties of level statistics at the energy scales of the order of a single level spacing, or, equivalently, at the longest time scale, the Heisenberg time $t_\mathrm H \sim e^{cL}$, below which the dynamics is not yet quasiperiodic. The implicit $t \to \infty$ limit makes $\overline r$ a quantity well suited for deciding whether the system is ergodic or MBL~\eqref{eq:MBL1}. Nevertheless, the significant shift of the ETH-MBL crossover with system size $L$, captured, e.g., by the disorder strengths $W^T(L)$ and $W^*(L)$, disallows unambiguous conclusions for the present-day numerical results for the disordered spin-1/2 chains. Other measures of spectral statistics, such as the level spacing distribution, are subject to similar flows with system size. 

The SFF quantifies two-point long-range spectral correlations, and its characteristic ramp-plateau structure in disordered many-body systems allows to define the Thouless time $t_{\mathrm{Th}}$, beyond which the approach to equilibrium state is over. Investigations of system size and disorder strength scaling of $t_{\mathrm{Th}}$ lead to ambiguous conclusions for the ETH-MBL crossover, similarly to the average gap ratio studies. The approximately exponential increase of $t_{\mathrm{Th}}$ with the disorder strength $W$ reflects the dramatic slow down of the many-body dynamics with increase of the disorder strength. These results, in the regime of system sizes $L$ accessible to the present-day numerics, could be consistent both with a slow approach to thermal equilibrium as well as with a complete freezing of the dynamics in the $t,L\to \infty$ limit.

\subsection{Properties of eigenstates}

Besides the eigenvalues $E_m$, the exact diagonalization algorithms also output the corresponding eigenstates $\ket{m}$, allowing for a more detailed understanding of the physics of investigated many-body Hamiltonian $\hat H$. In the following, we focus on the structure of many-body states in a specified basis of the Hilbert space and on the quantum information measures as ergodicity-breaking indicators.

\subsubsection{Wave function properties}
\label{subsec:wave-fun}
The eigenstates $\ket{m}$ are determined as linear combinations of $\mathcal N$ vectors of the many-body basis of the Hilbert space. The eigenbasis $\ket{\vec{\sigma}}$ of $\hat{S}^z_i$ operators, which we refer to also as the computational basis,  is a natural basis for the investigation of MBL in the disordered XXZ spin chain~\eqref{Hxxz} or the $J_1$-$J_2$ model~\eqref{eq:J1J2}. The eigenstates of these disordered spin chains coincide with the basis states $\ket{\vec{\sigma}}$ in the $W\to \infty$ limit at any fixed system size $L$. For finite $W$, the participation entropy, i.e., the R\'enyi entropy with index $q$ associated with the probability distribution  $|\braket{ \vec{\sigma}|m}|^2 $, defined as
\begin{equation}
  S_q = \frac{1}{1-q} \log_2\left( \sum_{\vec \sigma } | \braket{ \vec{\sigma} | m} |^{2q}\right), 
 \label{eq:PE}
\end{equation}
allows to study the spread of the state $\ket{m}$ in the many-body basis $\ket{\vec{\sigma}}$ which, for $U(1)$ symmetric models, spans a subspace with fixed magnetization (in the following, we focus on the zero magnetization sector). 
Participation entropy is related to the inverse participation ratio, which played a significant role in the analysis of Anderson transitions~\cite{Thouless74, Kramer93, Evers08}. Moreover, the participation entropies of many-body ground states distinguish between various quantum phases~\cite{Stephan09, Stephan10, Stephan14, Luitz14, Luitz14improving, Pino17, Lindinger19, Pausch21} and have been proposed as an ergodicity breaking measure in \cite{de2013ergodicity}. Quantum algorithms for measuring the participation entropies were proposed by~\cite{Liu24ipr}. 

The system size dependence of the participation entropy can be conveniently parameterized as 
\begin{equation}
 S_q = D_q \log_2(\mathcal N) + c_q,
 \label{eq:PEL}
\end{equation}
where $D_q$ is fractal dimension~\cite{Halsey86FractalMeasures},  $c_q$ is a sub-leading term, and $\mathcal N$ is the dimension of Hilbert space in the relevant symmetry sector, i.e., $\mathcal N = \binom{ L}{  L/2 }$ in the $\sum_i \hat{S}^z_i=0$ sector. In general, \eqref{eq:PEL} may be expected to apply only in a narrow interval of system sizes.
When the participation entropy $S_q$ is available for a wider interval of system sizes, interpolating the dependence of $S_q$ on $\log_2(\mathcal N)$ enables calculation of the fractal dimension as the first derivative, $D_q=\frac{\partial S_q}{\partial (\log_2(\mathcal N))}$.
If the state $\ket{m}$ is supported on a fixed number of basis states, then $S_q$ is independent of $L$ and consequently, the fractal dimension vanishes, $D_q=0$. In contrast, a Haar-random state\footnote{A Haar-random state can be obtained as $\ket{\psi} = U \ket{\psi_0}$, where $\ket{\psi_0}$ is a fixed state and $U$ is a matrix drawn with Haar measure on a selected matrix group. The orthogonal group, $U \in O(\mathcal N)$, is relevant for time-reversal invariant systems.}, which can often be used as an approximation of eigenstates of an ergodic many-body system, if fully delocalized in the Hilbert space, has $D_q=1$ (and $c_q<0$)~\cite{Backer19}. Multifractality~\cite{Stanley88} is the intermediate case when the fractal dimension $0<D_q<1$ depends non-trivially on the R\'enyi index $q$.

\begin{figure}
    \centering
    \includegraphics[width=0.95\columnwidth]{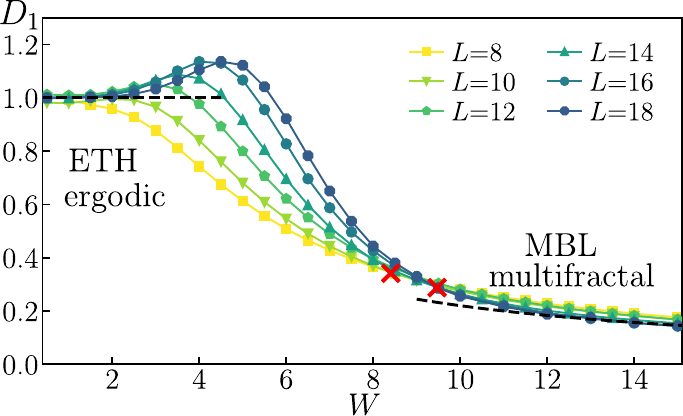}
    \caption{Fractal dimension $D_q$ for R\'enyi index $q=1$ at the ETH-MBL crossover in the $J_1$-$J_2$ model. The wave function $\braket{\vec{\sigma}|m}$ of mid-spectrum eigenstate $\ket{m}$ is fully delocalized, $D_q=1$, in the ETH regime, and multifractal, $0<D_q<1$ (with non-trivial dependence on the  $q$) in the MBL regime at the strong disorder $W$. The crossing point (denoted by the red cross) shifts with the system size $L$, hindering the location of the boundary of the putative MBL phase.
\label{fig:Dq}
}
\end{figure}

The behavior of the fractal dimension $D_q$ across the ETH-MBL crossover in disordered spin chains is illustrated in Fig.~\ref{fig:Dq} for the $q \to 1$ limit of Shannon entropy, on the example of the $J_1$-$J_2$ model \eqref{eq:J1J2}. We compute the participation entropy $S_q$ \eqref{eq:PE} of mid-spectrum eigenstates $\ket{m}$ of the $J_1$-$J_2$ model at system sizes $L$ and $L+\Delta L$ (where $\Delta L \ll L$), and we use \eqref{eq:PEL} to obtain the fractal dimension $D_q$ characterizing the spread of the states $\ket{m}$ in the eigenbasis $\ket{\vec{\sigma}}$ of $S^i_z$ operators. This way of calculating $D_q$, equivalent to taking the forward difference approximation of the derivative of $S_q(L)$ with respect to $L$, allows to estimate the extent of the ergodic regime more accurately than the typically assumed procedure of calculating $D_q$ as $S_q/ \log_2(\mathcal N)$, see, e.g., \cite{Mace19Multifractal}. The former procedure leads to weaker finite size effects in the ergodic and crossover regimes, yielding smaller values of $D_q$ for the disorder strengths $W$ just above the boundary of the ergodic regime $W^T(L)$. 
The difference between the two methods of $D_q$ calculation, see, e.g.,~\cite{Pino20} and \cite{Sierant23RRG}, was one of the factors leading to claims about the non-ergodic extended phase in the Anderson model on random regular graphs~\cite{Kravtsov18NonErgodic}, which is now believed to be shrinking to a single point in the $L\to \infty$ limit, see also~\cite{Tikhonov19stat, Tikhonov21eigen}.

For the $J_1$-$J_2$ model, at system size $L$, the fractal dimensions $D_q$ delineate a pronounced ETH regime, in which $D_q=1$, and the MBL regime, in which the states are multifractal and $0<D_q<1$, depending non-trivially on the R\'enyi index $q$, and decreasing approximately as $D_q \propto 1/W$ at $W \gg 1$~\cite{Mace19Multifractal}. Notably, this implies that the eigenstates in the MBL regime are \textit{not} localized in the computational basis $\ket{\vec{\sigma}}$, as already anticipated in \cite{Buccheri11Structure, Luca13}.
Considering many-body eigenstates in the Anderson insulator, by setting $\Delta=0$ in \eqref{Hxxz}, one finds that even in the absence of interactions, the states $\ket{m}$ are not localized in the $\ket{\vec{\sigma}}$ basis (in the sense of vanishing $D_q$), but rather are multifractal\footnote{To the best of our knowledge, a demonstration of this fact in large systems of free fermions is presently lacking.}. From this perspective, it is natural to expect that the eigenstates are still multifractal (in the computational basis) in the MBL regime upon the introduction of interactions to the Anderson insulator. For this reason, if an MBL \textit{phase} exists beyond a certain critical disorder strength $W_c$ in the considered spin chains, it is natural to expect that it is associated with multifractal eigenstates in the computational basis. In contrast, {\it localization} of the states in computational basis $\ket{\vec{\sigma}}$, denoted also as the Fock-space localization, occurring when $D_q=0$~\footref{footnoteFSL},
is a stronger form of ergodicity breaking, which may arise in disordered spin chains at disorder strengths larger than $W \propto L$~\cite{DeTomasi21}, and occurs in the quantum sun model~\cite{Suntajs24ultrametric}.

The fractal dimension $D_q$ exhibits a non-monotonous dependence on $W$, admitting a maximum at disorder strength at the boundary of the ETH regime at a given system size $L$, see Fig.~\ref{fig:Dq}. 
The fractal dimension $D_q$ can be bigger than unity since it reflects the local \textit{slope} of the  $S_q(L)$ curve. The interval of disorder strengths $W$, for which $D_q>1$, shrinks with the increase of $L$. 
This region of parameters coincides with the regime dubbed as \textit{maximally chaotic}, in which the sensitivity of the eigenstates to perturbations is maximal~\cite{Lim24DefiningChaos}, see also Sec.~\ref{subsec:other}. The maximum of $D_q$ is followed by its decrease with $W$ towards the value characteristic for the MBL regime. This behavior was interpreted as a transition to MBL phase~\cite{Mace19Multifractal}, leading to an estimate for the critical disorder strength consistent with the earlier study~\cite{Luitz15}. Significantly, however, the crossing point of the curves $D_q(L)$ for various system sizes drifts towards larger $W$ with increasing $L$. The drift of the crossing point for $D_q$, within the interval of available system sizes, is weaker than the drift for the average gap ratio, c.f.~Fig.~\ref{fig:ER}. Nevertheless, this finite size drift of the crossover prevents one from reaching unambiguous conclusions about the fate of the ETH-MBL crossover in the $L\to\infty$ limit.

The multifractal structure of the many-body wave function of eigenstates in the MBL regime can be employed to develop decimation schemes that discard the irrelevant parts of the Hilbert space, facilitating their numerical description~\cite{Pietracaprina21}, see also~\cite{Prelovsek18reduced, Prelovsek21percolation}. The properties of many-body wave function do depend on the choice of the basis. However, using another choice of the eigenbasis, namely, localized Anderson orbitals obtained by diagonalization of \eqref{Hxxz} in the single particle sector, the results for $D_q$ were reported to remain qualitatively the same~\cite{Mace19Multifractal}. Nevertheless, \cite{Luitz20a} suggests that the dynamical properties of many-body systems do not depend strongly on the degree of multifractality of many-body eigenstates. We note that the participation entropies at the ETH-MBL crossover may not be \textit{self-averaging}~\cite{Solorzano21}, so that the extraction of a representative value of $S_q$ at disorder strengths in that regime is possible only upon averaging the results over multiple disorder realizations of the system. Similar reservations also apply to other quantities at the ETH-MBL crossover~\cite{TorresHerrera20}.

Non-equilibrium problems for the disordered XXZ spin chain \eqref{Hxxz} can be viewed as a problem of a \textit{fictitious particle} hopping on a Fock space graph~\cite{Logan19local, Roy19percolation, Tarzia20}, whose vertices are the spin configurations~$\ket{\vec{\sigma}}$ 
corresponding to eigenstates of the $\hat{S}^z_i$ operators. The Fock space graph has an extensive local connectivity, and the on-site energies are correlated, which may prevent the wave function of the fictitious particle from trivial delocalization~\cite{Roy20correlations, Roy20LocalizationGraphs}. We note that the effects of Fock space correlations on finite system size drifts of the ETH-MBL crossover in several models was analyzed in~\cite{Scoquart24role}. 
The Fock space perspective enables application of insights and approximations developed for single particle problems to analyze properties of many-body system~\cite{Roy21anatomy}.
Still, the essential properties and system size drifts of the obtained measures are similar~\cite{Sutradhar22FockSpace} to the features typically observed at the crossover, as captured, e.g., by the participation entropy \eqref{eq:PE}.

\begin{figure}
    \centering
    \includegraphics[width=0.91\columnwidth]{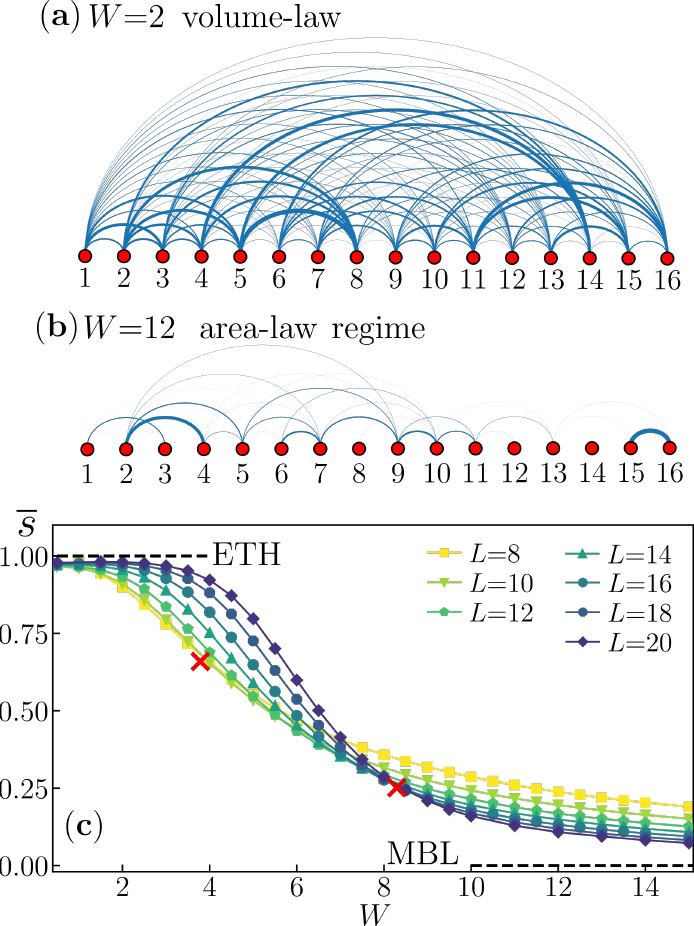}
    \caption{Entanglement of mid-spectrum eigenstates of the $J_1$-$J_2$ model at the ETH-MBL crossover. Quantum mutual information $I_2$ between pairs of the spins in chain of $L=16$ sites in the ETH regime, (a), and in the MBL regime, (b). The thickness of the line is proportional to the value of $I_2$. Rescaled entanglement entropy $\overline s$ crossovers between a volume-law scaling $\overline s \to 1$ at small disorder strength and an area-law scaling $\overline s \to 0$ deep in the MBL regime. Shift of the crossing point (denoted by the red cross) with system size $L$ is significant. 
\label{fig:I2}
}
\end{figure}
\subsubsection{Quantum information measures}
\label{subsec:quant_inf}

Quantum information theory provides measures of quantum correlations, which can be useful in characterization of many-body systems~\cite{Amico08} and are often framed in a broader context of quantum resource theories~\cite{Chitambar19resource}.  Characterization of entanglement in eigenstates of the disordered spin chains has been one of the dominating methods of investigation of  ergodicity and its breaking in many-body systems. In the following, we will focus on two measures of entanglement: entanglement entropy and quantum mutual information. 

The von Neumann entanglement entropy of state $\ket{ \psi }$ for a bipartition of the system into subsystems $A$ and $B$ is a measure of distillable quantum information~\cite{Karol}, defined as
\begin{equation}
    S_A = - \mathrm{tr}_A (\hat{\rho}_A \log_2 \hat{\rho}_A)
    \label{eq:entdef0},
\end{equation}
where $\hat{\rho}_A = \mathrm{tr}_{B} \ket{\psi}\bra{\psi}$ is the reduced density matrix of subsystem $A$, obtained by tracing out the degrees of freedom of subsystem $B$. Quantum mutual information (QMI) \cite{nielsen00} measures correlation between subsystems $A_1$ and $A_2$ of the system, and can be calculated as
\begin{equation}
     I_2 = S_{A_1} + S_{A_2} - S_{A_1 \cup A_2}.
    \label{eq:qmi}
\end{equation}

A basic idea is that the eigenstates of a system that obeys the ETH are well approximated by random Haar states~\cite{Dalessio16} and have a \textit{volume-law} entanglement entropy~\cite{Page93average, Bianchi22}. Consequently, the QMI calculated for mid-spectrum eigenstates is close to the RMT prediction, independently of the separation between the subsystems $A_1$ and $A_2$, see Fig.~\ref{fig:I2}(a). In contrast, eigenstates in the MBL regime are characterized by an \textit{area-law} of entanglement entropy, similarly to the  ground states of gapped Hamiltonians~\cite{Eisert10}. The QMI decreases, on average, exponentially with the distance between the subsystems $A_1$ and $A_2$~\cite{Tomasi17Mutual}, as illustrated in Fig.~\ref{fig:I2}(b).

Entanglement entropy has been employed in the studies of MBL from its early days~\cite{Znidaric08} and area-law for the entanglement entropy of eigenstates was proposed~\cite{Bauer13, Serbyn13b}.
It predicts that the entanglement entropy $S_A$~\eqref{eq:entdef0} is proportional to the boundary of the subsystem $A$, which, for 1D systems, implies $S_A \sim \mathrm{const}$ independently of the size $\ell_A$ of the subsystem $A$, provided that $\ell_A \gg 1$. This is consistent with the LIOMs picture, as discussed in Sec.~\ref{subsec:modelingANDphenom}. In contrast, the mid-spectrum eigenstates in ergodic regime are nearly maximally entangled~\cite{Page93average, Bianchi22}, leading to volume-law $S_A \sim \ell_A$. Global conservation laws, such as $U(1)$ symmetry of \eqref{Hxxz}, affect the constant term in the volume-law scaling of $S_A$~\cite{Vidmar17}.

For a meaningful comparison of entanglement entropy for different system sizes, it is convenient to rescale the entanglement entropy by the average $S_{\mathrm{RMT}}$ for Haar random states in the considered subspace of the Hilbert space, and consider $s=S_A/S_{\mathrm{RMT}}$. Averaging this quantity over mid-spectrum eigenstates $\ket{m}$ and disorder realizations, leads to $\overline s$ which may serve as an ergodicity breaking indicator. The rescaled entanglement entropy, $\overline s$,  across the ETH-MBL crossover in the $J_1$-$J_2$ model converges, with increasing the system size $L$, to unity in the ergodic regime and it decreases towards $0$ in the MBL regime, as shown in Fig.~\ref{fig:I2}(c). Even deep in the ergodic regime, we observe a small correction of the order of $\mathcal{O}(1)$ to the value of the averaged entanglement entropy\footnote{For $\ket{\psi} = U \ket{\psi_0}$ where $\ket{\psi_0}$ is fixed and $U$ is drawn with Haar measure from the orthogonal group in the zero-magnetization sector, $\sum_i \hat{S}^z_i=0$, and the symmetric bipartition of the chain, one finds $S_{\mathrm{RMT}} = L/2- (1/(4\log(2))+1/2)$~\cite{Vidmar17}. The average entanglement entropy $S_{\mathrm{ERG}}$ of the mid-spectrum eigenstates of $J_1$-$J_2$ model in the ETH regime, e.g., at $W=1$, is well fitted by $S_{\mathrm{ERG}} = S_{\mathrm{RMT}} - c/\log(2)$, consistently with the arguments of~\cite{Haque22ent}, and $c\approx 0.10$, in approximate agreement with calculations of~\cite{Huang19entmid, Huang21entmid}.
}. The magnitude of the correction depends on the choice of the parameters of the model~\cite{Kliczkowski2023midspectrum}. Notably, we observe a significant drift of the crossing point of $\overline s(W)$ curves with the increase of system sizes. This drift is analogous to the behavior of the averaged gap ratio $\overline r$ and the fractal dimension $D_q$ discussed in the preceding sections, disallowing a single parameter scaling and preventing one from reaching unambiguous conclusions about the fate of the crossover in the $L \to \infty$ limit.

Interestingly, the extent of the drift of the crossing point may depend on the considered observable. For instance, within the range of system sizes considered by us, the drifts of the crossing points are larger for the average gap ratio $\overline r$ and the rescaled entanglement entropy $\overline s$ than for the fractal dimension $D_q$, c.f. Fig.~\ref{fig:ER}, Fig.~\ref{fig:Dq}, and Fig.~\ref{fig:I2}, respectively. Similarly, the QMI calculated for extensive subsystems $A_1$ and $A_2$ of the Kicked Ising model in \cite{Sierant22floquet} exhibits a weaker drift of the crossing point than the entanglement entropy. A consistency of the extrapolations to the large $L$ limit of crossing points $W^*(L)$ yields consistent results for various observables~\cite{Sierant20polfed, Sierant21constraint, Sierant22floquet}.
The consistency between extrapolations of the crossing points for various observables may suggest their relevance. Nevertheless, such extrapolations remain uncontrolled in the absence of solid arguments for the form of the extrapolating functions.

Distribution of the entanglement entropy \cite{Yu16} and its fluctuations~\cite{Khemani17a} provide additional perspective on the ETH-MBL crossover. The fluctuations of entanglement entropy between disorder realizations are enhanced in the crossover region. Notably, the standard deviation of entanglement entropy $\sigma_S$ across the samples increases superlinearly, $\sigma_S \propto L^{\alpha}$ (with $\alpha>1$), within the interval of available system sizes. This trend cannot be sustained for $L\to \infty$, since the maximal standard deviation is proportional to $L$. Similar effects are visible in the analysis of sample-to-sample fluctuations of the average gap ratio~\cite{Sierant19level, Schliemann21, Krajewski22sample, Siegl23}. These strong pre-asymptotic trends in the results are related to the drifts of the crossing points, leading to the failure of the single-parameter scaling analysis, which contradicts the Harris-Chayes criterion $\nu \geq 2$, as suggested by~\cite{Khemani17a}.
 
The ETH-MBL crossover has been studied with other measures of entanglement such as the concurrence~\cite{Bera16concurrence}, total correlations~\cite{Goold15total}, Schmidt gap~\cite{Gray18}, entanglement spectrum~\cite{Geraedts16, Serbyn16power}, or entanglement negativity \cite{west2018global, Gray19negativity, Wybo20entanglement}, leading to similar system size trends and ambiguities in extrapolation of the results to the $L\to \infty$ limit. Notably, the calculation of entanglement entropy can be performed directly in the thermodynamic limit with a numerical linked cluster expansion~\cite{Devakul15}. However, in this case the predictions about the ETH and MBL regimes systematically drift with the order $n$ of the expansion, disallowing a clear extrapolation to $n\to\infty$ limit.

\subsubsection{Perspective}
Eigenvectors in the middle of the spectrum of many-body systems allow to directly probe the system's properties in the $t\to \infty$ limit and to investigate the ETH-MBL crossover. The differences between the featureless, Haar random, eigenstates of ergodic systems and area-law entangled eigenstates in the MBL regime allow to propose multiple quantities that distinguish between the two regimes. Participation entropies, motivated by the studies of Anderson transition, determine the values of fractal dimensions $D_q$, quantifying the spread of a many-body wave function in a selected basis of the Hilbert space. Entanglement entropy, in turn, depends on the type of bipartition of the system, but not on the choice of basis. Eigenstates of the analyzed disordered spin chains are fully extended, $D_q=1$, whenever entanglement entropy saturates to the maximal Page value. However, volume-law scaling of entanglement entropy may coexist when the states are multifractal~\cite{Sierant22MF}, or, entanglement entropy may saturate to the Page value even when $D_q < 1$~\cite{DeTomasi20multi}. The sub-leading term $c_q$ in the system size dependence of the participation entropy may be more closely related to the scaling of entanglement entropy, as suggested in \cite{Sierant22MF}, in analogy with the findings for ground state physics~\cite{Zaletel11, Luitz14universal}. Higher order correlations between the amplitudes on the Fock space graph provide another link between entanglement and many-body wave function properties~\cite{Roy22correlations}.

The eigenvectors encode much more information about the physics of the considered many-body model than its spectrum. Nevertheless, the derived ergodicity breaking indicators experience similar drifts with system size as the probes of level statistics, preventing one from reaching unambiguous conclusions about fate of disordered spin chains in the asymptotic limit $L,t\to \infty $, relevant for the MBL phase.

\subsection{Matrix elements of local observables}
\label{subsec:matrix_elements}

Matrix elements $\langle n|\hat A|m\rangle$ of local observables $\hat A$ in eigenstates $\ket{m}$ of many-body Hamiltonian $\hat H$ provide a direct link between the features of system's spectrum and measurable dynamical properties. The analysis of properties of matrix elements of local observables is  an important way of quantifying ergodicity and its breakdown in many-body systems.

\subsubsection{Spectral function}
\label{subsec:spectralFUN}
The ETH ansatz for the matrix elements $\langle n|\hat A|m\rangle$ \eqref{eq:ETH1} consists of the diagonal term ($m=n$) and the off-diagonal terms ($m \neq n$). We employ a spectral function of a local observable $\hat{A}$ as an organizing principle for studies of both diagonal and off-diagonal matrix elements of $\hat{A}$. The spectral function is defined as the Fourier transform of the autocorrelation function $\left \langle A(t) A  \right \rangle $, i.e., 
\begin{equation}
    f^2(\omega) = \frac{1}{2\pi } \int_{-\infty}^{\infty}dt e^{i\omega t}
    \left \langle e^{i \hat{H}t} \hat{A} e^{-i \hat{H}t} \hat A  \right \rangle 
    \label{eq:f2},
\end{equation}
where the average $\langle \cdots \rangle$ is taken with respect to a selected ensemble of states. 
In the following, to stay in a direct correspondence with the quantities analyzed so far in this Section, we focus on the microcanonical ensemble of states $\mathcal M_{E_c}$ around the middle of the spectrum.
In that case, calculating the Fourier transform in \eqref{eq:f2}, we arrive at  
\begin{equation}
    f^2(\omega) = \frac{1}{|\mathcal  M_{E_c}|}\sum_{m\in \mathcal  M_{E_c} }\sum_{n=1}^{\mathcal N} \delta( \omega-|E_m-E_n| ) |\bra{n} \hat A \ket{m}|^2  
    \label{eq:f2a},
\end{equation}
where $\mathcal M_{E_c}= \{ m : |E_m-E_c| < \delta E \}$ defines the microcanonical energy shell around the target energy $E_c$ (taken to be the arithmetic average of the ground state and highest excited state energy), and $\delta E$ decreases with $L$, so that the number of states $|\mathcal  M_{E_c}|$ in the energy shell is no bigger than a fixed number $n_\mathrm{ev}$ of states (we take $n_\mathrm{ev} = \min\{400, \mathcal N/20 \}$), where $\mathcal N$ is the Hilbert space dimension.
\footnote{ We note that various definitions of $f^2(\omega)$ can be found in the literature. The average  $\langle \cdots \rangle$ may be taken with respect to the infinite temperature ensemble of states in~\cite{Sels20obstruction, Vidmar21phenomenology}. The spectral function considered by \cite{Mondaini17, Jansen19polaron, Schonle21lens} involves contributions of eigenstates $\ket{m}, \ket{n}$ with the average energy fixed at a selected target value, while \cite{Beugeling15OffDiagonal} assumes that both eigenstates $\ket{m}, \ket{n}$ belong to a selected energy interval.}

The $f^2(\omega)$ function is directly linked with the dynamics of the system as the Fourier transform of the autocorrelation function. The spectral function enters the fluctuation-dissipation relations~\cite{Khatami13Fluctuation}, and, if the operator $\hat{A}$ is a density of a conserved quantity, e.g., a spin current operator~\cite{Prelovsek23}, $f^2(\omega)$ is related to the conductivity of the system \cite{Agarwal15}. Comparison of \eqref{eq:f2a} with the ETH ansatz provides a link between the spectral function and the structure function $f_\mathcal{A}(\bar E, \omega_{mn})$ in the ETH ansatz \eqref{eq:ETH1}. Moreover, the small frequency behavior of the spectral function, $f^2(\omega)\stackrel{\omega \ll 1}{=}C_0 \delta(\omega)$, reflects the averaged square of the diagonal matrix element of $\hat{A}$, $C_0 = \frac{1}{|\mathcal  M_{E_c}|}\sum_{m\in \mathcal  M_{E_c} } |\bra{m} \hat A \ket{m}|^2$, since the length of the energy interval corresponding to $ \mathcal  M_{E_c}$ vanishes in the large $L$ limit.  

\begin{figure}
    \centering
    \includegraphics[width=0.94\columnwidth]{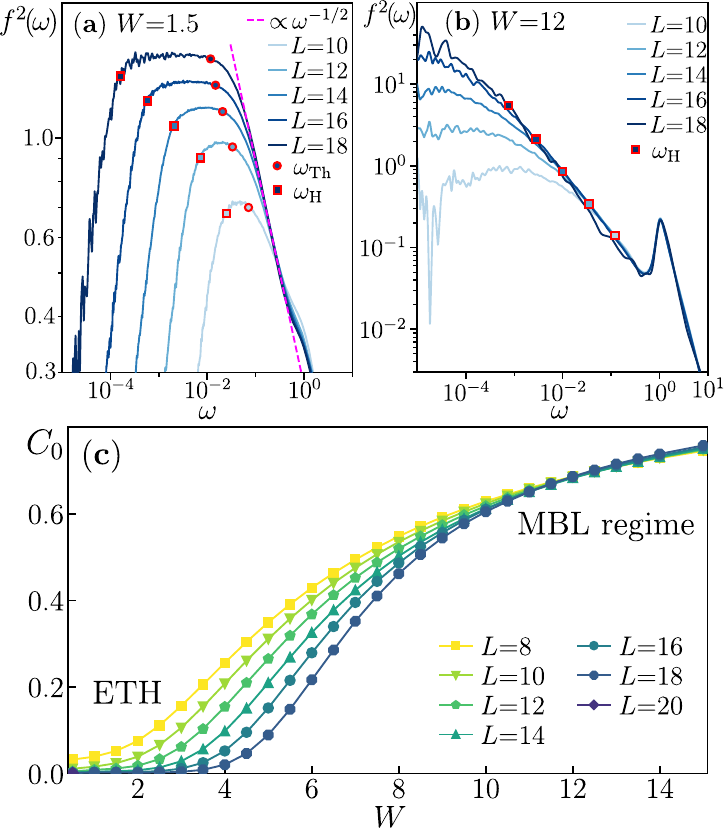}
    \caption{The spectral function $f^2(\omega)$ calculated for $\hat A=\hat{S}^z_j$ in the $J_1$-$J_2$ spin chain of size $L$ in the ergodic regime, (a), for $W=1.5$ and in the MBL regime, (b), for $W=12$. The Heisenberg energy $\omega_{\mathrm H} \sim e^{-cL}$ is denoted by red squares, while the Thouless energy $\omega_{ \mathrm{Th}}$ by red circles. The dashed line in (a) denotes a diffusive tail $f^2(\omega)\propto \omega^{-1/2}$. (c) The stiffness $C_0$ as function of disorder strength $W$ for various $L$ distinguishes ETH and MBL regimes.
\label{fig:f2}
}
\end{figure}

The spectral function $f^2(\omega)$ evaluated for the operator $\hat A = \hat{S}^z_j$, where $j$ is one of the lattice sites, and averaged over disorder realizations deep in the ETH regime of the $J_1$-$J_2$ model, is shown in Fig.~\ref{fig:f2}(a). The spectral function possesses a characteristic structure of a plateau, which extends from the Heisenberg energy $\omega_{\mathrm H}$ at small frequencies to the Thouless energy $\omega_{\mathrm{Th}}$ at large $\omega$\footnote{Interested in scaling of $\omega_{Th}$ with system size $L$ and disorder strength $W$, we identify the notions of energy, angular speed and frequency.}. These features motivated \cite{Serbyn17} to define the Thouless energy as the value of $\omega$ at which the plateau develops. Here, the Thouless energy is extracted as a point at which $\ln\left( f^2_0/f^2(\omega)) \right) > \epsilon$ where $\epsilon$ is a fixed threshold (we put $\epsilon=0.02$) and $f^2_0$ is the average of $f^2(\omega)$ over the plateau. Recalling that $f_{\mathcal A} (\bar E, \omega) = 1$ for the RMT, we note that the appearance of the plateau marks the onset of RMT behavior of the spectral function $f^2(\omega)$. The large frequency tail of the spectral function, $f^2(\omega) \propto \omega^{-1/2}$, corresponds to a diffusive decay of the autocorrelation function $\braket{\hat{S}^z_j(t)\hat{S}^z_j}$, which occurs deep in the ergodic regime of the $J_1$-$J_2$ model and lasts until the time scale proportional to $\omega_{\mathrm{Th}}^{-1}$. In principle, the Thouless time $t_{\mathrm{Th}}$ extracted from the behavior of SFF should scale in the same manner as $\omega_{\mathrm{Th}}^{-1}$ extracted here, reflecting the transport properties of the system. Approximately exponential scaling of $\omega_{\mathrm{Th}}$ with the disorder strength, $\omega_{\mathrm{Th}} \propto e^{-\kappa W}$ \cite{Sels20obstruction}, is consistent with the behavior of the Thouless time extracted from the SFF~\cite{Suntajs20e}. 


For a fixed system size $L$, the plateau of the  spectral function $f^2(\omega)$ becomes narrower and the ratio $\omega_{\mathrm{Th}}/\omega_{\mathrm{H}}$ decreases with the growth of the disorder strength $W$. At the same time, the spectral function decays algebraically at high frequencies, $f^2(\omega)\propto \omega^{-\alpha}$, with a power $\alpha$ increasing with disorder strength $W$, reflecting the slow-down of system's dynamics. At disorder strengths corresponding to the level statistics intermediate between the ETH and MBL regimes, the exponent $\alpha$ approaches values close to unity~\cite{Mierzejewski16}, which correspond either to a power-law decay of the autocorrelation function $\left \langle \hat{S}^z_j(t) \hat{S}^z_j  \right \rangle \propto t^{-\beta}$, with a very small exponent $\beta$, or when $\alpha \to 1$, to a logarithmic decay time~\cite{Sels20obstruction}. Upon further increase of the disorder strength, the averaged spectral function $f^2(\omega)$ starts to differ qualitatively from its \textit{typical} counterpart. In the MBL regime, there are no traces of the plateau in $f^2(\omega)$, as shown in Fig.~\ref{fig:f2}(b).

Increase of the disorder strength leads to the accumulation of weight $C_0$ of the spectral function at zero frequency. In the limit $\omega \to 0$, the spectral function reproduces the infinite time average of the autocorelation function $\left \langle \hat{S}^z_j(t) \hat{S}^z_j  \right \rangle$, reflecting the memory of the system about its initial state. The stiffness $C_0$ is shown in Fig.~\ref{fig:f2}(b). In the ETH regime, $C_0$ decays exponentially with the system size $L$ indicating the ergodicity of system's dynamics. In contrast, in the MBL regime, $C_0$ remains, to a good approximation, independent of $L$. Hence, $C_0$ follows the trends schematically represented in Fig.~\ref{fig:flow2}(b). 
This behavior follows the general trend illustrated in Fig.~\ref{fig:flow2}~(b) and,
similarly to the other examples discussed in this Section, does not allow for unambiguous conclusions about the fate of the system in the $L \to \infty$ limit.  

Understanding of evolution of the spectral function $f^2(\omega)$ with the disorder strength $W$, equivalent to comprehending the dynamics of autocorrelation functions $\left \langle \hat{A}(t) \hat{A}  \right \rangle $, remains an outstanding challenge for the non-equilibrium physics of disordered many-body systems. A recent attempt~\cite{Vidmar21phenomenology} proposes to use the non-interacting limit  ($\Delta=0$ in \eqref{Hxxz} or \eqref{eq:J1J2}) as the starting point. In the presence of interactions, Anderson LIOMs acquire finite relaxation times $t_R$. The distribution of $t_R$, relatable to the dynamics of $\left \langle \hat{A}(t) \hat{A}  \right \rangle $, becomes broad already for $W > W^T(L)$, i.e., directly above the ETH regime. At these disorder strengths, the off-diagonal elements of local observables show deviations from Gaussianity~\cite{Luitz16b}, while a fraction of Anderson LIOMs relaxation times $t_R$ exceeds the Heisenberg time $t_H$, see Sec.~\ref{sub:proximity} for details. The latter operators appear to be exactly conserved at the considered system size, and the interplay between the broad distribution of $t_R$ and the exponentially increasing Heisenberg time determines the fate of the ETH-MBL crossover in the system.

\subsubsection{Sensitivity to local perturbations}
\label{subsec:sensitivity}

\begin{figure}
    \centering
    \includegraphics[width=0.99\columnwidth]{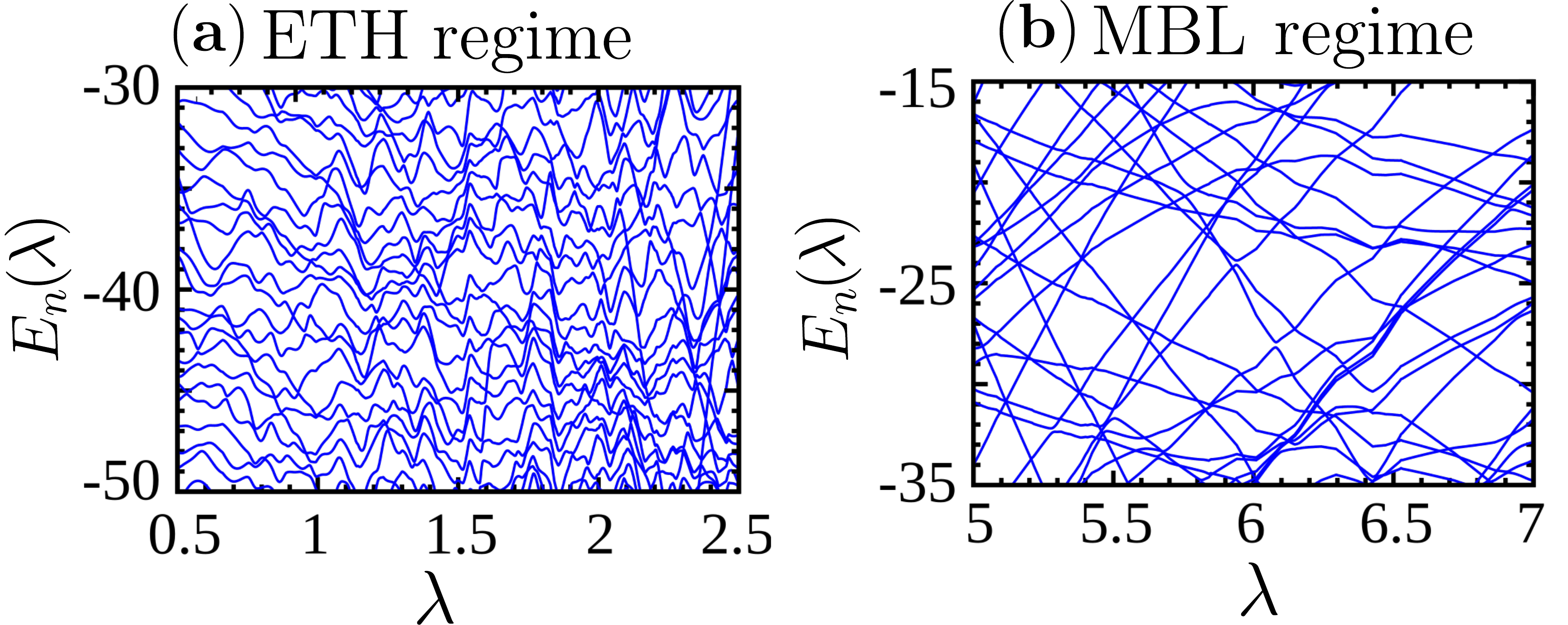}
    \caption{Parametric motion of eigenvalues $E_n(\lambda)$ for the disordered XXZ spin chain. The Hamiltonian reads $\hat{H}(\lambda) = H_{XXZ} + \lambda \sum_{i=1}^L u_i \hat{S}^z_i$, where $H_{XXZ}$ is given by Eq.~\eqref{Hxxz} with $W=0$ and $u_i$ are independent random variables distributed uniformly in the interval $[-1,1]$. In ergodic regime, (a), we observe avoided level crossings due to global correlations in the system. In the MBL regime, (b), the regular motion of $E_n(\lambda)$ and the presence of avoided crossing indicate the appearance of constants of motion localized in the real space. Reprinted and adapted from \cite{Maksymov19}.
\label{fig:enl}
}
\end{figure}

Consider a situation when a many-body system is affected by a local perturbation whose strength is determined by a parameter $\lambda$, leading to a Hamiltonian $\hat{H}(\lambda)$ that is functionally dependent on $\lambda$. In that case, the eigenvalues $E_n(\lambda)$ fulfill $\hat{H}(\lambda) \ket{n(\lambda)}= E_n(\lambda) \ket{n(\lambda)}$, becoming functions of $\lambda$. For a local perturbation of an ergodic system, the eigenvalues may be expected to perform a highly irregular motion~\cite{Zakrzewski23Dynamics} during which the energy levels $E_n(\lambda)$ repel each other, see Fig.~\ref{fig:enl}(a), and exhibit multiple avoided crossings due to the global correlations throughout the system~\cite{Zakrzewski93}. In contrast, for a system deep in the MBL regime, the energy levels likely cross each other as the local perturbation couples neighboring eigenvalues only in rare instances, as exemplified in Fig.~\ref{fig:enl}(b). Hence, investigations of the level dynamics, i.e., the parametric motion of the energy levels $E_n(\lambda)$, provide a means of assessing whether the system is ergodic.

To statistically analyze the properties of the level dynamics, it is convenient to define level velocities, $\partial_\lambda E_n$~\cite{Simons93}, and level curvatures, $\partial^2_\lambda E_n$, and to compare their distributions with RMT predictions~\cite{Zakrzewski93, vonOppen94}. The Hellmann–Feynman theorem allows for immediate identification of the level velocities $\partial_\lambda E_n$ and curvatures $\partial^2_\lambda E_n$ with expressions involving matrix elements of the perturbing operator $\partial_{\lambda}\hat{H}$. The study of level velocities and curvatures reproduces the position of the ETH-MBL crossover identified in exact diagonalization examinations of the disordered spin chains described in the preceding Sections, but, interestingly, certain deviations from the RMT predictions can already be observed deep in the ETH regime~\cite{Maksymov19}. 

The parametric dependence of the Hamiltonian on $\lambda$ can be introduced with various types of perturbations $\partial_\lambda \hat{H}$. Introducing a magnetic flux of strength determined by $\lambda$ is one notable choice. It corresponds to a "twist" of boundary conditions~\cite{Filippone16, Pouranvari21twisted}. In that case, the properties of level curvatures can be linked to the notion of Thouless time~\cite{Thouless74, Akkermans92}\footnote{The intuition from single-particle physics is that the sensitivity to the "twist" of boundary conditions determines the extent of delocalization/localization of eigenstates of the system.}. Another option is to subject the system to an \textit{imaginary} magnetic flux~\cite{Hatano97}, a method that has been explored for interacting models in the context of MBL~\cite{Hamazaki19}. This choice breaks down the hermiticity of $\hat{H}(\lambda)$ and defines a characteristic length scale, which can be identified with localization length in the system. This approach yields results consistent with the exact diagonalization results at the system sizes available to both methods~\cite{Heussen21, Obrien23}.

Investigations of level dynamics under local perturbations, such as $\partial_\lambda H = \hat{S}^z_j$, have been attempted to provide a link into the structure of integrals of motions that arise in the MBL regime. The relatively rare avoided level crossings deep in the MBL regime can be understood as resonances between LIOM configurations~\cite{Garratt21Local}, which occur on increasing length scales as the disorder strength is decreased towards the ETH-MBL crossover~\cite{Villalonga20,Villalonga20a}.

To obtain insights beyond the level dynamics, one may study sensitivity of eigenstates $\ket{n(\lambda)}$ to the perturbation $\partial_{\lambda}\hat{H}$. The sensitivity can be quantified by the fidelity $F$ given as $F\equiv \braket{ n(\lambda) | n(\lambda+\epsilon) } = 1-\frac{1}{2} \chi_n \epsilon^2 + \mathcal{O}(\epsilon^3)$, where $\epsilon \ll 1$. The latter equation defines the \textit{fidelity susceptibility} $\chi_n$, which, using perturbation theory, can be expressed as
\begin{equation}
 \chi_n = \sum_{m \neq n} \frac{| \bra{m}\partial_{\lambda} H \ket{n}|^2}{ (E_m-E_n)^2}.
 \label{eq:chiDEF}
\end{equation}
Analytic formulas for distributions of the fidelity susceptibility for $\hat{H}$ and $\partial_{\lambda}\hat{H}$ being GOE or GUE random matrices were derived in \cite{Sierant19f}, where $\chi_n$ was proposed as a dimensionless measure of ergodicity in many-body systems. Subsequently, the fidelity susceptibility averaged over eigenstates, 
\begin{equation}
\chi_{\mathrm{av}} = \frac{1}{\mathcal N} \sum_{n=1}^{\mathcal N} \chi_n,
\label{eq:chiAV}
\end{equation} 
where  $\mathcal N $ is the Hilbert space dimension, equals to the norm of an adiabatic gauge potential $\mathcal A_{\lambda}$, i.e., the generator, $\mathcal A_{\lambda} \ket{n(\lambda) } = i \partial_\lambda \ket{n(\lambda) }$, of the adiabatic deformation of the eigenstates. The latter was shown to be a particularly sensitive probe of quantum chaos~\cite{pandey_claeys_20}. 
Denoting $\omega_{mn} = E_m-E_n$ and regularizing the energy denominator in \eqref{eq:chiDEF} as $1/\omega_{mn}^2 \to \omega_{mn}^2/( \omega_{mn}^2 + \mu^2)^2$, where $\mu$ is a small energy cut-off, \eqref{eq:chiAV} can be rewritten as
\begin{equation}
\chi_{\mathrm{av}} = \int_{-\omega}^{\omega} d\omega \frac{\omega^2}{ (\omega^2 + \mu^2)^2} \left(f^2(\omega)-C_0 \delta(\omega)\right).
\label{eq:chiavSP}
\end{equation} 
The latter formula relies on the form of the spectral function $f^2(\omega)$~\eqref{eq:f2a} of the operator $\hat{A} = \partial_{\lambda} H$ with $\mathcal M_{E_c}$ containing the full spectrum of the system and with the contribution $C_0 \delta(\omega)$ stemming from $m=n$ in~\eqref{eq:f2a} subtracted. When the energy cut-off is taken to be of the order of inverse level spacing, $\mu \propto \omega_{\mathrm{H}} \propto 2^{-L}$, the average fidelity susceptibility allows for distinguishing different dynamical regimes. For ergodic system, the spectral function $f^2(\omega)$ has a plateau at $\omega_{\mathrm{H}}<\omega<\omega_{\mathrm{Th}}$, which, through~\eqref{eq:chiavSP}, implies an exponential scaling $\chi_{\mathrm{av}} \propto 2^L$. Vanishing of the spectral function at small frequencies for integrable models results in an algebraic scaling of $\chi_{\mathrm{av}}$ with $L$, while for non-interacting models, $\chi_{\mathrm{av}}$ is approximately system-size independent (as long as the perturbation $\partial_{\lambda} H $ preserves the class of the model).
This hierarchy of scalings motivated \cite{Lim24DefiningChaos} to propose the fidelity susceptibility as the quantity \textit{defining} the chaotic behavior, both in quantum systems and in their classical limit.
This classification distinguishes also a \textit{maximally chaotic} regime, in which the sensitivity of eigenstates to the perturbation $\partial_\lambda H$ is maximal, $\chi_{\mathrm{av}} \propto 4^L$. In disordered spin chains, the maximally chaotic regime is found at disorder strengths $W$ above the ergodic regime when the slow down of the dynamics results in the behavior $f^2(\omega) \propto \omega^{-\alpha}$ with $\alpha \to 1$.

Ref.~\cite{Sels20obstruction} quantified the sensitivity of eigenstates in the disordered XXZ spin chain~\eqref{Hxxz} to a perturbation $\partial_\lambda H = \hat{S}^z_j$ by means of a typical fidelity susceptibility $\chi_{\mathrm{typ}} =  \exp\left( \langle \langle \ln( \chi_n ) \rangle \right)$, where $\langle \langle . \rangle \rangle $ denotes the average over eigenstates and disorder realizations.
The typical fidelity susceptibility 
follows the ergodic scaling $\chi_{\mathrm{typ}} \propto 2^L$ in the ETH regime of the XXZ spin chain ($W < W^T(L) \approx 1.5 -2.1$ for $L \in [12,18]$). 
The rescaled fidelity susceptibility $\chi_{\mathrm{typ}}/2^L$ shows a crossing point at $W^*\approx 4.5$ which slowly drifts to a larger disorder strength with an increase of $L$. 
The 
maximally chaotic regime, in which the typical fidelity susceptibility scales as $\chi_{\mathrm{typ}} \propto 4^L$, is observed between the ETH and MBL regimes. 
For a fixed $L$, the maximum of $\chi_{\mathrm{typ}}$ as a function of disorder strength occurs at $W=W_m$, which drifts as $W_m \propto L$ in the regime of considered system sizes.

The described behavior of $\chi_{\mathrm{typ}}$ is consistent with system size drifts of the other ergodicity breaking indicators discussed in this Section, recall, e.g., the linear drift of $W^T(L) \propto L$. Hence, on its own, the analysis of the fidelity susceptibility does not bring up any qualitatively new information about the fate of the ETH-MBL crossover. The crossover features various system size drifts, including the linear drift with system size $L$ and the drifts that slow down with increasing $L$, exemplified by the crossing point of the $\chi_{\mathrm{typ}}/2^L$ curves. Beyond the crossing point, the rescaled fidelity $\chi_{\mathrm{typ}}/2^L$ decreases with system size, and the system is in the MBL regime. Extracting the Thouless energy from the spectral function, \cite{Sels20obstruction} reproduces the linear drift $W^T_{Th} \sim L$ implied by the scaling \eqref{eq:ThSc} from the analysis of SFF~\cite{Suntajs20e}, indicating consistency between the two approaches.
Moreover, when approaching the non-interacting limit $\Delta\to 0$ in the XXZ model at a fixed disorder $W$, Ref.~\cite{LeBlond21} also showed a tendency towards restoring ergodicity for arbitrary weak interaction.

Rescaling the fidelity susceptibility \eqref{eq:chiDEF} by the Hilbert space dimension $\mathcal N$, we obtain a variable $x_n = \chi_n / \mathcal{N}$, whose exact distribution for the GOE ($\beta=1$) and the GUE ($\beta=2$) is given at $\mathcal N \gg 1$ \cite{Sierant19f} as
\begin{equation}
P(x_n) = C_{\beta} \,f_{\beta}(x_n) \, x^{-\frac{\beta+3}{2}} \exp\left(-\frac{2}{\beta x}, \right),
    \label{eq:chiDIST}
\end{equation}
where $C_{\beta}$ is a normalization constant and $f_{\beta}(x) \stackrel{x \gg 1}{\longrightarrow} 1$ (specifically, $f_1(x) =1+\frac{1}{x} $ and $f_2(x) =\frac{3}{4}+\frac{1}{x}+\frac{1}{x^2}$). For $x\gg 1$, the distribution $P(x)$ has a power-law tail $x^{-\frac{\beta+3}{2}}$, dependent on the level repulsion exponent $\beta$, c.f. \eqref{eq:spacRMT}. This observation dates back to \cite{Gaspard90, Monthus17} and extends to the case of arbitrary $\beta \geq 0$. Indeed, for a bounded numerator in \eqref{eq:chiDEF},  large $\chi_n$ is obtained when the energy denominators $E_n-E_m$  are small. The probability of the latter is controlled by the level repulsion exponent $\beta$. 

The numerator of \eqref{eq:chiDEF}, i.e., the matrix element of the perturbation $| \bra{m}\partial_{\lambda} H \ket{n}|^2$, is another ingredient which controls the behavior of the tail of $P(x)$ at large $x$. In the ergodic regime, the numerator is bounded, similarly as in the RMT case. However, at disorder strengths above the ETH regime, there is a divergence of the matrix elements of local operators at small energy differences, as shown in Fig.~\ref{fig:f2}(b). Assuming that: i) the spectral function $f^2(\omega)$ diverges as $\omega^{-\alpha}$ at small $\omega$, ii) the numerator and the denominator in \eqref{eq:chiDEF} are not correlated, one obtains that $P(x)\propto x^{-\frac{3+\beta+\alpha}{2+\alpha}}$ at $x \gg 1$\footnote{Note that \cite{Sels20obstruction} consider a variable $z\equiv \log(x)$ and, consequently, the exponent analyzed there differs by unity with respect to the discussion here.}. Comparison of the latter expression with the RMT prediction $P(x) \propto x^{-\frac{3+\beta}{2}}$ for the tail of $P(x)$ suggests that the effect of the divergence of the spectral function $f^2(\omega)$ on the tail of $P(x)$ may be interpreted as altering the effective level repulsion exponent. For instance, the Poisson level statistics ($\beta=0$) combined with divergence of the spectral function ($\alpha= 0$) yields  a tail $P(x) \propto x^{-\frac{3}{2}}$. In contrast, $\beta=0$ accompanied with a divergence of spectral function $\alpha=1$ results in a slower decay $P(x) \propto x^{-\frac{4}{3}}$. The latter tail of the fidelity susceptibility distribution was observed in \cite{Sels20obstruction}. Within the RMT expression \eqref{eq:chiDIST}, an exponent $-\frac{4}{3}$ would be obtained for $\beta < 0$, suggesting that the observed behavior of the fidelity susceptibility distribution is reminiscent of an emergent level \textit{attraction} - 
lack of thereof is a crucial assumption of the mathematical analysis suggesting stability of the MBL phase~\cite{Imbrie16}. However, as shown by \cite{Garratt22Resonant}, the assumption about the lack of statistical correlation between the matrix element in the numerator of \eqref{eq:chiDEF} and the energy denominator is invalid. Moreover, while the distribution $P(x)$ is decaying according to a power law $x^{-\frac{4}{3}}$ in a broad regime, a crossover to the tail $P(x) \propto x^{-\frac{3}{2}}$ was observed for very large $x$~\cite{Maksymov19}. Nevertheless, as suggested in \cite{Sels20obstruction}, the divergence of the matrix elements of local operators at small energy differences,
together with the statistical correlations between the matrix elements and energy denominators~\cite{Garratt22Resonant}, should be taken into account  in the phenomenological attempts for a description of the MBL phase. Further insight into behavior of the fidelity susceptibility may be possible within the Lanczos based algorithm proposed in~\cite{Takahashi23shortcuts, Bhattacharjee23lanczos}.

\subsubsection{Other approaches and outlook}
\label{subsec:other}
Studies of ergodicity breaking have also focused on other properties of the matrix elements, among others, on the distribution of the diagonal matrix elements of local observables~\cite{Colmenarez19, Corps21kurtosis} or on the decay of the difference between diagonal matrix elements in neighboring eigenstates~\cite{Luitz16longTail}. The latter does not decay with system size $L$ in the MBL regime, in sharp contrast 
to the exponential in $L$ decay found in the ETH regime~\cite{Beugeling15}, which also differs from the algebraic decay of diagonal matrix element fluctuations in integrable systems~\cite{Ikeda13Finite, lydzba_swietek_24}.
System size drifts towards the ergodic behavior with increasing the system size $L$ are visible in the properties of distributions of matrix elements of local observables as seen, e.g., by~\cite{Colmenarez19}. 

A study~\cite{Panda20} of the Binder cumulant $\mathcal B$~\cite{Binder81}, a quantity proposed as a fine probe for quantum phase transitions, highlights the significance of the finite system size drifts at the ETH-MBL crossover. The Binder cumulant $\mathcal B$ was calculated from the off-diagonal matrix elements of a local observable ($\hat{S}^z_j$ operator) between eigenstates of \eqref{Hxxz} at a \textit{fixed} finite energy difference $\omega$. Hence, $\mathcal B$ probes properties of the system at times which do not scale with $L$, in contrast, e.g., to \cite{Luitz16b}, which focuses at energy scales with the corresponding times of the order of $t_H \propto e^{cL}$. The system size and disorder strength behavior of $\mathcal B$ does not show any clear signatures of a transition to the MBL phase. However, the observed behavior does not exclude a transition to the MBL phase at sufficiently large $L$ or $W$. Extrapolation of the scaling of $\mathcal B$ with $L$ and $W$ allows to obtain a lower bound on the system size $L_0$ needed to observe a transition to the MBL phase under an assumption that the transition occurs at a disorder strength $W_c$. For instance, \cite{Panda20} finds that the MBL transition at $W_c=10$ would be observable in the Binder cumulant behavior for system sizes larger than $L_0 \approx 57$.

A different perspective was pursued in studies of a single-particle density matrix, whose matrix elements are defined as $\rho_{ij} = \braket{ m | \hat{c}^\dag_i \hat{c}_j | m}$, where $\ket{m}$ is an eigenstate of a many-body system. Diagonalization of this matrix yields eigenvalues $n_{\alpha}$, interpreted as occupations of single-particle orbitals, which are expected to exhibit a step-like behavior in the MBL regime~\cite{Bera15}. 
Nevertheless, indicators of ergodicity breaking derived in this approach~\cite{Buijsman18natural, Hopjan19, Hopjan21opdm} are characterized by finite size drifts similar to the other quantities considered here. Analysis of properties of single-particle orbitals, i.e., the eigenvectors of $\rho_{ij}$, and the related dynamical quantities, such as the single-particle Green’s functions, provides a further outlook on the ergodicity and its breakdown in disordered many-body systems \cite{Villalonga18, Jana21LocalDensity, Prasad23SingleParticle, Jana23excitations, Roy23DiagnosticNEE}. 

To summarize, matrix elements of local observables provide a deep link between the spectral properties of many-body systems and their dynamics. Hence, understanding the properties of the matrix elements of local observables is vital for assessing the ergodicity of the considered system. Focusing on the spectral function $f^2(\omega)$, we have shown how features of ergodic and MBL regimes are encoded in the properties of matrix elements of local observables. In particular, the onset of the universal delocalization dynamics at large times manifests itself as a plateau of spectral function at small frequencies, leading to yet another definition of Thouless energy in many-body systems. We have argued that the essential trends observed in the analysis of spectral properties and of eigenstates, which prevent unambiguous conclusions about the fate of the ETH-MBL crossover in the $L\to\infty$ limit, are also present in the properties of matrix elements of local observables. 

We would like to note that the direct links between the notions of Thouless time $t_{\mathrm{Th}}$ and Thouless energy $\omega_{\mathrm{Th}}$, and the dynamics in disordered many-body systems, remain presently unclear.
Moreover, it has been suggested in~\cite{Dymarsky22BoundOnETH} that the Thouless energy $\omega_{\mathrm{Th}}$ should be \textit{parameterically larger} than the onset of the applicability of RMT, $\Delta E_{RMT}$, which is the energy scale below which $R_{mn}$ in the ETH ansatz \eqref{eq:ETH1} can be treated, for physical purposes, as being independent random variables.
The problem of consistency of various definition of Thouless time (see also \cite{Serbyn15}) in disordered many-body systems is an important open question relevant also for the field of mesoscopic conductors~\cite{Altland93Spectral, Altland96Thouless}. 

Therefore, despite the non-trivial insights stemming from investigations of the spectral functions or the level dynamics and sensitivity of eigenstates to local perturbations, the fundamental questions about non-equilibrium properties of disordered many-body systems, outlined in Sec.~\ref{sec:open}, remain open.

\section{Slow dynamics in disordered many-body systems}
\label{sec:dynamics}
This Section concentrates on the time evolution of disordered many-body systems, highlighting the abrupt slowdown of the dynamics in the MBL \textit{regime}. Focusing on the disordered XXZ model, we discuss the ubiquity of the regime of slow dynamics and outline numerical results for time evolution obtained with sparse matrix time propagation approaches and tensor network algorithms. We show how different scenarios for the fate of the ETH-MBL crossover discussed in Sec.~\ref{sec:numerical} translate to the properties of the dynamics in many-body systems. Considering local observables as well as entanglement entropy, we argue that the present-day numerical, as well as experimental, time evolution results allow for multiple and mutually inconsistent extrapolations to the limit of large times, $t \to \infty$, and system sizes, $L \to\infty$, relevant for the MBL \textit{phase}~\eqref{eq:MBL1}.

\subsection{Persistent thermalization of local observables}
\label{sec:density}
Average values of local observables are directly measurable in out-of-equilibrium experiments with synthetic quantum matter. Lack of thermalization of local observables is a distinct feature of the MBL regime and defines the MBL phase~\eqref{eq:MBL1} once the limit $L,t \to \infty$ is taken. Therefore, investigations of the time evolution of local observables constitute the fundamental aspect of studies of ergodicity and its breakdown in disordered many-body systems.

\subsubsection{Density correlation functions}
One approach to probe thermalization in an isolated many-body system is to perform a quench experiment studying the time evolution of an autocorrelation function 
\begin{equation}
C(t) = \frac{1}{c_0}\sum_{j\in \mathcal J} \left \langle \hat{S}^z_j(t)  \hat{S}^z_j \right \rangle,
\label{eq:den}
\end{equation}
where $c_0$ is a normalization constant ensuring that $C(0) \equiv 1$, and $\mathcal J= \{l_0,l_0+1,\ldots, L-l_0 \}$ is a set of lattice sites at which the autocorrelation function is measured,
with $0 \leq l_0\leq L/2$ being a constant. The average $\left \langle \cdots \right \rangle$ is taken for a selected ensemble of states and over multiple disorder realizations. For $U(1)$ symmetric spin-1/2 chains and $l_0=0$, the function $C(t)$ is, up to a constant shift, equal to the autocorrelation function of the fermion number operator  $\hat{n}_j= \hat{S}^z_j + 1/2$. In the following, we consider the time evolution of spin chains with open boundary conditions. To reduce the influence of the boundaries, we take $l_0=2$, which introduces a slight, non-trivial difference between $C(t)$ defined in~\eqref{eq:den} and the autocorrelation function of fermionic density $\hat{n}_j$. Since the latter difference is quantitatively small and does not change the overall behavior of the results, we keep referring to $C(t)$ as the density autocorrelation function.

When the mean $\left \langle \cdots \right \rangle$, besides the disorder average, involves an average over an ensemble $\{ \ket{ \vec{\sigma }} \}$ of product states in the eigenbasis of $\hat{S}^z_j$ operators, the autocorrelation function $C(t)$ measures the relaxation of the initial spin configuration $\ket{ \vec{\sigma } }$ throughout time evolution of the system. Using $\hat{S}^z_j(t) = e^{i \hat{H} t}\hat{S}^z_j e^{-i \hat{H} t}$, where $\hat{H}$ is the Hamiltonian of the system,  and denoting the time-evolved state as $\ket{\psi(t)} = e^{-i \hat{H} t} \ket{ \vec{\sigma } }$, 
the autocorrelation function \eqref{eq:den} can be written as 
\begin{equation}
C(t) = \frac{1}{c_0}\sum_{j\in \mathcal J} \left \langle  s_j \braket{\psi(t)|\hat{S}^z_j|\psi(t)} \right \rangle,
\label{eq:den2}
\end{equation}
where $s_j=\pm 1/2$ is the eigenvalue $\hat{S}^z_j \ket{ \vec{\sigma } } = s_j \ket{ \vec{\sigma } }$, and $\left \langle \cdots \right \rangle$ still involves the average over disorder realizations and the ensemble of initial states
\footnote{In particular, if the initial state is taken as the N\'{e}el state, $\ket{ \uparrow \downarrow \uparrow \downarrow  \ldots}$, the autocorrelation function $C(t)$ is sometimes called an \textit{imbalance}. We do not follow this distinction here as the time evolution of the N\'{e}el state follows the same patterns as for typical product states with energies close to the middle of the many-body spectrum.}.
The ETH \eqref{eq:ETH2} is valid in the ergodic regime. For the disordered XXZ spin chain (and the $J_1$-$J_2$ model), the microcanonical average of $\hat{S}^z_j$ vanishes in the large system size limit, c.f., Fig.~\ref{fig:f2}(c). Therefore, the autocorrelation function $C(t)$ decays to zero in the ergodic regime, and the information about the initial spin configuration $\ket{ \vec \sigma}$ is scrambled during the system's time evolution. In contrast, in the MBL regime, the approximate LIOMs preserve some fraction of the information about the initial spin configuration $\ket{\vec \sigma}$ indefinitely, leading to a non-vanishing autocorrelation function $C(t)$ in the $t \to \infty$ limit. The memory of the initial state retained at sufficiently strong disorder in many-body systems has been the basis of experimental observations of the MBL regime, see e.g.~\cite{Schreiber15}.

The decay of the autocorrelation function $C(t)$ to its long-time average contains information about the dynamics in the system. Diffusive dynamics is one possible scenario in systems with a conserved quantity, e.g., conservation of the total magnetization $\sum_j \hat{S}^z_j$ in the considered disordered spin chains may lead to diffusive spin dynamics. Then, at sufficiently large time and length scales, an effective hydrodynamic description of the dynamics is possible \cite{Kadanoff63}, which predicts a decay $C(t) \propto t^{-1/2}$ for 1D systems. Such a hydrodynamic behavior is inevitably accompanied by scaling corrections significant at intermediate time and length scales. The scaling corrections may be constrained due to additional symmetries in the system \cite{Crossley17fluids} and have to be taken into account when assessing the class of dynamics of the considered system \cite{Michailidis23CorrectionsToDiffusion}. Another scenario for systems with a conserved density is a subdiffusion with dynamical exponent $z>2$, which leads to a hydrodynamic description with an algebraic decay of the density autocorrelation function, $C(t) \propto t^{-1/z}$~\cite{Delacretaz23nonlinear}.

 \begin{figure}
    \centering
    \includegraphics[width=0.98\columnwidth]{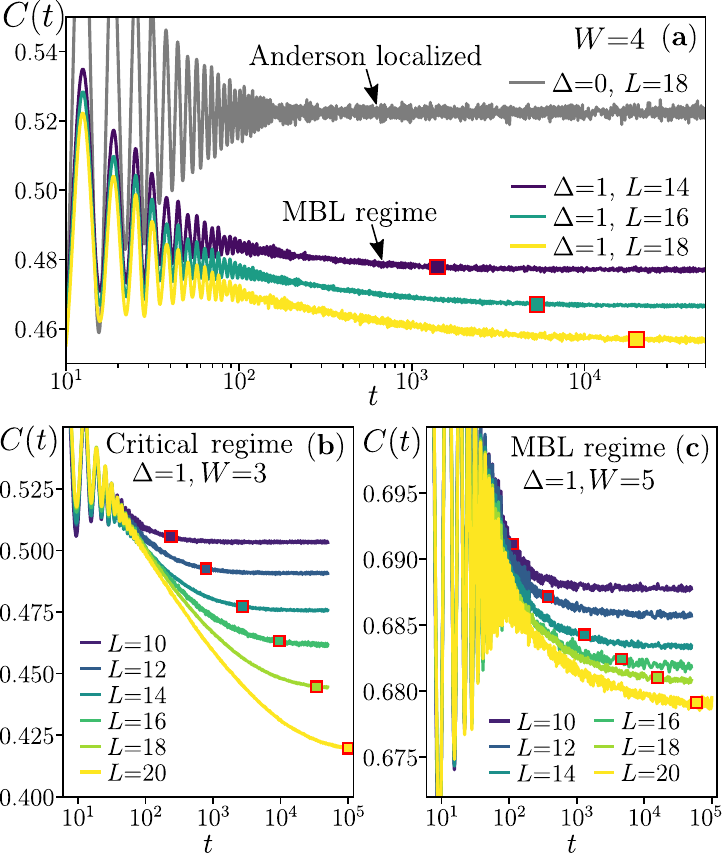}
    \caption{Time evolution of density autocorrelation function $C(t)$ in the disordered XXZ spin-1/2 chain \eqref{Hxxz}. (a) For $W=4$, the interactions ($\Delta=1$) induce a slow decay of $C(t)$ at long times, which continues approximately up to the Heisenberg time $t_H$ (denoted by the red squares). For $\Delta=0$, in the Anderson localized phase, $C(t)$ does not decay after the initial oscillations. (b) For disorder strength $W=3$, a strong decay of $C(t)$ is observed as the disordered XXZ spin-1/2 chain is in the crossover regime. (c) For $W=5$ the system is in the MBL regime, $W > W^*(L)$ for $12\leq L\leq 20$. The decay of $C(t)$ is persistent, but very slow, and an understanding of whether the system thermalizes in the $L,t\to\infty$ limit is missing. Reprinted and adapted from \cite{Sierant22challenges}. 
\label{fig:tevol}
}
\end{figure}

With increasing disorder strength, the dynamics of disordered spin chains abruptly slow down. The disordered XXZ spin chain \eqref{Hxxz}, in the absence of interactions ($\Delta = 0$), is in the Anderson localized phase. The arrest of dynamics in the Anderson insulator is manifested by the saturation of the density autocorrelation function $C(t)$ shown in Fig.~\ref{fig:tevol}(a) for $W=4$. The saturation is clearly visible after the initial transient period of oscillations, and extrapolation of the result to the asymptotic limit $L, t \to \infty$ is straightforward.

In the presence of interactions, $\Delta = 1$, and at the same disorder strength $W=4$, the autocorrelation function $C(t)$ slowly decays in time and ceases to decay only at long times approximately given by the Heisenberg time $t_H \propto e^{cL}$, see Fig.~\ref{fig:tevol}(a). The persistent decline of $C(t)$ is the effect of interactions in the system, and, despite being slow, the decay may have a strong impact on the fate of the system in the asymptotic $L,t\to \infty$ limit. The system's behavior illustrates this for $W=3$, shown in Fig.~\ref{fig:tevol}(b). At considered system sizes $12 \leq L \leq 20$, the value of $C(t)$ at any $t$ is no smaller than $0.4$, far from the ETH prediction $C(t) \stackrel{t\to\infty}{\longrightarrow} 0$ (relevant for $L\gg 1$). Hence, we observe a clear breakdown of ergodicity at considered $L$, which is consistent with the value of the average gap ratio at $W=3$, c.f. Sec.~\ref{subsec:spectral}, according to which the system is in the crossover regime $W^T(L)<W<W^*(L)$. Nevertheless, the saturation values of $C(t)$ are, to a good approximation, decreasing proportionally to $L$. Extrapolation of this trend indicates that in the limit of large $L$ and $t$, the autocorrelation function $C(t)$ decays to $0$, and the system thermalizes. 

The observation that power-law fits accurately describe the results in the \textit{crossover regime} of the disordered XXZ spin chain, at disorder strengths roughly fulfilling $W^T(L)<W<W^*(L)$, was attributed, by the early works~\cite{Gopalakrishnan16, Gopalakrishnan15, Agarwal17, Pancotti18}, to heterogeneities in disordered systems. Spatially uncorrelated disorder necessarily leads to Griffiths regions, where the disorder is locally stronger. Such regions play an essential role in the physics of quantum phase transitions~\cite{Vojta10}, and may act as bottlenecks hindering the thermalization of the surrounding spins and generating the slow power-law decays~\cite{Luitz16, Luitz17b} of local correlators observed experimentally by \cite{Luschen17}. The influence of Griffiths regions on the stability of the MBL phase will be revisited in Sec.~\ref{sec:aval}. 
In passing, we note that the power-law decays were also observed in the dynamics of survival probabilities~\cite{Tavora16}, which are discussed in more detail in Sec.~\ref{sec:survival}. Resonant regions, intimately connected to disorder fluctuations, play also an important role in the mathematical analysis of strongly disordered quantum Ising chains~\cite{deRoeck24absence}. The demonstration of the MBL 
in that work is limited only to rare realizations of the 1D chain, indicating the absence of diffusion at strong disorder, but allowing for subdiffusive behavior. 

The extrapolation of the results to the $L, t \to \infty$ limit is less straightforward when the disorder strength $W$ in an interacting system is increased beyond $W\gtrapprox3$. The power-law decay of $C(t)$, which accurately describes the results at $W=3$, is no longer a good fit for $W=5$ and system sizes up to $L=20$. Indeed, as visible in Fig.~\ref{fig:tevol}(c), the decay is slower than logarithmic in time. Subsequent studies realized that an increase of the system size beyond the capabilities of ED, possible due to Chebyshev time evolution method~\cite{Bera17} (up to $L=24$) or TEBD algorithm~\cite{Doggen18} (up to $L=100$), increases the prominence of the power-law decays in the XXZ spin chains at disorder strengths larger than $W^*(L=20)\approx 4$. The tensor network studies \cite{Doggen18, Chanda20t}, employing various types of threshold on the exponent $\beta$ governing the observed power-law decay of $C(t)$ (see also \cite{Popperl2021, Doggen2021manybody}), interpreted the observed behavior as a manifestation of the MBL transition occurring at disorder strengths above the early ED estimates~\cite{Luitz15}, c.f.~\cite{Nandy21}. Similar approaches in which the average in \eqref{eq:den2} is taken over specific initial states \cite{Prasad22InitialStateDependent} show that the onset of the MBL regime is dependent on the energy of the initial product state, which was interpreted as a premise for many-body mobility edge in disordered spin chains~\cite{Naldesi16, Chanda20m}.

While the latter observations of power-law decays reveal the essential features of the dynamics of disordered spin chains, formulation of conclusions for the system's fate in the $L,t\to \infty$ limit is not simple. Since non-trivial dynamics may be expected to occur in many-body systems at any time scale below the exponentially large Heisenberg time $t_H\propto e^{cL}$, even a very slow decay of $C(t)$ may lead to eventual thermalization of the system. Suppose that the decay of $C(t)$ can be approximated by an algebraic dependence, $C(t) \propto t^{-\beta}$, where $\beta$ is a fixed exponent. The persistence of this decay until the Heisenberg time would yield $C(t_H) \propto e^{-\beta c L}$. The latter result would suggest thermalization of the system in the $L, t \to \infty $ limit, irrespective of how small the value of the exponent $\beta$ was. Therefore, time evolution consistent with the MBL phase \textit{must necessarily} feature decay of $C(t)$ \textit{slower} than any power-law.

\begin{figure}
    \centering
    \includegraphics[width=1\columnwidth]{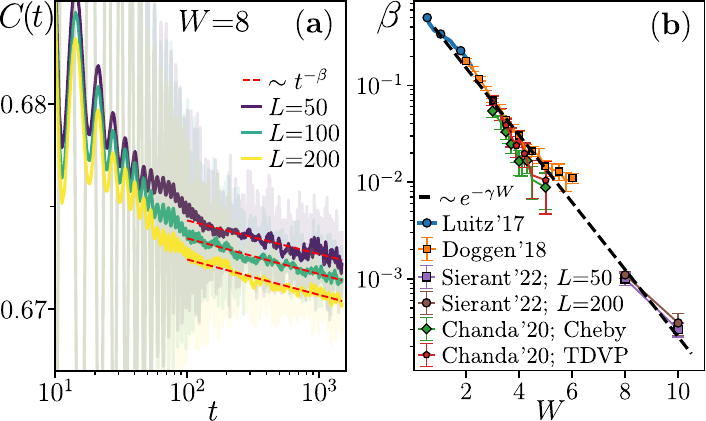}
    \caption{Persistent slow dynamics in the disordered XXZ spin-1/2 chain \eqref{Hxxz}. (a) Deep in the MBL regime, at $W=8$, the density autocorrelation function $C(t)$, calculated with TDVP algorithm, slowly decays, the decay can be fitted with a power-law $C(t) \propto t^{-\beta}$, with $\beta \approx 10^{-3}$.  (b) The dynamical exponent $\beta$ from: \cite{Luitz17b} [Luitz '17], governing the power-law decay of  $\left \langle \hat{S}^z_j(t) \hat{S}^z_j  \right \rangle $ obtained with typicality approach for $L=24$; \cite{Doggen18} [Doggen '18] governing the power-law decay of $C(t)$ for initial N\'{e}el state, calculated for $L=100$ with TDVP; \cite{Chanda20t} for the decay of $C(t)$ for initial N\'{e}el state, calculated with Chebyshev time propagation for $L=26$ (see Sec.~\ref{subsec:numtevol})  [Chanda '20; Cheby], and with TDVP for $L=50$ [Chanda '20; TDVP]; \cite{Sierant22challenges} [Sierant '22] for the decay of $C(t)$ for initial N\'{e}el state calculated with TDVP for $L=50, 200$. We observe an exponential decrease $\beta\propto e^{-\gamma W}$, where $\gamma=0.8$ is a constant. Panel (a) is reprinted and adapted from \cite{Sierant22challenges}.
\label{fig:beta}
}
\end{figure}

\subsubsection{Assessing the breakdown of ergodicity}
Tensor network investigation \cite{Sierant22challenges} of the disordered XXZ spin chain revealed a slow decay of the density autocorrelation function $C(t)$ at time scales of the order of $10^3$ tunneling times \textit{deep} in the MBL regime, which in ED studies at $L=20$ emerges at $W > W^*(L=20) \approx 4$. The decay, accurately fitted by a power-law with a small exponent, was found at $W=8$, as shown in Fig.~\ref{fig:beta}(a), and at $W=10$. Notably, since the value of $C(t)$ changes by a few per milles throughout the examined time interval, the autocorrelation function can be fitted equally accurately with a logarithmic in time dependence, $C(t) \propto \mathrm{const} -\log(t)$. Nevertheless, in this Section, we stick to the power-law fitting of $C(t)$, which describes the dynamics in the disordered spin chains in the broad regime of disorder strengths and time scales. The exponent $\beta$ governing the decay of $C(t)$, extracted with a variety of computational approaches and across several literature references for the disordered XXZ spin chain, decreases \textit{exponentially} with disorder strength $W$, $\beta \propto e^{-\gamma W}$, in a wide interval of $W \in[1,10]$, as shown in Fig.~\ref{fig:beta}(b). This result is another manifestation of the abrupt slowdown of the dynamics of disordered spin chains with the disorder strength $W$. 
Arguments for classical subdiffusive systems relate the exponent $\beta$ with the dynamical exponent $z$~\cite{Luitz17} via $\beta = 1/z$. The exponential decay $\beta \propto e^{-\gamma W}$ is qualitatively consistent with the rapid increase of the exponent $z$ extracted from the analysis of Thouless time with the subdiffusive ansatz $t_{\mathrm{Th}} \propto L^z$~\cite{Suntajs20e}.

The difficulties in extrapolating the time evolution results at larger disorder strengths to the asymptotic limit $L, t \to \infty$ forbid reaching unambiguous conclusions about the presence of the MBL phase in the system. Indeed, each of the Scenarios 1.-2. put forward in the analysis of ED results in Sec.~\ref{subsubsec:gapratio} can be seen as consistent with the time evolution results discussed here.

Scenario 1. assumes that both $W^T(L)$ and $W^*(L)$ converge in the $L\to\infty$ limit to a finite critical disorder strength $W_C$, above which the disordered spin chains are in the MBL phase, according to the definition \eqref{eq:MBL1}. Assuming that $W_C < 8$, consistently with the ED results at the available system sizes $L \sim 20$, the slow decay of $C(t)$ observed between times $t=10^2$ and $t=10^3$ in Fig.~\ref{fig:beta}(a) would be just a transient effect, replaced by a decay slower than an algebraic decay at longer times. In that case, for $L\gg1$, the autocorrelation function $C(t)$ would saturate at long times to a non-zero value, and the system would be in the MBL phase.

In contrast, Scenario 2. assumes that both $W^T(L)$ and $W^*(L)$ diverge, for instance, proportionally to system size $L$, as suggested by~\cite{Suntajs19, Sels20obstruction}, in the $L\to \infty$ limit. In that case, no MBL phase exists in the considered disordered spin chains, and ETH \eqref{eq:ETH2} applies at any disorder strength $W$ for $L\to \infty$. Within this Scenario, the decay of $C(t)$ observed at times $t > 10^2$ for $W=8$ in Fig.~\ref{fig:beta}(a), and for $W=10$ in \cite{Sierant22challenges}, could be a stable feature of the dynamics that persists until the time scales of the order of Heisenberg time, leading to thermalization of the system at sufficiently large system size $L$.

Presently available numerical results do not allow us to distinguish the two Scenarios or point toward any other type of behavior. The decay of $C(t)$ at $W=8$ and $10$ could remain algebraic at longer times, leading to the thermalization of the system. However, we cannot exclude that $C(t)$ decays \textit{slower} than the power-law on longer time scales and is only well fitted by a power-law in relatively narrow intervals of times $10^2 < t < 10^3$. Faced with an alternative of two competing ideas, we could ask which of the Scenarios is simpler given the numerical results at our disposal. Examining the trends in the decay of $C(t)$ for system sizes $12\leq L \leq 20$ at times much smaller than the Heisenberg time $t_H$ and $W=3, 5$, shown in Fig.~\ref{fig:tevol}(b), \ref{fig:tevol}(c), respectively, we may conclude that the increase of system size always leads to a quicker decay that persists to a longer time. From that perspective, demanding that the decay of $C(t)$ presented in Fig.~\ref{fig:beta}(a) slows down at a time larger than $t\approx 10^3$ but still much smaller than $t_H$, seems equivalent to requiring an emergence of a new trend in the data, hence placing an additional assumption on the dynamics of the system. On the other hand, the $C(t)$ decays only by a small fraction of its value in the available time interval $10^2 < t < 10^3$. Moreover, the shift of the crossing point $W^*(L)$ does slow down with $L$ for system sizes available to ED studies. Demanding that the slowdown of that drift ceases to occur at sufficiently large $L$ is yet another assumption about the system in a regime that we cannot presently probe.

The above discussion illustrates that the attempts to answer the question of the existence of the MBL phase based solely on the presently available numerical results belong to the realm of speculation. The exponential slowdown of the dynamics of spin chains with increasing disorder strength is the root of the problems, preventing us from deciding whether the dynamics slow down but the thermalization process continues indefinitely or whether the dynamics get ultimately arrested at sufficiently large disorder strengths and system sizes.
Irrespective of the answer to the question of the MBL phase, the numerically observed MBL regime is very robust in the sense that the dynamics at large $W$ are extremely slow. For instance, if the decay of $C(t)$ continued to be an algebraic decay with the exponent $\beta\approx 10^{-3}$ observed for $W=8$, the $C(t)$ would decay to $10\%$ of its initial value after time  $t \approx 10^{1000}$ tunneling times. In that case, the dynamics would effectively be arrested for all practical purposes.

\subsubsection{Spatial decay of correlations}
\label{subsec:spatial}
The density autocorrelation function $C(t)$ discussed above provides a scalar quantity that probes the approach to thermal equilibrium of the disordered spin chains. A connected two-point correlation function $\Phi(x,t) \equiv \langle \hat{S}^z_{j+x}(t) \hat{S}^z_{j}(0)\rangle - \langle \hat{S}^z_{j+x}(t) \rangle \langle \hat{S}^z_{j}(0)\rangle$ allows to probe the spatial correlations in the system, yielding much more complete information about the dynamics.  A hydrodynamic description of diffusive dynamics~\cite{Michailidis23CorrectionsToDiffusion} predicts that 
\begin{equation}\label{eq_nn_noph}
\Phi(x,t)
	= 
\frac{d_0}{\sqrt{t}}\left[F_{0,0}(y) + \frac{1}{\sqrt{t}} F_{1,0}(y) + O \left(\frac{\log t}{t}\right)\right] ,
\end{equation}
where $F_{0,0}$, $F_{1,0}$ are the scaling functions of the scaling variable $y\equiv x/\sqrt{Dt}$ with the diffusion constant $D$, and $d_0$ is a constant. The leading scaling function $F_{0,0}(y) = e^{-y^2/4}$ solves the diffusion equation, while the sub-leading terms include perturbative corrections \cite{ChenLin19}, derived within the effective field theory of dissipative fluids \cite{Crossley17fluids}. For subdiffusive classical dynamics the scaling functions can present significant subleading corrections, as shown recently in \cite{McRoberts23}. Even in classical systems, where the system size can be pushed to the thousands of spins, the determination of the diffusion coefficient, or the subdiffusion exponent, must pass through the correct modellization of the subleading corrections.

The two-point correlation function $\Phi(x,t)$ was probed with the Chebyshev time propagation algorithm in \cite{Bera17} in the disordered XXZ spin chain of size $L \leq 24$. The analysis found a pronounced non-Gaussian shape of $\Phi(x,t)$, which decays nearly exponentially in $x$ even in the ergodic regime at $W=1.5 <W^T(L)$ \cite{Weiner19}. Analysis of the time dependence of the width of the $\Phi(x,t)$ profile yields a dynamical exponent $z$ significantly larger than the diffusive value $z=2$. However, at longer times, the value of the exponent $z$ becomes strongly dependent on the system size, decreasing towards $z=2$ with an increase of $L$. This prevents us from concluding whether the observed subdiffusive dynamics are a stable property of the disordered XXZ spin chains or whether the features of subdiffusion are transient and disappear at sufficiently large times and system sizes, being replaced by diffusive dynamics. We revisit this question in Sec.~\ref{sec:trans}.

The finite time and finite size effects become even more pronounced when the disorder strength increases and the system reaches the MBL regime. An exponential shape of $\Phi(x,t)$ is still observed in that case. However, despite the slowness of the dynamics, a monotonous decrease of the dynamical exponent $z$ with an increase of the system size $L$ is clearly observed. The latter finding is consistent with the trends towards thermalization found for the autocorrelation function $C(t)$.

\subsubsection{Overview}
The dynamics of local observables is a direct approach to assessing the ergodicity and the class of dynamics in interacting many-body systems. Probing of time evolution is not only directly relevant for experiments with synthetic quantum matter but also numerically feasible, in limited time intervals, for system sizes larger than those accessible in ED studies. 

Studies of thermalization of local observables in disordered spin chains allow us to distinguish the ergodic regime, in which the systems thermalize at the accessible time scales, from an MBL regime in which the dynamics are nearly arrested. Investigations of the time evolution over experimentally relevant time scales show that an increase of the disorder strength $W$ leads to an exponential slowdown of the dynamics. Nevertheless, as we have argued, the presently available results for time evolution lead to ambiguous conclusions about the system's fate in the asymptotic limit $L,t \to \infty$.

\subsection{Entanglement growth}
\label{sec:entanglement}
The growth of entanglement in quench protocols provides another perspective on ergodicity and MBL in disordered many-body systems. While probing the entanglement entropy is more challenging than measurements of correlation functions of local observables, see, e.g., \cite{Elben23}, entanglement entropy growth in disordered many-body system has been experimentally observed in \cite{Lukin19}. In the following, we describe the properties of slow entanglement spreading in the MBL regime. We highlight the trends toward thermalizing behavior, which prevent us from pinpointing the status of the MBL phase, similarly as in the case of the autocorrelation function of local observables.

\subsubsection{Logarithmic growth of entanglement entropy}
\label{sec:entanglementSC}
Understanding the growth of entanglement entropy under unitary dynamics of many-body systems has been a subject of intense efforts over the last two decades. In non-interacting systems~\cite{Calabrese05Evolution}, the entanglement entropy $S(t)$, defined by Eq.~\eqref{eq:entdef0}, increases linearly with time $t$. Such a ''ballistic'' growth of $S(t)$ is also found in generic ergodic systems \cite{Kim13}. The introduction of symmetries may alter this behavior. For instance, the R\'{e}nyi entanglement entropy grows diffusively, proportionally to $\sqrt{t}$, in ergodic models with $U(1)$ symmetry \cite{Rakovszky19SubBallistic}\footnote{This behavior was substantiated in \cite{Huang20} and confirmed numerically in \cite{Han2023entanglement}.}, which, however, is specific to qubit systems in one dimension~\cite{Znidaric20entanglement}. General properties of the entanglement growth in higher dimensional systems, including the leading term quantifying the average value of $S(t)$ and sub-leading terms describing the entanglement fluctuations, are predicted by the minimal membrane picture conjectured in~\cite{Nahum17} and substantiated numerically beyond one dimension in \cite{Sierant23Membranes}.

The power-law growth of entanglement entropy in ergodic systems can be readily contrasted with the slow spreading of entanglement in strongly disordered spin chains~\cite{DeChiara06, Znidaric08}. In the MBL regime, an unbounded logarithmic growth of entanglement entropy $S(t) \propto \log(t)$ is expected~\cite{Bardarson12, Serbyn13a} (see also~\cite{Huang21extensive}), and has been found even in simulations directly in the thermodynamic limit \cite{Andraschko14Purification}. The logarithmic growth of $S(t)$ is consistent with the picture of LIOMs~\cite{Serbyn13b} and in line with the phenomenological arguments of \cite{Vosk13, Ruggiero22}. 

In the crossover regime of the disordered XXZ spin chain, for $W^T(L)<W<W^*(L)$, the entanglement entropy grows sub-ballistically, $S(t) \propto t^{\gamma}$~\cite{Luitz16}. The exponent $\gamma$ characterizing this growth is relatively stable when $L$ is increasing~\cite{Enss17infinite}. Nevertheless, a slight systematic increase of the exponent $\gamma$ [which may be initial state dependent, see~\cite{Prasad22InitialStateDependent}] can be observed when time and length scales grow~\cite{Chanda20t, Evers23Internal}. This opens up a possibility of a gradual emergence of slow thermalization in the disordered spin chains when the logarithmic increase of $S(t)$ is replaced by a slow algebraic growth of the entanglement entropy. Moreover, distinguishing between the algebraic and logarithmic growth becomes not trivial when the dynamics slow down with an increase of $W$, especially when the results are available only in a limited time interval.

\begin{figure}
    \centering
    \includegraphics[width=1\columnwidth]{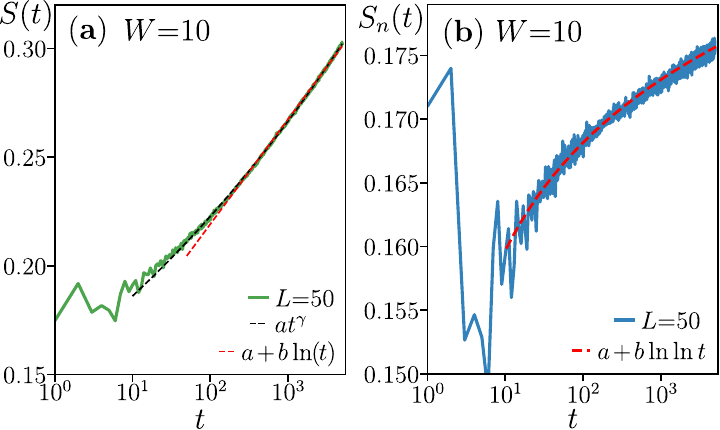}
    \caption{Slow growth of entanglement in the disordered XXZ spin-1/2 chain. (a) Entanglement entropy $S(t)$ deep in the MBL regime, for $W=10$, calculated with TDVP for system size $L=50$, for bipartition of the chain into two contiguous subsystems each containing $L/2$ sites, averaged over more than $10^3$ disorder realizations. Power-law fit $S(t) \propto t^{\gamma}$ with $\gamma \approx 0.08$ describes the results accurately in a longer time interval than a logarithmic fit $S(t) \propto \log(t)$. (b) Number entropy $S_n(t)$ deep in the MBL regime is accurately captured by the double logarithmic fit $S(t) \propto \ln \ln(t)$. Reprinted and adapted from \cite{Sierant22challenges}.
\label{fig:ent}
}
\end{figure}

Deep in the MBL regime of the disordered XXZ spin chain, as exemplified in Fig.~\ref{fig:ent}(a) for $W=10$, a logarithmic time dependence $S(t) \sim \log(t)$ accurately describes the entanglement entropy growth in the interval of times $t \in [200,4000]$. However, examining the entanglement entropy behavior in a broader time regime starting at $t \approx 10$, we see an evident curvature of the $S(t)$ on the log-linear scale. Consequently, a power-law increase $S(t) \sim t^{\gamma}$ with exponent $\gamma \ll 1$ describes the results rather well in such broad time interval. This behavior appears on the same time scales and system sizes as the slow decay of the density autocorrelation function $C(t)$ shown in Fig.~\ref{fig:beta}(a). Hence, the better adherence of $S(t)$ to power-law fit is another manifestation of the tendency towards slow thermalization in the system. This conclusion must be, however, treated with caution. Firstly, we cannot exclude that the logarithmic time dependence could describe the increase of $S(t)$ better than the power-law at time scales longer than $t>4000$. Moreover, a comparison with predictions of the LIOM model of~\cite{Chavez2023ultraslow} reveals an apparent curvature of $S(t)$ on the log-linear scale in time interval $t \in[10, 10^4]$. This behavior is entirely consistent with the results presented in Fig.~\ref{fig:beta}(a), while the LIOM model considered in~\cite{Chavez2023ultraslow} possesses, by construction, all features ascribed to the MBL phase, including the exponentially localized LIOMs and exponentially decaying interactions between them. 

A better understanding of the dynamics of LIOM models similar to the one considered by~\cite{Chavez2023ultraslow} will allow us to formulate more quantitative predictions on the form of the growth of the entanglement entropy $S(t)$, consistent with the emergence of the MBL phase in the system. The asymptotic logarithmic growth, $S(t)\sim \log(t)$, of entanglement entropy, which is expected to be a signature of the MBL phase, is not easily distinguishable from an algebraic growth of $S(t)$, which would signify thermalization of the system, again disallowing clear conclusions about the presence and the extent of the MBL phase. While we have concentrated here on entanglement entropy, the time evolution of different, more accessible entanglement measures, such as concurrence, has been investigated in the MBL regime~\cite{Iemini16signatures, Campbell17}. Concurrence provides a spatiotemporal characterization of the entanglement dynamics in strongly disordered spin chains, revealing striking similarities with the dynamical heterogeneity in classical glasses~\cite{Artiaco22heterogeneity}.

\subsubsection{Symmetry resolved entanglement: number entropy}
\label{sec:entanglementSN}
The presence of globally conserved quantities allows a  decomposition of the density matrix of a subsystem of the many-body system into orthogonal sectors, giving rise to the notion of symmetry resolved entanglement~\cite{Goldstein18SymmetryResolved, Murciano20symmetryResolved, Turkeshi20}, and uncovering counter-intuitive behaviors such as the quantum Mpemba effect~\cite{Ares23}. In particular, the conservation of the total  spin $z$ component, $\hat{S}^z \equiv \sum_{i=1}^{L} \hat{S}^z_i$, i.e., the $U(1)$ symmetry of the disordered XXZ spin chain \eqref{Hxxz}, opens up the possibility of studying a number entropy at the ETH-MBL crossover, providing another perspective on the dynamics of the disordered spin chains.

We consider a bipartition of the chain into subsystems $A$ and $B$, and denote $\hat{S}^z_A = \sum_{i\in A}\hat{S}^z_i$, $\hat{S}^z_B = \sum_{i\in B}\hat{S}^z_i$. The $U(1)$ symmetry implies that the total spin $z$ component commutes with the density matrix  $\rho =\ket{\psi}\bra{\psi}$ corresponding to the state of the system $\ket{\psi}$, $[ S^z, \ket{\psi}\bra{\psi}]=0$. Taking the trace over the subsystem $B$ of the latter equation, and using the identities $\mathrm{tr}_B \left( \hat{S}^z_A \rho \right)=\mathrm{tr}_B \left( \rho \right) \hat{S}^z_A$ and  $\mathrm{tr}_B \left( \hat{S}^z_B \rho \right)=\mathrm{tr}_B \left( \rho \hat{S}^z_B  \right)$, we get 
\begin{equation}
 [ \hat{S}^z_A,\hat{\rho}_A] = 0,   
\end{equation}
implying that the reduced density matrix $\hat{\rho}_A$ of the subsystem $A$ has a block diagonal structure, with blocks indexed by the eigenvalues $s\in[-L_A/2,\ldots,L_A/2]$ of the $\hat{S}^z_A$ operator. Denoting by $\Pi_s$ a projector into the subspace  of $\hat{S}^z_A$ to eigenvalue $s$, we have
\begin{equation}
  \hat{\rho}_A = \sum_s {p_s} \hat{\rho}_{A,s},
  \label{eq:rhoBLOCK}
\end{equation}
where $p_s = \mathrm{tr}_A\left( \Pi_s \hat{\rho}_A \Pi_s \right) $ and $\hat{\rho}_{A,s} \equiv \Pi_s \hat{\rho}_A \Pi_s/p_s $. The block structure in \eqref{eq:rhoBLOCK} allows us to rewrite the entanglement entropy \eqref{eq:entdef0} as $S = S_n +S_c$, where
\begin{eqnarray}
  S_n &=& - \sum_s {p_s} \ln(p_s),
  \nonumber \\
  S_c &=& - \sum_s {p_s} \mathrm{tr}_A \left( \hat{\rho}_{A,s} \ln(\hat{\rho}_{A,s}) \right)\;.
\label{eq:number}
\end{eqnarray}
The quantity $S_n$ is known as the number entropy~\cite{Schuch04,Schuch04b,Donnelly12,Lukin19}, since it is related to the distribution of probabilities of finding the subsystem $A$ in various eigenspaces of the $\hat{S}^z_A$ operator, which is, upon Jordan-Wigner transformation, equal to the particle number operator in the subsystem $A$, up to a constant shift. The quantity $S_c$, called configurational entanglement entropy,  stems from superpositions of states from sector with fixed value of $\hat{S}^z_A$ but with different spin configurations.

For systems of free fermions, the number entropy $S_n$ can be used to bound the entanglement entropy $S$ from below and above, as derived in~\cite{Kiefer20bounds}. The bounds imply an asymptotic scaling $S \propto \exp\left( S_n \right)$, showing that the growth of the entanglement entropy always implies a growth, although logarithmically slower, of the number entropy, and vice versa\footnote{The free fermion bounds in~\cite{Kiefer20bounds} were derived for R\'{e}nyi entropy with index $q=2$, but may be expected to hold for any $q$.}. These observations prompted an investigation of the number entropy in the strongly disordered XXZ spin chain~\cite{Kiefer20}.

For a many-body system with dynamical critical exponent $z \geq 2$, the number entropy is expected to grow as $S_n(t) \propto \frac{1}{z}\ln(t)$~\cite{Kiefer20}. In that case, the free fermion bounds extended to the interacting case would predict a power-law growth of the entanglement entropy $S(t)$, consistently with the expectations for ergodic systems. In contrast, in the MBL phase, one expects that the particle number fluctuations between the subsystems cease to increase at sufficiently large times~\cite{Bardarson12, Singh16number} and that the number entropy $S_n(t)$ saturates. 

Numerical results of \cite{Kiefer20} for the disordered XXZ spin chain indicate, instead, a double logarithmic increase of the number entropy, $S_n(t) \propto \ln \ln (t)$. This behavior, consistent with the relation for non-interacting models, $S \propto \exp\left( S_n \right)$,  was confirmed by a TDVP study~\cite{Sierant22challenges} for $L=50$, deep in the MBL regime of the disordered XXZ spin chain at $W=10$, see Fig.~\ref{fig:ent}(b). The double logarithmic growth of $S_n(t)$ was interpreted~\cite{Kiefer20} as a manifestation of persistent particle transport at a very slow rate, signifying the absence of the MBL phase. This leads to the proposition of an alternative picture of entanglement growth in MBL regime~\cite{Kiefer21, Kiefer21Unlimited}, in which $S(t) \propto \ln(t)$ and $S_n(t) \propto \ln \ln(t)$.

A subsequent study of the number entropy~\cite{Luitz20} suggested that the double logarithmic growth of $S_n(t)$ is a transient feature associated with a single particle fluctuating across the boundary between the two subsystems, ruling out the slow particle transport deep in the MBL regime. Similarly, \cite{Ghosh22} argued that the double logarithmic growth of the number entropy may be compatible with the MBL phase in the system. This triggered a comment with counter-arguments~\cite{Kiefer22comment} and, subsequently, a response~\cite{Ghosh22response}. The latter concluded that presently available numerical results could agree with both hypotheses, and access to larger system sizes and time scales would be needed to confirm or convincingly rebuke the arguments of \cite{Kiefer20, Kiefer22particle}.

The observation of the double logarithmic growth of the number entropy prompted a study of $S_n(t)$ in a LIOM circuit model~\cite{Chavez2023ultraslow}, which is, by construction, in MBL phase in the $L,t\to\infty$ limit. For parameters of the LIOM circuit that lead to the time dynamics parametrically similar to the disordered XXZ spin chain considered in~\cite{Luitz20}, the LIOM circuit shows a clear growth $S_n(t) \propto \ln \ln (t)$. Notably, the conclusion about the double logarithmic growth of number entropy persists to the largest system size $L=28$ considered in \cite{Chavez2023ultraslow}, larger than the system size examined in \cite{Kiefer20}. The slow growth of $S_n(t)$ persists until a time that scales exponentially with $L$, similarly to the Heisenberg time $t_H$. This finding suggests that the ultraslow growth of the number entropy growth may be a feature inherent to the MBL phase, even though the presently available results for the LIOM circuits model do not fully exclude the possibility that the growth of $S_n(t)$ may saturate in LIOM circuits at larger times and system sizes. Moreover, as suggested by~\cite{Kiefer20}, the $S_n(t) \propto \ln \ln (t)$ is reminiscent of the dynamics for which the time scale $t_{sp}$ for spreading of particles over length $\ell$ is exponentially large, $t_{sp} \propto e^{\ell}$. Depending on the factors in that relation, the time scale for reaching ergodicity, i.e., spreading particles over the entire system, $t_{sp}$ for $\ell=L$, may scale faster or slower than the Heisenberg time depending on the prefactors in the scaling. This may or may not enable a transition between ergodic and MBL phases in the disordered spin chains, and its resolution remains an open question.

\subsection{Extracting universal features of the slow dynamics}
\label{sec:extracting}

The results discussed in this Section demonstrate that comprehending the putative ETH-MBL transition in the time domain is an outstanding challenge. The regime of slow (glassy) dynamics, characterized by persistent decays of local correlators and a gradual growth of entanglement entropy, extends to the largest system sizes and time scales available in present-day numerical simulations. In the following, we outline various approaches aimed at providing a thorough understanding of the slow dynamics in disordered spin chains.

In contrast to the broad regime of glassy dynamics in disordered spin chains, transitions in the eigenstate structure of quantum systems with a well-defined transition point are typically associated with an evident change of character of the dynamics, at least sufficiently far away from the transition. For instance, the diffusive dynamics at small disorder strength $W$ in the 3D Anderson model can be plainly contrasted with localized dynamics at large $W$ already at moderate system sizes (see also Sec.~\ref{subsec:unitTEV}). While a direct observation of the subdiffusive dynamics at the transition requires larger time and length scales \cite{Ohtsuki97, Sierant20thouless}, a proper rescaling of the time evolution results~\cite{Hopjan23survival} allows to observe the universal features of the transition even at times much shorter than the Heisenberg time $t_H$~\cite{Hopjan23dynamics,Jiricek23observables}, see also~\cite{Bhakuni23dynamic}. Consequently, a comprehensive identification of the features of the slow dynamics in disordered spin chains may be conceptually and practically more relevant than only limiting the discussion to the question about the existence of the MBL phase in the $L, t\to \infty$ limit.

\subsubsection{Proximity to Anderson Insulator}
\label{sub:proximity}
Whether the Anderson localization may persist in the presence of interactions was the original formulation of the MBL problem~\cite{Basko06}. The possibility of investigating MBL in the disordered XXZ spin chain \eqref{Hxxz} from the perspective of this model's non-interacting limit ($\Delta=0$) allows for non-trivial insights into the physics of strongly disordered spin chains.

Even in the ergodic regime of the disordered XXZ spin chain, e.g., at $W=1.5$ and $L=16$, the spectral function  $f^2(\omega)$ (defined in Eq.~\eqref{eq:f2a}) of the non-interacting model ($\Delta=0$) provides, at sufficiently large frequencies $\omega$, a good approximation of $f^2(\omega)$ in the presence of interactions ($\Delta=1$). This demonstrates the resemblance of the dynamics on relatively short time scales in the interacting and non-interacting cases. Significant differences between $\Delta=0$ and $\Delta=1$ appear only at small frequencies $\omega$, corresponding to evolution on longer time scales, c.f. Fig.~\ref{fig:tevol}(a). The presence of the Anderson localized phase at $\Delta = 0$ allowed \cite{Vidmar21phenomenology} to propose a phenomenological model describing the behavior of the spectral function $f^2(\omega)$ at \textit{small} frequencies. 

The main premise is that the Anderson LIOMs $\hat{Q}_j$, corresponding to the localized single-particle orbitals of the Anderson insulator, may determine the memory of the initial state kept in the system at late times, and hence give rise to anomalously slow dynamics.
In the strict non-interacting case, $\Delta=0$, the Anderson LIOMs contribute to the $\omega\to 0$ limit of the spectral function as  $f^2(\omega)\stackrel{\omega \ll 1}{=}C_0 \delta(\omega)$. When the interactions are introduced, $\Delta=1$, the Anderson LIOMs are no longer stable and decay in time. Calculation of the correlator $\braket{\hat{Q}_j(t) \hat{Q}_j}$ allows to extract a time scale $\tau$ at which the Anderson LIOM $\hat{Q}_j$ decays. The key finding of \cite{Vidmar21phenomenology} is that the distribution 
$w(\tau)$ of the numerically extracted relaxation time scales
$\tau$ becomes extremely broad with increasing $W$, and is well captured, in a wide range of $\tau$, by a power-law dependence 
$w(\tau)\propto \tau^{-\mu}$,
with an exponent $\mu$ dependent on the disorder strength $W$. The finite lifetime of Anderson LIOMs amounts to the broadening of the Delta peaks in $f^2(\omega)$ at $\omega=0$ present in the non-interacting case. Assuming a Lorentzian broadening of the Delta peaks,  the following low-frequency behavior of the spectral function was obtained \cite{Vidmar21phenomenology},
\begin{equation}
    f^2(\omega) \propto \int_{\tau_{\mathrm{min} } }^{\tau_{\mathrm{max} } } d\tau
    \frac{1}{\tau^{\mu-1}} \frac{1}{(\omega \tau )^2+1},
\label{eq:f2phenom}
\end{equation}
where $\tau_{\mathrm{min}}$ and $\tau_{\mathrm{max}}$ correspond to minimal and maximal time scales at which the weight of the distribution 
$w(\tau)$
becomes negligible. For $\omega \ll \tau_{\mathrm{max}}^{-1}$, the spectral function has a plateau, $f^2(\omega) \propto \mathrm{const}$, which can be used to define the Thouless energy, see Sec.~\ref{subsec:spectral}. The plateau appears in the ergodic regime, $W<W^T(L)$, and shrinks with the growth of $W$ as $\tau_{\mathrm{max}}$ increases. When $\tau_{\mathrm{max}}$ approaches the Heisenberg time $t_H$ at $W\approx W^T(L)$ (for instance, in the disordered XXZ spin chain at $L=16$ and $W\approx 2$), the plateau disappears. At that point, the exponent $\mu$ is close to unity, which corresponds to the $f^2(\omega) \propto 1/\omega$ behavior observed in \cite{Mierzejewski16, Sels20obstruction} for the imbalance or the local $\hat S_j^z$ operator. The exponent $\mu$ increases upon a further increase of $W$, corresponding to the behavior $f^2(\omega) \propto 1/\omega^{2-\mu}$ observed in Fig.~\ref{fig:f2}(b), and the regime of the slow persistent decays of local autocorrelation functions emerges. 

The main consequence of the phenomenological model of the spectral function \eqref{eq:f2phenom} is that the dynamics of disordered spin chains may be perceived as being governed, in a wide interval of $W$, by the broad distribution of the relaxation times of Anderson LIOMs. Even in the MBL regime, $W>W^*(L)$, $\tau_{\mathrm{min}}$ is significantly smaller than $t_H$, while $\tau_{\mathrm{max}} $ exceeds the Heisenberg time. Accordingly, in that regime, a certain fraction of Anderson LIOMs remain stable and the remaining fraction delocalize, corresponding to the non-trivial dynamics observed in the MBL regime, see, e.g., Fig.~\ref{fig:beta}(a).

The presence of the Anderson localized phase at $\Delta=0$ motivates the question of how the interactions affect the dynamics of the Anderson insulator. To answer this question quantitatively, we note that the Hamiltonian of the disordered XXZ spin chain \eqref{Hxxz}, can be expressed as $ \hat{H} = \hat{H}_0 +\hat{H}_{\mathrm{int}} $, where
\begin{equation}
 \hat{H}_{0} = \sum_{\alpha=1}^{L} \epsilon_\alpha \hat{Q}_\alpha 
\end{equation}
is diagonal in the basis of Anderson orbitals, and the interaction part $H_{\mathrm{int}} \equiv \sum_{j=1}^{L} \Delta \hat{S}^z_j \hat{S}^z_{j+1}$ is a perturbation of the Anderson model\footnote{We note that we express the XXZ Hamiltonian in terms of spin-1/2 operators, while the notion of single-particle Anderson orbitals is usually invoked in therms of spinless fermion Hamiltonians, such as the one in Eq.~\eqref{eq:Hand}, onto which the spin-1/2 Hamiltonians can be mapped via the Jordan-Wigner transformation.}. The important observation of \cite{Krajewski22Restoring} is that the interaction term can be split into two orthogonal parts as 
\begin{equation}
 \hat{H}_{\mathrm{int}  } = \hat{H}^{\parallel}_{\mathrm{int}  } + \hat{H}^{\perp}_{\mathrm{int}  },
\end{equation}
where the orthogonality is defined by the Hilbert-Schmidt product, $\braket{A,B} = \frac{1}{\mathcal N} \mathrm{tr} \left( A^{\dag}B \right)$, and ${\cal N}$ is the corresponding Hilbert space dimension. The division of the interaction term is performed in such a way that $[\hat{H}^{\parallel}_{\mathrm{int}  }  ,  \hat{Q}_j]= 0$, i.e., the Anderson LIOMs commute with $\hat{H}^{\parallel}_{\mathrm{int}  }$. Consequently, only the term $\hat{H}^{\perp}_{\mathrm{int}  }$ is a genuine non-trivial perturbation to the Anderson insulator. 

In the limit of large disorder strength $W$, the Anderson LIOMs $\hat{Q}_j$ become identical to the local spin operators $\hat{S}^z_j$. In that limit, the interaction term $\hat{H}_{\mathrm{int}}  =\sum_j \Delta \hat{S}^z_j \hat{S}^z_{j+1}$ already commutes with Anderson LIOMs, and $\hat{H}^{\perp}_{\mathrm{int}  }=0$ for  $W\to \infty$. The Anderson LIOMs get broadened for finite disorder strengths, and the perturbation to Anderson insulator, $\hat{H}^{\parallel}_{\mathrm{int}  }$, is non-vanishing. Nevertheless, at strong disorder $W$, the interaction term, $\Delta \hat{S}^z_j \hat{S}^z_{j+1}$, is still largely parallel to the Anderson LIOMs due to their exponential localization. Quantitatively, \cite{Krajewski22Restoring} showed that $|| \hat{H}^{\parallel}_{\mathrm{int}  }||\propto 1/W$ (where $||A|| = \sqrt{\braket{A,A}}$), i.e., the norm of the true perturbation of the Anderson insulator decays as $1/W$. In contrast, the norm of the non-interacting Hamiltonian increases as $||\hat{H}_{0} || \propto W$. 

Therefore, at sufficiently large $W$, the perturbation $\hat{H}^{\perp}_{\mathrm{int}  }$ becomes rather weak, and the disordered XXZ spin chain may be viewed as a weakly perturbed Anderson insulator. A weak but non-vanishing perturbation results in slow dynamics towards the thermal equilibrium, sharing certain similarities with the weakly perturbed integrable models~\cite{Mierzejewski22multiple, Surace23weak}, posing a challenge to understanding the fate of the disordered spin chains in the $L, t \to \infty$ limit. Instead, when the perturbation term $\hat{H}^{\perp}_{\mathrm{int}  }$ is rescaled to match the norm of $\hat{H}_0$, the ensuing model remains ergodic at any disorder strength $W$~\cite{Krajewski22Restoring}. Similarly, the ergodicity is restored upon an introduction of a strong pair hopping term~\cite{Krajewski23twobody}, which introduces non-trivial perturbation to the Anderson insulator with norm $|| \hat{H}^{\perp}_{\mathrm{int}  }|| $ of the similar value as the norm of $\hat{H}_0$.

The discussed results outline the perspective on the disordered spin chains provided by the presence of the non-interacting, Anderson-localized limit. Further investigations of the interplay of Anderson LIOMs $\hat{Q}_\alpha$ with interactions that induce their gradual decay may lead to new insights into the physics of slow dynamics in strongly disordered spin chains.

\subsubsection{Stretched exponential decays}
\label{sub:stretched}
For disorder strengths roughly corresponding to the edge of the ergodic regime at $W\approx W^T(L)$, e.g. at disorder strengths $W \approx 2.5$ for the disordered XXZ spin chain of size $L=20$, the decay of the autocorrelation functions of local observables, such as $C(t)$ defined in \eqref{eq:den}, is manifestly faster than the power-law decay found at larger $W$. Instead, the faster decay in that regime can be well-fitted~\cite{Lezama19, Haldar23} by a stretched exponential dependence 
\begin{equation}
    C(t)  = A e^{-(t/\tau)^\kappa},
    \label{eq:stretched}
\end{equation}
with exponent $\kappa < 1$ and a characteristic time scale $\tau$ which increases with $W$.

The power-law decay of local correlators may be attributed to rare fluctuations in the disorder, i.e., to Griffiths regions, which act as bottlenecks for transport~\cite{Agarwal17}. The observed stretched exponential decay at $W\approx W^T(L)$ may be explained by the resonances between many-body states with locally differing magnetization patterns~\cite{Crowley21}. Such resonances manifest themselves as large oscillations of local correlators and average out when the disorder average is taken, c.f. Fig.~\ref{fig:tevol}. This motivated \cite{Long22Prethermal} to propose, without making any predictions about the value of $W_C$ nor about the mechanism of the possible transition to MBL phase,  
that many-body resonances characterize thermalization in most of the prethermal MBL regime, i.e., at disorder strengths $W^T(L) \lessapprox W \lessapprox W_C$ at which the system ultimately thermalizes in the $L,t \to \infty$ limit.
In contrast, while the rare-region effects may play an important role at larger disorder strengths, ~\cite{Long22Prethermal} suggested that the rare-region effects are not relevant at those intermediate disorder strengths.
The argument relies on the absence of the fat power-law tails in the resistance distribution [fat tails are consistent with the presence of Griffiths rare regions in the disordered XXZ spin chain~\cite{Schulz20}], and on the fact that power-law decays are also observed in quasiperiodic systems~\cite{Doggen19} in which the rare-regions are absent, see also Sec.~\ref{subsec:QPD}. Importantly, the many-body resonances are proposed as an intermediate step towards thermalization associated with establishing a local equilibrium. At longer times, the hydrodynamic behavior characterized by the power-law decays of local observables is expected to take over.

The claim about stretched exponential decay is substantiated numerically by the Jacobi algorithm for iterative matrix diagonalization \cite{Golub00}. The algorithm, see also~\cite{Long23beyond}, identifies the largest off-diagonal matrix element of the currently considered rotated Hamiltonian $\hat{H}(j)$. It performs a unitary transformation $\hat{R}$ to eliminate the identified element from the rotated matrix, $\hat{H}(j+1) = \hat{R}^{\dag} \hat{H}(j) \hat{R}$, so that the off-diagonal weight of $\hat{H}(j)$ decreases at each step $j$. Resonances between respective many-body states are found when the decimated off-diagonal matrix element is much larger than the corresponding difference in diagonal elements. This observation allowed \cite{Long22Prethermal} to extract a broad, power-law distribution of resonance frequencies and argue that the hierarchy of successive resonances leads to the stretched exponential decay of autocorrelators, unless the system flows towards the MBL phase and the dynamics become arrested.

Nevertheless, the stretched exponential prediction breaks down, for $L=20$, already at $W=3$ for the disordered XXZ spin chain. Indeed, as shown in Fig.~\ref{fig:tevol}(b), the decay of $C(t)$ may be well fitted by a power-law or logarithmic dependence in time\footnote{The decay of $C(t)$ at $W=3$ is \textit{consistent}, for instance, with a power-law. The amount by which $C(t)$ decays in the available time interval precludes any strong statements about the functional form of the decay.}. Still, a comparison of the $W=3$ case with the behavior of $C(t)$ at $W=2$, which is well fitted by a stretched exponential \cite{Long22Prethermal}, indicates a possibility for the stretched exponential decay of $C(t)$ to be observed at $W=3$ for system size slightly larger than $L=20$. An implication of the assumed stretched exponential decay for $W=8$ would be that the slow decay of $C(t)$, well fitted by a power-law in Fig.~\ref{fig:beta}(a), could become a faster decay at longer times. However, even under that assumption, the time scale of thermalization for a stretched decay, consistent with the results in Fig.~\ref{fig:beta}(a), would be enormous, and the system would remain non-ergodic for any practical purpose. The broad distribution of time scales associated with the hierarchy of resonances uncovered by the Jacobi rotations is a property of strongly disordered spin chains similar to the broad distribution of the decay times of Anderson LIOMs~\cite{Vidmar21phenomenology}. Understanding the relation between the two approaches is an interesting direction for further investigations of the regime of slow dynamics in strongly disordered spin chains.

\subsubsection{Internal clock of many-body dynamics}
Searching for a paradigm to provide a universal description of the regime of slow dynamics in the disordered spin chains, \cite{Evers23Internal} made the following observation. Considering a fixed value of disorder strength $W$, the dynamics of a specified observable $A(t)$ for two particular disorder realizations differ due to sample-to-sample fluctuations in the properties of the system. Extraction of universal trends in the dynamics of the observable $A$ could be more straightforward if one introduces an \textit{internal clock} that measures the flow of a fictitious time $\Gamma$. If the quantity $\Gamma$ is appropriately chosen, the dynamics of $A$, expressed as a function of $\Gamma$, should show at least a partial synchronization between the samples, simplifying the time evolution analysis.

As the internal clock,~\cite{Evers23Internal} proposed to employ the entanglement entropy, 
\begin{equation}
    \Gamma = S(t).
\end{equation}
Deep in the ergodic regime of disordered spin chains, the entanglement spreads ballistically, $S(t)\propto t$, and the fictitious time measured by the internal clock is proportional to the real-time, $\Gamma \propto t$. With the increase in the disorder strength, the relation between the entanglement entropy and time ceases to be linear. Additionally, the fictitious time $\Gamma$ flows differently for separate disorder realizations, and the relation between the considered observable $A$ and $\Gamma$ becomes non-trivial. Focusing on the density autocorrelation function $C(t)$, and its fluctuation $\mathcal F$ in a time interval of a fixed length $\Delta t$, \cite{Evers23Internal} demonstrated that the sample-to-sample fluctuations of $C(\Gamma)$ and $\mathcal F(\Gamma)$ are significantly reduced compared to  $C(t)$ and $\mathcal F(t)$.

The main finding of~\cite{Evers23Internal} is that the disorder averages of $C(t)$ and $\mathcal F(t)$, plotted as functions of the disorder-averaged entanglement entropy $S(t)$, collapse onto a single master curve which depends only on the system size for different values of $W<8$ in the disordered XXZ spin chain. Thus,  $S(t)$ works well as the internal clock $\Gamma$, enabling the identification of a high degree of universality in the regime of slow dynamics in the disordered XXZ spin chain. Consequently, in the considered interval of system sizes and time scales, there are no visible signatures of the MBL phase. Independently of the disorder strength $W$, the autocorrelation function $C(t)$ decay is dictated by the flow of the fictitious time $\Gamma$, i.e., the increase of entanglement entropy $S(t)$. In contrast, an MBL phase would require the emergence of another branch of the master curve for $L\to \infty$, along which the fictitious time grows indefinitely (since $S(t) \propto \log(t)$), and the autocorrelation functions do not decay. The absence of such a branch of the master curve at the time scale $t<10^3$ considered in \cite{Evers23Internal} is consistent with the persistent decay of the autocorrelation function $C(t)$ shown in Fig.~\ref{fig:beta}(a). Therefore, a smoking gun evidence for the MBL phase, if it exists, should be looked for at larger time scales and system sizes.

\subsubsection{Operator growth}
\label{subsec:operatorGR}
Operator spreading, or operator growth, is a fundamental aspect of the dynamics of quantum many-body systems providing links between quantum chaos, entanglement, transport properties, and quantum information propagation~\cite{Roberts15fourpoint, Ho17Entanglement, Nahum18operator, Khemani18spreading, Keyserlingk18}. Out-of-time-order correlators (OTOC) provide one way of examining the information propagation in isolated quantum systems. Deep in the ergodic regime of the disordered XXZ spin chain, the information propagates behind a ballistically moving front, which can be linked to the ballistic growth of the entanglement entropy~\cite{Luitz17}. When the disorder increases towards the MBL regime, the information front spreads algebraically in time and crosses over to a logarithmic in time spreading at large $W$~\cite{Yichen17, Fan17, He17}.

Investigating the Krylov complexity is a possible approach to quantifying the operator spreading. Consider an autocorrelation function $C(t) = \frac{1}{\mathcal{N}}\mathrm{tr} \left( \hat{A}(t) \hat{A} \right )$, where $A$ is a local observable, and, for simplicity, we have considered the infinite temperature ensemble of states. Defining a Liouvillian superoperator $\mathcal L(\hat{A} ) \equiv [\hat{H}, \hat{A}]$ and using the Baker-Campbell-Hausdorff formula, the autocorrelation function can be written as
\begin{equation}
C(t) = \braket{\hat{A} , \exp(i \mathcal L t) \hat{A} };,
\end{equation}
where $\braket{ \hat{A} , \hat{B} } \equiv \frac{1}{\cal N} \mathrm{tr}(\hat{A}^\dag \hat{B})$ is the Hilbert-Schmidt inner product. The latter equation indicates that the spreading of the operator $A$ under the Hamiltonian dynamics can be analyzed by examination of the sequence of operators $\{ \mathcal L^n \hat{A} \} $ obtained by consecutive actions of the Liouvillian on $\hat{A}$. Employing the Lanczos algorithm \cite{Lanczos50} to construct the Krylov basis of the subspace $\{ \mathcal L^n \hat{A} \} $ of the operator space, one obtains a sequence of positive numbers, Lanczos coefficients, $\{ b_n \}$. According to the hypothesis conjectured in \cite{Parker19growth}, in $d$-dimensional chaotic systems, the growth of the Lanczos coefficients is as fast as possible, i.e., linear (up to logarithmic corrections for $d=1$),
\begin{equation}
    b_n = \alpha n + \gamma,
\end{equation}
with real constants $\alpha>0$ and $\gamma$.  In that case, the growth of operator complexity is maximal, see also~\cite{Suchsland23krylov}. An effective transfer of quantum information into highly non-local operators, corresponding to this behavior, constitutes a basis for time evolution algorithms tailored for ergodic systems~\cite{Banuls09, Perez22, Lerose23barier, Artiaco23efficient}. In some cases, these approaches may overcome the entanglement barrier of the standard tensor network algorithms for time propagation outlined in Sec.~\ref{subsec:tenNET}. When the dynamics of many-body systems cease to be chaotic, the operator spreading is hindered~\cite{Rabinovici22, Rabinovici22b}. Interestingly, the Lanczos coefficients in the MBL regime in $d=1$ dimensional systems scale as $b_n \propto n /\ln(n)$~\cite{Ballar22}, the same as in the ergodic systems.
However, an additional even-odd alteration of $b_n$ and an effective randomness of the obtained sequence $b_n$ in the MBL regime of disordered many-body systems lead to the localization of a local operator $A$ in the Krylov basis, preventing its spreading.

An alternative approach introduced in the context of quantum chaos in \cite{Avdoshkin20Euclidean} is to consider the time evolution of observable $A$ in Euclidean time $\tau$, $\hat{A}(\tau) = e^{\tau \hat{H} } \hat{A}e^{-\tau \hat{H} } = \exp( \mathcal L \tau) \hat{A}$. In contrast to the real-time evolution, the norm of $\hat{A}(\tau)$ is not conserved during the Euclidean time evolution, and its growth is related to the class of the dynamics generated by the Hamiltonian $\hat{H}$. To analyze the growth of $||\hat{A}(\tau)||$, it suffices to focus on the norms of the operators $\mathcal L^n A$ generated during the Euclidean time evolution. The following bound was introduced for the dynamics under local interacting Hamiltonian $\hat{H}$ in dimension $d=1$ \cite{Avdoshkin20Euclidean},
\begin{equation}
   || \mathcal L^n \hat{A}  || \leq ||\hat{A}|| B_n(2) (2J)^n,
   \label{eq:eucl}
\end{equation}
where $J$ characterizes the local couplings of $\hat{H}$, and $B_n(2)$ are Bell polynomials. For large $n$, the Bell polynomials scale like $B_n(2) \propto (n/\ln(n))^n$. For chaotic dynamics, the bound \eqref{eq:eucl} may be expected to be tight, i.e., for ergodic systems the growth of $ || \mathcal L^n \hat{A}  ||$ is expected to be faster than exponential and nearly factorial in $n$. 
Defining \textit{a localized} interacting system as a system in which the operator $ \mathcal{L}^n \hat{A}$ remains exponentially localized with a characteristic length scale $\xi_{\rm loc}$ (equivalently, the operator $\hat{A}$ remains localized when commuted $n$ times with Hamiltonian),~\cite{Weisse24OperatorSpreading} derived the following bound on the norm
\begin{equation}
   || \mathcal L^n \hat{A}  ||\leq ||\hat{A}|| \frac{2^{\xi_{\rm loc}}}{(\xi_{\rm loc}-1)!}(2\xi_{\rm loc}CJ)^n.
   \label{eq:euclLOC}
\end{equation}
The latter inequality shows that the growth of the norm  $|| \mathcal L^n \hat{A}  ||$ is at most exponential with $n$ in a localized interacting system. Similarly, for free-fermionic systems, the growth $|| \mathcal L^n \hat{A}  ||$ is also at most exponential in $n$. Performing symbolic calculations of the nested commutator  $|| \mathcal L^n \hat{A}  ||$ for the disordered XXZ spin chain, \cite{Weisse24OperatorSpreading} find that the ratio $\frac{|| \mathcal L^{n+1} \hat{A}  ||}{|| \mathcal L^{n} \hat{A}  ||}$ keeps increasing with $n$, even at disorder strength $W=25$. The largest considered order $n$ is equal to $16$, which corresponds to operators that spread over $L=33$ lattice sites, which allows \cite{Weisse24OperatorSpreading} to conclude that there are no convincing signatures of the MBL phase visible at the probed length scale. A comparison of the ratio $\frac{||\mathcal L^{n+1} \hat{A}  ||}{||\mathcal L^{n} \hat{A}  ||}$ between interacting and non-interacting cases shows that the growth is similar in the two cases at small $n \lesssim 8$, and it then increases its pace significantly in the interacting model. This finding parallels the observations that the non-interacting limit of the disordered XXZ spin chains provides a good proxy for the dynamics on relatively short time scales in the interacting case, provided that the disorder is sufficiently strong.

Finding quantitative relations of the Euclidean time operator spreading with the persistent decays of local autocorrelation functions observed in the time evolution approaches is an interesting avenue for further research. Understanding the properties of operator spreading in models which are by definition MBL, such as \cite{Chavez2023ultraslow}, or obtaining mathematical insights into operator spreading in disordered systems~\cite{Elgart2022localization, Elgart2023slow}\footnote{ Arguments for the stability of slow Hamiltonian dynamics to generic local perturbations are presented in~\cite{Toniolo24stabilityslow}.}, could shed new light on distinguishing the regime of slow dynamics in strongly disordered spin chains from the properties of the dynamics in MBL phase in interacting models.

\subsubsection{Dynamics of survival probabilities} \label{sec:survival}
Survival probability is another quantity that allows the assessment of the ergodicity and ergodicity breaking in many-body systems~\cite{Torres-Herrera14,Torres-Herrera15,Torres-Herrera18, Prelovsek18reduced}. It is defined as 
\begin{equation}
    P(t)  = |  \braket{\psi(t) | \psi(0)}|^2,
\end{equation}
where $\psi(t)$ is the state of the system at time $t$. In ergodic systems $P(t)$ decays to a value exponentially small in system size $L$, while in the MBL regime it saturates to higher values, signifying that the memory of the initial state $\ket{\psi(t)}$ is retained even at long times.  

Expressing the initial state $\ket{\psi(0)} = \sum_n c_n \ket{n}$ in the eigenbasis $\{ \ket{n} \}$ of the Hamiltonian $\hat{H}$, and performing an average $\braket{\cdots}$ over disorder realizations, the survival probability can be rewritten as 
\begin{equation}
P(t) = \left\langle \left| \sum_{n=1}^{\mathcal N} | c_n |^2 e^{-i E_n t}  \right |^2 \right \rangle,
\end{equation}
where the coefficients $| c_n |^2$ specify the distribution of the initial state in the eigenbasis of the Hamiltonian.
The latter expression can be made identical to the definition of the SFF in Eq.~\eqref{eq:Kt} when $| c_n |^2$ are replaced by the Gaussian filter $g(E)$. Consequently, in the ergodic regime, the behavior of the survival probability at long times resembles the ramp-plateau structure of the SFF. Before the increase of $P(t)$  occurs (the increase is associated with the onset of ramp behavior),  the survival probability has a broad minimum called a correlation hole~\cite{Leviandier86}. Pin-pointing the position of the correlation hole allows for an alternative definition of a Thouless time \cite{Schiulaz19}, at which the spread of the initial state in the many-body Hilbert space is complete. The Thouless time defined in this fashion was shown to scale exponentially with the system size in the ergodic regime of disordered spin chains, starkly contrasting the $L^2$ scaling found in the Thouless time from SFF. Notably, the correlation hole is associated with a "bulge" in the time evolution of the entanglement entropy $S(t)$ described by~\cite{Torres17bulge}. Both the correlation hole and the entanglement entropy "bulge" are dynamic manifestations of quantum chaos and are sensitive even to exponentially weak integrability terms~\cite{Santos20}. The correlation hole, measurable experimentally via survival probability or correlation functions~\cite{Das24proposal}, allows for the detection of the onset of chaotic dynamics in the regime of slow dynamics in the disordered spin chains.

Despite considerable activities in studies of survival probabilities, no sharp prediction for the onset of MBL (or its absence) has been formulated so far. In contrast, for the quantum sun model in Eq.~\eqref{eq:Havalanche}, in which the transition point is well established, \cite{Hopjan23survival} showed that the survival probability, when rescaled by its asymptotic values as routinely done in studies of the SFF, exhibits scale invariance at the transition.
The latter provides an intriguing perspective to be explored in other interacting models as well.

The survival probability can also be viewed as a return probability on the Fock-space graph. In this case it can be directly linked with the properties of correlation functions of local observables~\cite{Pain23return}. Notably, higher-order correlations on the Fock-space graph carry information about entanglement in the system~\cite{Roy22correlations}, providing an alternative perspective on the slow growth of entanglement in the disordered spin chains. The transport of probability on the Fock-space graph~\cite{Creed23transportFS} is an alternative framework for studying universal features of the slow dynamics in the disordered spin chains. For example, the slow sub-diffusive dynamics in strongly disordered spin chains can be interpreted as originating from heavy-tailed distributions of the escape times in the vicinity of the depinning transition, found in directed polymers in random media~\cite{Biroli20}.

\section{Relation of thermalization and transport}
\label{sec:trans}
This Section is devoted to the transport properties of disordered spin chains. As we have signaled in the preceding Section, assessing the unitary dynamics provides one venue for investigating transport in many-body systems. Different approaches to probe many-body coherent transport involve studies of the spectral function of the spin current operator and examination of the effects of attaching reservoirs of conserved density to the system's boundaries. Ref.~\cite{Bertini21review} provides a comprehensive review of transport properties in 1D lattice models, while \cite{Landi22nonequilibrium} considers nonequilibrium boundary-driven systems in a more general context. In the following, we focus on selected aspects of transport in disordered spin chains relevant for understanding the regime of slow dynamics and MBL.

\subsection{Transport in the ergodic regime}
A general expectation for quantum, as well as classical, many-body chaotic systems is that their dynamics are diffusive~\cite{Saito03, Mejia05}. 
The presence of diﬀusion in many-body classical systems can be linked to the positivity of some Lyapunov exponents, which necessarily arise when integrability is broken. Such arguments, which in a classical system can be made by a combination of analytical and large system size numerics (thousands of microscopic constituents), do not have direct quantum-mechanical counterparts. In quantum non-integrable systems, one has to necessarily resort to smaller system sizes, and investigate transport with almost no support from analytical calculations. Even in these circumstances, however, one might be led to believe that diffusive transport must be the most common behavior of an ergodic quantum system, 
with sub- and super-diffusion or ballistic transport relegated to a few counterexamples. In this Section we review this question in the context of systems exhibiting 
MBL regime, and we show that the counterexamples are more than a few.

The presence of diffusion in 1D non-integrable lattice models was substantiated by several complementary methodological approaches~\cite{Bertini21review}, including, e.g., tensor network investigations~\cite{Prosen09, Prosen12} as well as time evolution results~\cite{Kim13, Karrasch14, Steinigeweg16}. Moreover, diffusion is observed when weak interactions are introduced to localized quasiperiodic systems~\cite{Znidaric18diffusiveQP, Varma19diffusive}. However, an example of a non-integrable spin ladder, which supports a ballistic magnetization transport in its invariant subspaces and whose transport is otherwise diffusive, was put forward in \cite{Znidaric13}. Hence, there are examples of chaotic, ergodic systems with spectral statistics following RMT predictions, which support transport types that are different from diffusive. 

Integrable systems with a \textit{single} impurity exhibit spectral properties that adhere to predictions of RMT~\cite{Santos04defects, Barisic09impurity}.
However, as shown on the example of the XXZ spin chain with a single impurity, such system hosts a ballistic transport~\cite{Brenes18ballistic}.
Investigations of kinetic constraints in 1D many-body systems, such as the $PXP$ model [the paradigmatic model of quantum many-body scars~\cite{Serbyn21}] provide further example of transport that is different from diffusive in the ergodic phase. 
For example, the analysis of energy transport in the kinetically constrained $PXP$ model, using extensive tensor network investigations, indicates that the transport may be superdiffusive, and the dynamical critical exponent is $z\approx 3/2$~\cite{Ljubotina23Superdiffusive}. The appearance of a superdiffusive transport in the $PXP$ model is aligned with the view of its proximity to the integrable point~\cite{Khemani19}, and with the claim about the lack of MBL phase in this type of constrained spin chains~\cite{Sierant21constraint}. 
Similar findings about anomalous diffusion have been reported for ergodic systems with dipole and higher moment conserving lattice systems~\cite{Feldmeier20Anomalous},
in models with sparse interactions~\cite{de2020subdiffusion} and in kinetically constrained quantum circuits~\cite{Singh21kinetic}. 
From that perspective, understanding conditions in which non-diffusive transport may emerge in ergodic systems remains an interesting question for further research.

\subsubsection{Spin conductivity} 
\label{subsec:spinCON}
The standard approach to study transport in disordered spin chains is to consider the dynamical spin conductivity~\cite{Karahalios09finite, Barisic10, Agarwal15, Gopalakrishnan15}
 defined as
\begin{equation}
    \sigma(\omega) \propto \int_{-\infty}^{\infty}dt e^{i\omega t}
    \left \langle \hat{j}(t) \hat{j} \right \rangle = \sum_{n,m} \delta(\omega-E_n + E_n) |\braket{n|\hat j|m}|^2
    \label{eq:f2J},
\end{equation}
where $\hat{j} = \sum_l \hat{j}_l$, with $j_l$ defined in \eqref{eq:jl} is the spin current operator\footnote{If one considers spinless fermions, $\sigma(\omega)$ corresponds to charge conductivity.}, and the proportionality factor in Eq.~\eqref{eq:f2J} includes the temperature $T$, such that the non-trivial quantity in the high-temperature limit $T\to\infty$ is $\tilde \sigma(\omega) = T \sigma(\omega)$; in what follows, we set $\tilde\sigma\to\sigma$. The dynamical conductivity $\sigma(\omega)$ is the spectral function \eqref{eq:f2} of the current operator $\hat{j}$. However, the matrix elements $\braket{n|\hat j|m}$ of the current operator $\hat j$ in the disordered spin chains differ significantly from the matrix elements of $\hat{S}^z_j$, leading to the expected small frequency behavior 
\begin{equation}
    \sigma(\omega) \stackrel{\omega \ll 1}{=} \sigma_0 + \zeta \omega^{\bar\alpha},
\end{equation}
where $\sigma_0$ is the DC conductivity, $\zeta$ is a constant and the exponent $\bar\alpha$ can be related to the dynamical critical exponent as $\bar\alpha = 1-\frac{2}{z}$~\cite{Luitz17b}. To extract the DC conductivity, one must first consider the limit of $L\to \infty$, followed by $\omega \to 0$. This task is feasible in the ergodic regime, in which the dynamical conductivity $\sigma(\omega)$ at small frequencies behaves in a fashion independent of the broadening $\eta$ of the Dirac delta functions used for numerical evaluation of Eq.~\eqref{eq:f2J}, see also the discussions in \cite{Prelovsek17, Luitz17b}. Analysis of the disordered spin chains~\cite{Prelovsek23, Barisic10} finds that the diffusion constant $D_0$ (satisfying $D_0 \propto \sigma_0$ due to Einstein relation) in the ergodic regime decreases as $D_0\propto e^{-a W}$, i.e., exponentially with the disorder strength. Linking $D_0$ with a parameter that quantifies the sensitivity of energy levels to "twist" of boundary conditions (see Sec.~\ref{subsec:sensitivity}), Ref.~\cite{Prelovsek23} argued that the deviations from the RMT universality in disordered spin chains are observable when the diffusion constant $D_0$ starts to be comparable to the average level spacing. This occurs at disorder strength $W\propto L$, similar to $W^T(L) \propto L$ extracted in Sec.\ref{subsec:spectral}, due to the combination of the exponential decrease of $D_0$ with $W$, and the exponential decrease of the average level spacing with system size $L$.

\subsubsection{Unitary time evolution}
\label{subsec:unitTEV}
Spatiotemporal decay of correlation functions of local observables, as discussed in Sec.~\ref{sec:density}, can be used for probing the class of dynamics in many-body systems. 

Early studies of the dynamics in the disordered XXZ spin chain reported anomalous, sub-diffusive dynamics at disorder strengths below the MBL regime, $W \lessapprox W^*(L)$~\cite{BarLev15}. Similar conclusions have been obtained from analyzing the density autocorrelation function $C(t)$ in \cite{Luitz16}. Importantly, the subsequent studies~\cite{Bera17, Weiner19} of the two-point correlation function $\Phi(x,t)$ defined in Sec.~\ref{subsec:spatial} indicate, instead, a prominent decrease of the extracted dynamical exponent $z$ with system size $L$.
Denoting the $n$-th moment of $\Phi(x,t)$ by $\braket{x^n} = \sum_x x^n \, \Phi(x,t)$, the density spreading may be quantified by $\Delta x(t) = ( \braket{x^2} -\braket{x}^2 )^{1/2}\propto t^{1/z}$.
For instance, the maximal value of $z$ (extracted after the initial transient period and before the late time saturation) changes between $z\approx 4.2$ for $L=16$, through $z \approx 3.4$ at $L=24$  down to $z\approx 2.95$ for $L=32$ at $W=1.5$~\cite{Weiner19}. While at $L=32$ the dynamical exponent $z$ is still markedly larger from the diffusive limit $z=2$, the significance of its drift with system size prevents one from excluding a diffusive scenario at $W=1.5$. With increasing the disorder strength, the system size drift of $z$ becomes weaker but is still observable. For instance at $W=3$, \cite{Weiner19} found $z\approx 6.4$ at $L=16$ and $ z\approx 5.2$ at $L=28$. While the slowdown of the dynamics is evident, the available numerical results for $\Phi(x,t)$ disallow clear conclusions about the class of the ergodic dynamics at $W=3$ in the disordered XXZ spin chain\footnote{Similarly, at $W=3$, there is no stable power-law exponent $\beta$ governing the decay of $C(t)$. Instead, a monotonic increase of $\beta$ with system size $L$ is observed; see Fig.~\ref{fig:tevol}(b).}.

Similar difficulties are faced even in transport studies in non-interacting systems. For example, let us consider Anderson model \eqref{eq:Hand} on a 3D regular (cubic) lattice with on-site energies $\epsilon_i$ being random, independent variables uniformly distributed in the interval  $[-W/2,W/2]$. It is well established that this model has the Anderson transition at $W\approx 16.5$~\cite{Slevin99} and that the dynamics are subdiffusive at the transition, with the dynamical exponent $z=3$~\cite{Ohtsuki97}.

Due to the absence of interactions in the Anderson model, we consider a quantity related to $\Phi(x,t)$ by probing the time evolution of an initial state $\ket{\psi_0}$ with a single particle completely localized at a given lattice site. Calculating the time evolved state $|\psi_0(t) \rangle = e^{-i \hat H t} |\psi_0\rangle$, we compute the mean square displacement,
\begin{equation}
 \langle r^2(t)\rangle = \langle \psi_0(t) | \sum_{i=1}^{D}  (\hat r_i - \overline r_i )^2|\psi_0(t)\rangle, 
\end{equation}
where $r_i$ is $i$-th component of the position operator $\bf{ \hat r}$ and $\overline r_i = \langle \psi_0(t) | \hat r_i |\psi_0(t)\rangle$. For the dynamics characterized by a dynamical exponent $z$, we expect $ \langle r^2(t)\rangle \propto t^{2/z}$. While this behavior is expected to emerge in the $L\to \infty$ limit at sufficiently large times, in order to account for the finite size and time effects, we consider the derivative $\alpha(t) \equiv d \log \langle r^2(t) \rangle / d \log t$, such that $\alpha(t\to\infty) = 2/z$.

\begin{figure}
    \centering
    \includegraphics[width=0.99\columnwidth]{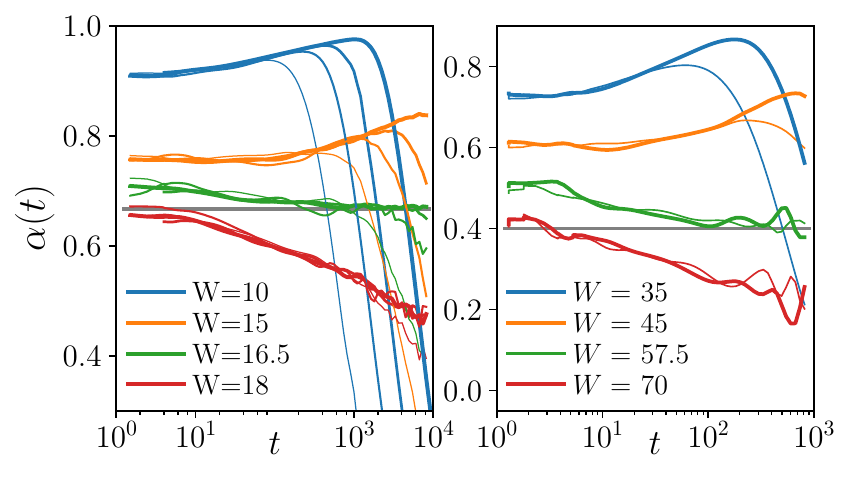}
    \caption{Subdiffusion at Anderson transition in $3$D (left) and $5$D (right) hypercubic lattices. Results show the time dependent $\alpha(t)$ function
  for 3D (left) and 5D (right) Anderson models for various
  disorder strengths $W$. In the 3D case, results for
  the system size {$L=80,120,160,240$} 
  are denoted by {progressively thicker} lines,
  whereas
  in the 5D case {thin (thick)} lines correspond to $L=20$ ($L=30$). For $t\to\infty$ the flowing exponent vanishes, $\alpha(t) \to 0$, when the wave-packet finishes spreading over the entire lattice. Reprinted from \cite{Sierant20thouless}. 
\label{fig:alpha}
}
\end{figure}

The results for the 3D Anderson model are shown in Fig.~\ref{fig:alpha}(a). Away from the transition in the 3D model, i.e., at $W=10$ ($W=18$), we see a progressive increase (decrease) of $\alpha(t)$ towards unity (zero), signaling the diffusive (localized) behavior. However, incorrect predictions about the character of the dynamics may be formulated without access to extensive time and length scales. For instance, results at $W=15 <W_C$ for $L=80$ suggest a stable subdiffusion with $\alpha \approx 0.75$. The trend towards diffusion, $\alpha \to 1$ and $z\to 2$, becomes apparent only for system size $L = 120$. Notably, the system size $L=80$ of the 3D Anderson model is already beyond the reach of the present-day full exact diagonalization. In the 5D Anderson model, the localization transition occurs at $W_C\approx 57.5$~\cite{Tarquini17}. Understanding the character of the dynamics is even more challenging than in the 3D case due to the more limited range of available system sizes, see Fig.~\ref{fig:alpha}(b).

The results for Anderson models indicate that the time and length scales required to assess the character of the dynamics correctly may be substantial. The same conclusion applies also to disordered \textit{classical} Heisenberg chains~ \cite{McRoberts23}. Hence, from 
this perspective it is not surprising to observe the finite size drifts of the exponent $z$ towards $z=2$, such as those found in the ergodic regime of the disordered XXZ spin chain~\cite{Bera17, Weiner19}, and their interpretation must be aligned with results from other transport measures obtained from complementary methods.

\subsubsection{Non-equilibrium steady states in open systems} 
\label{subsec:NESS}

Instead of probing the unitary dynamics, an alternative way to study coherent transport in interacting many-body systems is to mimic an experimental scenario in which reservoirs of a conserved density are attached to two ends of a chain. In that case, the Gorini-Kossakowski-Sudarshan-Lindblad (GKSL) equation~\cite{Gorini76, Lindblad76} describes the evolution of system's density matrix,
\begin{equation}
\frac{d \rho}{dt}=i [ \rho, \hat H ]+ 
\frac{1}{2}\sum_{k}  \left( [ L_k \rho,L_k^\dagger ]+[ L_k,\rho L_k^{\dagger} ] \right) \equiv \hat{\cal L}(\rho),
\label{eq:Lin}
\end{equation}
where  $\hat H$ is the Hamiltonian of the considered many-body system, and the Lindblad operators $L_k$ effectively account for driving by the Markovian baths attached to the end of the chain. The density matrix $\rho(t)$, initialized in a generic initial state $\rho(0)$, converges, at a sufficiently long time, to a non-equilibrium steady state (NESS) $\hat{\rho}_\infty = \lim_{t \to \infty}{\exp{(\hat{\cal L}t)}}\rho(0)$. Keeping track of the full density matrix $\rho(t)$ requires storing $\mathcal N^2$ complex numbers, much larger than $\mathcal N$ numbers required to store a pure state $\ket{\psi}$. For that reason, the exact numerical simulation of the GKSL equation \eqref{eq:Lin} is feasible on present-day computers only for small systems $ L \lessapprox 14 $, unless the quantum jumps method to simulate \eqref{eq:Lin} is employed~\cite{Daley14trajectories}.

A clear path forward was proposed by Ref.~\cite{Prosen09, Znidaric10},  which suggests employing a tensor network algorithm to simulate the evolution of $\rho(t)$, generated by the GKSL equation. A simulation of $\rho(t)$ expressed as an 
matrix product operator~\cite{Schollwoeck11} is limited by the amount of operator entanglement entropy\footnote{Mapping $\hat{\rho} = \sum_{i,j} \rho_{i,j} \ket{i}\bra{j} \longrightarrow \ket{ \rho} = \sum_{i,j} \rho_{i,j} \ket{i} \otimes \ket{j} $ allows to identify $
\hat{\rho}(t)$ with a state in a fictitious doubled Hilbert space with basis states $\ket{i,j}\equiv \ket{i} \otimes \ket{j}$,  interpreted as Hilbert space for $L$ lattices sites with on-site dimension equal to $d^2$ for qudits ($d=2$ for spin-1/2 systems). Operator entanglement entropy of $\hat{\rho}(t)$ is defined as the bipartite entanglement entropy~\eqref{eq:entdef0} of the state $\ket{ \rho}$, see e.g.~\cite{Prosen07}.} of $\hat{\rho}(t)$. The latter is a measure of correlations in many-body systems differing substantially from the entanglement entropy generated by unitary dynamics under $\hat H$. It allows for finding the NESS even in large systems, provided that the relaxation to the NESS $\hat{\rho}_\infty$ is sufficiently fast.

To probe spin transport in the disordered XXZ spin chain \eqref{Hxxz}, the Lindblad operators in \eqref{eq:Lin} can be chosen as $\hat{L}_1=\frac{1}{2}\sqrt{1+\mu}\,\hat{S}^+_1, \hat{L}_2= \frac{1}{2}\sqrt{1-\mu}\, \hat{S}^-_1$, and $\hat{L}_3 =  \frac{1}{2}\sqrt{1-\mu}\,\hat{S}^+_L, \hat{L}_4= \frac{1}{2} \sqrt{1+\mu}\, \hat{S}^-_L$. The asymmetry in driving between the two ends, i.e., $\mu \ne 0$, induces a non-vanishing value of the spin current $j_l = \mathrm{tr} \left(\hat{j}_l \hat{\rho}_\infty \right)$ in the NESS, where the spin current operator is defined as
\begin{equation}
    \hat j_l \equiv \hat{S}^x_{l} \hat{S}^y_{l+1}-\hat{S}^y_{l}\hat{S}^x_{l+1}.
    \label{eq:jl}
\end{equation}
The NESS obtained in this setting has, away from the edges of the system, a uniform spin current profile $j_l \equiv j$, and the value of the spin current scales as 
\begin{equation}
    j \propto \mu L^{-\gamma},
\end{equation}
where the exponent $\gamma$ is linked to the dynamical exponent $z$ via $\gamma=z-1$~\cite{Luitz17b}. Deep in the ergodic regime of the disordered XXZ spin chain at $0<W<0.5$, by examining large systems comprising up to $L=400$ sites, the 
matrix product operator simulations of the NESS~\cite{Znidaric16} find a clear diffusive behavior $j \propto L^{-1}$. This conclusion aligns with the general expectation of diffusive behavior in non-integrable ergodic systems.

With increasing the disorder strength, but still in the ergodic regime at $0.5<W<2$, \cite{Znidaric16, Schulz20} reported a subdiffusive behavior $j \propto L^{-\gamma}$, with $\gamma > 1$. However, the dynamics exhibit a slowing down with the growth of $W$, leading to an increase of the decay time of the slowest mode, which makes calculating the NESS $\hat{\rho}_\infty$ more challenging, even if one assumes that the operator entanglement does not significantly grow in this range of $W$.
Consequently, at $W=2$, the matrix product operator method employed in \cite{Znidaric16} provides access to results for relatively small system sizes up to $L\leq 20$, accessible also to the ED. Moreover, already at $W=0.65$, a small but noticeable trend for a decrease of the value of $\gamma$ towards unity, i.e., towards the diffusive transport, is visible when the system size $L$ is increased. Thus, the presently available numerical results do not allow for clear conclusions about those disorder strengths, similar to the investigations of the unitary time dynamics. Finally, we note that the spin transport properties on which we focus here may differ from the energy transport in disordered models~\cite{Mendoza19}.

The importance of finite size effects on the putative transition between diffusive and sub-diffusive dynamics in the disordered XXZ spin chains is further exemplified by a random dephasing model studied in \cite{Taylor21subdif}. The random dephasing model is a non-interacting 1D Anderson insulator~\eqref{eq:Hand} with randomly positioned on-site dephasing terms. 

In a disordered interacting spin chain, fluctuations of the disorder could create a complex landscape of regions with insulating and thermal regions. When a single excitation passes through a thermal region, it exchanges energy with its surroundings and acquires a phase due to interaction, which depends, in a complicated manner, on the state of neighboring particles. A simplified scenario in which an averaging over the acquired phase can be performed leads to the random dephasing model. In a genuinely interacting model, the dephasing process should be treated for each particle individually in a self-consistent manner. Despite simplifying this process by the random on-site dephasing, the model of ~\cite{Taylor21subdif} hosts a transition between diffusive and subdiffusive phases, of the kind depicted in \cite{Agarwal15}, providing insights into the possible changes in the dynamics in the ergodic phase of the disordered XXZ spin chain. The transition between diffusion and subdiffusion in the random dephasing model is a result of an interplay of three characteristic length scales: the localization length of the non-interacting model, the length over which the phase of the particle is effectively randomized, and lengthscale $s_\star$ associated with the percolation problem defining large, rare, insulating regions in the system. The latter length scale grows logarithmically in system size $L$. Therefore, observation of a significant change in $s_\star$, crucial for altering the transport properties, may require an enormous change in system size $L$, hindering the possibility of observing the effects of the rare insulating regions in studies of finite systems. The absence of the power-law tails of resistivity in the putative subdiffusive phase of disordered spin chains found in~\cite{Schulz20} may be attributed to this mechanism.

An alternative approach to studying the dynamical properties of a quantum many-body system with Hamiltonian $\hat{H}$ is to investigate the NESS that emerge when the system is weakly coupled to a non-thermal bath~\cite{Lenarcic20}. In this approach, one chooses the same set of Lindblad operators $\{L^{\nu}_i\}$ with $\nu=1,...,\nu_{\mathrm{max}}$ ($\nu_{\mathrm{max}} \geq 1$ is a constant) for each lattice site $i$, uniformly injecting energy to the system~\cite{Lange17, Lenarcic2018}. The local temperature profile of the resulting NESS $\hat{\rho}_{\infty}$, analyzed as a function of the coupling strength $\epsilon$ to the bath, can be used to quantify the dynamics of the underlying many-body Hamiltonian $\hat{H}$. A classical hydrodynamic description of an ergodic system, realized, for instance, by a spin chain, coupled to phonons and driven by light~\cite{Lenarcic18activating}, shows that the local temperature variations fulfill $\delta T \propto \epsilon^{ 1/(2z) }$, where $z$ is the dynamical critical exponent. In contrast, one may expect that the local variation of system properties in an MBL regime will lead to a non-uniform density profile even in the weak coupling limit $\epsilon \to 0$. These intuitions were employed in the analysis of the results of tensor-network simulations of the GKSL equation with the dissipative bulk driving of strength $\epsilon$~\cite{Lenarcic20}, yielding 
\begin{align}
\frac{\delta \beta}{\bar \beta}(\epsilon) \sim \left\{ 
\begin{array}{ll}
\epsilon^{1/(2z)}, & \mathrm{ergodic} \,\, \mathrm{regime} \nonumber \\
\frac{\delta \beta}{\bar \beta}\big|_{\epsilon \to 0} - b \, \epsilon + \mathcal{O}(\epsilon^2),  & \mathrm{MBL}  \,\,  \mathrm{regime}\;, \nonumber
\end{array}
 \right.
\end{align}
where $\delta \beta/\overline{\beta}$ is the inverse temperature variation. The non-analytical dependence of $\delta \beta/\overline{\beta}$ on the driving strength for $\epsilon \to 0$ enables the calculation of the dynamical critical exponent $z$ in the ergodic regime. The dynamical critical exponent $z$ increases abruptly beyond the diffusive value $z=2$ with the disorder strength $W$. At stronger disorder, a non-zero value of $\frac{\delta \beta}{\bar \beta}\big|_{\epsilon \to 0}$, extrapolated from the results at small but finite $\epsilon$, indicates the MBL regime. The MBL phase could be, strictly speaking, differentiated from the ergodic phase only in the weak-driving limit $\epsilon \to 0$ in the thermodynamic limit. The decrease of $\epsilon$ to $0$ implies a slower convergence to the NESS $\hat{\rho}_{\infty}$, which constitutes the main challenge for numerical simulations, especially at larger disorder strength $W$.

\subsubsection{Conclusion} 
While there is clear evidence for diffusive transport deep in the ergodic regime of the disordered XXZ spin provided by several complementary methods, the transport properties at larger disorder strength $W$ remain unclear. The increase of $W$ leads to a slowdown of the dynamics, hindering the assessment of the class of transport both in the NESS approach to the disordered XXZ spin chain, as well as in the direct examinations of the unitary dynamics of the model. Investigations of the dynamical spin conductivity run into a similar barrier when the extracted diffusion constant decreases and starts to be comparable to the many-body level spacing. 

In a broader perspective, the outlined problems leave the question about the relation of transport and thermalization in disordered spin chains at least partially open. This question is of particular conceptual relevance as both diffusion and subdiffusion may be expected, on the grounds of classical random walks considerations (see App.~\ref{app:dif-sub-Gibbs}), to be consistent with relaxation to thermal equilibrium. The uncertainty of the position of the putative MBL transition makes the question about the fate of the ergodic regime of the disordered spin chains even more exciting.

\subsection{Transport at large disorder strengths}
The difficulty of studying coherent transport in disordered spin chains can be directly linked to the slowdown of the dynamics at increasing disorder $W$. To show that,~\cite{Panda20} assumed that the disordered XXZ spin chain is initialized in a weakly polarized domain wall state, $\hat{\rho}(0) \propto 
\prod_{k=1}^{L/2} \left(  1 + \mu \hat{S}^z_k \right) \otimes \prod_{k=L/2+1}^{L} \left(  1 - \mu \hat{S}^z_k \right) $, where $\mu \ll 1$ describes the initial polarization of the system. The domain wall decay is quantified by $\Delta S(t) = \mu \frac{L}{2} - \sum_{k=1}^{L/2} \mathrm{tr} \left( \hat{\rho}(t) \hat{S}^z_k \right) $, which quantifies how many particles (when the XXZ spin-1/2 chain is expressed in terms of spinless fermions) have moved after a time $t$. 

At small disorder strength $W=0.5$, a power-law growth $\Delta S(t) \propto t^{1/z}$ with the dynamical exponent $z=2$ is observed, consistent with the emergence of diffusion. While the matrix product operator simulation employed in~\cite{Panda20} does not suffer from the effects of finite system size, it can access only relatively small times $t< 100$. At larger disorder strengths $W$, the exponent $z$, quantifying the behavior of $\Delta S(t)$, systematically decreases with increasing time $t$. For instance, at $W = 2$ the obtained exponent is $z \approx 3.8, 3.3, 3.1$, respectively, for the fitting windows $t \in [20, 40], [40, 60], [60, 80$]. Hence, the observed subdiffusion may be a transient effect, consistent with the conclusions of the earlier studies. 

An increase in the disorder strength leads to a dramatic growth of the time $t_N$ at which $N$ particles cross the position of the domain wall. The time $t_1$, after which a single particle is transported over the interface between the domains, increases exponentially with the disorder strength, $t_1 \propto e^{a W}$, for $W<3$, and the increase becomes even faster at larger disorder strengths. To make substantial statements about the nature of the transport in the system, the number of the particles transferred between the subsystems should scale with system size $L$, e.g., proportionally to $L$. The simulations of \cite{Panda20}, based on extrapolation of the short-time behavior, suggest that the transport time $t_N$ exceeds the Heisenberg time $t_H$ with increasing the disorder strength. For instance, the time required to transport $N=L/5=4$ particles exceeds the Heisenberg time for $L=20$ at $W=3$. This implies that attempts to understand the transport properties at $W=3$ must be performed at system sizes significantly exceeding $L=20$ and involve probing the dynamics at much larger time scales, e.g., at times larger than the Heisenberg time for $L=20$. The further slowdown of the dynamics at larger disorder strength $W$ results in progressively stronger conditions on the required system sizes and time scales.
 
Several works ~\cite{Berkelbach10, Barisic16, Herbrych22Relaxation} investigated the behavior of dynamical spin conductivity~\eqref{eq:f2J} in disordered spin chains at disorder strengths exceeding the boundary of the ergodic regime, $W>W^T(L)$. These studies unravel that the exponential decrease of the DC conductivity, $\sigma_0 \propto e^{-aW }$, persists to disorder strengths exceeding the onset of the MBL regime, $W>W^*(L)$. This behavior is another manifestation of the abrupt slowdown of the dynamics with increasing disorder strength. However, the validity of the assumptions about the independence of the broadening of the Dirac delta functions in~\eqref{eq:f2J}, required for the determination of $\sigma_0$ in this regime of disorder strengths, may not be taken for granted~\cite{Prelovsek17, Luitz17b}.

The results discussed in this Section show that the persistent slow dynamics at significant disorder strengths prevent not only reaching conclusions about the MBL phase but also about the character of transport in the disordered spin chains when the disorder strength increases. While the numerical and experimental studies do not reach a definite conclusion about transport in strongly disordered many-body systems, the mathematical analysis in the disordered quantum Ising model~\cite{deRoeck24absence} finds that the conductance of a chain of length $L$ vanishes faster than $1/L$, consistently with the dynamical critical exponent $z$ \textit{strictly larger} than $2$.

\section{Modelling MBL transition}
\label{sec:aval}
In this Section, we describe the possible mechanisms of the MBL transition.
In particular, we focus on the quantum avalanche mechanism, which is a non-perturbative scenario of growth of a small thermal region which, for specific parameters of the system, may destabilize the MBL phase. We outline ways of modeling the avalanche scenario, discuss its relevance for the dynamics of disordered spin chains, and briefly consider alternative frameworks providing insights into the physics behind the ergodic to MBL crossover.

\subsection{Rare thermal bubbles}

The uncorrelated nature of the on-site magnetic fields assumed in the disordered spin chains typically studied in the context of MBL, e.g., in the XXZ model \eqref{Hxxz} or in the $J_1$-$J_2$ model~\eqref{eq:J1J2}, necessarily leads to the appearance of inhomogeneities in the system. The probability $p_\ell$ that the on-site potentials $h_i$ fulfill $|h_i| < W_0$ on $\ell$ consecutive sites, where $W_0$ is a constant smaller than the disorder strength $W$, is given by $p_\ell = (W_0/W)^\ell = e^{-\ell \log(W/W_0)}$. The probability $p_\ell$ is exponentially suppressed in $\ell$, and the regions with locally weaker disorder strength are rare. Nevertheless, such rare regions of \textit{arbitrary size} $\ell$ exist in the chain with probability $1$ in the thermodynamic limit $L\to \infty$, and understanding their effects on the behavior of the disordered spin chains is of vital importance.

\subsubsection{Quantum avalanches}
\begin{figure}
    \centering
    \includegraphics[width=1\columnwidth]{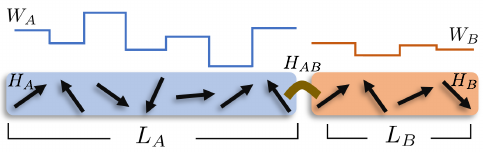}
    \caption{The avalanche scenario. Subsystem $A$ is subject to strong disorder and it is in the MBL regime. The subsystem $B$ is weakly disordered and it is in the ergodic regime. The two subsystems interact via the boundary term $\hat{H}_{AB}$, which may induce a gradual thermalization of the subsystem $B$. Reprinted and adapted from \cite{Szoldra24catching}. 
\label{fig:sketchAVA}
}
\end{figure}
The avalanche mechanism was introduced by \cite{DeRoeck17}, who proposed that the rare regions of anomalously weak disorder in the many-body system may become ergodic "bubbles" that destabilize the surrounding MBL regions. To describe this mechanism, let us consider a disordered spin chain in which a rare region with small local disorder strength $W_B$ and size $L_B$ is formed\footnote{The following exposition of the avalanche mechanism is similar to descriptions of \cite{Potirniche19} and \cite{Szoldra24catching}.}. The weakly disordered subsystem $B$ is in contact with subsystem $A$ of size $L_A$ subject to disorder strength $W_A$, see Fig.~\ref{fig:sketchAVA}. The Hamiltonian of the disordered spin chain, e.g., the XXZ model \eqref{Hxxz}, corresponding to this scenario, can be expressed as
\begin{equation}
    \hat{H} = \hat{H}_A + \hat{H}_B + \hat{H}_{AB},
    \label{eq:ava1}
\end{equation}
where $\hat{H}_A$ describes the MBL subsystem $A$, $\hat{H}_B$ is the Hamiltonian of the ergodic bubble, and $\hat{H}_{AB}$ involves local operators with support at the boundary between the subsystems $A$ and $B$.

The avalanche scenario describing interaction of an ergodic bubble $A$ with the MBL subsystem $B$ relies on the following simplifying assumptions. 
\begin{enumerate}
    \item Matrix elements $\braket{m_B| \hat{O} |n_B}$ of operators $\hat{O}$ with support in the subsystem $B$, i.e., in the ergodic bubble, between the eigenstates $\ket{m_B}$ of $\hat{H}_B$ ($\hat{H}_B\ket{m_B}=E_{m_B} \ket{m_B}$) follow the ETH~\eqref{eq:ETH1}, i.e.,
    \begin{eqnarray}\nonumber 
    \bra{m_B}\hat{O}\ket{n_B} =  O(\bar{E}) \delta_{n_B m_B} & \\  + 
      e^{-S(\bar{E})/2}  
     f_O(\bar{E}, &  \omega) \, R_{m_B n_B},  
     \label{eq:Obath}
    \end{eqnarray}
    where $ e^{-S(\bar{E})/2} = \rho_B^{-1/2} $, with $\rho_B$ the density of many-body states in the subsystem $B$, and $\bar{E} =(E_{m_B}+E_{n_B})/2 $. For simplicity, the energy $\bar{E}$ is assumed to be in the middle of the spectrum of $H_B$.
    \item The Hamiltonian of the MBL subsystem $A$ can be expressed in terms of the set of LIOM operators, $\lbrace \hat{I}_\alpha \rbrace$, $\alpha=1,\dots L_A$. Each of the LIOMs can be expanded in the microscopic spin operators $(\hat{S}^x_j, \hat{S}^y_j, \hat \hat{S}^z_j)$. The norm of the terms in the expansion decreases exponentially with characteristic lengthscale $\xi$ with the distance from the center of localization of the LIOM $\hat{I}_\alpha$.
    For simplicity, one may consider replacing $\hat{I}_\alpha$ with a simple spin operator, such as $\hat{I}_\alpha \to \hat S_j^z$.
    \item The coupling $\hat{H}_{AB}$ involves the spin operators $(\hat{S}^x_j, \hat{S}^y_j, \hat{S}^z_j)$ supported at the interface between the subsystems $A$ and $B$. Hence, it can be rewritten as $\hat{H}_{AB} = \sum_{j=1}^{L_A} \hat{H}_{AB}^{(j)}$, with
\begin{equation}
    \hat{H}_{AB}^{(j)} =  \left( V e^{-r/\xi }\hat{O}_i \hat{I}^{x}_j + \text{h.c.} \right),
    \label{eq:int}
\end{equation}
where $\hat O_i$ acts on the degrees of freedom of the ergodic subsystem $B$, 
$\hat{I}^{x}_j$  is a spin 
flip operator associated with LIOM $\hat{I}_j$ centered at site $j$, and $r=|i-j|$ is the distance between the LIOM and the ergodic bubble.
For simplicity, one may model $\hat{O}_i \to \hat{S}_i^x$ and $\hat{I}_j^x \to \hat{S}_j^x$, giving rise to the model similar to the one in Eq.~\eqref{eq:Havalanche}.
\end{enumerate}
At time $t=0$, the system is prepared in the state 
\begin{equation}
    \ket{\phi_{\mathrm{in}, n_B}} = \ket{I_1 \dots  I_j \dots I_{L_A}} \otimes \ket{n_B}\;,
    \label{phi0}
\end{equation} 
with a fixed configuration of LIOMs in the subsystem $A$, $I_j\in \{\uparrow,\downarrow\}$, and with the subsystem $B$ initialized in its eigenstate $\ket{n_B}$ with energy $E_{n_B}$.
Within the first order perturbation theory, the interaction term $\hat{H}^{(L_A)}_{AB}$ couples the initial state $\ket{\phi_{\mathrm{in}}}$ to the states of the form
\begin{equation}
    \ket{\phi_{\mathrm{fin},m_B} } = \ket{I_1 \dots  I_j \dots \overline{I}_{L_A}} \otimes \ket{m_B}\;,
\end{equation}
where $\overline{I}_{L_A}$ is the opposite spin to $I_{L_A}$ and $\ket{m_B}$ is an eigenstate $m \neq n$ of the bath. The first order perturbation theory correction to the state $\ket{\phi_\mathrm{in} }$ can be calculated as
\begin{equation}
 \ket{\phi_{\mathrm{in}, n_B}^{(1)}} = \sum_{m_B \neq n_B} \frac{ 
 \bra{\phi_{\mathrm{fin},m_B} } \hat{H}_{AB}^{(L_A)} \ket{ \phi_{\mathrm{in},n_B} }
 }{ E_{n_B}-E_{m_B}+ 2\epsilon_{L_A}  } \ket{{\phi_{\mathrm{fin},m_B}} },
    \label{eq:PER}
\end{equation}
where $2\epsilon_{L_A}$ is the energy associated with flipping the LIOM $\hat{I}_{L_A}$ and $\epsilon_{L_A}$ is of the order of $W_A$. The sum \eqref{eq:PER} may be dominated by a single term with minimal energy denominator, which is typically of the order of the level spacing $\rho_B^{-1}$ of the subsystem $B$. The matrix element of $\hat{H}_{AB}^{(L_A)}$ in the numerator can be evaluated with the help of \eqref{eq:int} and the ETH ansatz for subsystem $B$~\eqref{eq:Obath}. Hence, the contribution to the first order correction $\ket{\phi_{\mathrm{fin},m_B} }$ is given by
\begin{equation}
 h_1 \propto  \frac{V e^{- 1/\xi}}{ \rho_B^{1/2}} \frac{1}{\rho_B^{-1}} = V e^{-1/\xi} \rho_B^{1/2}.
    \label{eq:PER2}
\end{equation}
The quantum avalanche scenario of \cite{DeRoeck17} describes the following mechanism for thermalizing the MBL subsystem $A$ due to the interaction with the ergodic grain $B$. The spin in subsystem $A$ closest to subsystem $B$ will get thermalized if the correction $h_1$~\eqref{eq:PER2} is of the order of unity. When the spin thermalizes, it becomes a member of the ergodic bubble, increasing the level density of the bubble by a factor of $2$, $\rho_B \to 2\rho_B$. As a result, the denominator in Eq.~\eqref{eq:PER2} is decreased twice, showing that the enlarged bath is more potent to thermalize further spins. 
However, also the strength of the coupling is decreased as $e^{-k/\xi}$, where $k$ represents the distance of a spin in subsystem $A$ to the ergodic grain $B$.
This gives rise to a non-trivial competition between the decreasing interaction and the growing density of states.
After $k$ iterations of the process, the equation describing the thermalization of $k$-th spin away from the bath has the form
\begin{equation}
 h_k \propto  V e^{- k/\xi} \sqrt{2^k\rho_B} \sim  e^{ k[(\ln 2)/2-1/\xi]}.
    \label{eq:PER3}
\end{equation}
Moreover, if one assumes the 1D geometry in which the avalanche spreads in two directions (left and right), the density of states is multiplied by $4^k$ instead of $2^k$, giving rise to
\begin{equation}
 h_k \sim  e^{ k[\ln 2-1/\xi]}.
\end{equation}
Suppose the localization length $\xi$ is larger than the threshold value  $\xi^* = 1/\ln 2$. In that case, the growth of the density of states of the enlarged ergodic bubble compensates for the exponential decrease of the norm of the interaction term $H^{(j)}_{AB}$ in \eqref{eq:int}. Hence, the thermalizing avalanche propagates through the subsystem $A$, leading to its thermalization, and the LIOMs become destabilized by the ergodic inclusion. 
The avalanches seeded by rare ergodic grains propagate through the MBL regions, leading to the transition to the ergodic phase.
In contrast, if $\xi < \xi^*$, the avalanche does not propagate throughout the lattice and hence this theory predicts absence of thermalization.
A strong approximation involved in the latter statement is that the localization length $\xi$, introduced in the non-interacting picture, remains unchanged when adding interactions.

In passing, we note that the avalanche scenario relies crucially on the possibility of balancing the exponential decrease of the interaction matrix element (associated with the exponential decay of the LIOMs) and the exponential decrease of the level spacing of the ergodic grain. The balance may be found provided that the system is one dimensional, as argued above. In contrast, the avalanche mechanism predicts no MBL phase for higher dimensions. Additionally, the avalanches could not propagate in a 1D system with a superexponential decay of LIOMs~\cite{Foo23Stabilization}.

While the avalanche mechanism relies on expressions \eqref{eq:PER3} originating in the first-order perturbation theory, the process of enlarging the ergodic bubble by thermalizing neighboring spins is highly non-perturbative and, as such, is likely beyond the diagrammatic approach of \cite{Basko06}. Notably, the relevance of the avalanche mechanism for the dynamics of disordered spin chains remains to be determined. Numerical investigations \cite{DeRoeck17, Colmenarez23} did not provide entirely conclusive answers to this question, which we revisit in Sec.~\ref{subsec:avalREL}.

The avalanche model can be directly investigated numerically. To that end, one first rewrites $H_A$ in terms of LIOMs $\hat{I}_\alpha$ and neglects the dephasing terms involving more than a single LIOM operator. Subsequently, the ergodic bubble is assumed to be modeled by a random matrix $\hat{R}$. These suppositions, together with the form of the interaction term \eqref{eq:int}, lead to the Hamiltonian \eqref{eq:Havalanche}. The latter, dubbed a "quantum sun model", hosts an ergodicity-breaking phase transition~\cite{Luitz17bath, Suntajs22}. The position of the transition in the quantum sun model can be accurately pin-pointed with standard ergodicity breaking indicators such as the average gap ratio $\overline r$. For the quantum sun model, the gap ratio $\overline r$ follows the behavior illustrated in Fig.~\ref{fig:flow}(a), up to a very mild drift of the crossing point with system size $L$. This feature is noteworthy since the quantum sun model is a genuinely interacting many-body system, with Thouless time scaling exponentially with $L$ at the transition, as discussed in Sec.~\ref{subsec:thouless}. This makes the quantum sun model an exciting test bed for various phenomena that remain unclear in numerical studies of the disordered XXZ spin chains. One example is the many-body mobility edge whose signatures have been observed numerically in the disordered XXZ spin chain~\cite{Naldesi16, Villalonga18, Chanda20m, Brighi20} despite the theoretical arguments against its existence~\cite{DeRoeck16}. Numerical calculations in the quantum sun model provide clear evidence of the many-body mobility edge~\cite{Pawlik23}.
Another new development represents a detailed understanding of the fate of the observable matrix elements when the ergodicity breaking transition is approached from the ergodic side.
In the quantum sun model, the ETH ansatz from Eq.~(\ref{eq:ETH1}) breaks down while the short-range spectral statistics still comply with the GOE predictions~\cite{kliczkowski_swietek_24}.
This phenomenon, dubbed fading ergodicity, provides a theoretical framework to detect proximity of the ergodicity breaking transition despite the system being ergodic and the observables thermalize after a quantum quench.

Recently, \cite{Suntajs24ultrametric} argued that the corresponding RMT-like model description of the avalanche physics is the ultrametric model, defined on a Fock space.
Carrying out a detailed comparison between the later model and the quantum sun model~\eqref{eq:Havalanche}, the authors showed that both models exhibit a nearly identical value of the ergodicity breaking transition point.
Moreover, building on the results for the ultrametric random matrices~\cite{Fyodorov_2009}, it was argued that the eigenstates on the non-ergodic side of the quantum sun model are Fock-space localized, i.e., their fractal dimensions $D_q$ vanish in the non-ergodic phase at sufficiently large system size~\cite{Suntajs24ultrametric}. This is in contrast to the eigenstates in the MBL regime of the disordered XXZ and $J_1$-$J_2$ models, as discussed in Sec.~\ref{subsec:wave-fun}.

The ultrametric model on a Fock space defines a general class of interacting models that describe the interaction between the pinned impurities and the remaining (non-interacting) degrees of freedom. 
Beside the avalanche models (e.g., the quantum sun model), another subset of models that share similar geometry are the central spin models, see, e.g.,~\cite{Ponte2017}.
Future works should explore in more details the relevance of the ultrametric model on a Fock space for the physics of the disordered spin chains.

\subsubsection{Phenomenological investigations of MBL transition}
The avalanche mechanism is the leading theoretical proposal for the nature of the MBL transition. As a consequence, several phenomenological schemes relying on the real space renormalization group (RG) approach have been developed~\cite{Zhang16simplified, Dumitrescu17, Thiery18, Goremykina19, Dumitrescu19, Morningstar19, Morningstar20}. The main idea of RG approaches is to view a disordered system as a collection of ergodic and MBL blocks and assume specific simplified criteria to describe their evolution.

The RG approaches begin by dividing the disordered spin chains into "thermalizing" and "MBL" blocks of adjacent spins. Subsequently, the ergodic regions may cause thermalization of the neighboring MBL blocks provided that the relevant matrix elements are big enough compared to the level spacing of the block, according to criteria similar to Eq.~\eqref{eq:PER2}. If the MBL block is thermalized, it joins the thermalizing block, forming an enlarged ergodic region, which may destabilize MBL in its surroundings. As the RG progresses, the involved blocks grow, and finally, the system is globally identified as being ergodic or MBL.

It has been established that the renormalization group approaches relying on the avalanche mechanism lead to a BKT universality class for the MBL transition~\cite{Goremykina19, Dumitrescu19}. The associated length scale grows exponentially near the transition \eqref{eq:BKT}, hindering studies of properties of a BKT transition in numerical or experimental investigations. The analysis of~\cite{Morningstar20} indicates that the critical properties of such a transition, when analyzed in systems of size up to $L$, can be well captured assuming a power-law divergence of the correlation length~\eqref{eq:NU}, with exponent $\nu(L)$ depending on $L$. For the proposed BKT-like scaling (differing from the BKT scaling by the presence of non-analytic terms in the RG equations),~\cite{Morningstar20} showed that $\nu(L)$ grows indefinitely with $L$. However, $\nu(L)$ growth with system size is extremely slow, and results for the disordered spin chains of size as large as $L=10^6$  may be well described by a power-law scaling with $\nu(L=10^6) \approx 3$. Hence, distinguishing the BKT scaling or the BKT-like scaling from a power-law divergence is a complex task 
even within the RG schemes, which provide access to large system sizes due to their polynomial-in-$L$ complexity~\cite{Dumitrescu17, Dumitrescu19}. Consequently, telling apart the BKT-like scaling and power-law scalings in numerical studies of disordered spin chains is a formidable task due to the exponential increase of the dimension of the many-body Hilbert space, which restricts the available system sizes to $L \lessapprox 20$.

\subsubsection{Probing avalanches with planted thermal inclusions}
\label{subsec:avalREL}
The rare thermal bubbles emerging in regions of anomalously weak disorder are necessarily present in sufficiently large disordered many-body systems, even deep in the MBL regime. The ergodic bubbles amount to the structure of entanglement clusters at the crossover between the ergodic and MBL regimes~\cite{Herviou19}, and are characterized by an enhanced decay of correlation functions of local observables. Nevertheless, detecting such rare regions requires complex tools, for instance, based on machine learning approaches~\cite{Szoldra21} or analysis of pairwise correlation functions~\cite{Hemery22}. Hence, the relevance of the quantum avalanche mechanism for the dynamics of disordered spin chains is most conveniently studied in systems where a weakly disordered region is introduced intentionally in a controlled fashion.

To manually plant a thermal bubble, we enforce the disordered landscape shown in Fig.~\ref{fig:sketchAVA}. To that end, the on-site magnetic fields $h_i$ in the disordered XXZ spin chain~\eqref{Hxxz} are taken as uncorrelated random variables, \begin{equation}
    h_i \in \begin{cases}
  [-W_A, W_A]  & \text{ if }1\leq i \leq L_A, \\
  [-W_B, W_B] & \text{ otherwise}.
  \label{eq:disAB}
\end{cases}
\end{equation}
For $W_B$ deep in the ergodic regime and $W_A$ in the MBL regime, the subsystem $B$ plays the role of an ergodic bubble in contact with the MBL subsystem $A$, reproducing the situation considered in the avalanche scenario \eqref{eq:ava1}.

The avalanche scenario assumes that the ergodic bubble grows, thermalizing the neighboring spins, which enhances its capabilities to thermalize further spins, as reflected by Eq.~\eqref{eq:PER3}. This hypothesis may be contrasted with  an opposite scenario in which the ergodic bubble induces non-trivial dynamics of the neighboring LIOMs, which, however, do not influence their neighbors' thermalization process, i.e., they do not become part of the ergodic bubble. In that case, the impact of the ergodic bubble decreases exponentially with the distance $r$ and the MBL in the subsystem $A$ remains stable. In that case, a decay rate $\Gamma_j$ of LIOM $\hat{I}_j$ \eqref{phi0} at a distance $r$ from the ergodic bubble can be estimated by the Fermi golden rule~\cite{Tannoudji02quantum} as
\begin{equation}
     \Gamma_j \sim V^2 e^{-2 r/\xi}.
     \label{eq:FGR1}
\end{equation} 
The latter equation is valid only on timescales shorter than $\rho_B^{-1}$. At longer times, the discreteness of the spectrum is resolved, and, typically, the LIOM $\hat{I}_j$ does not decay.

Both the avalanche scenario and the above perturbative, non-thermalizing picture involve assumptions that may not be met by the dynamics of the disordered spin chains. Instead, the disordered spin chains may realize a scenario intermediate between the two situations. For that reason,~\cite{Leonard23} proposed to monitor the process of thermalization of the MBL subsystem using the two-point correlation function, 
\begin{equation}
   g^{(2)}(i) = \frac{1}{L_B}\sum_{j\in B} \left( \braket{\hat{S}^z_i \hat{S}^z_j} - \braket{\hat{S}^z_i} \braket{\hat{S}^z_j}\right).
   \label{g2def}
\end{equation}
Assuming that the initial state is $\ket{\phi_{\mathrm{in}, n_B}} $ \eqref{phi0} and that $W_A$ corresponds to the MBL regime, the correlation function $g^{(2)}(i)$  falls-off exponentially with a characteristic length $\xi_d$, 
\begin{equation}
    |g^{(2)}(i)|  = c \exp \left( -\frac{r}{\xi_d(t)} \right),
\label{eq:xi_d_definition}
\end{equation} 
at a fixed time $t$. The length scale $\xi_d(t)$ can be interpreted as a "thermalization distance," i.e., the depth at which the ergodic bubble penetrates the MBL subsystem. The avalanche scenario \eqref{eq:PER3} implies an unbounded growth of $\xi_d(t)$, logarithmic in time, or faster. In contrast, the perturbative decays of LIOMs \eqref{eq:FGR1} correspond to the growth of $\xi_d(t)$ that should be observed only up to a time scale proportional to $\rho_B^{-1}$.

Investigations of the dynamics of $g^{(2)}(i)$ in the disordered XXZ spin chain \cite{Szoldra24catching} confirm the expected trends. The thermalization distance $\xi_d(t)$ grows faster than logarithmically in time for disorder strengths $W_A$ corresponding both to the ergodic and the crossover regimes of the disordered XXZ spin chain, i.e., for $W_A \lessapprox W^*(L)$. In this regime of disorder strengths, the avalanches seeded by the ergodic bubble $B$ propagate throughout the subsystem $A$, causing its thermalization. In contrast, in the MBL regime, for $W_A \gtrapprox W^*(L)$, the saturation of the growth of thermalization distance $\xi_d(t)$ at times proportional to $\rho_B^{-1}$ is observed when the system is initialized in the state $\ket{\phi_{\mathrm{in}, n_B}}$, i.e., in an eigenstate of the MBL subsystem. Notably, the stop of the propagation of avalanches is observed at disorder strengths comparable to $W^*(L)$, even though the planted thermal inclusion would typically be found only in much larger spin chains. For instance, \cite{Szoldra24catching} considers thermal inclusion of size $L_B=6$ and with disorder strength $W_B=0.5$, while the disorder in the MBL subsystem is $W_A=6$. Without planting them manually, the density of such rare regions may be approximated as $(W_B/W_A)^{L_B}$, which is of the order of $10^{-6}$ for the chosen parameters. 

The results of \cite{Szoldra24catching} are consistent with the relevance of the avalanche scenario for the thermalization of MBL blocks by ergodic inclusions in the dynamics of the disordered spin chains. There are, however, two main caveats. The tendency towards ergodicity with increased time and length scales is similar to that of the uniformly disordered spin chains. This prevents distinguishing the dynamics of the MBL phase from a very slow thermalization. Moreover, the presence of two lengthscales $L_A$ and $L_B$ makes the extrapolation of the results to the thermodynamic limit even more complicated than in the case of the uniformly disordered XXZ spin chains. Nonetheless, the two-point correlation function \eqref{g2def} can be probed experimentally in many-body systems with planted thermal inclusions, as demonstrated in~\cite{Leonard23}.

Another approach to the investigation of the effects of planted regions of weak disorder was taken in~\cite{Peacock23}, which considers the dynamics of the autocorrelation function $C(t)$\footnote{Strictly speaking, \cite{Peacock23} considered the dynamics of the imbalance $I(t)$ from the initial N\'{e}el state.} in the disordered XXZ spin chain with on-site potential drawn according to Eq.~\eqref{eq:disAB}. The decay of $C(t)$ is strongly enhanced in the presence of thermal inclusion compared to the situation when $W_A=W_B$. Moreover, the results for $C(t)$ in the presence of the thermal inclusion collapse on a single curve when the time is rescaled according to $t\to t e^{-aW}$, where $a=2.1$ is a constant. The rescaling shows a significant degree of uniformity of the dynamics across a broad regime of disorder strengths, suggesting thermalization of the system in the presence of thermal inclusion. However, the system considered in \cite{Peacock23} is periodically driven by turning on and off the coupling $\hat{H}_{AB}$ between the ergodic inclusion and MBL region. The same system, but with coupling $H_{AB}$ present at all times, shows a much weaker tendency towards thermalization~\cite{Szoldra24catching}. While the coupling to thermal inclusion indeed increases the tendency towards thermalization, the rescaling $t\to t e^{-aW}$ does not provide a satisfactory collapse of the results for $C(t)$ in a broad regime of disorder strengths in the system without the periodic driving. This suggests disparate dynamics in the autonomous system at small and large $W_A$, leaving the question about its thermalization in the $L,t\to \infty$ limit open.

The results of the studies of disordered many-body systems with planted thermal inclusions 
are consistent with the predictions of the avalanche scenario. However, the approach of \cite{Leonard23, Szoldra24catching}, based on investigation of the thermalization length $\xi_d(t)$, asserts the mere consistence of the thermalization in the subsystem $A$ with the propagation of avalanches. Therefore, it is important to note that the results of \cite{Szoldra24catching} do not exclude the possibility that the thermalization of the subsystem $A$ occurs due to microscopic mechanism \textit{different} from the avalanche scenario, but with similar predictions for the behavior of $\xi_d(t)$. Whether such a mechanism could be formulated remains an open question. Nevertheless, for that reason, the problem of pin-pointing the microscopic mechanism responsible for the emergence of ergodicity in disordered spin chains remains a challenge for further investigations of the dynamics of disordered many-body systems.

\subsubsection{Open system approach}
The difficulties in finding direct numerical evidence pin-pointing the mechanism of the MBL transition motivate the search for alternative approaches. Ref.~\cite{Morningstar22, Sels21} proposed to examine an open system described by the GKSL equation \eqref{eq:Lin}, with the Lindblad operators 
\begin{equation}
    \hat{L}_k = \{ \hat{S}^x_1, \hat{S}^y_1, \hat{S}^z_1 \},
    \label{eq:Lks}
\end{equation} acting on one end of the chain. The solution of the GKSL equation is given as
\begin{equation}
    \hat{\rho}(t) = \sum_k e^{\lambda_k t } p_k \hat{\rho}_k,
\end{equation}
where $\lambda_k, \hat{\rho}_k$ are the eigenvalues and right eigenvectors of the Liouvillian superoperator $\hat{ \cal L}$ defined in \eqref{eq:Lin}, i.e., $\hat{ \cal L}(\hat{\rho}_k) = \lambda_k \hat{\rho}_k$, and the coefficients $p_k$ are the overlaps of the initial state with the left eigenvectors of $\hat{ \cal L}$. For the chosen Lindblad operators, the steady state $\hat{\rho}_\infty = \lim_{t\to\infty} \hat\rho(t)$ is unique and corresponds to an infinite temperature state, $\hat{\rho}_\infty \propto \mathbb{I}$, corresponding to the eigenvalue $\lambda_0=0$. At the same time, the slowest decaying mode is associated with the eigenvalue $\lambda_1$ with the second largest real part; its relaxation rate is given by $\Gamma = -\Re( \lambda_1 )$.

Ref.~\cite{Morningstar22, Sels21} proposed that insights into the stability of MBL in a disordered spin chain coupled to a large thermal inclusion can be obtained by the analysis of the GKSL equation with Lindblad operators  $\hat{L}_k$ given by Eq.~\eqref{eq:Lks}. According to this proposition, the MBL is stable to avalanches spreading throughout the system if the relaxation rate of the slowest decaying mode, $\Gamma$, decreases with system size $L$ faster than $4^{-L}$. This condition, in the perturbative limit of weak coupling to the Markovian bath described by \eqref{eq:Lin}, is fulfilled by the disordered XXZ spin chain at $W^{\mathrm{avch}}(L=14) \approx 20$~\cite{Sels21}, exceeding the value of $W^*(L=14)$ extracted from the standard indicators of ergodicity breaking available in exact diagonalization studies by a factor of more than $5$. The disorder strength $W^{\mathrm{avch}}(L)$ is compared by~\cite{Morningstar22} to other disorder strengths which are sensitive to the emergence of resonances between different spin configurations in eigenstates of \eqref{Hxxz}. The latter disorder strengths exceed the value of $W^*(L)$ by a factor of $2$ but are still well below the value of $W^{\mathrm{avch}}(L)$. We revisit their relation in Sec.~\ref{sec:reson}.

It is worth emphasizing that the connection between a GKSL equation with the Lindblad operators~\eqref{eq:Lks}, and the spreading of avalanches seeded by an ergodic inclusion, needs to be further clarified. In particular, it is even not apparent whether a large ergodic inclusion formed by a weakly disordered region in the XXZ spin chain fulfills the stringent assumptions behind the GKSL equation discussed here. One possible reconciliation of the value of $W^{\mathrm{avch}}(L)$ with the exact diagonalization results, which place the ETH-MBL crossover at significantly smaller disorder strengths, is that the effects of rare large thermal regions are absent at system sizes accessible to exact diagonalization~\cite{Sels21}.
This perspective is, however, discouraged by the results of~\cite{Szoldra24catching}. Despite the planted thermal inclusions are likely to be found only in systems as large as $L \approx 10^6$,~\cite{Szoldra24catching} do not observe any significant drift of the disorder strengths at which qualitative changes between the ergodic and MBL time evolutions are observed. Therefore, further investigations of microscopic mechanisms behind the avalanche propagation~\cite{Ha23} and the emergence of ergodicity in disordered spin chains are needed. Such studies should also consider the more complicated types of dynamics allowing for intermediate dynamical regimes~\cite{Crowley20, Crowley22meanfield}.

\subsection{Other approaches}
In the following, we briefly outline other propositions for the mechanism of the MBL transition.
We first highlight an alternative approach to the avalanche mechanism based on introducing strongly disordered impurities to an otherwise ergodic system, which enables investigation of the onset of the MBL-like dynamics.
We then focus on many-body resonances, which provide a distinct avenue for studies of the emergence of ergodicity in many-body systems.

\subsubsection{Thermalization of dilute impurities}
The idea of the avalanche scenario is to examine the stability of LIOMs to rare thermal regions. Let us assume that the conditions for the propagation of avalanches are fulfilled. In that case, the MBL regions are destabilized due to the coupling to rare thermal regions, and the whole system becomes ergodic. Ref.~\cite{Sels21dilute} proposed to reverse this perspective and look for a mechanism of MBL transition by examining the fate of a small density of strongly disordered sites, referred to as "impurities," in an otherwise ergodic system.

The impurity model is taken as the disordered XXZ spin chain \eqref{Hxxz} with on-site magnetic fields $h_i$ distributed uniformly in the intervals
\begin{equation}
    h_i \in \begin{cases}
  [V/2 - W_0, 3V/2+ W_0]  & \text{ if } i \in \cal I, \\
  [-W_0, W_0] & \text{ otherwise},
  \label{eq:disIMP}
\end{cases}
\end{equation}
where $V$ is the impurity potential strength, $W_0$ is a disorder strength (chosen as $W_0=1/4$ so that the system \eqref{Hxxz} is deep in the ergodic regime for $V=0$), and $\cal I$ is the set of sites at which the impurities are located. Introduction of the impurities, which at finite $L$ and $V \gg 1$ give rise to LIOMs, allows for insights into the finite size effects of the disordered spin chains and for understanding the mechanism of their thermalization in the thermodynamic limit.

Focusing on a single impurity case, when the set of the sites with impurities contains a single element, ${\cal I }= \{ (L+1)/2 \} $, Ref.~\cite{Sels21dilute}
demonstrated that the impurity is associated with a LIOM (which is adiabatically linked to the $\hat{S}^z_{ (L+1)/2}$ operator in the $V\to \infty$ limit) provided that the system size is smaller than a threshold value, which scales linearly with the impurity potential strength $V$ up to logarithmic corrections. Hence, the system is globally ergodic for $V$ below $V^*\propto L$ and the LIOM associated with the impurity becomes delocalized at $V<V^*$. Consequently, the average gap ratio $\overline r$ starts deviating from the GOE value at $V^*\propto L$. 
The subsequent increase of $V$ beyond $V^*$ leads to a further decrease of the value of $\overline r$, as the parts of the chain on the two sides of the impurity are connected at large $V$ only by the virtual terms $\propto 1/V$. 

Ref.~\cite{Sels21dilute} also considered the case of a finite density of impurities. In particular, by placing the impurities on every fourth lattice site, the impurity potential strength $V^*$, at which the deviation of $\overline r$ from the GOE value is visible, is again scaling linearly with system size,  $V^* \propto L$ (within the examined interval of system sizes $L<20$). However, an increasing number of lattice sites, associated with an increase of the number of impurities in the system, leads to a growth of the number of sectors into which the chain gets broken in the large $V$ limit. The increased number of blocks into which the Hilbert space is effectively fragmented causes an additional decrease of $\overline r$ at $V>V^*$ towards the Poisson value. This reduction of the value of $\overline r$ at $V>V^*$ leads to an appearance of crossing points of the  $\overline r(V)$ curves. However, these crossing points do not convey information about the ergodicity breakdown in the system, as the LIOMs are already destabilized at the weaker impurity potential $V=V^* \propto L$. 

These observations, substantiated by insights into the behavior of the spectral functions and operator spreading in the system, caused the authors of Ref.~\cite{Sels21dilute} to conclude that the system remains ergodic for finite impurity densities at system sizes beyond $L\propto V^*$. The appearance of the crossing points for the $\overline r$ is a result of the fragmentation of the Hilbert space into an increasing number of sectors rather than a signal of the breakdown of the ergodicity. The breakdown of the ergodicity occurs already at  $V^*\propto L$. The latter behavior is analogous to the linear drift $W^T(L) \propto L$ found in the results for the disordered spin chains as discussed in Sec.~\ref{subsubsec:gapratio}. If these linear system size drifts persist to $L\to \infty$, the spin chains are ultimately, in the thermodynamic limit, ergodic at any $V$. Regardless of whether that is indeed the case, the impurity approach enhances our understanding of finite-size drifts in disordered spin chains, providing a helpful perspective on MBL transition. Yet, in specific Floquet systems, the linear drift $W^T(L)\propto L$ is broken at sufficiently large system sizes and, instead, a sub-linear dependence of $W^T$ on $L$ is observed~\cite{Sierant22floquet}. Therefore, computational studies of linear drifts at large system sizes in autonomous disordered spin chains may shed further light on the status of MBL.

\subsubsection{Models of many-body resonances}
Broadly speaking, many-body resonances occur whenever an eigenstate of a many-body system can be expressed as a superposition of a few not entangled or weakly entangled states, e.g., the computational basis states, that substantially differ in extensively many local regions. Dynamically, the many-body resonances manifest as large oscillations in correlation functions of local observables.

Resonances models introduced in~\cite{Crowley21, Garratt21Local, Villalonga20} assume the presence of LIOMs in the sufficiently strongly disordered spin chains and investigate the statistics of many-body resonances upon local perturbations. The proliferation of many-body resonances leads to the destabilization of MBL and the onset of ergodicity. 

Within the model of \cite{Crowley21}, the probability that a given eigenstate finds a first-order resonance involving LIOMs within a range $r$ (defined as the distance between the perturbed site and the furthest LIOM flipped by the perturbation) is given by 
\begin{align}
        \label{eq:qr_intro}
        q(r) = \frac{\mathrm{e}^{-r/\xi}}{\lambda},
\end{align}
where the correlation length $\xi$ sets the typical range of resonances, while the resonance length $\lambda$ determines the density of resonances. The approach of \cite{Crowley21} employs perturbative arguments to obtain insights into the behavior of $\xi$ and $\lambda$ in disordered spin chains. These arguments suggest that $\xi$ diverges at the MBL transition and that the finite-size behavior in the vicinity of the transition is determined by the resonance length $\lambda$. Notably, the value of the latter parameter corresponding to the physics of the disordered XXZ spin chain is estimated as $15 \lessapprox \lambda \lessapprox 50$. For that parameter, the resonance model of \cite{Crowley21} reproduces many features of the results observed numerically for the disordered spin chains, including: the linear drift of the disorder strength $W^T(L)$ at the ETH-MBL crossover, slow dynamics, and the apparent $1/\omega$ behavior of the spectral function in the crossover regime, or the exponential increase of Thouless time $t_{Th}$ at small disorder strengths. Notably, all of those features are obtained within the resonance model, which, by construction, hosts an MBL phase. While the resonance model of \cite{Crowley21} relies on the first-order perturbation theory, which becomes invalid in the vicinity of the ergodic phase, it provides the basis for interpretation of the phenomenology of prethermal MBL proposed in \cite{Long22Prethermal}, as discussed in Sec.~\ref{sub:stretched}. 

The model of many-body resonances considered in~\cite{Garratt21Local} employs the perspective of parametric level dynamics induced by smooth variations of the disorder, see Sec.~\ref{subsec:sensitivity}, to understand the properties of resonances between the LIOM configurations in the MBL regime. This framework allowed to determine the distributions of off-diagonal matrix elements of local operators and their correlations with the level spectrum. A subsequent work~\cite{Garratt22Resonant} provided an extension of this framework, formulated without a direct reference to the picture of LIOMs. Ref.~\cite{Villalonga20} used numerically constructed LIOMs to investigate the many-body resonances, finding hybridizations at all considered length scales in the vicinity of the ETH-MBL crossover.

\subsubsection{Probing many-body resonances with exact diagonalization}
\label{sec:reson}
In addition to the open system approach and the corresponding disorder strength $W^{\mathrm{avch}}(L)$, Ref.~\cite{Morningstar22} proposed several probes of many-body resonances in eigenstates of disordered spin chains.

Quantum mutual information (QMI) $I^{\mathrm{e-e}}_2$~\eqref{eq:qmi} between the ends of the chain (assuming open boundary conditions), i.e., for subsystems $A_1=\{1\}$ and $A_2=\{L\}$ corresponding to the first ($j=1$) and the last ($j=L$) spin, allows to examine emergence of many-body resonances. Deep in the ergodic phase, the QMI $I^{\mathrm{e-e}}_2$ is small since the entanglement entropy follows the volume-law scaling and the contribution $S_{A_1}+S_{A_2}$ approximately cancels out with $S_{A_1 \cup A_2}$. Similarly, the QMI vanishes deep in the MBL regime when the end spins are not correlated. In contrast, $I^{\mathrm{e-e}}_2$  is enhanced in the intermediate crossover regime between the ergodic and MBL regimes when the eigenstates are superpositions of a limited number of basis states with substantially different spin configurations. For that reason, $I^{\mathrm{e-e}}_2$ is well suited for investigation of system-wide resonances appearing whenever an eigenstate is a Schr\"{o}dinger cat-like state composed of a few basis states. To probe the onset of such resonances, \cite{Morningstar22} considered the maximum $I^{\mathrm{e-e}}_{\mathrm{max}}$ of $I^{\mathrm{e-e}}_2$ over all eigenstates of the disordered XXZ spin chain, and take the median of $I^{\mathrm{e-e}}_{\mathrm{max}}$ over disorder realizations.

The resulting quantity, plotted as a function of $W$ for various $L$ has a crossing point $W^{\mathrm{swr}}(L)$, similarly to $\overline r$ shown in Fig.~\ref{fig:ER}. Notably, the crossing points $W^{\mathrm{swr}}(L)$ occur at significantly larger disorder strengths than $W^*(L)$. For instance, for the disordered XXZ spin chain, $W^*(L=16)\approx 3.5$, while $W^{\mathrm{swr}}(L=16) \approx 8.5$. The drift of $W^{\mathrm{swr}}(L)$  with an increase of system size $L$, relative to its value at small system sizes, is comparable to the drift of $W^*(L)$. In particular, between $L=11$ and $L=16$ the value of $W^{\mathrm{swr}}(L)$ increases by approximately $1/3$ of its value. Notably, however, the value of the QMI $I^{\mathrm{e-e}}_{\mathrm{max}}$ at the crossing point $W=W^{\mathrm{swr}}(L)$ is of the order of $10^{-7}$. Hence, the resonances found by this quantity are very weak. On the other hand, these resonances extend between the ends of the whole chain, indicating some form of breakdown of locality in the system. 
The maximum of $I^{\mathrm{e-e}}_2$ as a function of disorder strength also signals a characteristic disorder strength quantifying the MBL crossover. The maximum of $I^{\mathrm{e-e}}_2$ occurs at disorder strength comparable to the crossing point of gap ratio $W^*(L=16)\approx 3.5$~\cite{Morningstar22}. 

Another quantity considered by~\cite{Morningstar22} in the context of many-body resonances is the minimal energy gap $\delta$, and the minimal gap ratio $r_\mathrm{min}$. The latter quantity is obtained by minimizing \eqref{eq:er} over the entire many-body spectrum. Rescaling $r_\mathrm{min}$ by the expected behavior for Poisson statistics, ~\cite{Morningstar22} extract the resulting crossing points leading to yet another system size dependent disorder strength $W^{\mathrm{mg}}(L)$. While $W^{\mathrm{mg}}(L=16)\approx 5.7$ is significantly larger than $W^*(L=16)$, the drift of $W^{\mathrm{mg}}(L)$ with system size $L$ appears to be considerably weaker in comparison to the other disorder strengths.

Juxtaposing the values of $W^{\mathrm{swr}}(L)$, $W^{\mathrm{mg}}(L)$ and  $W^{\mathrm{avch}}(L)$ with the value of $W^*(L)$,~\cite{Morningstar22, Sels21} concluded that the MBL phase, if it exists at all in the disordered XXZ spin chain, can be present only at disorder strengths larger than $W \gtrapprox 20$. The latter inequality stems from $W^{\mathrm{avch}}(L=14)\approx 20$, which is considerably larger than the other characteristic disorder strengths, $W^{\mathrm{swr}}(L)$, $W^{\mathrm{mg}}(L)$, and $W^*(L)$.
This conclusion leads to two critical questions which remain open. The first question is whether $W^{\mathrm{avch}}(L)$, obtained in the open system approach, is relevant for the physics of isolated disordered spin chains. The second question is about understanding the relation of  $W^{\mathrm{swr}}(L)$ and $W^{\mathrm{mg}}(L)$, which are obtained as the extreme values of ergodicity breaking indicators, with the standard quantities analyzed in the context of ETH-MBL crossover such as $W^*(L)$. 
We can improve our understanding of this connection through computational studies of disordered spin chains and reference systems, such as the Anderson model on random regular graphs, or random matrix models \cite{Kutlin24resonances}, even at system sizes available for present-day classical computers. An investigation of extreme value statistics of an average of local observable in the disordered XXZ spin chains reported a sharp change of the properties of extreme value statistics at $W=4-7$~\cite{Colbois23}. These avenues constitute intriguing directions for further studies of MBL and may shed new light on the existence and critical disorder strength of the MBL transition. 

\subsubsection{Large-deviation approaches to resonances}
\label{sec:reson2}
Understanding many-body resonances as eigenstates that are superpositions of a small number of basis states reinforces the significance of the Fock space perspective on MBL~\cite{Logan19local, Roy19percolation, Tarzia20}. To enhance the effects of rare configurations corresponding to long-range resonances, Ref.~\cite{Biroli23largedeviation} introduced a biased sampling scheme that magnifies the impact of the tails of the probability distribution of the transmission amplitudes.

Considering the propagator $G_{j,0} = \braket{ j | H^{-1} | 0}$ at energy $E=0$ between a fixed state $\ket{0}$, and states $\ket{j}$ of the eigenbasis of $\hat{S}^z_j$ operators,~\cite{Biroli23largedeviation} examined the quantity
\begin{equation} \label{eq:TMBL}
{\cal T}_0 (\beta) = \sum_{j \in {\cal J}_n } |G_{j,0} |^\beta \,,
\end{equation}
where ${\cal J}_n$ denotes an exponentially large set of basis states differing by a fixed number of spin flips $n$ from the state $\ket{0}$. The function ${\cal T}_0 (\beta)$ is analogous to a 
partition function of statistical mechanics model defined on a Cayley tree~\cite{Derrida88}, with correlated energies $\omega_{j,0} = -\ln(G_{j,0})$ and the inverse temperature $\beta$. This analogy motivates the quantitative investigations of~\cite{Biroli23largedeviation}.
For $\beta=2$, the quantity ${\cal T}_0 (2)$, in terms of the single-particle hopping on the Fock space graph, can be interpreted as being proportional to the total Landauer transmission corresponding to the situation when the particles are injected at the point $\ket{0}$ and extracted at the node $\ket{j}$ for $j\in \cal{J}$. The basic idea is, however, to vary the auxiliary parameter $\beta$, tuning the relative contribution from members $\ket{j}$ of $\cal{J}$ that form strong resonances with the spin configuration $\ket{0}$, corresponding to significant values of $G_{j,0}$.

Analyzing dependence of a free energy $\phi(\beta) = \frac{1}{\beta n} \ln \braket{ {\cal T}_0 (\beta)  } $ of the value of $\beta$,~\cite{Biroli23largedeviation} distinguished four regimes of disorder strength characterized by different behavior of the propagator $G_{j,0}$. At disorder strength $W<W_{\mathrm{ergo}}$, all self-avoiding paths on the Fock space graph between $\ket{0}$ and the states $\ket{j}$ contribute to the dynamics. The system resembles a fully connected quantum dot in that regime, and the level statistics approach the RMT predictions. At larger disorder strengths $W_{\mathrm{ergo}} < W< W_g$, the number of self-avoiding paths contributing to the dynamics grows exponentially with $n$ but only a sub-extensive fraction of paths contributes, and the level statistics is no longer follows the RMT prediction. For $W_g<W<W_c$, the state $\ket{0}$ hybridizes only with a few, $O(1)$ states $\ket{j}$ far away in the Hilbert space. Those rare many-body resonances are unraveled thanks to tuning the value of $\beta$. Finally, at $W>W_c$ the probability to evolve away from the initial state $\ket{0}$ goes to zero exponentially with $n$ even when the system-wide resonances are accounted for. Interestingly, Ref.~\cite{Biroli23largedeviation} reported that the drift of the disorder strength $W_c$ with system size $L$ is very mild. Moreover, the same approach based on \eqref{eq:TMBL} to the Anderson model on a Cayley tree reproduces the known critical disorder strength $W_c$~\cite{parisi2019anderson, Tikhonov19} with a good accuracy. 
 
Interesting directions for future studies involve linking the four disorder regimes found in~\cite{Biroli23largedeviation} with properties of rare thermal inclusions, and understanding the relation of the four disorder regimes to the many-body resonance probes considered in \cite{Morningstar22}. Finally, it would be interesting to understand whether the biased sampling method of~\cite{Biroli23largedeviation} can be formulated for quantities that can be probed by the time evolution of many-body systems, increasing the relevance of the method both from theoretical and experimental standpoints. The described future directions may shed new light on the status of MBL in the disordered spin chains.

\section{Possible extensions}
\label{sec:ext}
In this Section, we briefly describe investigations of MBL in many-body systems different than the disordered spin chains on which we focused so far. 
We start with disordered one-dimensional Hamiltonians with short-range interactions, which share the gross features characterizing ergodicity breaking in the disordered XXZ spin chain, but also host a number of distinguishing features. We then proceed by concisely outlining the studies of MBL in systems with quasiperiodic potential, in disordered Floquet systems, in systems with the presence of long-range interaction, in systems beyond one spatial dimension, and in systems without disorder.

\subsection{Other models of MBL}
\label{subsec:otherModels}

\paragraph{Disordered quantum Ising models}
The one-dimensional transverse field Ising model (TFIM) is a spin-1/2 chain described by the Hamiltonian
\begin{equation}
\hat{H}_{\mathrm{TFIM} } =  \sum_{i=1}^L J_i \hat{S}_i^z \hat{S}_{i+1}^z + \sum_{i=1}^L (h_i \hat{S}_i^z + \gamma_i \hat{S}_i^x),
     \label{eq:TFIM}
\end{equation}
where $J_i$, $h_i$ and $\gamma_i$ are random couplings selected from certain distributions. The clean system with non-vanishing couplings $J_i=J$, $h_i=h$, and $\gamma_i = \gamma$ is an example of non-intergrable ergodic system with ballistic spreading of entanglement and diffusive energy transport~\cite{Kim13}. One choice of the couplings is $\gamma_i=1$, $J_i \in[0.8, 1.2]$, and $h_i \in[-W,W]$. In that case, tuning the disorder strength $W$ leads to a crossover between the ETH and MBL regimes~\cite{Abanin21, DeTomasi21}. The finite size drifts at the crossover appear analogous to the case of the disordered XXZ and $J_1$-$J_2$ models described in Sec.~\ref{subsec:spectral}. For instance, as shown in~\cite{Sierant22floquet}, the crossing point shifts to larger disorder strengths $W$ sub-linearly with the increase of $L$ and can be approximated as $W^*(L) \propto 1/L$. The boundary of the ergodic regime in the disordered TFIM drifts approximately linearly with the system size $L$, $W^T(L) \propto L$. The Hamiltonian~\eqref{eq:TFIM} belongs to the class of models considered in~\cite{deRoeck24absence}.

The disordered transverse field Ising model, in contrast to the disordered XXZ spin chain \eqref{Hxxz}, has no $U(1)$ symmetry. This leads to a larger dimension of the Hilbert space, $\mathcal N = 2^L$, at a fixed system size $L$. However, the geometry of the Fock-space graph is significantly simpler than in the disordered XXZ spin chain. Each of the states in the computational basis (i.e., the eigenbasis of the $\hat{S}^z_i$ operators) is linked by the Hamiltonian to exactly $L$ states differing by a single spin flip, simplifying the Fock-space perspective on MBL in the TFIM~\cite{Roy21anatomy, Creed23transportFS}.

Hamiltonian obtained as $\hat{H}'=\hat{H}_{\mathrm{TFIM} } + \sum_i g_i \hat{S}_i^z \hat{S}_{i+2}^z$ has a $Z_2$ symmetry given by the operator $\prod_{j=1}^L \hat{S}_i^x$, provided that $h_i = 0$ is set. Due to the breakdown of the ergodicity in the MBL regime, the $Z_2$ symmetry may be broken not only in the ground state, but also in highly excited eigenstates, as proposed by \cite{Huse13}\footnote{Analogously, the Anderson localization in non-interacting models leads to breakdown of the $Z_2$ symmetry at all 
energy densities as reviewed in \cite{Laflorencie22chapter}.}. Numerical evidence for the MBL regimes with paramagnetic and spin glass order at all energy densities was found in \cite{Kjall14}. 
Notably, the ergodicity was suggested~\cite{Moudgalya20, Sahay21} to be restored at crossovers between the different MBL regimes, and the wide extension of the intervening ergodic regime was verified by~\cite{Laflorencie22}\footnote{Tensor network investigations of the ground state of the disordered quanum Ising chain find that the single-particle localization is stable against finite interactions~\cite{Chepiga24}. This suggest presence of a crossover to ergodic states at finite energy density above the ground state.}. 
Signatures of topological MBL phases~\cite{Chandran14} were observed in a
variant of the disordered Ising spin chain in \cite{Laflorencie22}.

\paragraph{Disordered Fermi-Hubbard model}
The Fermi-Hubbard model~\cite{Gutzwiller63, Kanamori63, Hubbard64} is a fundamental tight-binding lattice model, offering insights into insulating, magnetic, and superconducting effects of interacting fermions~\cite{Lee06}, and may be realized in experiments with ultracold atoms~\cite{Esslinger10}. Hamiltonian of the Fermi-Hubbard model is given as
\begin{equation}
  \hat{H}_{\mathrm{FH}} = \sum_{i,\sigma=\uparrow, \downarrow} J( \hat{c}^{\dag}_{i\sigma}\hat{c}_{i+1,\sigma} +h.c)+
  \sum_i U \hat{n}_{i\uparrow}\hat{n}_{i\downarrow} +
  \sum_i \mu_i \hat{n}_{i},
 \label{eq:ham_FH}
\end{equation}
where $c^{\dag}_{i,\sigma}, \hat{c}_{i\sigma}$ are creation and annihilation operators for fermions of spin $\sigma = \uparrow, \downarrow$, $J=-1$ and $U$
are tunneling and interaction amplitudes, respectively, $\mu_i \in [-W;W]$ is a random on-site potential, and the sum over $i$ extends from $1$ to the system size $L$ . Defining $\hat{n}_{i,\sigma} = \hat{c}^{\dag}_{i,\sigma} \hat{c}_{i\sigma}$, we observe that $\hat{H}_{\mathrm{FH}}$ separately conserves the total number of fermions of spin up, $\hat{N}_\uparrow = \sum_i \hat{n}_{i,\uparrow}$, and spin down, $\hat{N}_\downarrow =\sum_i \hat{n}_{i,\downarrow}$. An optical lattice realizing the Fermi-Hubbard model with a quasiperiodic potential playing the role of the disorder $\mu_i$, see Sec.~\ref{subsec:QPD}, allowed for the first observation of signatures of the MBL regime \cite{Schreiber15}.

The on-site Hilbert space of the Fermi-Hubbard is four dimensional, which leads to richer physics than in the disordered spin-1/2 chains. The Hamiltonian \eqref{eq:ham_FH} possesses an additional parity symmetry, and the SU(2) symmetry that may be broken by an addition of weak magnetic field and chemical potential terms at the edges of the chain, enabling observation of the ETH-MBL crossover with the average gap ratio~\cite{Mondaini15}. Moreover, the disorder $\mu_i$ in Eq.~\eqref{eq:ham_FH} is a charge disorder, i.e., the fermions of different spins experience the same value of $\mu_i$. This leads to distinct dynamical behaviors of the charge degrees of freedom, $\hat{n}_i = \hat{n}_{i,\uparrow} +\hat{n}_{i,\downarrow}$, and spin degrees of freedom, $\hat{s}^z_i = (\hat{n}_{i,\uparrow} -\hat{n}_{i,\downarrow})/2$. It was proposed~\cite{Prelovsek16} that spin degrees of freedom remain ergodic, while charge degrees of freedom are in the MBL regime. This scenario has been  further investigated numerically in~\cite{Zakrzewski18,Kozarzewski18, Protopopov19, Johns19, Sroda19, Richter22}. Another possibility is to include a spin disorder upon replacing the term $\sum_i \mu_i \hat{n}_{i}$ in $\hat{H}_{\mathrm{FH}}$ by $\sum_i \mu_i \hat{s}^z_{i}$. In that case, the model has an  $\eta$-pairing symmetry \cite{Yang89} and it is possible to construct a large number of eigenstates of $\hat{H}_{\mathrm{FH}}$, which are characterized both by area-law entanglement entropy, as well as eigenstates with the entanglement entropy scaling logarithmically with system size $L$. Remarkably, the  $\eta$-pairing symmetry allows to numerically construct highly excited eigenstates of systems comprising up to few hundred lattice sites~\cite{Yu18}.

From the point of view of the currently understood trends towards ergodicity, observed in the disordered spin chains, it would be interesting to reconsider the case of charge disorder and understand to which degree the ergodic spin degrees of freedom may serve as a bath for the localized charge degrees of freedom. One step in this direction was done by~\cite{Krause21} and followed by~\cite{Brighi21a, Brighi21b} which consider a single mobile impurity, $ \hat{N}_\uparrow = 1$, in contact with $ \hat{N}_\downarrow = L/3$ particles of different species in Anderson localized state\footnote{Refs.~\cite{Brighi21a, Brighi21b} investigated a system of $L/3$ hard-core bosons interacting with a single boson from different species. The latter model can be mapped to Fermi-Hubbard Hamiltonian upon a generalization of Jordan-Wigner transformation considered in \cite{Shastry86, Prosen12}.}. In that case, the on-site potential term in $\sum_i \mu_i \hat{n}_{i}$ is replaced by $\sum_i \mu_i \hat{n}_{i, \downarrow}$, and the mobile impurity playes a role of an $L$-level heat bath. The Anderson localization regime is argued to remain stable in the regime of strong interaction between the two species of particles. In that case, the single impurity becomes immobile and the interactions effectively turn the Anderson insulator into an MBL system with non-trivial entanglement entropy dynamics. Subsequent~\cite{Sierant23slow} numerical simulations indicate that there is a wide regime of parameters in which the impurity spreads subdiffusively and induces a gradual delocalization of the Anderson insulator. Nevertheless, at sufficiently large interaction and disorder strength the dynamics remain non-ergodic at all accessible system sizes and time scales~\cite{Brighi23}.

\paragraph{Disordered Bose-Hubbard model}
The Bose-Hubbard model~\cite{Gersch63} is a fundamental model describing bosons hopping on a lattice, and hosting a superfluid to Mott insulator ground state phase transition \cite{Fisher89}, which was realized with ultracold bosons in optical lattices \cite{Greiner02}. The Bose-Hubbard Hamiltonian reads
\begin{equation}
  \hat {H}_{BH} = J \sum_{\langle i,j \rangle} \hat{a}^{\dag}_i\hat{a}_j +
 \frac{ U }{2} \sum_{i=1}^L \hat{n}_i (\hat{n}_i - 1) +
  \sum_{i=1}^L \mu_i \hat{n}_i,
 \label{eq:ham_BH}
\end{equation}
where $\hat{a}^{\dag}_i, \hat{a}_i$ are bosonic creation and annihilation operators, 
the tunneling amplitude $J=-1$ sets the energy scale, $U>0$ is the interaction strength, the chemical potential $\mu_i$ is distributed uniformly in an interval $[-W;W]$, and $\langle i,j \rangle$ is a sum over the neighboring sites of 1D lattice comprising $L$ sites. Defining the bosonic number operator as $\hat{n}_i= \hat{a}^{\dag}_i \hat{a}_i$, we observe that $ \hat {H}_{BH} $ conserves the total number of bosons $\hat{N} = \sum_i \hat{n}_i$.

Increase of the disorder strength $W$ in the disordered Bose-Hubbard model \eqref{eq:ham_BH} leads to a crossover between the ETH and MBL regimes~\cite{Sierant18, Orell19, Hopjan19}. The main difference between the Bose-Hubbard model and the disordered spin-1/2 chain is that the number of the bosons $n_i$ on a site $i$ is not restricted in the former model. At a fixed density of particles $\nu=N/L$, the states with atypically large particle occupation numbers lie at large energy densities. The density of states at these large energy densities is small, which leads to an inverted many-body mobility edge observed numerically in the disordered Bose-Hubbard model~\cite{Sierant18}, see also \cite{Koshkaki22}. The question whether this inverted many-body mobility edge persists in the asymptotic limit $L,t \to \infty$ remains open.

\paragraph{Disordered constrained spin chains}
Replacing the term $\sum_{i=1}^L J_i \hat{S}_i^z \hat{S}_{i+1}^z$ by $\sum_{i=1}^L J_i (\hat{S}_i^z+1/2) (\hat{S}_{i+1}^z+1/2) $ in the TFIM \eqref{eq:TFIM}, taking the limit of strong interactions, ${ J_i \to \infty }$, 
and reducing the description to the lowest energy subspace, we arrive at the (generalized) PXP Hamiltonian,
\begin{equation}
 \hat{H}_{\mathrm{PXP}}= \sum_{i=1}^{L} \hat{P}^{\alpha}_{i} \hat{S}^x_i \hat{P}^{\alpha}_{i+1+\alpha}+
\sum_{i=1}^{L} h_i \hat{S}^z_i,
 \label{Hpxp}
\end{equation}
where the projectors $\hat{P}^{\alpha}_i=\prod_{j=i-\alpha}^{i-1}(1/2-\hat{S}^z_j)$ assure that the dynamics are confined to a constrained subspace of the full Hilbert space, in which sites with state $\ket{\uparrow}$ are separated by $\alpha$ sites. The limit of strong interactions in the TFIM corresponds to the blockade radius $\alpha=1$, however, this model can be also considered for a more general $\alpha>1$.
In Eq.~\eqref{Hpxp}, the values of $h_i$ are independent and uniformly distributed
random variables in the interval $[-W/2,W/2]$ with $W$ being the disorder strength, and periodic boundary conditions (PBC) $\vec{S}_{L+i}\equiv \vec{S}_{i}$ are assumed. 

For $\alpha =1$, the model in Eq.~\eqref{Hpxp} is known as the PXP model. In the clean limit, $h_i=0$, the PXP Hamiltonian is a paradigmatic model of quantum many-body scars~\cite{Lesanovsky12, Turner18, Serbyn21}, describing the experiments with arrays of Rydberg atoms \cite{Bernien17, Browaeys20}, and can be mapped onto a U(1) gauge theory \cite{Surace20}. In presence of disorder, $W>0$, the PXP model hosts a crossover between the ETH and MBL regimes~\cite{Chen18}. However, the system size drifts at this crossover are significantly more pronounced than in the disordered XXZ spin chain, and both disorder strengths $W^T(L)$ as well as $W^*(L)$ shift linearly with $L$~\cite{Sierant21constraint}. 

The Hilbert space dimension of the PXP model \eqref{Hpxp} scales as $\mathcal N \propto (\Phi_\alpha)^L$, where $\Phi_\alpha<2$ decreases monotonously with the growth of the blockade radius $\alpha$. Hence, for larger $\alpha$ it is possible to access larger system sizes $L$ with exact diagonalization methods. Notably, at $\alpha>1$, the linear trends $W^T(L)\propto L$ and $W^*(L)\propto L$ are even more pronounced, suggesting that the PXP models \eqref{Hpxp} remain ergodic in the asymptotic limit. The nature of the kinetic constraints in $\hat{H}_{\mathrm{PXP}}$ is one factor responsible for this behavior. Indeed, the effective radius of the Fock space graph is much smaller in the PXP model than in the TFIM, enhancing the trends towards ergodicity in the former model~\cite{Sierant21constraint}. Notably, Ref.~\cite{Royen23EnhancedMBL} reported an opposite trend of enhancing the MBL regime due to the interplay between interaction and kinetic constraints that are of different type than those considered here.

The milder growth of the Hilbert space with increase of the number of lattice sites in constrained spin models has been also used in investigations of 2D models \cite{Theveniaut20, Pietracaprina20dimer}. Moreover, the local constraint resemble, to some degree, the local Gauss law in lattice gauge theories. The latter have been shown to host a pronounced region in which the ergodicity is broken. The boundary of this region, i.e., the disorder strength $W^T(L)$, is nearly independent of system size $L$ in the 1D Schwinger model with the disordered background charges~\cite{Giudici20}.

\subsection{Quasiperiodic potential}
\label{subsec:QPD}
The discussion above concentrated on the random disorder, i.e., on the situation in which the on-site potential strengths $h_i$ are independent and identically distributed random variables. 
Introduction of the so-called quasiperiodic potential is another way of breaking the translational invariance in many-body systems that leads to a similar phenomena of ergodicity breaking. One example of a quasiperiodic potential is
\begin{equation}\label{eq:qp}
  h_j=W\cos(2\pi \eta j +\varphi),  
\end{equation}
where $W$ is the amplitude, $\eta$ is the incommensurability factor and $\varphi$ a random phase
differentiating between realizations of the potential. The quasiperiodic potential Eq.~\eqref{eq:qp} is routinely realized in ultra-cold atom experiments when on top of the standard optical lattice one puts additional weak potential coming from the second laser of incommensurate frequency with the one creating the primary lattice \cite{Guidoni97,Fallani07}. In fact, the demonstration of single-particle localization in cold atom setting was performed in parallel for random potential obtained using the so-called speckle field \cite{Billy08} and the quasiperiodic potential \cite{Roati08} where, strictly speaking, the Aubry-Andr\'e model~\cite{Aubry80} was realized. 
In 1D systems, the Anderson localization occurs for arbitrary random disorder strength,  while the Aubry-Andr\'e localization occurs only for sufficiently strong disorder $W$.
The convenient access to quasiperiodic potentials in optical lattices resulted in a series of experiments 
unraveling signatures of the MBL regime, see Sec.~\ref{sec:exp}.

An early study \cite{Iyer13} considered the XXZ model with diagonal quasiperiodic potential given by \eqref{eq:qp}, finding a crossover between the ETH and MBL regimes with the increase of the potential strength $W$. The initial claim was that the crossover gives rise to the MBL transition, which is smoothly connected, at weak interaction strength $\Delta$, to the Aubry-Andr\'e localization threshold. This claim was disputed by \cite{Znidaric18diffusiveQP}, which demonstrated that transport properties discontinuously change from localization to diffusion upon switching on the interactions.
Nevertheless, the strong spatial correlations between on-site potentials $h_i$, present in the quasiperiodic case, eliminate the possibility of emergence of the large Griffiths regions of anomalously weak (or strong) disorder. This constitutes a major difference with respect to the MBL problem in the random disorder case.
Several aspects of the dynamics in quasiperiodic potentials (and their generalizations) has been addressed over the years in diverse systems~\cite{Naldesi16,Setiawan17,Khemani17,BarLev17,Bera17,Weidinger18,Doggen19,Weiner19,Mace19,Singh21, Agrawal22,Zhang22,Strkalj22,Xu19butterfly, Tu23spectrum, Vu22fermionic, Prasad23SingleParticle, Thomson23}. 
It is worth observing that the potential \eqref{eq:qp} depends on two parameters $W$ and $\eta$. It turns out that the position of the ETH to MBL crossover, $W^*(L)$, strongly depends on the value of the incommensurability factor $\eta$ as numerically observed \cite{Doggen19}. This has been partially explained in the single-particle case \cite{Fallani07}. Namely, the Aubry-Andr\'e
transition (i.e., the transition in the absence of interactions, at $\Delta=0$) is determined by the distribution of the neighboring differences $\delta_i=h_{i+1}-h_i$ whose amplitude depends on ${\eta}$. Nevertheless, 
a complete understanding of the impact of the value of ${\eta}$, or other parameters specifying the quasiperiodic potential, c.f.~\cite{Mace19, Agrawal20}, on the ETH-MBL crossover is still lacking.

The absence of Griffiths regions in the quasiperiodic systems translate into the features of the crossover between the ETH and MBL regimes. Recall that the cost function analysis of the mean gap ratio for the uniform random disorder, put forward in \cite{Suntajs20}, yields a linear growth of the characteristic crossover disorder value, $W_{cr}\sim L$, suggesting a BKT type of the transition. A similar analysis, performed for the quasiperiodic case \cite{Aramthottil21}, also pointed towards the BKT transition, however, it also suggested $W_{cr}\sim \ln L$ rather than $W_{cr} \sim L$, thereby exhibiting a weaker dependence on the system size $L$.
Extrapolation of the drift leads to $W_{cr}\approx 4$ for $L=50$. 
Interestingly, around this value of $W$, the density autocorrelation function $C(t)$, defined in Eq.~\eqref{eq:den2}, for the initial N\'e{e}l state reveals slow oscillations that are in phase 
for a wide interval of system system sizes $L\in[12,200]$.
As argued in \cite{Sierant22challenges}, this may indicate a truly localized character of the dynamics. This suggests that a
clear evidence for the MBL phase may be found in quasiperiodic systems at smaller time and length scales than in the case of random disorder. 
This expectation is further supported by the faster than exponential decay of the diffusion constant in the XXZ spin chain with quasiperiodic potential~\cite{Prelovsek23}. Finally, we note that the analysis of the MBL regime from the Fock-space perspective allows to distinguish quasiperiodic and random systems by their sample-to-sample fluctuations~\cite{Ghosh24FockPropagator}, similarly to the entropy measures~\cite{Khemani17} and level 
statistics \cite{Sierant19level}.

\subsection{Other sources/types of disorder}
\label{sec:otherDisorder}
Several other interesting types of disorder were used over the years to study aspects related to the MBL. One example, important from the experimental perspective, is realized by the speckle disorder used in the non-interacting case for demonstrating Anderson localization \cite{Billy08}.
It has been also applied in an early attempt \cite{Kondov15} to observe MBL in a 3D experiment. 
One may also use a  
speckle disorder on top of the optical lattice to 
 create a random diagonal potential with variable correlation length \cite{Maksymov19}, which affects the properties of the ETH-MBL crossover \cite{Samanta22, Shi22}. While in the latter case, the numerically observed properties may be traced back to the spatial correlations of the potential, \cite{Zhang22Optimizing} considered a problem of finding a distribution of the on-site potential that would enhance the MBL signatures in a maximal way.

Another interesting possibility
is to introduce disorder by selecting the values of the on-site fields $h_i$ from a discrete set.
To that end, \cite{Gavish04} proposed to use heavy, practically immobile atoms, interacting with light particles of a different species. In that case, the heavy particles, on the time scales on which their motion can be neglected, produce an effective random binary potential for the light particles. If the light particles are confined to one dimension and are themselves not interacting, the model leads to Anderson localization. 
The interactions between light particles together with the effective binary potential may give rise to the onset of an MBL regime. Nevertheless, even a slow motion of the heavy particles was argued to lead to thermalization of such a system at suffiently long time scales~\cite{Papic15}.

Nonetheless, one may consider a model with binary quenched disorder for which the on-site potentials are chosen from a two element set, $h_i \in \{ -W, W \}$. The model is attractive from the theoretical perspective, as the two possible values of disorder may be modelled by extending the Hilbert space, adding an auxiliary quantum 1/2-spin for each lattice site. Quantum evolution of the enlarged system, following the proposition of \cite{Paredes05}, reproduces, in a single run, the average over all $2^L$ realizations of the binary disorder, see also \cite{Smith17free}. 
This enables calculation of the autocorrelation function $C(t)$, for the initial N\'eel state, directly in the thermodynamic limit $L\to\infty$~\cite{Tang15, Enss17infinite}.
However, the time evolution in this approach is followed up to a relatively short time $t=30$. 
Hence, due to the observed oscillations and the decay of $C(t)$, the extrapolation of the results to the limit $t\to \infty$, relevant for the MBL phase, is not feasible. 
Exact diagonalization results of the XXZ spin chain with binary disorder leads to similar values of the exponent $\nu$ as for the typically studied unform random disorder~\cite{Enss17infinite}, suggesting similarities between the finite size drifts in the two types of the model.
Let us note that other discrete disorders may be studied in the context of the MBL transition \cite{Janarek18}. 
For uncorrelated discrete disorders with more than 2 values, the ETH-MBL crossover in the XXZ spin chain shows a large degree of universality, at least for the system sizes accessible to exact diagonalization studies.~\cite{Janarek18} showed that various ergodicity breaking indicators $\overline X$, e.g., the gap ratio $\overline r$, plotted as a function of the variance of the on-site potential $h_i$, collapse for different types of disorder.

All the models considered so far, with the exception of the PXP model, have one common feature.
In the absence of interactions the models reduce to disordered single-particle problems in one dimension, 
which are Anderson localized. In fact, the proximity of the Anderson localized limit provides the usual avenue for studies of the dynamics of strongly disordered many-body systems, as we argued in Sec.~\ref{sub:proximity}.
Nevertheless, can we observe signatures of MBL when the system is delocalized in the absence of interactions? Such a problem was studied for bosons in a lattice with random on-site {\it interactions} \cite{Sierant17, Sierant17b, Sierant18}. In that case, $\mu_i=0$ and the on-site interaction term in \eqref{eq:ham_BH} is modified as $\frac{ U }{2} \sum_{i=1}^L \hat{n}_i (\hat{n}_i - 1) \to \frac{ 1 }{2} \sum_{i=1}^L U_i \hat{n}_i (\hat{n}_i - 1)$, where $U_i$ is a site-dependent interaction amplitude, drawn independently at each site from a specified distribution.
Independently, similar problems were considered for spinful fermions with random on-site interactions \cite{BarLev16} as well as for spinless fermions~\cite{LiX17}.  In the absence of interactions, $U_i=0$, the systems are delocalized with Bloch waves forming the eigenstates. Turning on random interactions, $U_i>0$, which for ultracold bosons can be realized by placing the system close to a wave-guide~\cite{Gimperlein05}, may lead to the MBL that clearly has no localized single-particle limit. Despite that, the characteristics of MBL regime observed both in quantities derived from eigenvalues and eigenvectors (e.g., the gap ratio), as well as features of time dynamics, e.g., the density autocorrelation function $C(t)$, show significant similarity with the standard disordered spin chains. Another model with  similar properties was considered in \cite{Mace19}. Here, particles were considered on a Fibonacci chain and the non-interacting case revealed multifractal eigenstates. Better understanding of finite-size drifts in such models, as well as identification of their dynamical features distinct from the standard spin chains, remain interesting avenues for further research.

Apart for the on-site disorder, one may also encounter disorder in tunnelling terms in the models in Eqs.~\eqref{Hxxz}, \eqref{eq:ham_BH} and~\eqref{eq:ham_FH}. The standard formulation of Lieb-Robinson bounds~\cite{Lieb1972} is insufficient to capture the complexity of the dynamics induced by "weak links", i.e., bonds for which $J_i$ in \eqref{Hxxz} is anomalously small. Instead, capturing the rich dynamics of such models requires a distinction between the bounds that hold for all sites of the chain and bounds that hold for a subsequence of sites, as shown in~\cite{Baldwin23}. 
Ground state physics of random-bond Heisenberg spin, for which $J_i$ 
are taken to be random variables, was studied using the renormalization group~\cite{Ma79, Dasgupta80, Fisher94}. These approaches were extended to the non-equilibrium systems  \cite{Vosk13, Vosk14} and were followed by \cite{Turkeshi20, Ruggiero22}. Notably, the random bond XXZ spin chain at $\Delta =1$ has an SU(2) symmetry, which was argued to be incompatible with the MBL phase \cite{Potter16Symmetry, Protopopov17}. In spite of this, the random bond Heisenberg spin chain hosts a wide non-ergodic regime~\cite{Kozarzewski18, Protopopov20}, and a similar wide non-ergodic regime is found in the SU(3) symmetric spin chains~\cite{Dabholkar24su3}. Better understanding of the finite size drifts at this crossover could allow us to form more precise formulations about the system in which the MBL phase may be expected to occur. A similar class of models with bond disorder has been recently studied theoretically in the context of chains of Rydberg atoms~\cite{Signoles21,Schultzen22,Tan22}.

\subsection{Floquet MBL}
\label{sec:floq}
Periodically driven systems played historically a very prominent role in quantum chaos studies, with textbook models such as kicked rotator or kicked top, see, e.g.,~\cite{Haakebook}. Many-body generalizations of the kicked top model were studied theoretically by \cite{Das10exotic, Notarnicola18,Notarnicola20,Russomanno21, Fava20kicked} and realized experimentally~\cite{SeeToh2022, Cao2022, Guo23observation}.
In the kicked top model, the evolution operator $U$ over a single period of the driving can be expressed as a product of two unitary operators, one describing the dynamics between the kicks, and the second describing the kick itself. The product of these operators is the Floquet operator $U$ of the system~\cite{Eckardt17periodically}. Such a construction allows for a great flexibility in adjusting the dynamics. The other apparent advantage of Floquet operators is the fact that their eigenphases, in typical cases, are distributed uniformly over a unit circle, simplifying statistical analysis of eigenphases and eigenvectors.

The advantage of studying non-equlibrium many-body problems in Floquet systems was recognized by Tomaž Prosen long before the MBL era~\cite{Prosen98, Prosen99, Prosen02}. Remarkably, the Floquet systems became also the first system allowing for an analytical demonstration of quantum chaos in many-body systems~\cite{Bertini19,Kos18,Chan18a,Bertini18}\footnote{Soon, it was realized~\cite{Bertini19dual} that so-called dual-unitarity, introduced in~\cite{Akila16},  was the property allowing for analytical insights.}.

In the presence of strong disorder, the Floquet systems may host an MBL regime \cite{Ponte15,Lazarides15,Abanin16}. 
As an example, let us consider the disordered Kicked Ising model introduced by \cite{Prosen02, Prosen07} and defined by the Floquet operator for a spin-1/2 chain,
\begin{eqnarray}
 \hat{U}_F= e^{-i g \sum_{j=1}^L \hat{\sigma}^x_j} e^{ -i \sum_{j=1}^L (J \hat{\sigma}^z_j \hat{\sigma}^z_{j+1}  +  h_j \hat{\sigma}^z_j ) },
 \label{eq:KIM}
\end{eqnarray}
where $\hat{\sigma}^{x,y,z}_j$ are Pauli operators, $h_j\in[0,2\pi]$ are independent and uniformly distributed random variables, and periodic boundary conditions are assumed. Setting $g=J=1/W$, the parameter $W$ plays the role of the disorder strength in the system. 
Solving the eigenproblem $\hat{U}_F \ket{\psi_n} = e^{i\phi_n} \ket{\psi_n}$ yields  the eigenphases $\phi_n \in[0,2\pi]$ and eigenvectors. The eigenproblem can be solved with full exact diagonalization or with an approach relying on polynomial filter approach~\cite{Luitz21} analogous to the POLFED algorithm, as described in Sec.~\ref{subsec:ED}. The polynomial filter approach provides access to system sizes up to $L=20$ on present-day supercomputers, significantly beyond the scope of full exact diagonalization algorithms. The access to larger system sizes provides means to uncover quantitative changes in finite size drifts at the ETH-MBL crossover in the Kicked Ising model. 
 
In the ergodic regime of isolated periodically driven interacting systems~\cite{Alessio14}, the spectral properties 
are modelled by  Circular Orthogonal Ensemble of random matrices~\cite{Haakebook}. Hence, the ETH-MBL crossover in the Kicked Ising model can be investigated similarly to the autonomous\footnote{In this review an autonomous system is a many-body quantum system with time-independent Hamiltonian.} disordered spin chains, see Sec.~\ref{subsec:spectral}. Consequently, \cite{Sierant22floquet} analyzed the crossover using the system size dependent disorder strengths $W^T(L)$ and $W^*(L)$. The boundary of the ergodic regime $W^T(L)$ scales linearly with system size, $W^T(L) \propto L$, for $L \leq 15$, as in the
autonomous disordered XXZ spin chains. Notably, this trend is replaced by a sub-linear growth at $L \geq 15$, consistent with the possibility of a transition to the MBL phase at a sufficiently strong disorder. The crossing point $W^*(L)$ is drifting with $L$, however, the drifts slows down with the increase of system size, and $W^*(L)$ is well fitted by a first order polynomial in $1/L$. Moreover, extrapolation of the \textit{linear} scaling of $W^T(L)$, found in small systems, crosses the extrapolation of $W^*(L)$ at $L\approx 28$. The latter length scale is significantly smaller than $L\approx 50$ found for the disordered XXZ spin chain. The finite size drifts observed for several variants of kicked spin chains~\cite{Sierant22floquet} are similar to those in the Kicked Ising model, and also similar to the finite size drifts found in the Anderson model on random regular graphs~\cite{Sierant23RRG, vanoni2023renormalization}. 

Overall, these results provide premises suggesting the possibility of a stable  MBL phase in the Kicked Ising model. The presence of the MBL phase in Floquet systems is vital for the stability of Floquet time crystals \cite{Choi17, Bordia17p,  Mi22} and Floquet insulators \cite{Rudner20} by providing a mechanism to completely eliminate the heating due to periodic driving of the system. Comparison of the Kicked Ising model with autonomous TFIM carried out by~\cite{Sierant22floquet} suggests that the lack of energy conservation is the main factor responsible for the weaker finite size drifts in the Floquet models. Better understanding of the microscopic mechanisms responsible for the disparities between autonomous spin chains and Floquet models is needed, especially in view of the results of ~\cite{Hahn23localization}, which do not find strong premises for the MBL phase in a family of Floquet circuits. One possible route of future studies is the stroboscopic time dynamics in Floquet systems~\cite{Lezama19}, which can be examined numerically in a fashion simpler than time evolution of autonomous many-body systems.

\subsection{Long-range interactions}
\label{subsec:longrange}

A systematic analysis of non-equilibrium physics in models with power-law decaying couplings was carried our by \cite{Yao14}. The
Hamiltonian considered reads:
\begin{equation}
 \hat H = \sum_{i,j}\frac{t_{ij}}{|r_{ij}|^a}\left(
 \hat{S}^+_i\hat{S}^-_{j} + h.c.\right)+\sum_{i,j}\frac{V_{ij}}{|r_{ij}|^b}
 \hat{S}^z_i\hat{S}^z_{j}
 +  \sum_i h_i \hat{S}^z_i,
 \label{Hd}
\end{equation}
with $b \le a$, $h_i$ is randomly distributed and $r_{ij} = |i-j|$ is the distance between the sites $i$ and $j$.  Note that this model preserves the $z$-component of the total spin. Such Hamiltonians arise in descriptions of a variety of interacting physical systems such as trapped ions, Rydberg atoms, ultracold polar molecules or dipolar systems in optical lattices~\cite{Baranov2012}. The analysis of possible resonances between distant spins mediated by \eqref{Hd} leads \cite{Yao14} to the estimates for the MBL regimes depending on the relations between $a$, $b$ and the dimension $d$ of the system.
Independently,  similar results for systems that do not preserve the total spin, i.e., obtained by replacing $\left(
 \hat{S}^+_i\hat{S}^-_{j} + h.c.\right)$ term in \eqref{Hd} with  $\hat{S}^x_i\hat{S}^x_{j}$, were obtained for a more restricted class of systems by~\cite{Burin06,Burin15}. We note that the tendencies towards delocalization are enhanced when the coefficients $V_{ij}$ are random~\cite{Prasad21enhanced}.

Since then, a number of works considered different aspects of MBL-like behaviour in disordered long-range interacting systems, e.g., \cite{Hauke15,Li16, Gutman16, Singh17, Nandkishore17, Tikhonov18, Safavi19,DeTomasi19, Nag19, Roy19, Botzung19, MaksymovBurin20, Schiffer19, Kloss20spin,  Deng20, Yousefjani23, Thomson20}. Several signatures of MBL regime were discussed, and the decay of LIOMs in systems with long-range interactions was proposed to be algebraic as opposed to the exponential decay in short-range MBL systems. One of the characteristic features of the MBL regime in the presence of long-range (algebraic) interactions turns out to be algebraic (and not logarithmic, as expected for short-range interactions) growth of the entanglement entropy \cite{Safavi19,DeTomasi19}. Such an algebraic growth persists from deep inside the MBL regime down to the crossover to the ergodic regime. In fact, it has been numerically found \cite{Deng20} that the growth $S\sim t^\gamma$, with $\gamma=1/3$, corresponds to the characteristic disorder value designating the crossover in a broad range of the values of exponents $a$ and $b$ describing the interaction range in \eqref{Hd}. 
One may also consider a model of infinite-range, i.e., all-to-all, interactions as those mediated by photons in the cavity \cite{Sierant19c,Kubala21,Chanda22, Ge23cavity}. 
In that case, the entanglement entropy growth in the MBL regime is also well fitted by an algebraic dependence on time~\cite{Sierant19c}. Quantum dots provide another class of models with all-to-all interactions interesting from the perspective of ergodicity breaking phenomena, see e.g.~\cite{Herre23} and references therein.

The described results regarding the non-equilibrium physics of systems with long-range interactions were obtained numerically for systems of size up to $L=20$, or for larger systems with tensor networks but over short evolution times. If the interactions are decaying according to a power-law or slower, the avalanche scenario, see Sec.~\ref{sec:aval}, predicts the absence of MBL phase. Indeed, if the power-law decay of matrix elements was included in Eq.~\eqref{eq:int}, the balance between the decay of matrix element and the growth of the ergodic bath's density of states could not be attained, resulting in the thermalization of the system.  Nevertheless, long-range interactions, contrary to naive expectations, may lead to slow growths of entanglement \cite{Pappalardi18, Lerose20origin}, illustrating the need of a better understanding of microscopic origins of ergodicity in the presence of long-range couplings~\cite{Defenu24rev}. Solvable all-to-all Sachdev-Ye-Kitaev (SYK) models~\cite{SYK22} provide a reference point for these questions.
Further quantitative explorations of the drifts at the ETH-MBL crossover in the presence of long-range interactions are needed to better understand its relation to the crossover found in the disordered short-range systems.

\subsection{Higher Dimensions }

Studies of ergodicity breaking phenomena in more than one spatial dimension are quite rare by comparison with the abundant literature for the 1D problem. The first stimulus for the consideration of MBL problem in higher dimensions came with experiments \cite{Choi16}, in which  half of a 2D 
lattice was initially filled with bosonic particles. Then the particles were allowed to wander in the  presence of disorder, filling the lattice almost 
uniformly for a small disorder. In contrast, for a sufficiently strong disorder, the bosons remained in the vicinity of their initial positions and thermalization of the system was not achieved during the course of the experiment. The full scale simulation of the problem is beyond reach of the present classical computers. A first approach \cite{Yan17} used the time-dependent Gutzwiller approximation to tackle the problem. In this approach the system's wave function remains separable throughout the evolution time so no entanglement can build up. The results of the simulation gave, however, quite an accurate description of the experiment showing that its interpretation in terms of MBL may be questioned, and the results can be interpreted in terms of slow glassy dynamics. 
The attempt at the simulation of full many-body dynamics of this bosonic system was undertaken on a $6\times6$ lattice with tree tensor network states by \cite{Urbanek18}. The authors simulated the experiment tuning the parameters of the problem to match the experimental system and, despite following the evolution for 256 tunneling times, they did not observe a saturation of the imbalance. 

Even smaller 2D spinless fermion systems were studied for different configuration of lattices by means of exact diagonalization and gap ratio statistics \cite{Wiater18}. It was found that the crossover to the MBL regime depends on the geometry of the lattice and, in particular, on the number of nearest neighbors connected via direct tunnelings. This hints at a possible approximate mean-field description, in  agreement with~\cite{Yan17}.

The avalanche mechanism precludes the existence of MBL beyond one spatial dimension~\cite{DeRoeck17}. Nevertheless, understanding the relevance of the avalanche mechanism for the physics of disordered spin chains is incomplete, as discussed in Sec.~\ref{sec:aval}. Moreover, the numerical signatures of ergodicity breaking, as well as the experimental results, are characterized by a significant similarity between one spatial dimension and higher dimensional systems. Indeed, the numerical studies based on tensor networks~\cite{Wahl17, Wahl19, Venn22, Chertkov21,Doggen20} indicate the presence of an MBL regime in which the dynamics towards thermalization is very slow. Similar conclusions about the presence of a robust MBL regime were obtained in studies of small 2D many-body systems with kinetic constraints
~\cite{Theveniaut20, Pietracaprina20dimer}. Overall, the status of MBL phase in two and higher spatial dimensions still remains in several aspects unclear.

\subsection{Ergodicity breaking without disorder}
\label{subsec:clean}
Recent years brought a rich variety of interacting non-integrable systems that reveal non-ergodic dynamics even in the absence of the disorder. The generic feature linking these models is the existence of additional approximate constants of motion that prevent thermalization at accessible time scales. The examples include breakdown of ergodicity in:
\begin{itemize} 
\item{A.} Lattice gauge theory models where typically the Gauss law, or other local
conservation laws, provide restrictions for the dynamics \cite{Brenes18,Giudici20,Chanda20};
\item{B.} Strong interactions that slow down the dynamics and may lead to the Hilbert space fragmentation \cite{Sala20, Khemani20};
\item{C.} Tilted lattices where the emerging approximate conservation of the dipole moment leads to the so-called Stark MBL \cite{vanNieuwenburg19, Schulz19}.
\end{itemize}

The short-range models with the dynamical lattice gauge fields \cite{Giudici20, Chanda20}, after the elimination of the fields with the Gauss law, become effectively long-ranged. This leads to the MBL-like phenomenology. Understanding the possibilities of a breakdown of ergodicity in such models in the asymptotic limit $L,t \to \infty$ is an interesting direction for further research. 

While we have discussed models with long-range interactions above, it is worth mentioning that even without disorder they may show signatures of non-ergodic dynamics associated with the phenomenon of Hilbert space fragmentation. One of the examples is provided by the strongly interacting dipoles in optical lattices \cite{Li21}. 
In these systems, the long-range character of the couplings may be controlled by the optical lattice shape \cite{Korbmacher23}. Going beyond the hard-core bosons approximation may result in the enhanced importance of density dependent tunneling \cite{Dutta15}, which to some extent restores ergodicity~\cite{Aramthottil23}. 

The tilted lattice systems, i.e., the systems in which, instead of the on-site disorder, a \textit{linear potential} $F\sum_j j \hat{n}_{i}$ is added to the Hamiltonian, provide yet another platform to study ergodicity breaking phenomena.
The tile in optical lattices may be realized with an additional electric or space dependent magnetic field via Zeeman effect. Such a linear tilt of a 1D lattice leads, in the non-interacting limit, to the Wannier-Stark localization. Similarly, in the presence of interactions, 1D lattice models with tilted potential show signatures of non-ergodicity due to the Stark MBL~\cite{vanNieuwenburg19,Schulz19,Taylor20}. For strong lattice tilts, the observed ergodicity breaking may be linked to the fragmentation of the Hilbert space due to the emerging conservation of a dipole moment, $\mathcal D = \sum_j j \hat{n}_{i}$, which recently stimulated a vivid interest in the subject~\cite{Sala20,Rakovszky20,Yao20b,Chanda20c,Yao21a, Nandy24subdiffusionTilted}.  
Perturbing the titled potential with a weak disorder \cite{vanNieuwenburg19} or adding a small (e.g. quadratic) modification of the external potential \cite{Schulz19,Taylor20} results in the  phenomenology similar to the one observed in the MBL regime of disordered spin chains, e.g., logarithmic entanglement entropy growth from initial separable states. Nevertheless, for a pure linear tilt a different behavior is observed \cite{Yao20b,Yao21}. This may be linked to the emerging global dipole moment conservation in the system \cite{Doggen20s, Nandy24subdiffusionTilted}. Finally, we note that arguments for finite spin transport in tilted lattices were presented in~\cite{Kloss23Absence}, relying on a combination of mirror and spin-flip symmetries of such potential,  as well as on the assumption of the absence of degeneracies in the spectrum.

Experimental investigations of Rydberg atom arrays~\cite{Bernien17, Browaeys20} lead to the discovery of a form of weak-ergodicity breaking, defined via a strong dependence of the thermalization process on the system's initial state. For instance, the PXP model~\eqref{Hpxp}, for $\alpha=1$ and $h_i=0$, initialized in the N\'{e}el state $\ket{ \uparrow \downarrow \uparrow \downarrow  \ldots}$, features long-lived oscillations of local observables and revivals of wave function, while, for period-4 density wave state $\ket{ \uparrow \downarrow \downarrow \downarrow \uparrow  \ldots}$, the thermalization occurs quickly~\cite{Turner18prb}. This dynamical behavior of the PXP model is due to quantum many-body scar states~\cite{Turner18}. The quantum scar states form a set of equally spaced in energy eigenstates of $\hat{H}_{\mathrm{PXP}}$ characterized by low entanglement and atypically large overlaps with the N\'{e}el state. The set of scar states is embedded in the spectrum of ergodic eigenstates of $\hat{H}_{\mathrm{PXP}}$. Ideas to employ quantum many-body scars as a mechanism for maintaining coherence and the desire for a more complete understanding of dynamics of non-integrable many-body systems fueled the intensive investigations of weak-ergodicity breaking phenomena, as reviewed in~\cite{Serbyn21, Moudgalya22rev, Chandran23rev}. Non-thermal eigenstates can be found in multiple non-integrable quantum models, including Fermi-Hubbard model~\cite{Iadecola19, Moudgalya20eta}, Bose-Hubbard model~\cite{Hummel23Bose}, spin-1 systems~\cite{Iadecola19a, Chattopadhyay20}, spin-1/2 model with emergent kinetic constraints~\cite{Iadecola20}, and even in disordered spin chains~\cite{Shibata20}. The emergence of a decoupled subspace spanned by non-thermal eigenstates in an otherwise thermal spectrum of the system~\cite{Shiraishi17} is a common trait of the models of weak-ergodicity breaking.

\section{MBL in experiments}
\label{sec:exp}

In this Section we give a brief overview of the experiments related to MBL. We highlight the investigated system sizes and evolution times to illustrate the length and times scales relevant for the present-day experiments. 

\subsection{Early developments}
Early years of this millennium were rich in efforts to experimentally verify Anderson localization, when it was understood that the cold atomic settings provide a new and promising medium for explorations of disordered many-body systems~\cite{Damski03}. The early attempts to observe the Anderson localization were spoiled by weak interactions \cite{Clement05, Schulte05, Fort05, Schulte06}. Finally, diluting the cold atomic gas to make it non-interacting, led to an experimental verification of the Anderson localization \cite{Billy08,Roati08}.

Observation of MBL requires a preparation of the appropriate highly out-of equilibrium state with the finite energy density. In other words, the experiment should not be limited to low energy states as, e.g., in the first notable attempt to study the influence of strong disorder on the interacting bosons trapped in a 3D optical lattice~\cite{White09}. Using the speckle disorder potential, the authors were able to study the influence of the disorder on superfluid and Mott insulator phases of interacting bosons, as well as the possible emergence of the Bose glass phase. The latter study was extended later in \cite{Pasienski10}.

A subsequent experiment, \cite{Kondov15} realized the disordered 3D Fermi-Hubbard model and probed transport properties of the system. An external force was applied to displace the atoms from the centre of harmonic trapping potential and the velocity $v$ of their center-of-mass was measured. At weak disorder strength $W$, the velocity $v$ was manifestly non-vanishing, suggesting the presence of transport in the model. In contrast, large disorder strength $W$ lead to practical vanishing of $v$ suggesting absence of the transport. These observations are consistent with general expectations about the slowing down of the dynamics and the emergence of MBL regime with increasing the disorder strength $W$. However, their relation to the physics of thermalization and ergodicity breaking in quantum many-body system is not fully clear.

\subsection{MBL regime in ultracold atomic systems}
The breakthrough in studies of thermalization and MBL came with the experiment \cite{Schreiber15}, in which the 1D Fermi-Hubbard model~\eqref{eq:ham_FH} was realized by a 3D optical lattice with deep confining potential in two spatial dimensions leading to an array of decoupled 1D tubes. The disorder was added via a second weak laser, incommensurate with the one creating lattices, resulting in a quasiperiodic potential such as the one in Eq.~\eqref{eq:qp}. A preparation of the initial highly inhomogeneous state at high energy density, with every second lattice site occupied by a fermion and every second site empty, was followed by monitoring of the time evolution of the system. Measuring the density autocorrelation function (a generalization of $C(t)$ in Eq.~\eqref{eq:den} for spinful fermions), and varying the interaction strength $U$ and the amplitude of quasiperiodic potential $W$, the experiment~\cite{Schreiber15} demonstrated a crossover between the ergodic (thermalizing) behavior, for which vanishing of $C(t)$ is observed, and the MBL regime, in which $C(t)$ is significantly non-zero even at the longest probed time scale.

The longest probed evolution time reported ~\cite{Schreiber15} is $t_{\mathrm{max} }=50$ in the units of tunneling time $J^{-1}$. While the 1D lattices consisted of several tens of lattice sites, let us take $L=50$ as a rough approximation. The results ~\cite{Schreiber15} clearly demonstrate the dichotomy between thermalizing and non-ergodic behavior, which arises when the disorder strength in a many-body system is increased. Nevertheless, in light of the discussion in Sec.~\ref{sec:density}, we presently understand that there are good premises to believe that the autocorrelation function $C(t)$ may slowly decay at longer times, and hence, the experiment observed an MBL regime, without any significant implications for the MBL phase, i.e., for the fate of the system in the asymptotic limit $L,t\to \infty$. 

A subsequent experiment~\cite{Luschen17} extended the available time scale to $t_{\mathrm{max} }=100$, which allowed for observation of the slow, power-law like, decays of the density autocorrelation function $C(t)$ in the vicinity of the MBL regime.
The absence of the Griffiths regions in a quasiperiodic potential considered by~\cite{Luschen17} indicates that a different type of mechanism may be responsible for the decay of $C(t)$, see Sec.~\ref{sub:stretched}. Introduction of the coupling between the 1D tubes significantly enhances the tendencies towards thermalization, increasing the disorder strengths needed to observe the MBL regime~\cite{Bordia16,Bordia17}. Experiments in this line also considered the ETH-MBL crossover in the periodically driven MBL systems~\cite{Bordia17p} and showed how the MBL regime is destroyed in the presence of an engineered dissipation~\cite{Luschen17o}, following a theoretical analysis by~\cite{Levi16Dissipation, Fischer16Bath}.

The 1D Bose-Hubbard model~\eqref{eq:ham_BH} with quasiperiodic potential was realized in a line of experiments with a quite small number of ultracold bosons in optical lattices~\cite{Lukin19,Rispoli19,Leonard23}. The exquisite control of the setup enabled experimental studies of the slow growth of entanglement entropy from an initial product state in the MBL regime~\cite{Lukin19} (see Sec.~\ref{sec:entanglement}), as well as probing 2-point correlation functions at the ETH-MBL crossover~\cite{Rispoli19}. The single-site resolution enabled observation of signatures of quantum avalanches by planting thermal inclusions~\cite{Leonard23}, as described in Sec.~\ref{subsec:avalREL}. The typical lattice size considered in this experimental realization of the Bose-Hubbard Hamiltonian is $L=12$ and the probed time scales are of the order of $t_{\mathrm{max}}=100$ tunneling times. A separate experiment realized disordered Bose-Hubbard model~\cite{Choi16} on a 2D lattice comprising in total $L\approx 500$ sites. Studies of time evolution of the density autocorrelation function $C(t)$ for the initially prepared out-of-equilibrium state, in which bosons were occupying one half of the lattice, demonstrated a crossover between thermalization at weak disorder strengths and MBL regime at strong disorder. In the latter regime, the bosons remained in the vicinity of their initial positions even after $t_{\mathrm{max}}=300$ tunneling times.

For completeness, we mention realization of a titled Fermi-Hubbard model, which hosts non-ergodic regimes due to Hilbert space fragmentation~\cite{Scherg21, Kohlert23}. Here, up to $t_{\mathrm{max}}=700$ tunneling times have been reached. 
Experimental realization of quasicrystals~\cite{senechal1996quasicrystals} in systems of ultracold bosons in 2D optical lattices~\cite{Sbroscia20, yu2023observing} opens up a new testing ground for the MBL regime in two dimensions, especially interesting due to the absence of rare-region effects in these systems.

\subsection{MBL regime in other platforms}

Quantum Ising model \eqref{eq:TFIM} with additional long-range couplings was realized in Yb-ion quantum simulator with programmable random disorder~\cite{Smith16}. The experiment with $L=10$ spins has shown a crossover between thermalization at weak disorder and the presence of MBL regime at large disorder in which the density autocorrelation functions do not decay at the longest considered time scale $t_{\mathrm{max}}=10$. Assuming the relevance of the quantum avalanche mechanism, the observed MBL regime should eventually become ergodic with increasing time and length scales, see Sec.~\ref{subsec:longrange}.

Disordered 1D Bose-Hubbard model was also implemented on an $L=9$ superconducting qubit system~\cite{Roushan17}, and the ETH-MBL crossover was observed experimentally in the spectral properties of the system. In similar spirit, the experiment on up to $L=21$ superconducting qubits~\cite{Chiaro22}, where the tomographic reconstruction of single and double qubit density matrices helped to quantify temporal and spatial aspects of the entanglement growth in the MBL regime, explored the time scales up to $t_{\mathrm{max}} \approx 400$ tunneling times. An experiment monitoring the dynamics of $L=19$ superconducting qubits over about $t_{\mathrm{max}} \approx 100$ tunneling times~\cite{Guo21mbl} demonstrated that the disorder strength at which the ETH-MBL crossover occurs is dependent on the energy of the initial state~\cite{Chanda20m}. Fock-space perspective on the dynamics in ergodic and MBL regimes, see Sec.~\ref{subsec:wave-fun}, was investigated experimentally in a system of $L=24$ superconducting qubits~\cite{Yao23fockspace}.

Signatures of the MBL regime were observed in the conductivity of a disordered 2D electron system with power-law interactions and good insulation from thermal bath realized in Si metal-oxide-semiconductor ﬁeld-effect transistors, as reported in~\cite{Stanley23sceening}.
Finally, we note that Stark MBL, see Sec.~\ref{subsec:clean}, was observed in systems of trapped ions~\cite{Morong21} and of superconducting qubits~\cite{Guo20}.

\subsection{Perspective on MBL experiments}
The MBL regime, in which quantum many-body systems do not exhibit signatures of thermalization at the probed time scales $t< t_{\mathrm{max}}$, has been observed in several experiments across a variety of synthetic matter platforms. The time scales and system sizes probed in the experiments provide a reference point for the question of the MBL phase, Eq.~\eqref{eq:MBL1}, posed in this review. The limit of $L, t \to \infty$, relevant for the latter question, may seem immaterial from the experimental perspective since each synthetic matter platform is limited by finite coherence times. From that viewpoint, the very slow but persistent trends towards thermalization described in Sec.~\ref{sec:density} are irrelevant, as sufficiently strongly disordered many-body systems fail to thermalize at the experimentally considered times $t < t_{\mathrm{max}}$. This highlights the importance of identifying the universal features of the regime of slow dynamics, as described in Sec.~\ref{sec:extracting}.

The numerical results reviewed in this work typically concern time scales and system sizes similar to or larger than those accessible in present-day experiments. The notable exception is provided by cold atomic settings in which the system sizes significantly exceed the capabilities of exact numerical methods. Nevertheless, even in those cases, the finite coherence times allow us to reproduce the significant features of the experiments, at least at larger disorder strengths, with tensor network approaches. The numerical results, while indicating the possible benefits of increasing the coherence times and experimentally probed system sizes, at least partially discourage such an approach. Let us consider a \textit{gedanken} experiment in which the unitary time evolution of a disordered many-body system can be probed to $t_{\mathrm{max}}=10^4$ tunneling times in a lattice comprising $L=1000$ qubits. Can we expect the emergence of any new trend in the density autocorrelation function $C(t)$ in such an experiment? Sec.~\ref{sec:dynamics} suggests that the answer is \textit{no}. Fixing disorder strength $W$ and studying the time evolution of $C(t)$, we would likely find that $C(t)$ decays to smaller values at the longest evolution time than in an experiment with a smaller system and shorter probed times. Does it mean that no further progress is possible? Again, we believe that progress is indeed possible. Investigating microscopic mechanisms responsible for the emergence and breakdown of ergodicity, including, but not limited to, the scenarios considered in Sec.~\ref{sec:aval}, is experimentally feasible and may lead to more conclusive answers about the MBL phase.

\section{Conclusions}

\subsection{Summary} We close this review by revisiting the open questions outlined in Sec.~\ref{sec:open}. We comment on these problems in light of the numerical results for the disordered interacting quantum many-body systems reviewed above.

\noindent\textit{1. Definition of MBL.} 
The MBL regime, in which interacting many-body systems appear to avoid thermalization, is clearly observed for sufficiently large disorder strength $W$ in the systems of size $L$ studied numerically and experimentally. The breakdown of thermalization of local observables in the asymptotic limit $L, t \to \infty$, Eq.~\eqref{eq:MBL1}, is a natural, from the statistical physics perspective, way to define the MBL phase. 

However, taking the double limit $L, t \to \infty$ is a formidable task in numerical studies. Examination of eigenvalues and eigenvectors of the system allows to implicitly consider the $t \to \infty$ limit, but the unbiased state-of-the-art numerical approaches allow to consider modest system sizes no bigger than $L\approx20$, insufficient to reach conclusions about the $L \to \infty$ limit. Time evolution of strongly disordered many-body systems may numerically be probed for larger systems comprising more than a hundred sites. However, due to the persistent dynamical flow towards thermalization, the $t \to \infty$ limit is often infeasible. The definition relying on the $L, t \to \infty$ limit is also impractical, especially in the context of finite coherence times in experiments with synthetic quantum matter out-of-equilibrium, which may be used to probe the MBL regime.
On the other hand, providing an alternative definition of the MBL phase is non-trivial. 
It should allow for distinguishing a sharp boundary of the defined MBL phase in which the dynamics are markedly different than in the other dynamical phases. 

One option for an alternative definition of the MBL phase involves constraining the probed times $t$ by a certain threshold. For instance, one could consider only times scaling no faster than polynomially with the system size $L$. Restoration of the ergodicity at longer times, e.g., scaling exponentially with $L$, would be irrelevant from the point of view of such a definition, and systems that fail to thermalize at times scaling polynomially with $L$ would belong to some form of non-ergodicity, e.g., to the MBL phase. However, the abrupt slow-down of the dynamics of disordered systems makes such a definition of MBL not much more practical than the definition involving the $L, t \to \infty$ limit. Indeed, the decay of local correlators $C(t)$ at disorder $W=8$ in the disordered XXZ spin chain is well approximated by a power-law behavior, $C(t) \propto t^{-\beta}$, where $\beta \approx 10^{-3}$. If we allow the time to 
be a polynomial of $L$ of order higher than $\frac{1}{\beta}$, the disordered XXZ chain at $W=8$ may still be ergodic. However, to verify this, one would have to wait for an enormous amount of time, which is, again, irrelevant from the experimental perspective.
Therefore, the question of proposing a definition of the MBL phase that would be both practically relevant and satisfactory from the theoretical standpoint remains a largely open problem.

\noindent \textit{2. Understanding of the MBL phase.}  
Assuming that the MBL phase is defined by the failure of thermalization in the asymptotic limit $L, t \to \infty$, presently available numerical (and experimental) results do not provide a clear-cut understanding of the MBL phase. A broad consensus among mathematicians and physicists regarding rigorous results about the status of the MBL phase defined in that manner is yet to be achieved. The multi-scale Jacobi method considered in~\cite{Imbrie16a, deRoeck24absence} or other rigorous mathematical framework will be needed to finally resolve this question.

The phenomenology of the ETH-MBL crossover described in Sec.~\ref{sec:numerical} involves significant drifts of disorder strengths towards ergodicity with the increased system size $L$. Some drifts appear to be slowing down with an increase of $L$, providing premises for the presence of the MBL phase. Other drifts are significant and linear with $L$, suggesting that thermodynamically large, disordered many-body systems may eventually thermalize at any $W$. These drifts preclude analysis of the results at the crossover with the single-parameter scaling and make the extrapolation of the results to the asymptotic limit less reliable. This behavior hinders studies of the ETH-MBL transition, as the presently available numerical results do not determine whether and at what disorder strength the transition to the MBL phase occurs.
Consequently, our comprehension of the properties of the MBL phase still needs to be completed. For instance, the Fock-space localization in disordered many-body systems appears to be a more robust form of ergodicity breaking than the MBL phase. We cannot entirely exclude the possibility that the MBL phase and Fock-space localization, in the asymptotic limit, arise at disorder strengths that depend on $L$ in the same manner. Available numerical results, however, suggest that the latter is not a likely scenario. 

The exact diagonalization results provide a thorough picture of the MBL regime. Determining which features of the MBL regime are inherited, in the asymptotic limit, by the MBL phase, is an essential challenge for future research.

\noindent\textit{3. Relationship between thermalization and transport.} 
The exponential slow-down of the dynamics of many-body systems with increasing the disorder strength $W$, discussed in Sec.~\ref{sec:dynamics}, hinders understanding of the status of the MBL phase at large disorder strengths and investigations of the transport at intermediate disorder strengths. The disordered spin-1/2 chains host an evident diffusive regime at small disorder strengths. The transport significantly slows down with increasing $W$ within the ergodic phase. Whether it remains diffusive or becomes sub-diffusive is not fully understood, leaving the question of what types of transport are allowed in the ergodic phase of disordered many-body systems unanswered, as discussed in Sec.~\ref{sec:trans}. Further explorations of this subject are needed. Notably, the length and time scales needed to explore this problem may not be as extreme as the system sizes and times required to comprehend transport properties deep in the MBL regime.

\noindent\textit{4. Mechanism of the transition.}
The quantum avalanche scenario, which describes thermalization of the MBL region due to rare ergodic inclusions emerging in the regions of anomalously weak disorder, is the prevalent proposition for a microscopic mechanism of the MBL transition. As we argued in Sec.~\ref{sec:aval}, the implications of the avalanche mechanism can be studied numerically and experimentally, particularly by engineering conditions for the propagation of avalanches by planting thermal inclusions. Determination of the relevance of the avalanche mechanism for the thermalization in disordered many-body systems and exploration of its falsifiable consequences are important directions for further research. This includes comprehending how to model the effects of the rare thermal inclusions in a simplified manner, and what aspects of the dynamics of isolated disordered many-body systems may be captured by the open system approach.

Many-body resonances arise when eigenstates of disordered many-body systems are superpositions of a few product states that differ locally at extensively many lattice sites. The MBL transition may be associated with the appearance of system-wide many-body resonances that corroborate a breakdown of locality in the disordered spin chain. Establishing links between the quantities that probe rare instances of the many-body resonances, described in Sec.~\ref{sec:reson} and ~\ref{sec:reson2}, and the standard ergodicity breaking indicators involving averages of quantities derived from spectra of many-body systems, discussed in Sec.~\ref{sec:numerical}, is an essential direction for further explorations.

Pin-pointing the microscopic mechanisms behind the emergence of ergodicity with a decrease of $W$ in strongly disordered many-body systems, responsible for the MBL transition, would enhance our understanding of the transition. Hence, by extension, it would allow us to form more precise expectations about the properties of the MBL phase. Changing the perspective and exploring mechanisms destabilizing the ergodic phase, for instance, by placing strongly disordered inclusions in an otherwise thermalizing many-body system, is a promising avenue for studies of MBL transition.

\noindent\textit{5. Understanding of slow dynamics at finite times. } 
The regime of slow dynamics in strongly disordered many-body systems, described in Sec.~\ref{sec:dynamics}, is robust and relevant for experimentally verifiable questions. Hence, understanding the regime of slow dynamics is vital, regardless of the status of the MBL phase and the MBL transition. Approaches aimed at extracting the universal features of this regime, outlined in Sec.~\ref{sec:extracting}, constitute crucial steps in our understanding of the dynamics of disordered many-body systems.

Studying the MBL regime, even at the disorder strength at which it is only a prethermal phenomenon, is an important avenue of research. Questions about the MBL regime may be phrased in terms of the time evolution of specific quantities. Alternatively, such questions may also be stated in terms of spectral statistics. Are there universal spectral properties, beyond the standard random-matrix-theory predictions, which characterize the strongly disordered many-body systems at finite sizes?

\noindent\textit{6. Models of MBL and beyond.}
Disordered XXZ spin-1/2 chain and its close cousin, the disordered $J_1$-$J_2$ model, serve as paradigmatic models for studies of MBL. The majority of the numerical results reviewed here concern these models. Many disordered 1D many-body systems, mentioned in Sec.~\ref{sec:ext}, share similar dynamical features, including the finite size drifts at the ETH-MBL crossover and the regimes of persistent slow dynamics at strong disorder. Despite the similarities, details of the numerical results vary from model to model. For instance, the MBL phenomenology in the disordered spin-1/2 chains appears more stable at available system sizes in Floquet systems than in Hamiltonian systems. In contrast, when spin-1/2 chains' dynamics are subject to specific local constraints, numerical results indicate much stronger tendencies towards ergodicity, suggesting a lack of the MBL phase in such systems. These disparities indicate that the status of MBL may depend on the microscopic details of particular many-body Hamiltonians. Identifying and classifying microscopic features that enhance the ergodic phase or lead to more robust premises for the occurrence of the MBL phase is an important avenue for further studies. Such efforts can be facilitated by studies of interacting models in which the ergodicity-breaking transition is more accessible numerically, such as the quantum sun model introduced within the avalanche theory and the ultrametric model defined on a Fock space.

Irrespective of the status of MBL, the phenomenology of the MBL regime remains an important reference point for ergodicity-breaking phenomena in many-body systems, both in the presence and absence of disorder. Moreover, the stability of the MBL phase enables various exciting phenomena such as the emergence of localization-protected quantum order~\cite{Huse13}, its topological counterpart~\cite{Chandran14}, or non-trivial phase structure in disordered Floquet systems~\cite{Khemani16structure} and the discrete-time crystals~\cite{Sacha17, Zaletel23DiscreteTC}. The enormous time scales describing the dynamics deep in the MBL regime make these phenomena stable for many practical purposes. Nevertheless, understanding whether these phenomena may give rise to genuine stable dynamical phases of matter is inseparably tied to the status of the MBL phase.

\subsection{Outlook}
The intensive research of disordered many-body systems over the last two decades allowed us to theoretically identify and describe the MBL regime, and to observe this regime experimentally. Many features of the MBL regime have been established, including its manifestations in the spectra of many-body systems and the experimentally measurable aspects of their time evolution. Despite this progress, the status of the MBL phase and the MBL transition remain controversial. Phases and phase transitions, fundamental notions of statistical physics, are characterized by a significant degree of robustness and universality, and allow drawing sharp and essential distinctions between different types of behavior of complex systems. For these reasons, the MBL phase and MBL transition questions are of fundamental interest.

Presently, the status of the MBL field resembles, in some aspects, the situation of glassy systems. In spin glass models, the mean-field picture is understood through the mechanism of replica symmetry breaking \cite{Parisi80}. The value of the upper critical dimension, i.e., the spatial dimension above which the mean-field solution applies, however is not clear. Therefore, it is debated whether the mean-field picture explains the results in three dimensions better than the alternative phenomenology of {\it droplets} \cite{mcmillan1984scaling,bray1984lower,fisher1988nonequilibrium} [for a recent review of this debate see \cite{parisi2023nobel,Moore21Droplet} and references therein]. In configurational glasses, as reviewed by \cite{Berthier11}, simulations and experiments show very slow dynamics, and there is no clear evidence for either of the scenarios. 
Still, characterization of the microscopic mechanisms responsible for the different physical regimes in configurational glasses remains an important challenge. Similarly, studies of the microscopic mechanisms associated with the interplay of ergodicity and ergodicity breaking in the vicinity of the putative MBL transition may unravel novel physical phenomena, essential for our understanding of non-equilibrium physics of disordered many-body systems, irrespective of the fate of the ETH-MBL crossover in the asymptotic limit. 

The presence of strong disorder and interaction necessitates the use of numerical methods for studies of MBL. Paradoxically, our understanding that the available results for MBL may be interpreted in very distinct ways, commenced by the works described in Sec.~\ref{subsec:chall}, may indicate the maturing of our ideas about the MBL. The history of the drifts in the interpretation of numerical results about the MBL should serve as a reference and a cautionary tale for future studies of collective phenomena in complex systems.

The numerical methods described in Sec.~\ref{sec:num} allowed us to reach the present understanding of MBL. These methods are typically run on several CPUs (central processing units) of present-day classical supercomputers. Search for efficient and effective methods of investigating MBL with GPUs (graphical processing units) or TPUs (tensor processing units) \cite{Lewis22TPU, Morningstar22TPU} is an exciting challenge within the age of classical computing. Nevertheless, the simulation of a many-body system, even assuming it is in the MBL phase, is formally a hard problem if the evolution time is exponentially long~\cite{Ehrenberg22}, which is likely required to achieve the final answer for the MBL problem. Fault-tolerant quantum computers may open an entirely new chapter in MBL studies.

\section{Acknowledgements}

We acknowledge collaborations and insightful discussions with 
D. Abanin, B. Alsthuler, A.S. Aramthottil, C. Artiaco, 
F. Balducci, C. Baldwin, L. Barbiero, J. Bardarson, A. Bayat, S. Bera, K. Biedroń, J. Bonča, M. Bukov,
T. Chanda, A. Chandran, 
M. Dalmonte, B. De, X. Deng,
A. De Luca, W. De Roeck, G. De Tomasi,
F. Evers,
P. N. Falcao, R. Fazio, M. Feigelman,
A. M. García-García, A. Garg, S.J. Garratt, J. Goold,
F. Heidrich-Meisner, M. Hopjan, D. Huse,
J. Imbrie,
J. Janarek,
B. Krajewski, R. Kraus, I. Khaymovich, H. Korbmacher, K. Kottmann, V. Kravtsov, P. Kubala, M. Ku\'s,
C. Laumann, W. Li, E. Gonzalez-Lazo,
M. \L{}\c{a}cki, P. \L{}ydżba,
A. Maksymov, A. Mallick, M. Mierzejewski, R. Moessner, G. Morigi, M. Mueller,
R. Nandkishore,
V. Oganesyan, 
R. Panda, A. Pal, T. Parolini, S. Pascazio, S. Pappalardi, G. Parisi, K. Pawlik,  F. Pietracaprina, A. Polkovnikov, I. Protopopov, P. Prelovšek, T. Prosen, 
M. Rigol, V. Ros, S. Roy,
L. Santos, L.F. Santos, M. Schulz,  D. Sels, J. Sirker, S. Sondhi,  T. Szo\l{}dra, R. Świ\c{e}tek,
J. Šuntajs, K. Suthar,
L. Tagliacozzo, S. Taylor, X. Turkeshi,  
C. Vanoni,  V. Varma, 
D. Wiater, 
R. Yao,
M. Žnidarič.

P.S., L.V., J.Z. acknowledge the workshop "Dynamical Foundation of Many-Body Quantum Chaos" at Institute Pascal (Orsay, France) at which foundations for several Sections of this review were laid.
P.S. and L.V. acknowledge the program "Stability of Quantum Matter in and out of Equilibrium at Various Scales" (code: ICTS/SQMVS2024/01) at International Centre for Theoretical Sciences (Bengaluru, India) at which many useful discussions about ergodicity breaking phenomena took place.
P.S acknowledges the school "Quantum localization and Glassy physics" at Institut d'études scientifiques de Cargèse (Corsica, France) at which many useful conversations about MBL occurred.

The work of A.S. is funded under the National Recovery and Resilience Plan (NRRP), Mission 4 Component 2 Investment 1.3 funded by the European Union NextGenerationEU.
National Quantum Science and Technology Institute (NQSTI), PE00000023, Concession Decree No. 1564 of 11.10.2022 adopted by the Italian Ministry of Research, CUP J97G22000390007.
L.V. acknowledges  support by the Slovenian Research and Innovation Agency (ARIS), Research core funding numbers P1-0044, N1-0273, J1-50005, and N1-0369.
The research of J.Z. was
funded by National Science Centre (Poland) under grant 
No. OPUS18 2019/35/B/ST2/00034 and the OPUS call within the WEAVE programme
2021/43/I/ST3/01142.  
Support by 
Poland’s high-performance Infrastructure PLGrid  (HPC Centers: ACK Cyfronet AGH)
via providing computer facilities within computational Grant No. PLG/2023/016370 is acknowledged.
The research has been also supported by a grant from the Priority Research Area (DigiWorld)
under the Strategic Programme Excellence Initiative at Jagiellonian University (J.Z.).
P.S. and M.L. acknowledge support from
ERC AdG NOQIA; MCIN/AEI (PGC2018-0910.13039/501100011033, CEX2019-000910-S/10.13039/50110 0011033, Plan National FIDEUA PID2019-106901GB-I00, Plan National STAMEENA PID2022-139099NB-I00 project funded by MCIN/AEI/10.13039/501100011033 and by the ``European Union NextGenerationEU/PRTR'' (PRTR-C17.I1), FPI); QUANTERA MAQS PCI2019-111828-2);  QUANTERA DYNAMITE PCI2022-132919 (QuantERA II Programme co-funded by European Union's Horizon 2020 program under Grant Agreement No 101017733), Ministry of Economic Affairs and Digital Transformation of the Spanish Government through the QUANTUM ENIA project call - Quantum Spain project, and by the European Union through the Recovery, Transformation, and Resilience Plan - NextGenerationEU within the framework of the Digital Spain 2026 Agenda; Fundaci\'{o} Cellex; Fundaci\'{o} Mir-Puig; Generalitat de Catalunya (European Social Fund FEDER and CERCA program, AGAUR Grant No. 2021 SGR 01452, QuantumCAT \ U16-011424, co-funded by ERDF Operational Program of Catalonia 2014-2020); Barcelona Supercomputing Center MareNostrum (FI-2024-1-0043); EU Quantum Flagship (PASQuanS2.1, 101113690); EU Horizon 2020 FET-OPEN OPTOlogic (Grant No 899794); EU Horizon Europe Program (Grant Agreement 101080086 - NeQST), ICFO Internal ``QuantumGaudi'' project; European Union's Horizon 2020 program under the Marie Sk{\l}odowska-Curie grant agreement No 847648;  ``La Caixa'' Junior Leaders fellowships, ``La Caixa'' Foundation (ID 100010434): CF/BQ/PR23/11980043. Views and opinions expressed are, however, those of the author(s) only and do not necessarily reflect those of the European Union, European Commission, European Climate, Infrastructure and Environment Executive Agency (CINEA), or any other granting authority.  Neither the European Union nor any granting authority can be held responsible for them.

\section{Appendix: different kinds of transport and compatible equilibria}
\label{app:dif-sub-Gibbs}

When discussing localization, and in particular MBL, two clearly distinct but related phenomena are involved: transport (i.e., dynamical relaxation from an initial condition or non-equilibrium steady states) and equilibrium, either in terms of equilibrium distributions or, 
in more modern terms, eigenstate properties. Can one have different kinds of transport and the same equilibrium state?\footnote{The vice versa is probably easily excluded.} In principle, the answer is yes. While it is natural to associate diffusion to the Gibbs distribution, a diffusion coefficient to a temperature and mobility (like in Einstein's relation), it is possible to construct counterexamples of classical dynamical processes with anomalous transport but which nonetheless converge to Gibbs equilibrium distribution. So, if the question is: from the mere observation of ETH in the eigenstates, can we conclude that there is a non-zero conductivity/diffusion coefficient in the system, the answer is {\it no}. Let us give a couple of examples to show how this works.

We follow closely \cite{klages2008anomalous}, Chapter 5. Consider a continuous time random walk with independent random time and space jumps $\lambda(\xi)$ and $\psi(\tau)$. The dispersion relation in general connects the typical values of the time passed and the space covered,
\begin{equation}
    x\sim t^{\gamma},
\end{equation}
where $x=\sum_{i=1}^n \xi_i$ and $n$ is such that $\sum_{i=1}^n\tau_n=t$.

For $\gamma=1/2$, we have diffusion, and $\gamma<1/2$ is sub-diffusion. A more complete description is contained in the probability distribution $P(x,t)$ or equivalently its Fourier transform $P(k,t)$, or the Fourier-Laplace transform $P(k,s)$.

The same $\gamma$ can be obtained in two ways.

{\it a)} The waiting time $\psi(\tau)\sim 1/\tau^{1+\alpha}$, for $\alpha<1$, this distribution is long tailed (such that the average $\langle\tau\rangle$ diverges). The distribution $\psi(\xi)$ is Gaussian (or has at least two finite moments). In this case,
\begin{equation}
    P(k,t)\sim E_\alpha(-k^2 t^{\alpha}),
\end{equation}
where $E_\alpha(x)=\sum_{n\geq 0}(-x)^n/\Gamma(1+\alpha n)$. We have $\langle x^2\rangle=t^\alpha$, so $\gamma=\alpha/2$. In the Fokker-Plank equation, this corresponds to a fractional derivative in time,
\begin{equation}
    \frac{\partial^\alpha}{\partial t^\alpha}P=D_{\alpha}\frac{\partial^2}{\partial x^2}P.
\end{equation}
When a force $F(x)=-\partial V/\partial x$ is introduced, this modifies the right-hand side of the equation,
\begin{equation}
    \frac{\partial^\alpha}{\partial t^\alpha}P=-\mu\frac{\partial}{\partial x}(FP)+D_{\alpha}\frac{\partial^2}{\partial x^2}P.
\end{equation}
The stationary solution is however still of the Gibbs form, $P\propto e^{-\beta V}$, with $\beta$ related to $D,\mu$ by the Einstein relation.

{\it b)} The waiting time has a narrow distribution, while the jumping interval is broadly distributed. {\it Mutatis mutandis,} this leads to a Fokker-Planck equation of the form 
\begin{equation}
    \frac{\partial}{\partial t}P=-\mu\frac{\partial}{\partial x}(FP)+D_{\gamma}\frac{\partial^{2\gamma}}{\partial x^{2\gamma}}P,
\end{equation}
where $\gamma>1$ ($\gamma<1$) corresponds to subdiffusion (superdiffusion). The equilibrium distribution is not Gibbs anymore, since $e^{-\beta V}$ does not solve the equation with $\partial P/\partial t=0$. It must be mentioned, however, that the case with $\gamma<1$, which corresponds to Levi flights, is more natural than the case $\gamma>1$ since the latter requires fine-tuning to eliminate the $\partial^2/\partial x^2$ terms, which arise naturally and are dominant with respect to $\partial^{2\gamma}/{\partial x^{2\gamma}}$ in a long-wavelength expansion.

Summarizing, subdiffusion is compatible with the Boltzmann-Gibbs distribution, and although this is a statement in classical mechanics, it is most likely the case that subdiffusion is compatible with some version of the ETH as well, which explains why it is observed in transport in quantum spin chains at very small disorder \cite{Varma17}. Superdiffusion, instead, leads to non-Gibbsian equilibrium distribution, solution of the fractional derivative equation
\begin{equation}
    -\mu\frac{\partial}{\partial x}(FP_{eq})+D_{\gamma}\frac{\partial^{2\gamma}}{\partial x^{2\gamma}}P_{eq}=0,
\end{equation}
which can be achieved with the same techniques used before \cite{klages2008anomalous}.

%

\end{document}